\documentclass{elsart}
\usepackage{natbib}
\usepackage{slashed}
\usepackage{slashbox}
\usepackage{multirow}
\usepackage{graphicx}
\usepackage[dvips]{color}
\usepackage{epsfig,psfig}
\usepackage{colortbl}
\newcommand{\be}{\begin{equation}}
\newcommand{\ee}{\end{equation}}
\newcommand{\ba}{\begin{array}}
\newcommand{\ea}{\end{array}}
\newcommand{\bea}{\begin{eqnarray}}
\newcommand{\eea}{\end{eqnarray}}
\newcommand{\eqn}[1]{(\ref{#1})}
\newcommand{\pp}{~~~.}
\newcommand{\vv}{~~~,}
\newcommand{\bvec}{\mathbf}

\newcommand{\rt}{\rightarrow}
\newcommand{\pe}{{\phi_e}}
\newcommand{\lh}{\hat{L}}
\newcommand{\ov}{\overline}
\def\neb{\hbox{$\ov{\nu}_e \!$ }}
\def\ca{{C_{\scriptscriptstyle A}}}
\def\cv{{C_{\scriptscriptstyle V}}}

\newcommand{\tre}{\left( \cv^2 + 3 \ca^2 \right)}
\newcommand{\slp}{/\!\!\!p}
\newcommand{\PL}{{\it Phys. Lett.\,}}
\newcommand{\ApJ}{{\it Astrophys. J.\,}}
\newcommand{\ApJS}{{\it Astrophys. J. Suppl.\,}}
\newcommand{\NP}{{\it Nucl. Phys.\,}}
\newcommand{\PR}{{\it Phys. Rev.\,}}
\newcommand{\PRL}{{\it Phys. Rev. Lett.\,}}
\newcommand{\pth}{\texttt{PArthENoPE} }

\def\lesssim{\buildrel < \over {_{\sim}}}
\def\gtrsim{\buildrel > \over {_{\sim}}}
\def\hi2{{}^2{\rm H}}
\def\h3{{}^3{\rm H}}
\def\he3{{}^3{\rm He}}
\def\lisix{{}^6{\rm Li}}
\def\bers{{}^7{\rm Be}}
\def\bero{{}^8{\rm Be}}
\def\bern{{}^9{\rm Be}}
\def\cdod{{}^{12}{\rm C}}
\def\X{{\rm X}}

\def\slW{\slashed{W}}
\definecolor{Black}{named}{Black}
\definecolor{Red}{named}{Red}

\def\gsim{\;\raise0.3ex\hbox{$>$\kern-0.75em\raise-1.1ex\hbox{$\sim$}}\;}
\def\lsim{\;\raise0.3ex\hbox{$<$\kern-0.75em\raise-1.1ex\hbox{$\sim$}}\;}

\def\sw2{\sin^2 \theta_W}
\def\epsilon{\varepsilon}
\def\s132{\sin^2 \theta_{13}}
\newcommand{\neff}{N_{\rm eff}}

\def\yp{Y_{p}}
\def\He4{{}^4{\rm He}}
\def\he3{{}^3{\rm He}}
\def\Li6{{}^6{\rm Li}}
\def\li7{{}^7{\rm Li}}
\def\nue{\nu_e}
\def\num{\nu_\mu}
\def\nut{\nu_\tau}
\def\bnu{\bar{\nu}}
\def\Lag{{\mathcal L}}

\def\slW{\slashed{W}}
\begin{document}
\begin{frontmatter}
{\hfill \small DSF-20/2008, FERMILAB-PUB-08-216-A, IFIC/08-37}\\
\title{Primordial Nucleosynthesis: from precision cosmology to fundamental physics}
\author[Arcetri]{Fabio Iocco}\footnote{Present address: Institut d'Astrophysique de Paris, 98bis bd
Arago, 75014 Paris, France},
\author[Napoli]{Gianpiero Mangano},
\author[Napoli,Valencia]{Gennaro Miele},
\author[Napoli]{Ofelia Pisanti},
\author[FERMI]{Pasquale D.\ Serpico}\footnote{Present address: Physics Division, Theory Group, CERN, CH-1211
Geneva 23, Switzerland}
\address[Arcetri]{INAF, Osservatorio Astrofisico di Arcetri, \\Largo E. Fermi 5, I-50125 Firenze, Italy}
\address[Napoli]{Dip. Scienze Fisiche, Universit\`{a} di Napoli
{Federico II} \& INFN, Sez. di Napoli, Complesso Univ. Monte S.\ Angelo,
Via Cintia, I-80126 Napoli, Italy}
\address[Valencia]{
Instituto de F\'{\i}sica Corpuscular (CSIC-Universitat de Val\`{e}ncia),
Ed. Institutos de Investigaci\'on, Apartado de Correos 22085,
E-46071 Val\`encia, Spain}
\address[FERMI]{Center for Particle Astrophysics, Fermi National
Accelerator Laboratory, Batavia, IL~~60510-0500, USA}
\begin{abstract}
We present an up-to-date review of Big Bang Nucleosynthesis (BBN).
We discuss the main improvements which have been achieved in the
past two decades on the overall theoretical framework, summarize
the impact of new experimental results on nuclear reaction rates,
and critically re-examine the astrophysical determinations of
light nuclei abundances. We report then on how BBN can be used as
a powerful test of new physics, constraining a wide range of ideas
and theoretical models of fundamental interactions beyond the
standard model of strong and electroweak forces and Einstein's
general relativity.
\end{abstract}
\end{frontmatter}
\tableofcontents
\section{Introduction}
\label{sec:introduction}
A remarkable scientific achievement in the second half of the 20$^{\rm
th}$ century has been the establishment of ``Standard Models'' of Particle
Physics (SMPP) and Cosmology (SMC). In particular,  the latter has been
possible thanks to an incredibly fast growth of  the amount and quality of
observations over the last couple of decades. The picture revealed is at
the same time beautifully simple and intriguingly mysterious: on one hand,
known gauge interactions and Einstein's general relativity seem able to
explain a huge wealth of information in terms of a few free parameters
specifying the composition/initial conditions of the Universe; on the
other hand, these numbers are not explained in terms of dynamical
processes involving the known fields and interactions. This is the case of
the ``dark energy'' density (consistent with a cosmological constant), of
the non-baryonic dark matter, of the baryon-antibaryon asymmetry, the
flatness, homogeneity and isotropy of the universe on large scales, etc.

The very success of the cosmological laboratory is thus providing
much indirect evidence for physics beyond the SMPP. On the other
hand, advances in particle physics (a very recent example being
the phenomenology of massive neutrinos) have an impact at
cosmological level. This interplay has proven extremely fertile
ground for the development of `astroparticle physics', especially
since many theories beyond the  SMPP predict new phenomena far
beyond the reach of terrestrial laboratories, but potentially
testable in astrophysical and cosmological environments. In this
respect, the nucleosynthesis taking place in the primordial plasma
plays a twofold role: it is undoubtedly one of the observational
pillars of the hot Big Bang model, being indeed known simply as
``Big Bang Nucleosynthesis'' (BBN); at the same time, it provides
one of the earliest direct cosmological probes nowadays available,
constraining the properties of the universe when it was a few
seconds old, or equivalently at the MeV temperature scale.
Additionally, it is special in that all known interactions play an
important role: gravity sets the dynamics of the ``expanding
cauldron'', weak interactions determine the neutrino decoupling
and the neutron-proton equilibrium freeze-out, electromagnetic and
nuclear processes regulate the nuclear reaction network.

The basic framework of the BBN emerged in the decade between  the
seminal Alpher-Bethe-Gamow (known as $\alpha\beta\gamma$) paper in
1948 \citep{Alp48} and the essential settlement of the paradigm of
the stellar nucleosynthesis of elements heavier than $^7$Li with
the B$^2$FH paper \citep{Bur57}. This pioneering period---an
account of which can be found in \citep{Kra96}---established the
basic picture that sees the four light-elements $^2$H, $^3$He,
$^4$He and $^7$Li as products of the early fireball, and virtually
all the rest produced in stars or as a consequence of stellar
explosions.

In the following decades, the emphasis on the role played by the
BBN has evolved significantly.  In the simplest scenario, the only
free parameters in primordial nucleosynthesis are the baryon to
photon ratio $\eta$  (equivalently, the baryon density of the
universe) and the neutrino asymmetry parameters,
$\eta_{\nu_\alpha}$ (see Section~\ref{nuasymm}). However, only
neutrino asymmetries larger than $\eta$ by many orders of
magnitude have appreciable effects. This is why the simple case
where all $\eta_{\nu_\alpha}$'s  are assumed to be negligibly
small (e.g., of the same order of $\eta$) is typically denoted as
Standard BBN (SBBN). Since several species of `nuclear ashes' form
during BBN, SBBN is an over-constrained theory whose
self-consistency can be checked comparing predictions with two or
more light nuclide determinations. The agreement of predicted
abundances of the light elements  with their measured abundances
(spanning more than nine orders of magnitude!) confirmed the
credibility of BBN as cosmological probe. At the same time,  the
relatively narrow range of $\eta$ where a consistent picture
emerged was the first compelling argument in favor of the
non-baryonic nature of the ``dark matter'' invoked for
astrophysical dynamics.

The past decade, when for the first time a redundancy of
determinations of $\eta$ has been possible, has stressed BBN as a
consistency tool for the SMC. Beside BBN, one can infer the
density of baryons from the Lyman-$\alpha$ opacity in quasar
spectra due to  intervening high redshift hydrogen clouds
\citep{Mei93,Rau96,Wei97}; from the baryon fraction in clusters of
galaxies, deduced from the hot x-ray emission \citep{Evr97}; most
importantly, from the height of the Doppler peak in the angular
power spectrum of the cosmic microwave background anisotropy (see
\citep{Dun08} for the latest WMAP results). These determinations
are not only mutually consistent with each other, but the two most
accurate ones (from the CMB and BBN) agree within $5\div$10\%.
While losing to CMB the role of ``barometer of excellence'', BBN
made possible a remarkable test of consistency of the whole SMC.

It comes without surprise that this peculiar `natural laboratory'
has inspired many investigations, as testified by the numerous
reviews existing on the subject, see e.g.
\citep{Mal93,Cop95,Oli00,Sar96,Sch98,Ste07}. Why then a new
review? In the opinion of the authors, a  new BBN review seems
worthy because at present, given the robustness of the
cosmological scenario, the attention of the community is moving
towards a new approach to the BBN. On one hand, one uses it as a
precision tool in combination with other cosmological information
to reduce the number of free parameters to extract  from
multi-parameter fits. On the other hand, BBN is an excellent probe
to explore the very early universe, constraining scenarios beyond
the SMPP. The latter motivation is particularly intriguing given
the perspectives of the forthcoming LHC age to shed light on the
TeV scale. A new synergy with the Lab is expected to emerge in the
coming years, continuing a long tradition in this sense. Finally,
a wealth of data from nuclear astrophysics and neutrino physics
have had a significant impact on BBN, and it is meaningful to
review and assess it. In particular, the  recent advances in the
neutrino sector  have made obsolete many exotic scenarios popular
in the literature still a decade ago and improved numerous
constraints, providing a clear example of the synergy we look
forward to in the near future.\\
This review is structured as follows: in Section
\ref{sec:cosm_overview} we summarize the main cosmological notions
as well as most of the symbols used in the rest of the article.
Section \ref{sec:bbn_overview} is devoted to the description of
the Standard BBN scenario. Section \ref{sec:obsabund} treats the
status of observations of light nuclei abundances, which in
Section \ref{sec:BBNanalysis} are compared with theoretical
predictions. The following Sections deal with exotic scenarios:
Section \ref{sec:BBN_nuphys} with neutrino properties, Section
\ref{sec:IBBN} with inhomogeneous models, Section
\ref{sec:constr_fundam} with constraints to fundamental
interactions and Section \ref{s:massive-particle} with massive
particles. In Section \ref{sec:conclusions} we report our
conclusions. Although this article is a review, many analyses have
been implemented ex-novo and some original results are presented
here for the first time. Due to the large existing literature and
to space limitation, we adopt the criterion to be as complete as
possible in the post-2000 literature, while referring to previous
literature only when pertinent to the discussion or when still
providing the most updated result. Also, we adopt a more
pedagogical attitude in introducing arguments that have rarely or
never entered previous BBN reviews, as for example extra
dimensions or variation of fundamental constants in Section
\ref{sec:constr_fundam}, while focusing mainly on new results (as
opposed to a `theory review') in subjects that have been
extensively treated in the past BBN literature (as in SUSY models
leading to cascade nucleosynthesis, the gravitino `problem',
etc.). Other topics, which for observational or theoretical
reasons have attracted far less interest in the past decade in
relation to BBN bounds, are only briefly mentioned or omitted
completely (this is the case of technicolor or cosmic strings).
Older literature containing a more extensive treatment of these
topics can be typically retraced from the quoted reviews. In the
following, unless otherwise stated, we  use natural units
$\hbar=c=k_B=1$, although conventional units in the astronomical
literature (as parsec and multiples of it) are occasionally used
where convenient for the context.

\section{Standard Cosmology} \label{sec:cosm_overview}
To keep this review self-contained, and fix the notation which we
will be using in this paper, we summarize here the main aspects of
the cosmological model which are relevant for our analysis. The
standard hot Big Bang model is based on three fundamental
astronomical observations: the Hubble law, the almost perfect
black body spectrum of the background photon radiation, and the
homogeneity and isotropy of the universe on large scales, see e.g.
\citep{Pee80,Pee93}, The latter, also known under the spell of
\textit{Cosmological Principle} implies that the metric itself
should be homogeneous and isotropic, and singles out the
Friedmann-Lema\^{i}tre-Robertson-Walker (FLRW) models. In comoving
spherical coordinates one has: \be ds^2 = g_{\mu\nu} dx^\mu dx^\nu
= dt^2 - a^2(t) \left[ \frac{dr^2}{1-kr^2} + r^2 (d\theta^2 +
\sin^2 \theta d\phi^2)\right] \vv \label{e:FLRW} \ee where $a(t)$
is the cosmic scale-factor and $k=1,0,-1$ the rescaled spatial
curvature signature for an elliptic, euclidean or hyperbolic
space, respectively.

The Einstein field equations relate the energy-momentum tensor of
the perfect fluid representing the matter-energy content of the
universe,
\be
T_{\mu\nu} = - P\, g_{\mu\nu} + (P+\rho) u_\mu u_\nu \vv
\ee
with the space-time curvature $R_{\mu\nu\rho\sigma}$,
\be R_{\mu\nu} - \frac{R}{2} \, g_{\mu\nu} = 8\, \pi\, G_N\,
T_{\mu\nu} + \Lambda g_{\mu\nu} \vv \label{e:Einstein} \ee
where $R_{\mu\nu}$ is the Ricci tensor, $R_{\mu\nu} \equiv
g^{\rho\sigma} R_{\rho\mu\sigma\nu}$, $R$ the scalar curvature, $R
= g^{\mu\nu} R_{\mu\nu}$, $G_N$ the Newton gravitational constant,
and $\Lambda$ the cosmological constant. Substituting the FLRW
metric \eqn{e:FLRW} in the Einstein's equations \eqn{e:Einstein}
gives the Friedmann-Lema\^{i}tre (FL) equation for the Hubble
parameter $H$, \be H^2 \equiv \left( \frac{\dot a}{a} \right)^2 =
\frac{8\, \pi\, G_N}{3}~ \rho - \frac{k}{a^2} \pp \ee

The equation of state of the fluid filling the universe,
$P=P(\rho)$, specifying the pressure as a function of the energy
density, along with the covariant conservation of the energy
momentum tensor (which accounts for the entropy conservation if
the fluid corresponds to a thermal bath of particle excitations),
\be
\frac{d(\rho a^3)}{da} = -3\,P\, a^2 \vv \label{conservation}
\ee
allows one to get the evolution of $\rho$ as function of $a$,
\bea
\rho_M &\propto& a^{-3} \vv \\
\rho_R &\propto& a^{-4} \vv\\
\rho_\Lambda &\propto& const \vv
\eea
for matter (both baryonic and dark matter, $P_B,P_{DM}\sim 0$),
radiation ($P_R=\rho_R/3$), or cosmological constant
($P_\Lambda=-\rho_\Lambda$), respectively.

As usual, the present values of radiation, baryon matter, dark
matter and cosmological constant energy densities will be
expressed in terms of the parameters $\Omega_i=
\rho^{0}_i/\rho_{cr}$, $i=R,B,DM,\Lambda$, with $\rho_{cr}=3
H_0^2/(8 \pi G_N)$ the critical density today and $H_0= 100 \,h$
km s$^{-1}$ Mpc$^{-1}$, with $h=0.73^{+0.04}_{-0.03}$
\citep{Yao06}. To quantify the baryon density parameter we will
also use $\omega_b\equiv \Omega_B h^2$ and the baryon to photon
density ratio, $\eta= n_B/n_\gamma$. The latter is also
proportional to the initial baryon-antibaryon asymmetry per
comoving volume produced at some early stage of the universe
evolution. This ratio keeps constant after the $e^+-e^-$
annihilation phase taking place at a value of the photon
temperature $T \sim 0.3$ MeV (see later). Moreover, at low energy
scales there are no baryon violating interactions at work, thus
the value of $\eta$ can be simply related to $\Omega_B$, see e.g.
\citep{Ser04b}
\be \!\!\!\!\!\!\!\!\!\!\!\! \eta_{10} \equiv \eta \cdot 10^{10} =
 \frac{273.45\, \Omega_B h^2}{1- 0.007 Y_p} \left( \frac{2.725 \,
\textrm{K}}{T_0} \right)^3 \left( \frac{6.708 \cdot 10^{-45}
\textrm{MeV}^{-2}}{G_N} \right), \label{eta10def} \ee
where $Y_p$ stands for $^{4}$He mass fraction (see
Section~\ref{sub_overview}) and $T_0$ the photon temperature
today. Note that the numerical factor multiplying $Y_p$ takes into
account the effect of the ${{}^4{\rm He}}$ binding energy on the
whole energy budget in baryonic matter.

Matter and radiation fluids can be usually described in terms of a
bath of particle excitations of the corresponding quantum fields.
In particular, at high temperatures rapid interactions among them
ensures thermodynamical equilibrium and each particle specie is
described by an equilibrium (homogeneous and isotropic) phase
space distribution function,
\be f_i (|\bvec{p}|,T) = \left[ \exp \left(
\frac{E_i(|\bvec{p}|)-\mu_i}{T} \right) \pm 1 \right]^{-1} \vv \ee
where $E_i(|\bvec{p}|) = \sqrt{|\bvec{p}|^2+m_i^2}$ is the energy,
$+/-$ corresponds to the Fermi-Dirac/Bose-Einstein statistics, and
$\mu_i$ the chemical potential, which is zero for particles which
can be emitted or absorbed in any number (like photons).

In the comoving frame, the number density, energy density and
pressure can be expressed as follows
\bea
n_i (T) &=& g_i \int \frac{d^3 \bvec{p}}{(2\, \pi)^3}~ f_i (|\bvec{p}|,T) \vv \\
\rho_i (T) &=& g_i \int \frac{d^3 \bvec{p}}{(2\, \pi)^3}~
E_i(|\bvec{p}|)~
f_i (|\bvec{p}|,T) \vv \\
P_i (T) &=& g_i \int \frac{d^3 \bvec{p}}{(2\, \pi)^3}~
\frac{|\bvec{p}|^2}{3\, E_i(|\bvec{p}|)}~ f_i (|\bvec{p}|,T) \vv
\eea
where $g_i$ is the number of internal degrees of freedom. The BBN
takes place in the radiation dominated phase, hence
non-relativistic particles contribute negligibly to the total
energy density, which therefore can be conveniently written in
terms of the photon energy density $\rho_\gamma = \pi^2 T^4/15$,
\be \rho \sim \rho_R = g_* \frac{\rho_\gamma}{2} \vv \ee which
defines $g_*$, the total number of relativistic degrees of
freedom,
\be
g_* = \sum_{B_i} g_i \left( \frac{T_i}{T} \right)^4 +
\frac{7}{8} \sum_{F_i} g_i \left( \frac{T_i}{T} \right)^4 \vv
\ee
where the first and second terms are due to all boson and fermion
species, respectively. The possibility is left in the previous
formula of different $T_i$ for different species, accounting for
pseudo-thermal distributions of decoupled fluids (like
relativistic neutrinos at BBN times).

Finally, we will also exploit in the following the definition of
the entropy density, $s(T)$, in terms of the phase space
distribution function. For a given specie $i$ one has:
\be
s_i (T) = \frac{\rho_i+P_i}{T}= g_i \int \frac{d^3
\bvec{p}}{(2\, \pi)^3}~ \frac{3\, m_i^2 + 4\, |\bvec{p}|^2}{3\,T\,
E_i(|\bvec{p}|)}~ f_i (|\bvec{p}|,T) \pp \label{entropy}
\ee
The total entropy density is conventionally written as
\be
s(T) = \frac{\pi^4}{45\, \zeta (3)}~ g_{*s} (T)~ n_\gamma =
\frac{2\, \pi^2}{45}~ g_{*s}(T)~ T^3 \vv
\ee
where $n_\gamma = (2\, \zeta (3)/\pi^2)~ T^3$ is the number
density of photons and
\be
g_{*s}(T) = \sum_{B_i} g_i \left( \frac{T_i}{T} \right)^3 +
\frac{7}{8} \sum_{F_i} g_i \left( \frac{T_i}{T} \right)^3 \pp
\ee
Use of Eq. (\ref{conservation}) implies that entropy per comoving
volume is a conserved quantity, $s(t) a^3 = const$.

\section{Big Bang Nucleosynthesis} \label{sec:bbn_overview}
\subsection{Overview}\label{sub_overview}
Extrapolating the present universe back in the past, we infer that
during its early evolution, before the epoch of nucleosynthesis,
it was hot and dense enough for electrons, positrons, photons,
neutrinos and nucleons, as well heavier nuclei, to be in kinetic
and chemical equilibrium due to the high (weak, strong and
electromagnetic) interaction rates. In particular, the initial
values of all nuclear densities are set by Nuclear Statistical
Equilibrium (NSE). NSE implies that they constitute a completely
negligible fraction of the total baryon density, which is all in
the form of free neutrons and protons. As expansion proceeds, weak
process rates become eventually smaller than the expansion rate
$H$ at that epoch, hence some particle species can depart from
thermodynamical equilibrium with the remaining plasma. This is the
case of neutrinos which only interact via weak processes and {\it
freeze out} at a temperature of about 2-3 MeV. Soon after, at a
temperature $T_D\sim 0.7$ MeV, neutron-proton charged-current weak
interactions also become too slow to guarantee neutron-proton
chemical equilibrium. The n/p density ratio departs from its
equilibrium value and freezes out at the value n/p $= \exp(-\Delta
m /T_D) \sim 1/7$, with $\Delta m = 1.29$ MeV the neutron--proton
mass difference, and is then only reduced by subsequent neutron
decays. At this stage, the photon temperature is already below the
deuterium binding energy $B_D \simeq 2.2$ MeV, thus one would
expect sizable amounts of $^2$H to be formed via $n+p \rightarrow
^2$H + $\gamma$ process. However, the large photon-nucleon density
ratio $\eta^{-1}$, which is of the order of $10^9$, delays
deuterium synthesis until the photo--dissociation process become
ineffective (deuterium {\it bottleneck}). This takes place at a
temperature $T_N$ such that $\exp(B_D/T_N) \eta \sim 1$, i.e. $T_N
\sim 100$ keV, which states the condition that the high energy
tail in the photon distribution with energy larger than $B_D$ has
been sufficiently diluted by the expansion.

Once $^2$H starts forming, a whole nuclear process network sets
in, leading to heavier nuclei production, until BBN eventually
stops, see Section \ref{sec:obsabund}. An estimate of the main BBN
outcome, i.e. $^4$He, can be obtained with very simple arguments,
yet it provides quite an accurate result. Indeed, the final
density $n_{^4{\rm He}}$ of $^4$He is very weakly sensitive to the
whole nuclear network, and a very good approximation is to assume
that all neutrons which have not decayed at $T_N$ are eventually
bound into helium nuclei, see e.g. \citep{Kol90,Sar96}. This leads
to the famous result for the helium mass fraction $Y_p \equiv 4 \,
n_{^4{\rm He}}/n_B$ \be Y_p \sim \frac{2}{1+ \exp(\Delta m/T_D)
\exp(t(T_N)/\tau_n)} \sim 0.25 \vv \label{ypqualitative} \ee with
$t(T_N)$ the value of time at $T_N$ and $\tau_n$ the neutron
lifetime.

On the other hand, the determination of all light nuclei produced
during BBN, and a more accurate determination of $^4$He as well,
can be only pursued by a simultaneous solution of a set of coupled
kinetic equations which rule the evolution of the various nuclei,
supplemented by Einstein's equations, covariant conservation of
total energy momentum tensor, as well as conservation of baryon
number and electric charge. This is typically obtained
numerically, although nice semi-analytical studies have been also
recently performed \citep{Muk04}.

To summarize the general BBN setting, we start with some
definitions. We consider $N_{nuc}$ species of nuclides, whose
number densities, $n_i$, are normalized with respect to the total
number density of baryons $n_B$, \be X_i=\frac{n_i}{n_B}
\quad\quad\quad i=p,\,^2{\rm H}, \,^3{\rm He},\, ... \pp \ee The
list of nuclides which are typically considered in BBN analysis is
reported in Table \ref{t:nuclnumb}. To quantify the most
interesting abundances, those of $^2$H, $^3$He, $^4$He and $^7$Li,
we also use in the following the short convenient definitions \be
\!\!\!\!\!\!\!\!\!\!\! ^2{\rm H}/{\rm H} \equiv X_{^2{\rm
H}}/X_p,\,\,\,\,\, ^3{\rm He}/{\rm H} \equiv X_{^3{\rm
He}}/X_p,\,\,\,\,\,Y_p \equiv 4 X_{^4{\rm He}}, \,\,\,\,\,^7{\rm
Li}/{\rm H} \equiv X_{^7{\rm Li}}/X_p, \ee i.e. the $^2$H, $^3$He
and $^7$Li number density normalized to hydrogen, and the $^4$He
mass fraction, $Y_p$. Notice that, although the above definition
of $Y_p$ is widely used, it is only approximately related to the
real helium mass fraction, since the $^4$He mass is not given by 4
times the atomic mass unit. The difference is quite small, of the
order of $0.5 \%$ due to the effect of $^4$He binding energy.
However, in view of the present precision of theoretical analysis
on $^4$He yield, this difference cannot be neglected, and one
should clearly state if one refers to the conventional quantity
$Y_p$ (as we do here, too)  or the ``true'' helium mass fraction.
\begin{table}[t]
\begin{center}
\caption{Nuclides which are typically considered in BBN numerical
studies.}\label{t:nuclnumb}
\begin{tabular}{|c|c|c|c|c|c|c|c|c|c|}
 \hline\hline
\backslashbox{$Z$}{$N$} & 0 & 1 & 2 & 3 & 4 & 5 & 6 & 7 & 8\\
\hline 0 & & \cellcolor[gray]{.8} n & & & & & & & \\
\hline
1 & \cellcolor[gray]{.8}H & \cellcolor[gray]{.8}$^2$H & \cellcolor[gray]{.8}$^3$H & & & & & & \\
\hline
2 & & \cellcolor[gray]{.8}$^3$He & \cellcolor[gray]{.8}$^4$He & & & & & & \\
\hline
3 & & & & \cellcolor[gray]{.8}$^6$Li & \cellcolor[gray]{.8}$^7$Li& \cellcolor[gray]{.8}$^8$Li& & & \\
\hline
4 & & & & \cellcolor[gray]{.8}$^7$Be & & \cellcolor[gray]{.8}$^9$Be& & & \\
\hline
5 & & & & \cellcolor[gray]{.8}$^8$B & & \cellcolor[gray]{.8}$^{10}$B& \cellcolor[gray]{.8}$^{11}$B& \cellcolor[gray]{.8}$^{12}$B & \\
\hline
6 & & & & & & \cellcolor[gray]{.8}$^{11}$C& \cellcolor[gray]{.8}$^{12}$C & \cellcolor[gray]{.8}$^{13}$C& \cellcolor[gray]{.8}$^{14}$C\\
\hline
7 & & & & & & \cellcolor[gray]{.8}$^{12}$N& \cellcolor[gray]{.8}$^{13}$N & \cellcolor[gray]{.8}$^{14}$N& \cellcolor[gray]{.8}$^{15}$N\\
\hline
8 & & & & & & & \cellcolor[gray]{.8}$^{14}$O & \cellcolor[gray]{.8}$^{15}$O& \cellcolor[gray]{.8}$^{16}$O\\
 \hline\hline
\end{tabular}
\end{center}
\end{table}
In the (photon) temperature range of interest for BBN, $10 \,{\rm MeV}\gsim T
\gsim10\, {\rm keV}$, electrons and positrons are kept in thermodynamical
equilibrium with photons by fast electromagnetic interactions and
are distributed according to a Fermi-Dirac function $f_{e^{\pm}}$,
with chemical potential $\pm \mu_e$, parameterized in the
following by the ratio $\phi_e \equiv \mu_e/T$. The chemical
potential of electrons is very small, due to the universe charge
neutrality \citep{Lyt59,Sen96},
\be \frac{\mu_e}{T} \sim \frac{n_e}{n_\gamma} =
\frac{n_p}{n_\gamma} \sim 10^{-10} \pp
\ee
To follow the neutrino-antineutrino fluid in detail, it is
necessary to write down evolution equations for their distribution
in phase space, rather than simply using their energy density.
This is due to the fact, as we will illustrate in details in the
following, that they are slightly reheated during the $e^+-e^-$
annihilation phase and develop non--thermal momentum--dependent
features. We denote these distributions by
$f_{\nu_e}(|\bvec{p}|,t)$, $f_{\bar{\nu}_e}(|\bvec{p}|,t)$ and
\be
f_{\nu_\mu}=f_{\nu_\tau} \equiv f_{\nu_x}(|\bvec{p}|,t) \vv
~~~ f_{{\bar \nu}_\mu}=f_{{\bar \nu}_\tau} \equiv f_{{\bar
\nu}_x}(|\bvec{p}|,t) \pp
\ee
In the standard scenario of no extra relativistic degrees of
freedom at the BBN epoch apart from photons and neutrinos, the
neutrino chemical potential is bound to be a small fraction of the
neutrino temperature. This bound applies to all neutrino flavors,
whose distribution functions are homogenized via flavor
oscillations \citep{Dol02a,Aba02,Won02}. In this Section we focus
on non-degenerate neutrinos, namely $f_{\nu_e}=f_{\bar{\nu}_e}$
and $f_{\nu_x}=f_{\bar{\nu}_x}$, while we will consider in details
the effect of neutrino--antineutrino asymmetry in Section
\ref{nuasymm}.

The set of differential equations ruling primordial
nucleosynthesis is the following, see for example
\citep{Wag67,Wag69,Wag73,Esp00b,Esp00a}:
\bea
&&\frac{\dot{a}}{a} = H = \sqrt{\frac{8\, \pi G_N}{3}~ \rho} \vv
\label{e:drdt} \\
&&\frac{\dot{n}_B}{n_B} = -\, 3\, H \vv
\label{e:dnbdt} \\
&&\dot{\rho} = -\, 3 \, H~ (\rho + P) \vv
\label{e:drhodt} \\
&&\dot{X}_i = \sum_{j,k,l}\, N_i \left(  \Gamma_{kl \rt ij}\,
\frac{X_k^{N_k}\, X_l^{N_l}}{N_k!\, N_l !}  \; - \; \Gamma_{ij \rt
kl}\, \frac{X_i^{N_i}\, X_j^{N_j}}{N_i !\, N_j  !} \right) \equiv
\Gamma_i \vv
\label{e:dXdt} \\
&& n_B~ \sum_j Z_j\, X_{j} = n_{e^-}-n_{e^+}\equiv L \left(\frac{m_e}{T},
\pe\right)  \equiv T^3~ \lh \left(\frac{m_e}{T}, \pe\right) \vv
\label{e:charneut}\\
&&\left(\frac{\partial}{\partial t} - H \, |\bvec{p}|
\frac{\partial}{\partial |\bvec{p}|} \right) \,
f_{\nu_\alpha}(|\bvec{p}|,t) = I_{\nu_\alpha} \left[
f_{\nu_e},f_{\bar{\nu}_e},f_{\nu_x},f_{\bar{\nu}_x},f_{e^-},f_{e^+}
\right] \vv \label{e:nuboltz} \eea where $\rho$ and $P$ denote the
total energy density and pressure, respectively, \bea
\rho &=& \rho_\gamma + \rho_e + \rho_\nu + \rho_B \vv \\
P &=& P_\gamma + P_e + P_\nu + P_B\pp
\eea

Eq. \eqn{e:drdt} is the definition of the Hubble parameter $H$, whereas
Eq.s \eqn{e:dnbdt} and \eqn{e:drhodt} state the total baryon number and
entropy conservation per comoving volume, respectively. The set of
$N_{nuc}$ Boltzmann equations \eqn{e:dXdt} describes the density evolution
of each nuclide specie, Eq. \eqn{e:charneut} states the universe charge
neutrality in terms of the electron chemical potential, with $L
\left(m_e/T, \pe \right)$ the charge density in the lepton sector in unit
of the electron charge, and finally Eq.s (\ref{e:nuboltz}) are the
Boltzmann equations for neutrino species, with $I_{\nu_\alpha} \left[
f_{\nu_e},f_{\nu_x} \right]$ standing for the collisional integral which
contains all microscopic processes creating or destroying the specie
$\nu_\alpha$.

Since electromagnetic and nuclear scatterings keep the
non-relativistic baryons in kinetic equilibrium their energy
density $\rho_B$ and pressure $P_B$ are given by
\bea
\rho_B & = & \left[M_u + \sum_i \left( \Delta M_i + \frac{3}{2} \,
T \right)~ X_i \right] n_B\,\,\, , \label{neupress}\\
P_B & = & T \, n_B \, \sum_i X_i \,\,\,, \label{barpress}
\eea
with $\Delta M_i$ and $M_u$ the i-th nuclide mass excess and the
atomic mass unit, respectively.

The pressure and energy density of the electromagnetic plasma
($e^{\pm}$ and $\gamma$) receives a contribution at first order in
the fine structure coupling $\alpha$ when considering QED finite
temperature corrections which change the electromagnetic plasma
equation of state via the appearance of a thermal mass for both
photons and $e^\pm$, which in turn change the particle dispersion
relation $E_{e,\gamma}(p)$. This has been studied e.g. in
\citep{Hec94,Lop99,Man02}. These corrections also enter Eq.
~(\ref{e:drhodt}), the expression of the expansion rate $H$, as
well as the thermal averaged n-p weak conversion rates, which
depend upon electron-positron distribution function (see later).
It has been shown that at the time of BBN, all these effects
slightly influence the $^4$He abundance, at the level of per mille
\citep{Lop99}.

In Eq. \eqn{e:dXdt} $i,j,k,l$ denote nuclear species, $N_i$ the
number of nuclides of type $i$ entering a given reaction (and
analogously $N_j$, $N_k$, $N_l$), while the $\Gamma$'s denote
symbolically the reaction rates. For example, in the case of decay
of the species $i$, $N_i=1$, $N_j=0$ and $\sum\Gamma_{i\to kl}$ is
the inverse lifetime of the nucleus $i$. For two--body collisions
$N_i=N_j=N_k=N_l=1$ and $\Gamma_{ij\to kl}=\langle \sigma_{ij\to
kl}\,v\rangle$, the thermal averaged cross-section for the
reaction $i+j\to k+l$ times the $i-j$ relative velocity. In Eq.
\eqn{e:charneut}, $Z_i$ is the charge number of the $i-$th
nuclide, and the function $\lh(\xi,\omega)$ is defined as \be
\lh(\xi,\omega) \equiv \frac{1}{\pi^2} \int_\xi^\infty
d\zeta~\zeta\, \sqrt{\zeta^2-\xi^2}~ \left(
\frac{1}{e^{\zeta-\omega}+1} - \frac{1}{e^{\zeta+\omega}+1}
\right) \pp \label{lfunc} \ee

Eq.s (\ref{e:drdt})-(\ref{e:nuboltz}) constitute a set of coupled
differential equations which have been implemented in numerical
codes since the pioneering works of Wagoner, Fowler and Hoyle
\citep{Wag67} and Kawano \citep{Kaw88,Kaw92}. Before discussing
some details of this implementation and its present status, we
describe two crucial phenomena which take place before the BBN,
the freezing out of neutrino distribution functions and of the
neutron--proton density ratio, when the weak charged current n-p
processes become too slow to ensure chemical equilibrium between
the two nucleon species. At the time these two phenomena occur the
synthesis of deuterium, and thus of the whole nuclear chain, is
still strongly forbidden by photo--dissociation processes, and all
baryon density is in the form of free neutrons and protons. This
means that in principle they could be treated
\textit{independently} of the whole set of nuclear processes which
leads to the proper BBN phase. In particular, once obtained the
shape of neutrino distribution functions, they can be used as
given inputs when solving the set of equations
(\ref{e:drdt})-(\ref{e:dnbdt}), thus significantly simplifying the
problem.

\subsection{The role of neutrinos in BBN and neutrino decoupling}
\label{sec:neutrinodecoupling}

At early epochs neutrinos were kept in thermal contact with the
electromagnetic primordial plasma by rapid weak interactions with
electrons and positrons, controlled by a rate
$\Gamma_w\simeq\langle\sigma_w\,v\rangle n_{e^{\pm}}\sim G_F^2\,T^2\times
T^3$. When the temperature dropped below a few MeV, these weak processes
became too slow compared to the Hubble expansion rate ($\Gamma_w<
H\sim\sqrt{G_N}T^2$) and the process of neutrino decoupling took place. An
accurate estimate for the neutrino decoupling temperature is 2.3 MeV for
the electron neutrino and slightly larger for $\nu_{\mu,\tau}$, 3.5 MeV
\citep{Enq92b}, since the latter only interact with the electromagnetic
plasma via neutral current processes.

Shortly after, the $e^{\pm}$ pairs began to annihilate almost entirely
into photons, thus producing a difference between the temperatures of the
relic photons and neutrinos. The MeV to 0.1 MeV range is crucial for BBN
physics, and in particular for the neutron-proton fraction, so it comes
without surprise that BBN is sensitive to the properties of neutrinos: it
derives from the basic facts that the neutrino decoupling, the deuterium
binding energy, the $n-p$ mass difference and the electron mass all fall
in the MeV range. In more detail, neutrinos enter BBN equations in three
ways:
\begin{enumerate}
\item[I] the momentum distributions of the $\nue$ and $\bnu_e$
entering the n-p inter--conversion weak rates;
\item[II] their overall energy density content $\rho_\nu$, which
determines the Hubble expansion rate;
\item[III] their overall pressure $P_\nu$, which enters the
energy-momentum conservation law. {\it Assuming} the equation of state of
relativistic species $P_\nu=\rho_\nu/3$, this effect is not independent
from the previous one, and we shall not discuss it further.
\end{enumerate}

The effect (I) is clearly model-dependent, in the sense that it
can be quantified and parameterized only on the basis of a
physically-motivated hypothesis. One popular example where the
effect of neutrinos is dominated by this distortion is the case of
a $\nu_e-\bar\nue$ asymmetry, see Section \ref{nuasymm}.
Concerning the effect (II), it is customary to parameterize it via
an effective number of neutrinos $\neff$ \citep{Shv69,Ste77},
defined by the relation \be \rho_\nu \equiv
\frac{\neff}{3}\rho_{\nu,0} \vv \label{neffdef} \ee where
$\rho_{\nu,0}$ is the energy density in neutrinos in the limit of
instantaneous decoupling from the $e^{+},e^{-},\gamma$ plasma and
no radiative or plasma corrections (see Eq. (\ref{e:rhonu0})
below). Nowadays, precision electroweak measurements at the
$Z^0$-resonance pin down the number of {\it light active} neutrino
species with high accuracy, $N_{\nu}=2.9840\pm 0.0082$
\citep{Ale05}, consistent within $\sim 2\,\sigma$ with the known
three families of the SMPP. However, well before these
measurements were available, BBN was already invoked to favor 3
light, thermalized (and thus probably active with respect to weak
interactions) neutrino families, with a range not wider than 2-4
(see Ref.s in the reviews \citep{Sar96,Sch98,Oli00}). While this
connection between collider physics and cosmology has been
historically important, we would like here to clarify some common
misunderstanding and ambiguities on the BBN bound on $\neff$. If
the bound is derived by changing only $\neff$, but without
including any effect of the type (I), it is incorrect to refer to
it as a bound on neutrino properties. It is rather a statement on
the rate of expansion of the universe at the BBN time, which may
be indicative e.g. of the presence of additional
(semi)relativistic species in the plasma, or of exotic thermal
histories. Any reasonable BBN bound on the number of neutrino
species $N_{\nu}$ requires a modification of the type (I) as well.
Since no universal parametrization exists, this is often
neglected. Yet, this should not diminish the importance of the
point. For example, if we accompany the rescaling  of
Eq.~(\ref{neffdef}) by an identical rescaling of the $\nue$ and
$\bnu_e$ distribution function (i.e. adopting a gray-body
parametrization, keeping the temperature of the spectrum unchanged
but altering the normalization), the range of $Y_p$ that for
standard parameters corresponds to $2.8 \leq\neff\leq 3.6$ (see
Section \ref{sec:BBNanalysis}) translates into a more severe
constraint, $2.85\leq N_{\nu}\leq 3.12$. It is worth noting that
no role is played in BBN by possible heavy fourth generation
neutrinos, whose mass must be larger than $\sim$45 GeV to avoid
the $Z$-width bound. Even if such a neutrino exists, it would be
natural to expect it to be unstable and rather short-lived, for
example due to mixing with the light neutrinos (neutrino
oscillations prove that family lepton-numbers are violated). Even
if some protective symmetry prevents the decay (so that it
accounts for a subleading fraction of the cold dark matter), no
effect on the BBN is present,  and we do not discuss it further.

A major consequence of the settlement of the neutrino oscillation
issue is that a very refined calculation of the neutrino
decoupling is possible. The standard picture in the instantaneous
decoupling limit is very simple (see e.g.\ \citep{Kol90}): coupled
neutrinos had a momentum spectrum with an equilibrium Fermi-Dirac
(FD) form\footnote{It was noted in \citep{Cuc96} (and more
recently the topic reanalyzed in \citep{Dol05}) that the $^4$He
abundance in the early Universe is sensitive to the difference
between FD or an exotic Bose-Einstein (BE) distribution of the
(quasi)thermal neutrino bath, a conclusion which is expected
according to the spin-statistics theorem but otherwise difficult
to prove in laboratory experiments. They found indeed a good
sensitivity to the quantum nature of the statistics,
$Y_p$(BE)-$Y_p$(FD)$\simeq -(3\div 4\%) Y_p$(FD) (or equivalent to
$\neff\simeq 2.4$), with the present range of $Y_p$ thus
disfavoring a BE spectral shape.} with temperature $T$, \be f_{\rm
eq}(|\bvec{p}|,T)=\left
[\exp\left(\frac{|\bvec{p}|}{T}\right)+1\right]^{-1}\,, \label{FD}
\ee which is preserved after decoupling. Shortly after neutrino
decoupling the photon temperature drops below the electron mass,
$m_e$, and $e^{\pm}$ annihilations heat the photons. If one
assumes that this entropy transfer did not affect the neutrinos
because they were already completely decoupled (instantaneous
decoupling limit), using conservation of entropy per comoving
volume it is easy to calculate the difference between the
temperatures of relic photons and neutrinos and thus the eventual
neutrino energy density $\rho_{\nu,0}$ introduced before.

At high temperatures, $T \geq 2-3$ MeV, one can write the conservation of
entropy per comoving volume in the form (we are considering a temperature
range well below muon annihilation epoch)
\be
\left( s_{e^\pm, \gamma}+ s_\nu \right) a^3 = const \pp
\ee
After $\nu$-decoupling, one has instead two separate conservation
conditions for neutrinos and for the electromagnetic plasma. Nevertheless,
until photons are reheated by $e^\pm$ annihilation, both photon and
neutrino temperatures redshift by the same amount and keep equal. If one
specifies the entropy conservation laws at the two different epochs,
$a_{in}$ well before $e^\pm$ annihilation, and $a_{end}$ well after this
phase, one obtains
\bea
&& s_\nu(a_{in}) a_{in}^3 = s_\nu(a_{end}) a_{end}^3 \vv \nonumber \\
&& s_{e^\pm, \gamma}(a_{in}) a_{in}^3 = s_{\gamma}(a_{end})
a_{end}^3 \vv \eea where in the second equation one takes into
account that both photon and $e^\pm$ degrees of freedom contribute
at $a_{in}$, while only photons are present (and reheated) at
$a_{end}$. The ratio of these two equations using the expression
of entropy density (\ref{entropy}), gives the well-known
asymptotic ratio of neutrino/photon temperatures after $e^\pm$
annihilation phase, \be \frac{T_\nu}{T} = \left(\frac{2}{2+4
\times 7/8} \right)^{1/3}= \left(\frac{4}{11} \right)^{1/3} \simeq
1.401 \vv \ee and the instantaneous decoupling expression of the
neutrino energy density in terms of $\rho_\gamma$, \be
\rho_{\nu,0} = 3 \frac{7}{8} \left(\frac{4}{11} \right)^{4/3}
\rho_\gamma \pp \label{e:rhonu0} \ee

However, the processes of neutrino decoupling and $e^{\pm}$
annihilations are sufficiently close in time so that some relic
interactions between $e^{\pm}$ and neutrinos exist. These relic
processes are more efficient for larger neutrino energies, leading
to non-thermal distortions in the neutrino spectra (larger for
$\nue$ than for $\nu_{\mu,\tau} $, since $\nue$ also feel
charged-current interactions) and a slightly smaller increase of
the comoving photon temperature. Even in absence of mixing, a
proper calculation of the process of non-instantaneous neutrino
decoupling requires the solution of the momentum-dependent
Boltzmann equations for the neutrino spectra (\ref{e:nuboltz}), a
set of integro-differential kinetic equations that are quite
challenging to attack numerically. In the last two decades a
series of  works has been devoted to solving this system in an
increasingly general and precise way, ultimately also including
finite temperature QED corrections to the electromagnetic plasma
(a full list of the pre-2002 works is reported in \citep{Dol02b}).
To give a feeling of the overall effect, it suffices to say that
the combination of these corrections leads to an effective
neutrino number $\neff\simeq 3.046$ , while asymptotically one
finds $T_\nu/T = 1.398$, slightly smaller than the instantaneous
decoupling value \citep{Man05}.

The existence of neutrino oscillations imposes modifications to
these corrections. The effect of $f_\nu$ distortions on the $\yp$
yield in the simplified case of two-neutrino mixing, averaged
momentum, and Maxwell-Boltzmann statistics was estimated in~
\citep{Khl81,Han01}. All three approximations were relaxed in
\citep{Man05} and a full calculation was performed with a density
matrix formalism. The neutrino ensemble is described by the
momentum-dependent density matrices $\varrho_p$
\citep{Dol81,Raf92,Sig93,McK94}. The form of the neutrino density
matrix for a mode with momentum $p$ is \be \varrho_p(t) = \left
(\matrix{\varrho_{ee} & \varrho_{e\mu}& \varrho_{e\tau}\cr
\varrho_{\mu e}& \varrho_{\mu\mu}& \varrho_{\mu\tau}\cr
\varrho_{\tau e} &\varrho_{\tau \mu}& \varrho_{\tau \tau}
}\right). \label{3by3} \ee The diagonal elements correspond to the
usual number density of the different flavors, while the
off-diagonal terms are non-zero in the presence of neutrino
mixing. There exists a corresponding set of equations for the
antineutrino density matrix $\bar{\varrho}_p$. In absence of a
neutrino asymmetry (or of additional couplings flipping $\nu$'s
into $\bar\nu$ and vice versa) antineutrinos follow the same
evolution as neutrinos, and a single matrix suffices to describe
the system. The equations of motion for the density matrices are
\be i\dot\varrho_p= \left[ H_0+H_1,\varrho_p\right]
+{C}[\varrho_p]~, \label{eq:3by3evol} \ee where the first
commutator term includes the free Hamiltonian $H_0$ and the
effective potential of neutrinos in medium $H_1$, while the last
term is a collisional term of order $G_F^2$ describing the
breaking of coherence induced by neutrino scattering and
annihilation as well as neutrino production by collisions in the
primeval plasma. In a FLRW universe,
$\dot\varrho_p=\left(\partial_t-Hp\,\partial_p\right)\varrho_p$.
The Hamiltonian can be written explicitly in the flavor basis as
\be H_0=\mathcal{U}\frac{M^2}{2\,p}\mathcal{U}^\dagger,\:\:\:M^2=
{\rm diag}(m_1^2,m_2^2,m_3^2)\,, \ee where $\mathcal{U}$ is the
mixing matrix which, assuming vanishing CP-violating phases,
writes \be \mathcal{U}=\left(
    \ba{ccc}
        c_{12} c_{13}
        & s_{12} c_{13}
        & s_{13} \\
        -s_{12} c_{23} - c_{12} s_{23} s_{13}
        & c_{12} c_{23} - s_{12} s_{23} s_{13}
        & s_{23} c_{13} \\
        s_{12} s_{23} - c_{12} c_{23} s_{13}
        & -c_{12} s_{23} - s_{12} c_{23} s_{13}
        & c_{23} c_{13}
    \ea
    \right).
\ee
Here $c_{ij}=\cos \theta_{ij}$ and $s_{ij}=\sin \theta_{ij}$ for $ij=12$,
23, or 13. Apart for CP-violating phases, there are five oscillation
parameters: $\Delta m^2_{21} = m^2_2 - m^2_1, \Delta m^2_{32} = m^2_3 -
m^2_2, \theta_{12},\theta_{23}$ and $\theta_{13}$. Best-fit values and
uncertainties for these parameters (and upper bound for $\theta_{13}$) can
be found in \citep{Yao06}.

Neglecting non-diagonal components of the effective potential, the
matter term writes
\be
H_1= {\rm diag}(V_e,V_\mu,V_\tau)~,
\ee
where, assuming $T\ll m_\mu$,
\bea
V_e  & =& - \frac{8 \sqrt{2} G_F\, p}{3} \left( \frac{\rho_{\nue + \bar\nue}}{M_Z^2}
+ \frac{\rho_{e^- + e^+}}{M_W^2}  \right)\,,\\
V_\mu & =&  - \frac{8 \sqrt{2} G_F\,p}{3 M_Z^2} \rho_{\num +\bar\num}\,,  \\
V_\tau &=&    - \frac{8 \sqrt{2} G_F\,p}{3 M_Z^2}
\rho_{\nut+\bar\nut} \,.
\eea

Finally, the collisions of neutrinos with $e^{\pm}$ or among
themselves are described by the term $C[{\cdot}]$, which is
proportional to $G_F^2$. For the off-diagonal terms it is
sufficient to adopt simple damping coefficient as reported in
\citep{Dol02a}. Instead, for the diagonal ones, in order to
properly calculate the neutrino heating process one must consider
the exact collision integral $I_{\nu_\alpha}$, that includes all
relevant two--body weak reactions of the type $ \nu_\alpha(1) + 2
\rightarrow 3 + 4$ involving neutrinos and $e^{\pm}$, see e.g.
\citep{Dol97}. The kinetic equations for the neutrino density
matrix are supplemented by the covariant conservation equation for
the total energy-momentum tensor. Given the order of the effects
considered, one should also include the finite temperature QED
corrections to the electromagnetic plasma \citep{Man02}.

\begin{figure}[t]
\begin{center}
\includegraphics[width=0.8\textwidth]{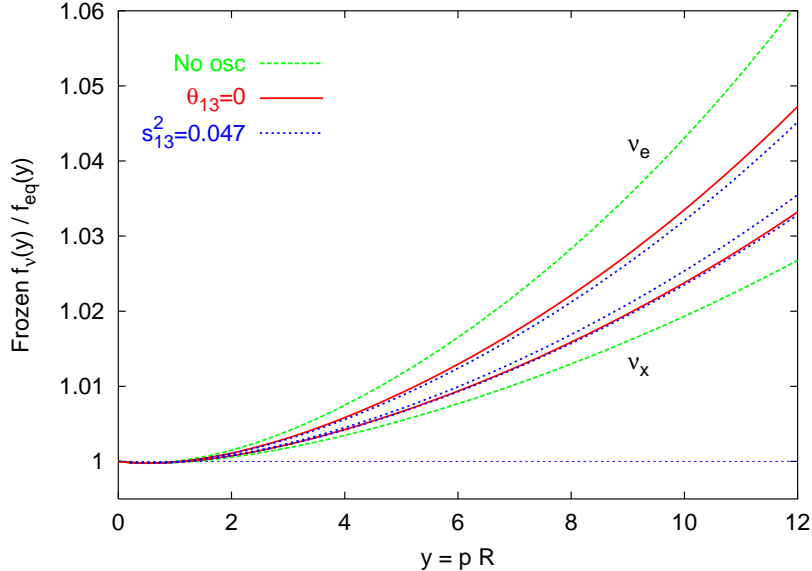}
\end{center}
\caption{Frozen distortions of the flavor neutrino spectra as a
function of the comoving momentum, for the best fit solar and atmospheric
mixing parameters. R is the scale factor. In the case where we allow for
$\theta_{13}\neq 0$ consistently with present bounds (blue dotted lines),
one can distinguish the distortions for $\nu_\mu$ and $\nu_\tau$ (middle
and lower, respectively). From \citep{Man05}.}
\label{fig:finalfnu}
\end{figure}

We show in Figure \ref{fig:finalfnu} the asymptotic values of the
flavor neutrino distribution, both without oscillations and with
non-zero mixing. The dependence of the non-thermal distortions in
momentum is well visible, which reflects the fact that more
energetic neutrinos were interacting with $e^{\pm}$ for a longer
period. Moreover, the effect of neutrino oscillations is evident,
reducing the difference between the flavor neutrino distortions.
Fitting formulae for these distributions are available in
\citep{Man05}.
\begin{table}[t]
\begin{center}
\caption{$\neff$ and $\Delta Y_p$ obtained for different cases,
with and without neutrino oscillations, as reported in
\citep{Man05}.}\label{tab:neffyp}
\begin{tabular}{l|l|l}
\hline\hline
Case      & $\neff$         & $\,\,\,\,\,\Delta Y_p$ \\
\hline\hline
No mixing (no QED) & 3.035 & $1.47{\times} 10^{-4}$\\
\hline
No mixing  & 3.046 & $1.71{\times} 10^{-4}$\\
\hline
Mixing, $\theta_{13}=0$ & 3.046 & $2.07{\times} 10^{-4}$\\
\hline
Mixing, $\sin^2(\theta_{13})=0.047$ & 3.046 & $2.12{\times} 10^{-4}$\\
\hline Mixing, Bimaximal, $\theta_{13}=0$ & 3.045 & $2.13{\times}
10^{-4}$ \\
\hline\hline
\end{tabular}
\end{center}
\end{table}
In Table \ref{tab:neffyp} we report the effect of
non-instantaneous neutrino decoupling on the radiation energy
density, $\neff$, and on the $^4$He mass fraction. By taking also
into account neutrino oscillations, one finds a global change of
$\Delta Y_p\simeq 2.1{\times} 10^{-4}$ which agrees with the
results in \citep{Han01} due to the inclusion of QED effects.
Nevertheless the net effect due to oscillations is about a factor
of 3 smaller than what was previously estimated, due to the
failure of the momentum-averaged approximation to reproduce the
true distortions.

It is worth remarking that the precise computation of the effect
of particular neutrino decoupling scenario on primordial yields
can only be performed numerically. The neutrino distribution
functions, once obtained by the solution of Eq.
(\ref{eq:3by3evol}), have to be substituted in Eq.s
(\ref{e:drdt})-(\ref{e:charneut}) which will predict the
primordial abundance. The process is particularly involved, it is
in fact important to follow the single neutrino distribution as a
function of time because of the particular role played by the
different neutrino flavors. In particular, the $\neff$ reported in
Table \ref{tab:neffyp} is the contribution of neutrinos to the
whole radiation energy budget, but only at the very end of
neutrino decoupling. Hence, not all the $\Delta \neff$ there
reported will be really contributing to BBN processes. In order to
clarify this subtle point, we report in Table \ref{tab:nuclides}
the effect on all light nuclides, of the non-instantaneous
neutrino decoupling in the simple scenario of no neutrino
oscillation, and compare this column with the simple {\it
prescription} of adding a fix $\Delta \neff = 0.013$ contribution
to radiation. Even though $Y_p$ is reproduced (by construction),
this is not the case for the other nuclear yields.
\begin{table}[t]
\begin{center}
\caption{Comparison of the exact BBN results with a fixed-$\Delta
\neff$ approximation. From \citep{Man05}.}\label{tab:nuclides}
\begin{tabular}{l|l|l}
\hline\hline
Nuclide        & Exact (No $\nu$-oscillations)          & Fixed $\Delta \neff = 0.013$\\
\hline\hline
$\Delta Y_p$   & $1.71{\times} 10^{-4}$ & $1.76{\times} 10^{-4}$\\
\hline
$\Delta$($^2$H/H) & $-0.0068{\times} 10^{-5}$ & $+0.0044{\times} 10^{-5}$\\
\hline
$\Delta$($^3$He/H) & $-0.0011{\times} 10^{-5}$ & $+0.0007{\times} 10^{-5}$\\
\hline
$\Delta$($^7$Li/H) & $+0.0214{\times} 10^{-10}$ & $-0.0058{\times} 10^{-10}$\\
\hline\hline
\end{tabular}
\end{center}
\end{table}

\subsection{The neutron-proton chemical equilibrium and the role of weak rates}
\label{sec:weakrates}
Neutrons and protons are kept in chemical equilibrium by charged current
weak interactions,
\bea
(a)~~~ \nu_e + n \rightarrow e^- + p &~~~,~~~~~~& (d)~~~ \neb + p
\rightarrow e^+ + n \vv \nonumber \\
(b)~~~e^- + p \rightarrow \nu_e + n &~~~,~~~~~~& (e) ~~~n \rightarrow e^-
+ \neb + p \vv \nonumber \\
(c)~~~ e^+ + n \rightarrow \neb + p &~~~,~~~~~~& (f)~~~ e^- + \neb + p
\rightarrow n \vv
\label{e:reaction}
\eea
which enforce their number density ratio to follow the equilibrium value,
n/p =$\exp(-\Delta m/T)$. Shortly before the onset of BBN, processes
$(a)-(f)$ become too slow, chemical equilibrium is lost and the ratio n/p
freezes out for temperatures lower than the decoupling temperature
$T_D\sim 1$ MeV. Residual free neutrons are partially depleted by decay
until deuterium starts forming at $T_N$ and neutrons get bound in nuclei,
first in deuterium and eventually in $^4$He.

The leading role of $(a)-(f)$ in fixing the neutron fraction at
the BBN, and thus $Y_p$, simply means that to get an accurate
theoretical prediction for $^4$He abundance requires a careful
treatment of the weak rates. Large improvements on this issue have
been obtained in the last decade, which we summarize in the
following. Extensive analysis can be found in e.g.
\citep{Lop99,Esp99}.

\begin{figure}[t]
\begin{center}
\epsfig{file=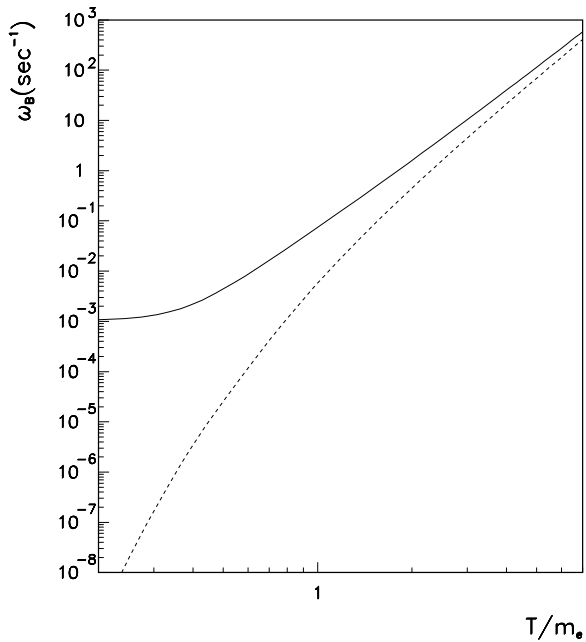,width=0.6\textwidth}
\end{center}
\caption{The total Born rates, $\omega_B$, for $n \rightarrow p$
(solid line) and $p \rightarrow n$ transitions (dashed line). From
\citep{Esp99}}
\label{f:born}
\end{figure}

At the lowest order, the calculation is rather straightforward,
and is obtained by using V-A theory and in the limit of infinite
nucleon mass (we will refer to this as the Born limit), see
\citep{Wei72}. The latter approximation is justified in view of
the typical energy scale of interest, of order $T \sim $ MeV, much
smaller than the nucleon mass $M_N$. For example, the neutron
decay rate takes the form (neglecting the very small neutrino
masses)
\bea \omega_B ( n \rightarrow e^- + \neb + p ) \, &=& \,
\frac{G_F^2 }{2 \pi^3} \tre\, \int d |\bvec{p}_e| \,
|\bvec{p}_e|^2 \, |\bvec{p}_\nu|^2  \,
\nonumber \\
&{\times}& \Theta(|\bvec{p}_\nu|) \, \left[ 1 - f_{\bar{\nu}_e}
(|\bvec{p}_\nu|)\right] \left[ 1 - f_e(|\bvec{p}_e|) \right] \vv
\label{e:cb15} \eea
where $\cv$ and $\ca$ are the nucleon vector and axial coupling,
and $|\bvec{p}_\nu| = \Delta m - \sqrt{|\bvec{p}_e|^2 + m_e^2}$.
The rates for all other processes $(a)-(d), (f)$ can be simply
obtained from (\ref{e:cb15}) by changing the statistical factors
and the expression for neutrino energy in terms of the electron
energy. An average can be performed at this level of approximation
over equilibrium Fermi-Dirac distribution for leptons, i.e.
neglecting the effects of distortion in neutrino-antineutrino
distribution functions, but taking into account the time evolution
of the neutrino to photon temperature ratio $T_\nu/T$. In Figure
\ref{f:born} we report the Born rates, $\omega_B$, for n-p
processes. The accuracy of Born approximation results in being, at
best, of the order of $7\%$. This can be estimated by comparing
the prediction of Eq. \eqn{e:cb15} for the neutron lifetime at
very low temperatures, with the experimental value $\tau_n^{ex} =
(885.7 {\pm} 0.8)$ s \citep{Yao06}.

The Born calculation can be improved by considering four classes
of effects:

{\it Electromagnetic radiative corrections}. These are typically split
into $outer$ and $inner$ terms (for a review see e.g. \citep{Wil82}). The
first ones involve the nucleon as a whole and consist of a multiplicative
factor to the squared modulus of transition amplitude of the form
\be
1+ \frac{\alpha}{2 \pi} g(E_e,E_\nu) \vv
\label{e:outer}
\ee
where analytic expression for $g(E_e,E_\nu)$, can be found in Ref.
\citep{Sir67}. On the other hand, the inner corrections are deeply related
to the nucleon structure. They have been estimated in Ref. \citep{Mar86},
and applied in the BBN context in \citep{Lop99,Esp99}. Furthermore, when
electron and proton are both either in the initial or final state, one
should also add the effect of {\it Coulomb correction}
\citep{Dic82,Cam82,Bai90}, due to rescattering of the electron in the
proton electromagnetic field and leading to the Fermi factor
\be
1 \, + \, \alpha \pi \, \frac{E_e}{{|\bvec{p}_e|}} \pp
\label{e:coulomb}
\ee

{\it Finite nucleon mass corrections}. For finite nucleon mass $M_N$ and
at order $1/M_N$, the weak hadronic current receives a contribution from
the weak magnetic moment coupling,
\be
J_\mu^{wm} = i \frac{G_F}{\sqrt{2}} \frac{f_2}{M_N} \, \ov{u}_p(p) \,
\sigma_{\mu \nu} \, (p-q)^\nu u_n(q) \vv
\ee
where, from conservation of vector current (CVC), $f_2 = V_{ud} (\mu_p -
\mu_n)/2 = 1.81 V_{ud}$. Both scalar and pseudoscalar contributions can be
shown to be much smaller and negligible for the accuracy we are interested
in. At the same order $1/M_N$ the allowed phase space for the relevant
scattering and decay processes gets changed, due to nucleon recoil.
Finally, one has to consider the effect of the initial nucleon thermal
distribution in the comoving frame. All these effects are proportional to
$m_e/M_N$ or $T/M_N$, and in the temperature range relevant for BBN, can
be as large as radiative corrections. This has been first pointed out in
\citep{Sec93,Lop97} and then also numerically evaluated in
\citep{Lop99,Esp99}.

\begin{figure}[t]
\begin{center}
\epsfig{file=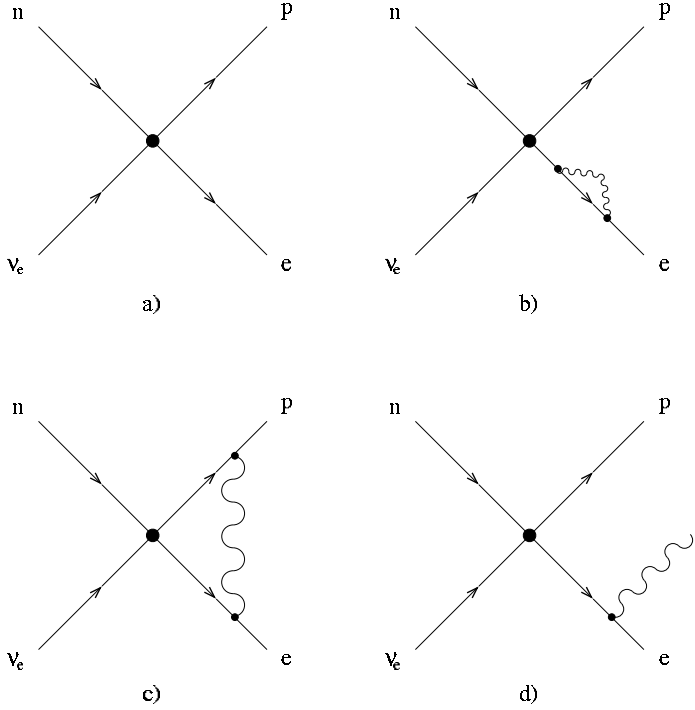,width=0.7\textwidth}
\end{center}
\caption{The tree level Born (a), the one-loop (b),(c), and the
photon emission/absorption diagrams (d) for $n \rightarrow p$ processes.}
\label{f:grafici}
\end{figure}

{\it Thermal-Radiative corrections}. The n-p rates get slight
corrections from the presence of the surrounding electromagnetic
plasma. To compute these corrections one may use Real Time
formalism for Finite Temperature Field Theory \citep{LeB96} to
evaluate the finite temperature contribution of the graphs of
Figure \ref{f:grafici}, for the $n \rightarrow p$ processes.
Inverse processes $p \rightarrow n$ are obtained by inverting the
momentum flow in the hadronic line. The first order in $\alpha$ is
given by interference of one-loop amplitudes of Figure
\ref{f:grafici} b) and c) with the Born result (Figure
\ref{f:grafici} a)). Photon emission and absorption processes
(Figure \ref{f:grafici} d)), which also give an order $\alpha$
correction, should be included to cancel infrared divergences.
Notice that photon emission (absorption) amplitudes by the proton
line are suppressed as $M_p^{-1}$ and can be neglected.

All field propagators get additional on-shell contributions
proportional to the number density of that particular specie in
the surrounding medium. For $\gamma$ and $e^{\pm}$ (neglecting the
small electron chemical potential) we have \bea i
\Delta_\gamma^{\mu \nu}(k) &=& - \left[\frac{i}{k^2} + 2 \pi \,
\delta(k^2)~f_\gamma(k) \right] g^{\mu \nu}~~~,
\label{e:a1} \\
i \, S_e(p_e)&=& \frac{i}{\slp_e - m_e} - 2 \pi~\delta({p_e}^2 -
m_e^2)~ f_e(E_e)~(\slp_e + m_e)~~~, \label{e:a2} \eea with
$f_\gamma$ the photon distribution function. The whole set of
thermal/radiative corrections have been computed by several
authors
\citep{Dic82,Cam82,Don83,Don84,Don85,Joh86,Kei89a,Bai90,Kei89b,LeB90,Alt89,Kob85,Kob86,Saw96,Cha97,Esp98}.
Though they agree on the order of magnitude---which is quite
small---there is nevertheless no consensus on the detailed value,
due for example to different ways of treating the wave function
renormalization at finite temperature. Finally, it was correctly
pointed out in \citep{Bro01} that a (small) contribution to
neutron-proton chemical equilibrium is also provided by
photon-proton interactions (and inverse processes), \be \gamma +p
\rightarrow e^+ + \nu_e + n ~~~,~~~~~~e^+ + \nu_e + n \rightarrow
\gamma + p \pp \label{newpgamma} \ee The corresponding thermal
averaged rates can be found in e.g. \citep{Ser04b}.

{\it Non-instantaneous neutrino decoupling effects}. Distortion of
neutrino distribution functions changes the weak rates $(a)-(f)$
which are enhanced by the larger mean energy of electron
neutrinos. On the other hand, there is an opposite effect due to
the change in electron-positron temperature. Finally, since the
photon temperature is reduced with respect to the instantaneous
decoupling value $(11/4)^{1/3}$, the onset of BBN, via $^2$H
synthesis, takes place earlier in time. This means that fewer
neutrons decay from the time of freezing out of weak interactions,
and this in turn corresponds to a larger $^4$He yield.

The effect of all corrections to the weak rates discussed so far
has been considered in details in \citep{Lop99,Esp99,Ser04b}. The
leading contribution is given by electromagnetic radiative
corrections, which decrease monotonically with increasing
temperature for both $p \rightarrow n$ and $n \rightarrow p$
processes, and by finite nucleon mass effects. Their sum changes
the Born estimate for a few percent correction at the freeze out
temperature $T\sim $ MeV. Comparing the theoretical prediction for
the neutron lifetime at zero temperature using $G_F=(1.16637{\pm}
0.00001){\cdot} 10^{-5}\, GeV^{-2}$, $C_V=0.9725 {\pm} 0.0013$ and
the ratio $C_A/C_V=-1.2720 {\pm} 0.0018$, \citep{Yao06}, one finds
$\tau_n^{th} = 886.5~s$, quite an accurate result when compared
with the experimental value (agreement is at the $0.1 \%$ level).
It is worth commenting here on the fact that a recent measurement
of the neutron lifetime exists \citep{Ser05a}, which gives
$\tau_n=878.5\pm0.7_{\rm stat}\pm0.3_{\rm syst}$ and results in a
5.6$\sigma$ discrepancy from the previous most precise result,
which in turn is consistent with the other six determinations used
in \citep{Yao06} to determine the best fit. According to the
Particle Data Group, {\it [The result of \citep{Ser05a}] is so far
from other results that it makes no sense to include it in the
average. It is up to workers in this field to resolve this issue.
Until this major disagreement is understood our present average of
$885.7\pm 0.8$ s must be suspect.} While implications for BBN of
this different lifetime have been explored \citep{Mat05a}, in the
following we shall assume that the best fit provided by the PDG is
correct. Note that the PDG value also appears to be in better
agreement with global electroweak fits~\citep{Severijns:2006dr}.
Modulo this caveat, one might be confident that n-p weak rates are
presently quite accurately computed, at per mille precision.
Plasma corrections and finite temperature radiative effects are
sub-leading, changing the rates at the level of $(0.3 \div 0.6)
\%$ only. Their effect is to slightly increase $Y_p$ by a very
small amount, $\Delta Y_p \sim 1 {\cdot} 10^{-4}$, \citep{Fie93}.

To conclude, we report a fit of the n-p rates which include all effects
described in this Section as a function of $z=m_e/T$, \citep{Ser04b},
accurate at the $0.01 \%$ level, which the reader might find useful:
\bea
\omega(n \rightarrow p) &=& \frac{1}{\tau_n^{ex}} \exp \left( -q_{np}
/z\right) \,\sum_{l=0}^{13} a_l ~z^{-l}~~~~~ 0.01\leq
T/{\rm MeV} \leq 10\vv
\label{e:fitnp} \\
\omega(p \rightarrow n) &=& \left\{
\ba{cc}
\frac{1}{\tau_n^{ex}} \exp
\left( -q_{pn} z\right) \,
\sum_{l=1}^{10} b_l ~z^{-l} & 0.1 \leq T/{\rm MeV}  \leq 10 \\
0 & 0.01\leq T/{\rm MeV}  < 0.1
\ea \vv \right.
\label{e:fitpn}
\eea
with
\bea
\ba{lll}
a_0= 1 & a_1= 0.15735 & a_2= 4.6172 \\
a_3= -0.40520 {\cdot} 10^2 & a_4= 0.13875 {\cdot} 10^3 & a_5=
-0.59898 {\cdot} 10^2 \\
a_6= 0.66752 {\cdot} 10^2 & a_7= -0.16705 {\cdot} 10^2 & a_8=
3.8071 \\
a_9= -0.39140 & a_{10}= 0.023590 & a_{11}= -0.83696 {\cdot}
10^{-4} \\
a_{12}= -0.42095 {\cdot} 10^{-4} & a_{13}= 0.17675 {\cdot} 10^{-5} &
q_{np}= 0.33979~~~, \\
\ea \label{e:coeffnp}
\eea
\bea
\ba{lll} b_0= -0.62173 & b_1= 0.22211 {\cdot} 10^2 & b_2=
-0.72798 {\cdot} 10^2 \\
b_3= 0.11571 {\cdot} 10^{3} & b_4= -0.11763 {\cdot} 10^2 & b_5= 0.45521
{\cdot} 10^2
\\  b_6=
-3.7973 & b_7= 0.41266 & b_8= -0.026210 \\
b_9= 0.87934 {\cdot} 10^{-3} & b_{10}= -0.12016 {\cdot} 10^{-4} &
 q_{pn}= 2.8602 ~~~.
\ea
\label{e:coeffpn}
\eea

\subsection{Nuclear Reaction Network}
\label{NucNet}
Nuclear processes during the BBN proceed in an environment very
different with respect to the perhaps more familiar stellar
plasmas, where stellar nucleosynthesis takes place. The latter is
a dense plasma where species are mostly in chemical equilibrium,
the former is a hot and low density plasma with a significant
population of free neutrons, which expands and cools down very
rapidly, resulting in peculiar  ``out of equilibrium''
nucleosynthetic yields. The low density of the plasma at the time
of BBN is responsible for the suppression of three--body reactions
and an enhanced effect of the Coulomb--barrier, which as a matter
of fact inhibits any reaction with interacting nuclei charges $Z_1
Z_2\gtrsim$6. The most efficient categories of reactions in BBN
are therefore proton, neutron and deuterium captures --$(p,
\gamma)$ $(n, \gamma)$, $(d,\gamma)$--, charge exchanges
--$(p,n)$--, and proton and neutron stripping --$(d, n)$, $(d,
p)$.

From a more technical point of view, accurate  BBN predictions
require a detailed knowledge of the nuclear rates entering the set
of Equations~(\ref{e:dXdt}): a physical understanding of the basic
reaction chains helps however to implement a code with a good
compromise between reliability and computational time. In this
Section we start with the phenomenology of nuclear processes
related with the choice of such nuclear network, whereas a
chronology of the codes devoted to solve the whole set of
equations, and related numerical issues, is reported in Section
\ref{sec:numsol}.

\begin{figure}[t]
\begin{center}
\includegraphics[width=0.7\textwidth]{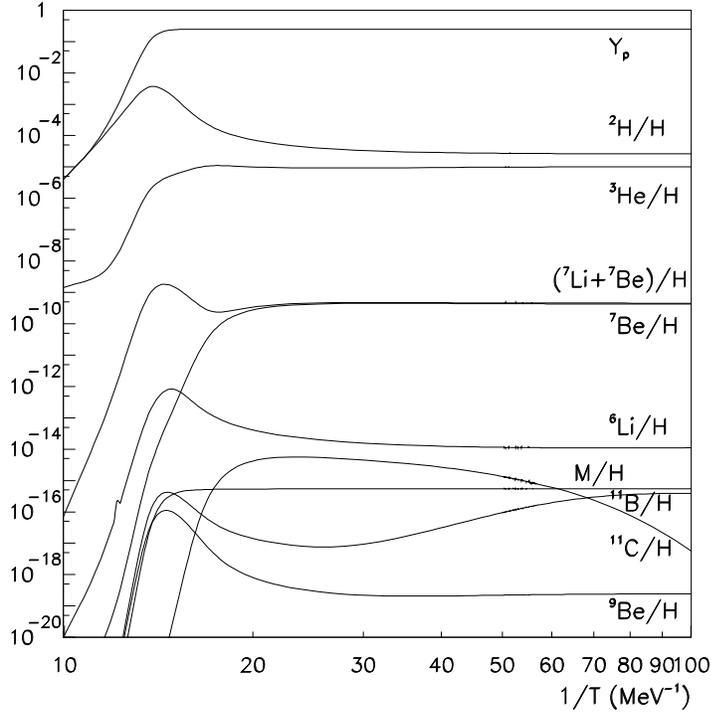}
\end{center}
\caption{The evolution of some element abundances
produced during BBN. The line M/H refers to the abundance of ``metals"
(see text).}
\label{fig:bbnevol}
\end{figure}

After the freeze--out of weak interactions, $p(n,\gamma)\hi2$ is
the only reaction able to synthesize sensible amounts of nuclei,
since it involves the only two nuclear species with non-vanishing
abundances, protons and neutrons. The very high entropy of the
primordial plasma keeps this reaction in equilibrium down to
energies much lower than its $Q$--value, which corresponds to the
deuterium binding energy of $B_D\simeq 2.2$ MeV. The deuterium
{\it bottleneck} ends at a temperature of T$_D\sim$100~keV, given
roughly by the condition $\exp(B_D/T_D)\eta\sim 1$, which ensures
that the high energy tail of the photon distribution has been
sufficiently diluted by the expansion. At this point, almost
instantaneously all the available neutrons are locked into
deuterium nuclei. The effect of the bottleneck on the following
nuclear processes is two--fold: the ``delayed'' population of the
deuterium specie results in BBN taking place at relatively low
temperatures, with consequences on the efficiency of all the
following reactions. Second, but not less important, is the
intervening decay of the neutron, which alters the neutron to
proton ratio at the time of effective deuterium production.
Efficient production of deuterium marks the effective beginning of
the nuclear phase of BBN: the nuclear path towards heavier
elements, before all $\h3$, $\he3$, $^4$He, is enabled by the
interaction of deuterium nuclei with nucleons and other $\hi2$
nuclei. The bulk of the remaining $\li7$ is produced via tritium
and especially $\he3$ radiative capture on $\He4$. The latter path
leads to $\bers$, which eventually decays into $\li7$ by electron
capture. The evolution of light nuclide abundances as a function
of the temperature of the plasma is shown in Figure
\ref{fig:bbnevol}: they  undergo the nuclear phase of departure
from chemical equilibrium and, by the time the universe cools to
few keV, they reach their final values. It is worth noting the
dramatic change that all species undergo following the deuterium
synthesis at 1/T$\gtrsim$14 MeV$^{-1}$.

\begin{table}[t]
\begin{center}
\caption{The most relevant reactions for BBN.}\label{NucProc}
\begin{tabular}{|c|c||c|c|}
\hline
Symbol & Reaction & Symbol & Reaction \\
\hline
\hline
$R_0$ & $\tau_n$             & $R_8$ & $\he3(\alpha,\gamma)\bers$ \\
$R_1$ & $p(n,\gamma)d$       & $R_9$ & $\h3(\alpha,\gamma)\li7$   \\
$R_2$ & $\hi2(p,\gamma)\he3$ & $R_{10}$ & $\bers(n,p)\li7$        \\
$R_3$ & $\hi2(d,n)\he3$      & $R_{11}$& $\li7(p,\alpha)\He4$     \\
$R_4$ & $\hi2(d,p)\h3$       & $R_{12}$& $\He4(d,\gamma)\lisix$   \\
$R_5$ & $\he3(n,p)\h3$       & $R_{13}$& $\lisix(p,\alpha)\he3$   \\
$R_6$ & $\h3(d,n)\He4$       & $R_{14}$& $\bers(n,\alpha)\He4$    \\
$R_7$ & $\he3(d,p)\He4$      & $R_{15}$& $\bers(d,p)2\:\He4$      \\
\hline
\end{tabular}
\end{center}
\end{table}

\begin{figure}[t]
\begin{center}
\includegraphics[width=0.7\textwidth]{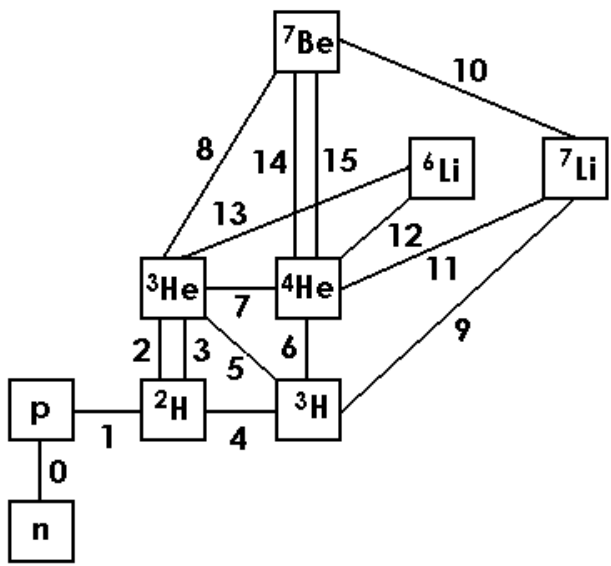}
\end{center}
\caption{The most relevant reactions for BBN.}
\label{fig:network}
\end{figure}

The most important nuclear processes for the BBN were identified
in \citep{Smi93a} (reactions $R_0$--$R_{11}$ in Table
\ref{NucProc}). Reaction $R_{9}$ provides actually only a
sub-leading contribution to $^7$Li for today's preferred value of
$\eta$, being more relevant at lower $\eta$ where $\li7$
production is direct rather than coming from $\bers$ synthesis.
Also, if one is concerned with the traces of $^6$Li produced in
BBN, the reactions $R_{12}:\, \He4(d,\gamma)\lisix$, and
$R_{13}:\, \lisix(p,\alpha)\he3$ are relevant. Due to their larger
uncertainties, even reactions as $R_{14}:\, \bers(n,\alpha)\He4$
and  perhaps $R_{15}:\, \bers(d,p)2\:\He4$ are important in the
$\li7$ error budget determination. All these reactions are
summarized in Table \ref{NucProc} and Figure \ref{fig:network}.
Both leading and sub-leading reactions of some interest have been
discussed e.g. in \citep{Ser04b}, which we mostly follow here.
Other compilations can be found in \citep{Ang99,Cyb04,Des04}.

These reactions are not only important to understand the nuclear
physics of the BBN, but also to assess an error budget for its
theoretical predictions. The first ``modern'' papers addressing
these issues are \citep{Kra90,Smi93a}. For example, in
\citep{Smi93a}, the authors performed a systematic analysis of the
nuclear network in BBN, as it had been implemented in the first
publicly released code \citep{Wag67}, with the--at the time--new
rates compiled in \citep{Cau88}. For the range of $\eta$ allowed
at the time, they identified the twelve leading reactions reported
in Table~\ref{NucProc}. A detailed study of their rates and
uncertainties led to a new code \citep{Kaw92}, implementing an
updated nuclear network.

Inferring the uncertainties on the light elements yields is
conceptually a three-step process. First, one has to determine the
uncertainties on the cross-sections $\sigma_i(E)$ (as functions of
energy) measured in the Lab or theoretically predicted. Since a
first principle computation is virtually impossible for almost all
processes, one has to rely on some form of theoretically motivated
fitting formulae and interpolate between and extrapolate beyond
the data. Indeed, the reactions in the early universe happen in a
thermal plasma, so the relevant nuclear input are the
temperature-dependent rates given by the convolution of the
cross-sections with the Maxwell-Boltzmann distribution of
nuclides, i.e. $\langle \sigma v\rangle\propto
T^{-3/2}\int_0^\infty{\rm d}E\, \sigma(E)\,E\,\exp(-E/T)$. These
can be approximated with analytical formulae in some simple
hypotheses for the functional form of $\sigma(E)$. Although no
modern compilation relies on such approximations, they are often
used to suggest a fitting formula for the numerical integration
results. Anyway, it is the uncertainties on these rates that
ultimately propagate onto the final errors on the nuclides. The
formalism/machinery to treat this problem can be found in
classical papers (as  \citep{Smi93a}), textbooks
\citep{Cla83,Rol88} and has also been reported in several
compilations of the last decade \citep{Ang99,Cyb04,Des04,Ser04b},
so we do not repeat it here. We want to remark, however, that: a)
In the last decade, in particular following \citep{Ang99}, a
world-wide effort in obtaining new measurements at low energy for
several important reactions involving light nuclei has been
undergone; b) in several cases, especially when new measurements
have become available, the different experimental data sets do not
seem to agree within the quoted uncertainties. In this situation,
it is a tricky business to assess uncertainties in a statistically
meaningful way, unless one has reason to believe that some of the
datasets are affected by unaccounted systematics and decides for
example to dismiss the older measurements. Here we follow the
prescription illustrated in \citep{Ser04b} and motivated on the
basis of the arguments presented in \citep{DAg94}. While in
agreement with the error estimates given in other compilations
\citep{Cyb04,Des04} when statistical errors dominate, we warn the
reader that our procedure tends to produce smaller errors than
other compilations when discrepancies among datasets exist.

An additional technical aspect arises in the way uncertainties are
accounted for in the BBN calculations. A lot of attention has been
paid in the last two decades to this problem. One may adopt Monte
Carlo simulations directly, with various degrees of sophistication
\citep{Kra90,Smi93a,Nol00}, or use an error matrix approach as in
\citep{Fio98} and its generalization in \citep{Cuo04}. Each one
has its own advantages and disadvantages, but the agreement
between the two is typically very good \citep{Fio98}. In
Table~\ref{uncertNuc} we report the light nuclide abundances for
WMAP 5-year result $\Omega_B h^2 = 0.02273 \pm 0.00062$
\citep{Dun08} (second column), showing in the third and fourth
columns, respectively, the uncertainties due to $\Omega_B h^2$
($\sigma_{\omega_b}$) and nuclear rates errors
($\sigma_{ii}$)\footnote{A recent analysis finds a larger value
for $^7$Li/H$=(5.24^{+0.71}_{-0.67}) 10^{-10}$, due to new
determinations of the rate $R_8$, see \citep{Cyb08} and references
therein. This further exacerbates the $^7$Li problem, see Sections
4.4-4.5.}. The last two columns report the rates mostly
contributing to the nuclear uncertainties, and their relative
contributions in percent in a quadrature sum, according to
\citep{Ser04b}.
\begin{table*}
\begin{center}
\caption{The light nuclide abundances for WMAP 5-year result
$\Omega_B h^2=0.02273 \pm 0.00062$ (second column). The
uncertainties due to $\Omega_B h^2$ ($\sigma_{\omega_b}$) and
nuclear rates errors ($\sigma_{ii}$) are also shown in the third
and fourth columns, respectively. The last two columns report the
rates mostly contributing to the nuclear uncertainties, and their
relative contributions in percent in a quadrature sum, according
to \citep{Ser04b}.} \label{uncertNuc}
\begin{tabular}{|c|c|c|c|c|c|}
\hline nuclide $i$ & central value &  $\sigma_{\omega_b}$ & $\sigma_{ii}$  & rate & $\delta \sigma^2/ \sigma^2 (\%)$\\
\hline
 \multirow{1}{*}{$Y_p$ }  &  0.2480  &  $^{+0.0002}_{-0.0003}$ & $\pm 0.0002$  & $R_0$& 98.5\\
\hline
 \multirow{3}{*}{$^2$H/H $\times 10^{5}$}  & &  &  &$R_2$& 49\\
 & 2.53 &  $\pm 0.11$ &${\pm} 0.04$  & $R_3$& 37\\
& & & & $R_4$& 14 \\
\hline
 \multirow{2}{*}{$^3$He/H $\times10^{5}$ } & 1.02 &  $^{+0.01}_{-0.02}$ & ${\pm} 0.03$  &$R_7$& 80.7\\
  & &  &   & $R_2$& 16.8\\
\hline
 \multirow{4}{*}{$^7$Li/H $\times 10^{10}$ }  & &  &  &  $R_{14}$& 40.9\\
 & 4.7 &  $\pm 0.3$ & $\pm 0.4$ &   $R_8$ & 25.1\\
& & & &  $R_{15}$& 16.2 \\
& & & &  $R_7$ & 8.6\\
\hline
 \multirow{1}{*}{$^6$Li/H $\times 10^{14}$ }  &1.1 & ${\pm} 0.1$  & $^{+1.7}_{-1.1}$  &  $R_{13}$& $\sim100$\\
\hline
\end{tabular}
\end{center}
\end{table*}

To illustrate the main dependence of the yields on the nuclear
rates, we report here the scaling relations introduced in
\citep{Cyb04} and normalized with respect to the predictions of
\texttt{PArthENoPE} \citep{Par08} around the fiducial value of
$\omega_b=0.02273$ (WMAP 5-years). The scalings are:
\bea \label{eqn:scal-d} \frac {^2{\rm H}}{\rm H} &=& 2.53\times
10^{-5}\,R_3^{-0.55} R_4^{-0.45} R_2^{-0.32}
R_1^{-0.20}\left(\frac{\omega_b}{0.02273}\right)^{-1.62}\,\left(\frac{\tau_n}{\tau_{n,0}}
\right)^{0.41} \\
\label{eqn:scal-he3} \frac{\he3}{\rm H} &=& 1.02\times
10^{-5}\,R_7^{-0.77} R_2^{0.38} R_4^{-0.25} R_3^{-.20} R_5^{-0.17}
R_1^{0.08} \nonumber\\
& \times & \left(\frac{\omega_b}{0.02273}\right)^{-0.59}\,\left(
\frac{\tau_n}{\tau_{n,0}}\right)^{0.15} \vv \\
\label{eqn:scal-he4} Y_p &=& 0.2480\, R_3^{0.006} R_4^{0.005}
R_1^{0.005} \left(\frac{\omega_b}{0.02273}
\right)^{0.39}\left(\frac{\tau_n}{\tau_{n,0}}\right)^{0.72} \vv \\
\label{eqn:scal-li7} \frac{\li7}{\rm H} &=& 4.7 \times
10^{-10}R_1^{1.34} R_8^{0.96} R_7^{-0.76} R_{10}^{-0.71}
R_3^{0.71} R_2^{0.59} R_5^{-0.27} \nonumber\\
& \times & \left(
\frac{\omega_b}{0.02273}\right)^{2.12}\,\left(\frac{\tau_n}{\tau_{n,0}}\right)^{0.44}
\pp \eea
It is clear, for example, that $Y_p$ prediction is dominated by
the neutron mean lifetime (see Section \ref{sec:weakrates}).  For
all the other reactions the relative weight of the different
cross-sections is consistent with what is reported in
Table~\ref{uncertNuc}, modulo a caveat: some important reactions
may have a minor impact on the uncertainty due to their better
determination or vice versa. This is the case of $R_1$, whose role
does not reflect in the error budget being theoretically well
under control (at the \% level), or conversely the case of
$R_{14}, R_{15}$, which contribute appreciably to the $\li7$ error
budget in Table~\ref{uncertNuc} due to the assumption done in
\citep{Ser04b} that they are only known at the order of magnitude.
In general, while improvements in the nuclear reaction rates would
still sharpen the BBN predictions (especially for $\li7$), it is
fair to conclude that these uncertainties are at the moment
negligible compared to the observational ones (see Section
\ref{sec:lithium7}). Thus an experimental campaign does not seem
mandatory for BBN purposes alone. Yet, a careful assessment of the
systematic uncertainties would be useful in reducing the present
discrepancies in data regression protocols. A realistic account
especially of scale uncertainties in the {\it already existing}
datasets, perhaps excluding unreliable measurements and correcting
underestimated error assignments would be certainly a more useful
input from the experimental community.

It is worth commenting here on an additional robust prediction of
BBN: elements heavier than $\li7$ are virtually absent from the
chemical composition of the early universe. This was realized from
the first pioneering studies on BBN \citep{Alp48,Fer50}, and
mostly explained in terms of the very high entropy and on the low
density of the primordial plasma. They result in extremely low
abundances $A\geq 2$ elements at the weak reaction freeze--out,
and an inhibition of three--body reactions\footnote{The leading
reaction for the synthesis of $\cdod$ in stars is the
3$\alpha\rightarrow\cdod$.}, respectively. To assess more
quantitatively the robustness of this result, the present authors
have recently studied the nuclear processes involved in the
production of $A\geq 8$ elements in BBN \citep{Ioc07}. It was
confirmed that their synthesis can only start at T$\lesssim$60~keV
(see Figure \ref{fig:bbnevol}), close to the freeze--out of
nuclear reactions, when they have a very small efficiency. Heavier
elements in BBN are in fact produced by an $\alpha$ capture over
$\li7$ and subsequent build-up of $\cdod$; their final abundance,
mainly produced by means of an $\alpha$ capture over $\li7$ and
subsequent proton capture, is to be well below $10^{-10}$ that of
hydrogen (see Figure \ref{fig:bbnevol} for details). While
processes till now neglected were identified and included, their
role does not alter this basic conclusion. An important
implication of these results follows for the first stars in the
universe, known as Population III stars, whose formation process
and initial mass function is thought to be critically dependent on
the carbon and oxygen content of the environment\footnote{For a
review we address the reader to the proceedings of the last
plenary conference on the topic \citep{FS308}.}. While the amount
of ``heavy'' element predicted in SBBN is safely below the
critical level to alter the standard (but yet to be tested)
scenario, in exotic models this might not be true. More in
general, even when roughly reproducing $\hi2$, $\he3$, $\He4$ and
$\li7$ yields of the SBBN, exotic scenarios may differ for the
predictions of $\lisix$, $^9$Be or elements with $A\geq 12$
(`metals' in the astrophysical jargon). Examples of this will be
provided in Section \ref{sec:BImodels} and \ref{Be9CBBN}.

\subsubsection{Numerical solution of the BBN set of equations}
\label{sec:numsol}
Despite of the fact that some simple estimate of BBN predictions
can be made {\textit{on the back of the envelope}}, detailed
theoretical estimates that can be compared with experimental data
require a careful numerical solution of BBN equations
\eqn{e:drdt}-\eqn{e:nuboltz}. This is particularly relevant when
one considers exotic scenarios with extra parameters or where a
different BBN dynamics is considered, with the aim of constraining
physics beyond the SMPP. Since the original Wagoner computer code
\citep{Wag69,Wag73} much effort has been devoted to develop
numerical tools and provide reliable numerical results
\citep{Kaw88,Kaw92,Fio98,Lop99,Esp99,Bur00,Esp00a,Oli00,Cyb01,
Cyb03b,Coc:2003ce,Cyb04,Ser04b,Pis08}. In 1988, Kawano modified the BBN
program of Wagoner \citep{Kaw88}, and in 1992 an updated and user
friendly public version has been released \citep{Kaw92}, which has
served as a reference tool for a wide number of research groups.
Nuclear reaction rates were updated in 1993 \citep{Smi93a} and
radiative and Coulomb corrections to the weak rates, to which the
$^4$He abundance is very sensitive, as well plasma effects, were
included as a constant multiplicative factor to n--p weak rates.
In 1993 Kernan discovered a relatively large time-step systematic
error in the $^4$He abundance prediction of the public Kawano
code, $\delta Y_p = 0.0017$, since then routinely added to its
result. This, strictly speaking, is not fully adequate because
different users use different step sizes and furthermore,
numerical error is machine dependent. In 1999 Lopez and Turner
wrote a new nucleosynthesis code \citep{Lop99}, which used the
same nuclear rates  but incorporated more accurately the
radiative, finite nucleon mass and finite temperature corrections
to the weak rates. The code used by Olive, Steigman and Walker
\citep{Oli00} agreed, over the whole range $1\leq \eta_{10} \leq
10$, better than the 0.1\% level, with the predicted $^4$He
abundance of the Lopez and Turner code. In 2000, the weak rate
corrections were  also calculated in \citep{Esp99} and included in
a BBN code. This code was developed and continuously updated over
almost a decade, giving particular care also to the treatment of
neutrino decoupling and the nuclear reaction chain which enters
the light abundances evolution \citep{Esp00a,Ser04b}. It was
eventually made public in 2008 with the name \texttt{PArthENoPE}
\citep{Par08}. Details can be found in \citep{Pis08}, where
particular emphasis is given to a comparison with \citep{Kaw92}.
To our knowledge, only the original Kawano code or improved
versions of it\footnote{See the link
http://www-thphys.physics.ox.ac.uk/users/SubirSarkar/bbn.html for
a version of the Kawano standard code, made Linux-friendly by S.
Dodelson.} and \pth are publicly available.

\begin{table}
\begin{center}
\caption{A comparison of light nuclei theoretical predictions in
some recent analyses. Results are either produced by public
numerical tools, or obtained by using fitting formulae made
available by the authors or finally, simply quoted in the papers.
The theoretical errors are estimated using \citep{Par08}, and
account for the effects of nuclear rate uncertainties. Results are
shown for a baryon fraction $\Omega_B
h^2$=0.0224.}\label{comparing-codes}
\begin{tabular}{|c|c|c|c|c|c|c|}
\hline & \citep{Kaw92} & \citep{Fio98} & \citep{Bur00} &
\citep{Cyb04}&
\citep{Ste07} & \citep{Par08} \\
\hline $Y_p$ & 0.2463 & 0.2479 & 0.2483 & 0.2485 & 0.2485 &
$0.2479 \pm 0.0002$ \\
$^2$H/H $\times 10^5$ & 2.57 &2.60 & 2.60 & 2.55 & 2.59 & 2.58 $\pm$ 0.04 \\
$^3$He/H $\times 10^5$ & 1.04 & 1.04 & 1.02 & 1.01 & 1.04 & 1.03 $\pm$ 0.03 \\
$^7$Li/H $\times 10^{10}$ & 4.53 & 4.42 & 4.91 & 4.26 & 4.50 & 4.57 $\pm
0.4$ \\
\hline
\end{tabular}
\end{center}
\end{table}

To examine the concordance of theoretical predictions we have
considered some recent results obtained using numerical BBN codes,
either publicly available or whose outputs have been made public
by the authors as fitting formulae versus the baryon density or
finally, simply quoted in their papers for some reference value of
$\Omega_B h^2$. These results are shown in Table
\ref{comparing-codes}. The list is of course largely incomplete,
but we think it is representative enough to give the idea of the
status of BBN theoretical accuracy. In the last column we also
quote the theoretical uncertainty due to nuclear rate
(experimental) errors, as estimated using \citep{Par08}. We see
that all results are in very good agreement, in particular $Y_p$
agrees at the $\pm 0.1\%$ level in the most recent calculations.
There is a somehow larger spread of $^7$Li estimates which are
however all compatible within the larger theoretical uncertainty
(Table \ref{comparing-codes}).
\begin{table}[h]
\caption{Coefficients of the fit of Eq. (\ref{e:abundfit}) for the
$Y_p$ abundance.}\label{t:He4fit}
\begin{tabular}{|c|cccccc|}
\hline\hline
\backslashbox{$m$}{$n$} & 0 & 1 & 2 & 3 & 4 & 5 \\
\hline 0 & 0.24307                 & -14.242 & 1418.4  & -65863. &
1.4856$\times
10^6$  & -1.3142$\times 10^7$ \\
\hline 1 & -3.6433$\times 10^{-2}$ & 14.337  & -1375.0 & 64741.  &
-1.4966$\times
10^6$ & 1.3601$\times 10^7$  \\
\hline 2 & 1.6132$\times 10^{-2}$  & -4.5189 & 444.13  & -21353. &
502610.
      & -4.6405$\times 10^6$ \\
\hline 3 & -1.6279$\times 10^{-3}$ & 0.43362 & -42.850 & 2069.4  &
-48890.
      & 452740.              \\
\hline\hline
\end{tabular}
\end{table}

\begin{table}[h]
\caption{Coefficients of the fit of Eq. (\ref{e:abundfit}) for the
$^2$H/H$\times 10^5$ abundance.} \label{t:Dfit}
\begin{tabular}{|c|ccccc|}
\hline\hline
\backslashbox{$m$}{$n$} & 0 & 1 & 2 & 3 & 4 \\
\hline 0 & 14.892                  & -1551.6 & 70488.  &
-1.5390$\times 10^6$ &
1.3164$\times 10^7$  \\
\hline 1 & 6.1888                  & -916.16 & 56639.  &
-1.6046$\times 10^6$ &
1.7152$\times 10^7$  \\
\hline 2 & -0.60319                & 118.51  & -8556.3 & 267580.
&
-3.0624$\times 10^6$ \\
\hline 3 & 4.5346$\times 10^{-2}$  & -8.7506 & 624.51  & -19402.
&
221200.              \\
\hline\hline
\end{tabular}
\end{table}

\begin{table}[h]
\caption{Coefficients of the fit of Eq. (\ref{e:abundfit}) for the
$^ 3$He/H$\times 10^5$ abundance.}\label{t:He3fit}
\begin{tabular}{|c|ccccc|}
\hline\hline
\backslashbox{$m$}{$n$} & 0 & 1 & 2 & 3 & 4 \\
\hline 0 & 3.1820                 & -298.88 & 15974.  & -422530. &
4.4031$\times
10^6$  \\
\hline 1 & 0.57549                & -91.210 & 6376.6  & -201070. &
2.3486$\times
10^6$  \\
\hline 2 & -0.15717               & 33.689  & -2651.2 & 89571.   &
-1.0998$\times
10^6$ \\
\hline 3 & 1.4594$\times 10^{-2}$ & -3.2160 & 256.66  & -8780.2  &
109100.
      \\
\hline\hline
\end{tabular}
\end{table}

\begin{table}[h]
\caption{Coefficients of the fit of Eq. (\ref{e:abundfit}) for the
$^7$Li/H$\times 10^{10}$ abundance.}\label{t:Li7fit}
\begin{tabular}{|c|ccccc|}
\hline\hline
\backslashbox{$m$}{$n$} & 0 & 1 & 2 & 3 & 4 \\
\hline 0 & 2.5274                  & -614.44 & 62186.  &
-1.5670$\times 10^6$ &
1.4339$\times 10^7$  \\
\hline 1 & 1.9384$\times 10^{-2}$  & 55.173  & -11365. & 492710.
&
-6.2826$\times 10^6$ \\
\hline 2 & -8.6994$\times 10^{-2}$ & 16.437  & -431.32 & -13313.
&
359980.              \\
\hline 3 & 2.2257$\times 10^{-2}$  & -4.2339 & 260.16  & -6277.3
&
55300.               \\
\hline\hline
\end{tabular}
\end{table}

We report the fit of the main nuclide abundances as function of
$\omega_b\equiv \Omega_B h^2$ and the number of effective degree
of freedom, $\neff$. The fit holds for $0.015\leq \omega_b\leq
0.029$ and $0\leq \neff\leq 7$. The fitting function for all
nuclei is
\be \sum_n \sum_m a_{nm}\, \omega_b^n\, \neff^m \vv
\label{e:abundfit}
\ee
and the coefficient are reported in Tables
\ref{t:He4fit}--\ref{t:Li7fit}. The fit accuracy is better than
0.13\% for $Y_p$, than 0.3\% for $^2$H/H and $^7$Li/H, and than
0.6\% for $^3$He/H.

\section{Observational Abundances}
\label{sec:obsabund}

The abundances of primordial elements are inferred from
measurements performed in a large variety of astrophysical
environments. What \textit{precision cosmology era} has meant in
this field is an increased number and precision of spectroscopic
data over the past two decades. Presently, the situation is still
quite involved due to the presence of relevant systematic errors
which are comparable to (where not dominant over) the statistical
uncertainties. Unfortunately, these errors are to a large extent
irreducible and intrinsic to the astrophysical determinations
themselves,  which rely on highly evolved systems where
reprocessing, astration, and contaminations from younger systems are
possibly present and difficult to correct for.

Since our nearby universe is far from reflecting its primordial conditions, the
methods proposed and developed in the last forty years to infer
primordial yields have focused on very old and hence little
evolved astrophysical regions, as well as on the capability to
correct for the effect of the galactic evolution on the pristine
abundances. For example, due to its very weak binding, any $^2$H
nucleus contained into pre--stellar nebulae is burned out during
their collapse. Hence, the post-BBN deuterium evolution is
expected to be a monotonic function of time and astrophysical
deuterium measurements can be assumed to represent lower bounds of
its primordial abundance.

Unfortunately, such a simple scheme cannot be applied to the more tightly
bound $\he3$ nucleus. In this case, in fact, in stellar interior it can be
either produced by $^2$H-burning or destroyed in the hotter regions. As a
consequence, all the $\he3$ nuclides surviving the stellar evolution phase
contribute to the chemical composition of the InterStellar Medium (ISM)
and thus stellar and galactic evolution models are necessary to track back
the primordial $\he3$ abundance from the post-BBN data, at least in the
regions where stellar matter is present. For this reason the inferred
primordial $\he3$ abundance is intrinsically a model-dependent quantity.

The situation of $\He4$ is reversed with respect to that of
deuterium, since in this case the hydrogen burning in the
successive stellar population has increased the amount of $\He4$
as well as "metals" (nuclei with $Z > 4$) such as C, N and O.
Usually, $\He4$ is measured in old and very little evolved systems
versus their metallicity, and extrapolating {\it linearly} to
zero-metallicity. While this is certainly a reasonable approach,
still it could lead to some systematic uncertainty.

Finally, $\li7$, whose bulk is also believed to be produced in the
primordial cauldron, is a very weakly bound nuclide which has an
extremely involved post-BBN chemical evolution. In fact, it is
easily destroyed in the interiors of stars but can survive in the
cooler outer layers of stars with shallow convective zones, where
it can be measured by means of absorption spectra. However, the
scenario is much more involved due to the observation of enhanced
lithium abundance in some red-giants. This suggests that $\li7$
formed in the interior of some stars may be transported by
convective modes to the cooler exteriors. In this case these stars
behave as lithium producers and thus enrich the ISM with lithium.
Furthermore, the cosmic rays scattering on ISM nuclei can
contribute to the total amount of $\li7$ and since CNO elements
are necessary for spallation processes, this would imply a
correlation between post-BBN $\li7$ and the metallicity, which may
help in tracking back this non-primordial component.

\subsection{Deuterium}
\label{sec:deuterium}

It is commonly believed that there are no astrophysical sources of
deuterium since it is destroyed by stellar evolution processes
\citep{Eps76} and non-thermal production channels have been
constrained to be negligible (see e.g. \citep{Pro03}). Thus, any
astrophysical observation can provide a lower bound for the
primordial abundance. Therefore, the local ISM in the Milky Way
can provide an order of magnitude more determinations of $^2$H/H
than high redshift Quasar Absorption Systems (QAS) (see Ref.
\citep{Lin06} for the most recent compilation), given its easier
observational accessibility. By using the Far-Ultraviolet
Spectroscopic Explorer (FUSE) \citep{FUS99}, a large database of
Galactic $^2$H/H measurements has been compiled. However, despite
FUSE and other satellite observations, providing measurements of
$^2$H/H in almost 50 lines of sight, the picture of Galactic
deuterium abundances remains puzzling. In particular, inside the
Local Bubble ($<$100 pc from the Sun) the deuterium to hydrogen
ratio seems roughly constant at $^2\textrm{H}/\textrm{H}|_p =
(1.56 \pm 0.04) \times 10^{-5}$ \citep{Woo04}. However, beyond
this bound, an unexpected scatter of a factor of $\sim 2$ in
$^2$H/H values is observed \citep{Jen99,Woo04,Lin06,Ell07} as well
as correlations with heavy element abundances, which suggest that
ISM deuterium might have suffered stellar processing, namely
astration (see for example \citep{Rom06}), but also that it may
reside in dust particles which evade gas-phase observations. This
is supported by a measurement in the lower halo \citep{Sav07}
which indicates that the Galactic $^2$H abundance has been reduced
by a factor of only $1.12 \pm 0.13$ since its formation. As an
alternative explanation, it is worth reporting the possible
existence of a strong late infall of pre-galactic material with
primordial composition (high $^2$H) (see for example
\citep{Pra07,Pro08}) which has some observational evidence via the
study of kinematics of high latitude gas regions. Finally, it is
important to mention the analysis of Infrared Space Observatory
spectra of H$_2$, H-$^2$H, CH$_4$, and CH$_3$$^2$H in Jupiter's
atmosphere which led authors of Ref. \citep{Lel01} to infer the
value of deuterium to hydrogen ratio for the protosolar cloud
$^2$H/H$_{psc} = (2.1 \pm 0.4) \times 10^{-5}$, which corresponds
to our galaxy value when its age was two thirds of the present
one.

The astrophysical environments which seem most appropriate to
obtain reliable measurements of the primordial deuterium fraction
are the hydrogen-rich clouds absorbing the light of background
QSOs at high redshifts, as recognized already in \citep{Ada76}.
Conventional models of galactic nucleosynthesis (chemical
evolution) predict a small contamination of pristine $^2$H/H
\citep{Fie96}. However, a successful implementation of this method
requires: (i) neutral hydrogen column density in the range
$17\lesssim \log[{\rm N(H}_I{\rm)/cm}^{-2}]\lesssim 21$; (ii) low
metallicity [M/H] to reduce the chances of deuterium astration;
(iii) low internal velocity dispersion of the atoms of the clouds,
allowing the isotope shift of only 81.6 km /s to be resolved
\citep{Pet08}. For this reason, only a handful of determinations
have been obtained since the advent of the $>8$m class telescopes
in the 1990s in damped Lyman-$\alpha$ systems and Lyman limit
systems:

\begin{itemize}
\item[i)] {\bf Q1009+2956}, with the absorber placed at $z =
2.504$, yielding
$^2\textrm{H}/\textrm{H}=\left(3.98^{+0.59}_{-0.67} \right)\times
10^{-5}$ ($\log{^2\textrm{H}/\textrm{H}} = -
4.40^{+0.06}_{-0.08}$) \citep{Bur98,Kir03}.

\item[ii)] {\bf PKS1937-1009 (I)}, with the absorber placed at
$z=3.572$, yielding $^2\textrm{H}/\textrm{H}=(3.3 \pm 0.3)\times 10^{-5}$
($\log{^2\textrm{H}/\textrm{H}}=-4.49 \pm 0.04$) \citep{Bur98}.

\item[iii)] {\bf HS 0105+1619}, with the absorber placed at $z=2.536$
with metallicity [Si/H]$\sim 0.01$, yielding
$^2\textrm{H}/\textrm{H}=(2.54\pm 0.23) \times 10^{-5}$
($\log{^2\textrm{H}/\textrm{H}}=-4.596 \pm 0.040$) \citep{OMe01}.

\item[iv)] {\bf Q2206-199}, with the absorber placed at $z=2.0762$
with metallicity [Si/H]$=-2.23$, yielding $^2\textrm{H}/\textrm{H}=(1.65
\pm 0.35) \times 10^{-5}$
($\log{^2\textrm{H}/\textrm{H}}=-4.78^{+0.08}_{-0.10}$) \citep{Pet01}.

\item[v)] {\bf Q0347-3819}, with the absorber placed at $z=
3.024855$ with metallicity [Si/H]$=-0.95 \pm 0.02$, yielding
$^2\textrm{H}/\textrm{H} = (3.75 \pm 0.25) \times 10^{-5}$
($\log{^2\textrm{H}/\textrm{H}}=-4.43 \pm 0.03$) \citep{Lev01}.
Note that the previous analysis in \citep{DOd01} yielded
$^2\textrm{H}/\textrm{H} = (2.24 \pm 0.67) \times 10^{-5}$ due to
an incorrect velocity distribution function (see \citep{Lev01}).
Similar considerations apply for the value discussed in vii)
below).

\item[vi)] {\bf Q1243+3047}, with the absorber placed at
$z=2.525659$ with metallicity [O/H]$=-2.79 \pm 0.05$, yielding
$^2\textrm{H}/\textrm{H}=\left(2.42^{+0.35}_{-0.25} \right) \times
10^{-5}$
($\log{^2\textrm{H}/\textrm{H}}=-4.617^{+0.058}_{-0.048}$)
\citep{Kir03}.

\item[vii)] {\bf PKS1937-1009 (II)}, with the absorber placed at
$z=3.256$ with metallicity [Si/H]$=-2.0  \pm 0.5$, yielding
$^2\textrm{H}/\textrm{H}=\left( 1.6^{+0.25}_{-0.30} \right) \times
10^{-5}$ ($\log{^2\textrm{H}/\textrm{H}}=-4.80^{+0.06}_{-0.09}$)
\citep{Cri04}. Since not all the ionized deuterium (D$_I$)
components are resolved, it is often claimed that this value of
$^2\textrm{H}/\textrm{H}$ is more dependent on the precise
description of the kinematics of the gas and thus less robust
against systematics (see \citep{Pet08}).

\item[viii)] {\bf SDSS 1558-0031}, with the absorber placed at
$z=2.70262$ with metallicity [O/H] $= -1.49$, yielding
$^2$H/H$=\left(3.31^{+0.49}_{-0.43} \right)\times 10^{-5}$
($\log{^2\textrm{H} /\textrm{H}} = -4.48 \pm 0.06$) \citep{OMe06}.

\item[ix)] {\bf Q0913+072} shows six well-resolved D$_I$ Lyman
series transitions placed at $z = 2.61843$ and recently observed
with the ESO VLT \citep{Pet08}. With an oxygen abundance of about
1/250 of the solar value the authors of Ref. \citep{Pet08} deduce
a value of the deuterium abundance $^2\textrm{H}/\textrm{H} =
\left(2.75^{+0.27}_{-0.24} \right) \times 10^{-5}$
($\log{^2\textrm{H}/\textrm{H}} = -4.56 \pm 0.04$).

\item[x)] {\bf Q0014+813}, with the absorber placed at $z= 3.32$,
yielding $^2\textrm{H}/\textrm{H} \sim (1.9 \pm 0.5) \times
10^{-4}$ \citep{Rug96,Car94}. Very old and no more used after
\citep{Bur99a}.

\item[xi)] {\bf Q0420-388}, with the absorber placed at $z=3.08$,
implying, if the deuterium identification is correct, that the $^2$H/H
ratio could have any value $\leq 2 \times 10^{-5}$. Whereas, if the
O$_I$/H$_I$ ratio is constant throughout the complex, then
$^2\textrm{H}/\textrm{H} \sim 2 \times 10^{-4}$ \citep{Car96}.

\item[xii)] {\bf BR1202-0725}, two Lyman-$\alpha$ systems placed
at $z=4.383$ and $z=4.672$, yielding $^2\textrm{H}/\textrm{H} \leq 1.5
\times 10^{-4}$ \citep{Wam96}.

\item[xiii)] {\bf PG1718+4807}, with the absorber placed at $z=0.701$
$^2\textrm{H}/\textrm{H} \in (1.8-3.1) \times 10^{-4}$
\citep{Web97,Tyt99,Kir01,Cri03}, but having been criticized in
past by Ref. \citep{Kir01}.

\item[xiv)] {\bf Q0130-4021}, with the absorber placed at $z~2.8$
with metallicity [Si/H]$ \leq -2.6$, yielding
$^2\textrm{H}/\textrm{H} \leq 6.7 \times 10^{-5}$
($\log{^2\textrm{H}/\textrm{H}}\leq -4.17$) \citep{Kir00},
considered not very interesting and thus typically neglected.
\end{itemize}

\begin{figure}[t]
\begin{center}
\epsfig{file=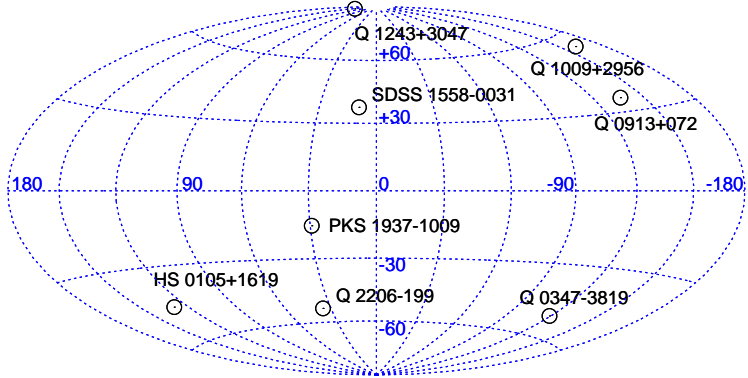,width=0.8\textwidth}
\end{center}
\caption{Hammer-Aitoff projection in Galactic coordinates of the
positions of the eight quasars along which nine measurements of deuterium
abundance have been reported (see i)-ix), in the text).}
\label{figDmap}
\end{figure}

Recently, three different analysis of high-redshift systems appeared:
\begin{itemize}
\item [1)] in \citep{OMe06,Ste07} quasars used are: HS 0105+1619,
Q2206-199, PKS1937-1009 (I), Q1009+2956, Q1243+3047,  SDSS
1558-0031, and the cumulative value reported is
$$ ^2\textrm{H}/\textrm{H}=\left(2.68^{+0.27}_{-0.25} \right)\times 10^{-5},
$$
when averaging over $^2\textrm{H}/\textrm{H}$ determinations
\citep{Ste07}, or rather
$$
\log{^2\textrm{H}/\textrm{H}} = -4.55 \pm 0.04 \Longrightarrow
^2\textrm{H}/\textrm{H}=\left(2.82^{+0.27}_{-0.25} \right)\times
10^{-5},
$$
when the log's are used;
\item [2)] the authors of \citep{Fie07} use: HS 0105+1619,
Q2206-199, PKS1937-1009 (I), Q1009+2956, Q1243+3047 and the cumulative
value reported is
$$
^2\textrm{H}/\textrm{H}=(2.78 \pm 0.29) \times 10^{-5},
$$
averaging over the $\log[^2\textrm{H}/\textrm{H}]$ determinations;
\item [3)] finally, in the very recent analysis of \citep{Pet08}
the following sample is exploited: HS 0105+1619, Q0913+072,
Q1009+2956, Q1243+3047, SDSS 1558-0031, PKS1937-1009 (I),
Q2206-199. An average consistent with all the data is
$\log[^2\textrm{H}/\textrm{H}]=-4.55\pm0.03$ or, equivalently,
$$
^2\textrm{H}/\textrm{H}=(2.82^{+0.20}_{-0.19}) \times 10^{-5}.
$$
\end{itemize}

\begin{figure}[t]
\begin{center}
\epsfig{file=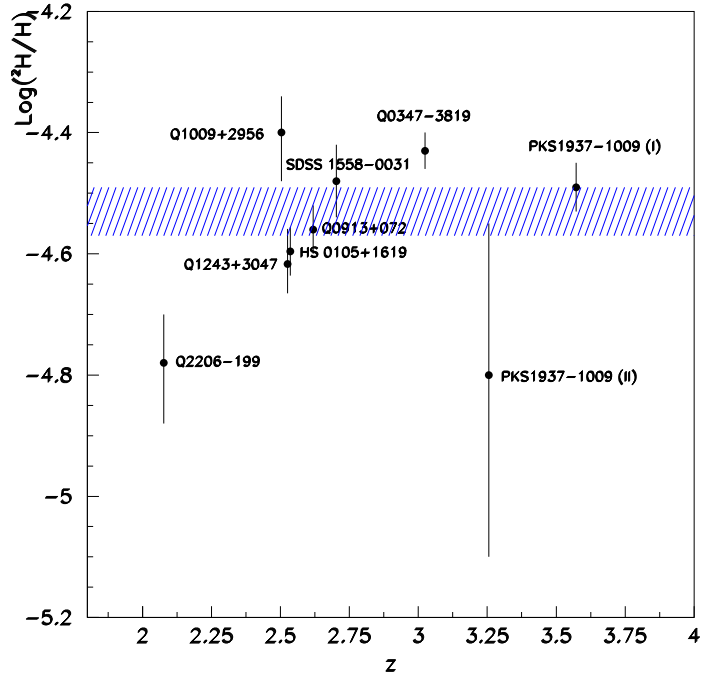,width=0.7\textwidth}
\end{center}
\caption{The nine measurements of i)-ix) QSA's used in our analysis.
The horizontal band represents the value of Eq. (\ref{Dfin}).}
\label{figD}
\end{figure}

The positions in the sky of the reported measurements i)-ix), are
given in galactic coordinates in Figure \ref{figDmap}, and in
redshift space in Figure \ref{figD}. We have performed a
re-analysis of $^2\textrm{H}/\textrm{H}$ using these results,
averaging over the values of $\log[^2\textrm{H}/ \textrm{H}]$ and
using the method described in Ref. \citep{Bar04a} which has the
particular advantage to give a continuous $\chi^2$ function even
in the case of asymmetric errors. In particular, let us denote
with ${\overline{x}_i}\, ^{+\sigma_i^+}_{-\sigma_i^-}$ the generic
measurement of $\log[{^2\textrm{H}/\textrm{H}}]$ corresponding to
the i-th QSA. According to Ref. \citep{Bar04a} one can define the
following quantities:
\be
\sigma_i \equiv \frac{\sigma_i^+ +
\sigma_i^-}{2}\,\,\,\,\,\,\,\,\,\, A_i \equiv \frac{\sigma_i^+ -
\sigma_i^-}{\sigma_i^+ + \sigma_i^-} \vv \label{chi2-D-param}
\ee
and define a total $\chi^2(\mu)= \sum_{i} \chi^2_i(\mu)$, where
each contribution $\chi_i^2$ is expanded up to $A_i^2$ terms
\be
\chi^2_i(\mu) = \left(\frac{\overline{x}_i -
\mu}{\sigma_i}\right)^2 \left( 1 - 2\, A_i
\left(\frac{\overline{x}_i - \mu}{\sigma_i}\right) + 5 \, A_i^2 \,
\left(\frac{\overline{x}_i - \mu}{\sigma_i}\right)^2 \right) \pp
\label{chi2-D-i} \ee
Using this procedure we find a value of the reduced $\chi^2$,
$\sqrt{\chi^2_{\rm min}/(9-1)}=2.715$, which shows the effect of
some systematic effects and that one or more uncertainties have
been underestimated. If one chooses to treat all the data on the
same footing, one can account for this by inflating each
uncertainty by the multiplicative factor $2.715$. In this case,
after repeating the procedure, the new minimization leads to the
result
\be
\log{^2{\rm H/H}} = -4.53 \pm 0.04 \Longrightarrow ^2{\rm
H/H}=\left(2.98^{+0.29}_{-0.23}\right) \times 10^{-5} \label{Dfin}
\pp
\ee
In \citep{Pet08} it was argued that the determinations v) and vii)
of our list are less robust against systematics in the modeling of
the cloud, due to the minor number of resolved deuterium lines. To
be more conservative, one can thus exclude these two data from the
regression. In this case, with the choice for central values and
symmetric error bars done in \citep{Pet08}, we reproduce their
results. However, we find that it is not irrelevant to take into
account the asymmetric errors in the regression procedure. When
using the published asymmetric errors in $\log$[$^2$H/H] and
applying the same procedure as above, we find a significantly
lower multiplicative factor $\sqrt{\chi^2(-4.539)/6}=1.897$,
showing indeed that the dispersion of the measurements for this
dataset is more consistent with a statistical one. To be
conservative we multiply the error bars by the factor 1.897 and
find \be \log{^2{\rm H/H}} = -4.54 \pm 0.03 \Longrightarrow ^2{\rm
H/H}=\left(2.87^{+0.22}_{-0.21} \right) \times 10^{-5}
\label{Dfin2} \vv \ee which is the value we will be using in the
following.

To conclude, it is worth mentioning the recent proposal to use the
fluctuations in the absorption of cosmic microwave background photons by
neutral gas during the cosmic dark ages, at redshifts $z \approx 7 - 200$,
to reveal the primordial deuterium abundance of the Universe. This method
is based on the strength of the cross-correlation of
brightness-temperature fluctuations due to resonant absorption of CMB
photons in the 21-cm line of neutral hydrogen, with those due to resonant
absorption of CMB photons in the 92-cm line of neutral deuterium. This
results to be proportional to the ratio $^2\textrm{H}/\textrm{H}$ fixed
during BBN. Although technically challenging, this measurement could
provide the cleanest possible determination of $^2{\rm H/H}$
\citep{Sig06}. A difficulty which has been pointed out---that may prevent
the viability of the method at redshifts when the first UV sources turn
on, $z\lsim 40$---is that when including Ly$\beta$ photons in the
analysis, the inferred ratio $^2$H/H would not be constant, but depend
sensitively on the UV spectrum \citep{Chu06}.

\subsection{Helium-3}

Like deuterium, whose primordial yield is extremely sensitive to the
baryon density parameter, $\eta$, $\he3$ is a crucial test of the standard
BBN scenario as well. From the observational point of view, several
environments are studied in order to derive its primordial abundance.
Terrestrial determinations yield e.g. the ratio $\he3/\He4 \sim 10^{-6}$
from balloon measurements or $\sim 10^{-8}$ from continental rock
\citep{And93,Roo02}. These observations, which show a large spread of
values, confirm the idea that the terrestrial helium has no cosmological
nature. In fact, most of it is $\He4$ produced by the radioactive decay of
elements such as uranium and thorium. No natural radioactive decay
produces $\he3$, hence its observed terrestrial traces can be ascribed to
unusual processes such as the testing of nuclear weapons or the infusion
of extraterrestrial material.

In the solar system, $\he3$ is measured through the solar wind and
meteorites \citep{Glo00}. The most accurate value was measured in
Jupiter's atmosphere by the Galileo Probe \citep{Mah98}. These
measurements support the idea of the conversion of the deuterium
initially present in the outer parts of the Sun into $\he3$ via
nuclear reactions. One can infer that in the ProtoSolar Material
(PSM) out of which the Sun formed, $\he3/\He4 = (1.66 \pm 0.05)
\times 10^{-4}$ \citep{Mah98,Roo02} and simultaneously
$^2$H/H$=(2.6 \pm 0.7)\times 10^{5}$ \citep{Mah98}. Note that in
order to transform the above mentioned ratios $\he3/\He4$ into
$\he3/\mbox{H}$ one can use the value $\He4/\mbox{H}\sim 0.1$ and
thus derive the important ratio $\he3/^2$H$= 0.6 \pm 0.2$ which
allows one to constrain electromagnetically decaying particles
(see Section~\ref{s:massive-particle}).

The previous observation of $\he3/\He4$  is compatible with the
same measurement performed in meteoritic gases, yielding
$\he3/\He4 = (1.5 \pm 0.3) \times 10^{-4}$
\citep{Bla72,Ebe74,Gei72,Gei93}. Further information comes from
Local Inter-Stellar Medium (LISM). Occasionally, LISM atoms
crossing the termination shock region that separates the solar
system from interstellar space get ionized. By counting the helium
ions in this particular component of the solar wind, the Ulysses
spacecraft has measured a $\he3/\He4$ ratio of
$2.48^{+0.68}_{-0.62} \times 10^{-4}$ \citep{Glo98}, which is not
inconsistent with the idea that $\he3$ at our galaxy's location
might have grown in the last 4.6 billion years since the birth of
the Sun.

\begin{figure}[t]
\begin{center}
\epsfig{file=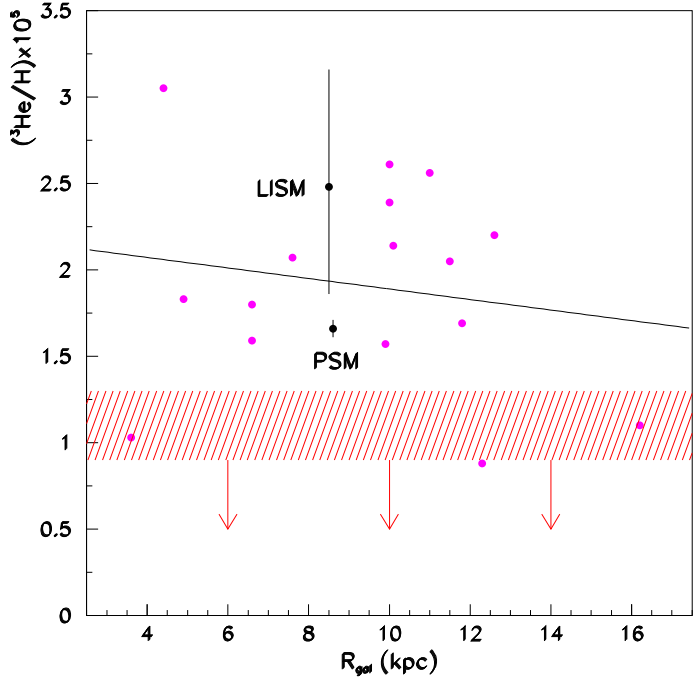,width=0.7\textwidth}
\end{center}
\caption{The 16 purple spots represent the HII regions studied in
Ref. \citep{Ban02} versus their distance from galactic center. Also the
PSM and LISM measurements are reported. The black solid line stands for
the linear fit of purple data, whereas the red band represents the upper
bound obtained in \citep{Ban02}.}
\label{he3dist}
\end{figure}

\begin{figure}[t]
\begin{center}
\epsfig{file=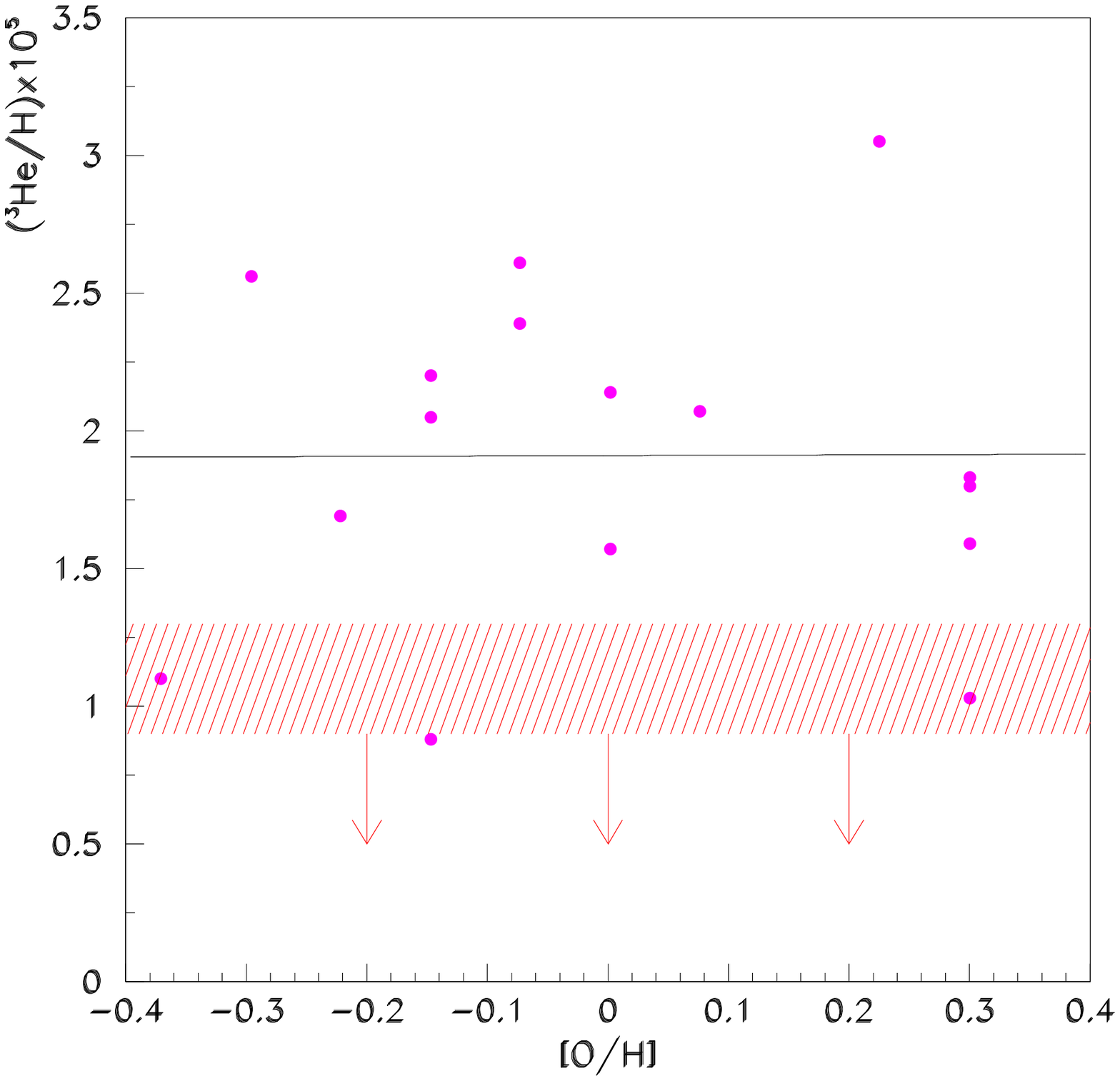,width=0.7\textwidth}
\end{center}
\caption{The data of Ref. \citep{Ban02} versus metallicity are reported.
The black solid line stands for the linear fit of purple data, whereas the
red band represents the upper bound obtained in \citep{Ban02}.}
\label{he3met}
\end{figure}

Far beyond the LISM, only one spectral transition allows the
detection of $\he3$, namely the 3.46 cm spin-flip transition of
$\he3^+$, the analog of the widely used 21-cm line of hydrogen;
this is a powerful tool for the isotope identification, as there
is no corresponding transition in $\He4^+$. The emission is quite
weak, hence $\he3$ has been observed outside the solar system only
in a few HII regions and Planetary Nebulae (PN) in the Galaxy. The
values found in PN result one order of magnitude larger than PSM
and LISM determinations (for example $\he3/\mbox{H} = (2 - 5)
\times 10^{-4}$ is measured in NGC3242 \citep{Ban02}), confirming
a net stellar production of $\he3$ in at least some stars. From
the expected correlation between metallicity of the particular
galactic environment and its distance from the center of the
galaxy, one would expect a gradient in $\he3$ abundance versus
metallicity and/or distance. In Ref. \citep{Ban02} the
$\he3/\mbox{H}$ abundance ratios are reported for the sample of
simple HII regions.

Figures (\ref{he3dist}) and (\ref{he3met}) report the data as
functions of the distance from galactic center and metallicity,
respectively, together with the two determination from PSM and
LISM. No significant correlation between the $\he3$ abundance and
location (or metallicity) in the Galaxy is revealed. The linear
fits reported in Figures (\ref{he3dist}) and(\ref{he3met}) as
black solid lines correspond respectively to \bea (\he3/\mbox{H})
\times 10^{5}& = & 2.194 - 0.030 \, R_{gal}(\mbox{kpc})\,\,\,,
\\(\he3/\mbox{H}) \times 10^{5}& = & 1.910 +
0.014 \left[\mbox{O/H}\right]\,\,\,,
\eea
where $[\mbox{O/H}] \equiv \log$(O/H)-$\log$(O/H)$_{\odot}$. Note that we
adopt (O/H)$_{\odot}= 4.2 \times 10^{-4}$ \citep{Asp05}, differently from
\citep{Ban02} where $\log$(O/H)$_{\odot}=6.3 \times 10^{-4}$ was used.
Qualitatively, these data suggest a remarkable compensation between
stellar production and destruction of $\he3$.

The failure in observing a galactic $\he3$ dependence on
metallicity, typically predicted by a chemical evolution model of
the Galaxy (see \citep{Rom03} for a review) has been referred to
as the ``$\he3$ problem". However, in the last years this subject
received new insight \citep{Egg06} by the study of 3D mixing
models which seem to reconcile the predictions with the data. In
this scenario, by assuming a more conservative approach, the
authors of Ref. \citep{Ban02} prefer to report an upper limit to
the primordial abundance of $\he3$ by using the observations of a
peculiar galactic HII region, \be \he3/\mbox{H} < (1.1 \pm
0.2)\times 10^{-5} \pp \label{upperhe3} \ee This upper bound is
reported as the red band in Figures (\ref{he3dist}) and
(\ref{he3met}) \citep{Ste07}.

A different approach could be based on the observation that the
ratio $(^2{\rm H}+ \he3)/\mbox{H}$ shows a high level of stability
during the galactic evolution. This is partially supported by
observations and chemical evolution models (see for example
\citep{Gei07}). In this case, by using the value reported in Ref.
\citep{Gei98} for PSM, namely $(^2{\rm H}+\he3)/\mbox{H}=(3.6 \pm
0.5) \times 10^{-5}$, and considering it as a good estimate for
the primordial value one gets
\be
\frac{\he3}{\mbox{H}} = (0.7 \pm 0.5) \times 10^{-5}
\ee
by using the primordial deuterium abundance discussed in the previous
Section, whose upper bound is consistent with the one derived from Eq.
(\ref{upperhe3}).

\subsection{Helium-4}

The post-BBN evolution of $\He4$ can be simply understood in terms
of nuclear stellar processes which, through successive generations
of stars, have burned hydrogen into $\He4$ and heavier elements,
hence increasing the $\He4$ abundance above its primordial value
\citep{Ste07}. Since the history of stellar processing can be
tagged by measuring the {\it metallicity} ($Z$) of the particular
astrophysical environment, the primordial value of $\He4$ mass
fraction $\yp$ can be derived by extrapolating the $\yp - $O/H and
$\yp - $N/H correlations to O/H and N/H$\rightarrow 0$, as
proposed originally in Ref.s \citep{Pei74,Pei76,Pag86}. However,
heavy elements like oxygen are produced by short-lived massive
stars whereas $\He4$ is essentially synthesized in all stars, so
one has to minimize model-dependent evolutionary corrections. The
key data for inferring $\He4$ primordial abundance are provided by
observations of helium and hydrogen emission lines generated from
the recombination of ionized hydrogen and helium in
low-metallicity extragalactic H$_{II}$ regions \citep{Ste07}. Many
attempts to determine $\yp$ have been made, constructing these
correlations for various samples of Dwarf Irregular (DIrrs) and
Blue Compact Galaxies (BCGs) \citep{Izo98}. These systems are the
least chemically evolved known galaxies. Plausibly, they contain
very little helium synthesized in stars after the BBN, minimizing
the chemical evolution problems that affect e.g. the determination
of $\he3$ \citep{Izo04}.

Uncertainties in the determination of $\yp$ can be statistical or
systematic. Statistical uncertainties can be decreased by
obtaining very high signal-to-noise ratio spectra of BCGs. These
BCGs are undergoing intense bursts of star formation, giving birth
to high-excitation supergiant H$_{II}$ regions and allowing an
accurate determination of the helium abundance in the ionized gas
through the BCG emission-line spectra. The theory of nebular
emission is understood well enough not to introduce additional
uncertainty. According to the standard scenario, the universe was
born with zero metallicity; hence, $\yp$ can be determined
extrapolating to $Z \rightarrow 0$ the relationship between $Z$
and the $\He4$ abundance for a sample of objects. This procedure
relies on the determination of the individual $\yp$ and $Z$ values
and of the slope $d\yp/dZ$, which is assumed to be linear. The
uncertainty affecting $\yp$ depends directly on the uncertainties
on $d\yp/dZ$ and the ensemble of the $(\yp, Z)$ pairs. For this
reason, it has long been thought that the best results are
obtained from the analysis of extremely low metallicity objects
like DIrrs and BCGs, since their use minimizes the uncertainty
associated with $d\yp/dZ$. However, the authors of Ref.s
\citep{Pei03a, Pei03b} have noted that this advantage is
outweighed by the relatively higher uncertainty on the
$\yp$-values, which derives from the (unknown) collisional
contribution to the Balmer line intensities, an uncertainty
especially affecting these objects since collisional contribution
is quite important at high temperatures and rapidly fades away at
intermediate and low temperatures \citep{Lur03}. Mostly following
the analysis reported in  Ref. \citep{Pei07} we list below the
most recent estimates of $\yp$.

\begin{itemize}
\item[i)] In Ref. \citep{Izo04} is reported the estimate $\yp =
0.2421 \pm 0.0021$. According to Ref. \citep{Pei07} the
differences with the previous determination are mainly systematic.
One is due to the use in Ref. \citep{Pei07} of $He_{I}$
recombination coefficients studied in Ref.s \citep{Por05,Por07},
which yield $\yp$ values about 0.0040 higher than the previous
ones. Moreover, in Ref. \citep{Pei07} they use some recent $H_{I}$
collisional data, which further increase the $\yp$ values over the
older $H_{I}$ collisional corrections by about 0.0025.

\item[ii)] The value quoted in Ref. \citep{Oli04} is $ \yp =
0.249 \pm 0.009$. The small sample size used and the large
uncertainty affecting the parameters derived from the $H_{II}$
regions considered in the analysis are responsible in this case
for the very conservative error estimate. Also in this case, the
systematic differences with Ref. \citep{Pei07} are due to the
$He_{I}$ recombination data used by both groups and to the
estimation of the collisional contribution to the H Balmer lines.

\item[iii)] In Ref. \citep{Fuk06}, based on a reanalysis of a
sample of 33 $H_{II}$ regions from Ref. \citep{Izo04}, the authors
determined a value of $\yp = 0.250 \pm 0.004$. In addition to a different
treatment of the underlying $H$ and $He_{I}$ absorption there are few
systematic effects discussed in detail in Ref. \citep{Pei07}.

\item[iv)] In Ref. \citep{Pei07}, the authors present a new $\He4$ mass fraction
determination, yielding $\yp= 0.2477 \pm 0.0029 $. This result is based on
new atomic physics computations of the recombination coefficients of
$He_{I}$ and of the collisional excitation of the $H_{I}$ Balmer lines
together with observations and photoionization models of metal-poor
extragalactic $H_{II}$ regions.

\item[v)] Finally, in Ref. \citep{Izo07} is reported the estimate $\yp =
0.2516 \pm 0.0011$ when using the $He_{I}$ emissivity of Ref.
\citep{Por05}.
\end{itemize}

All recent analyses of $\yp$ agree on the fact that the systematic
error is the main responsible for the spread of the $\yp$
determinations. For example, in Ref. \citep{Pei07} three of the
four main sources of error ($\Delta Y_p\geq 0.01$) in estimating
$Y_p$ are reported to be of systematic nature. Different authors
however, report different error budgets (sometimes analyzing the
same objects). Nevertheless, the use of new $He_{I}$ recombination
coefficients studied in Ref.s \citep{Por05,Por07} has sensibly
changed the predictions by pushing them up a few percent as shown
in Ref.s \citep{Pei07,Izo07}. For this reason we will use their
determinations only to derive an estimate of primordial $\yp$.

In particular, in Ref. \citep{Pei07} the uncertainty quoted on the
value of $\yp$ seems to provide a more realistic estimate (0.003)
of the residual indetermination affecting the $\He4$ mass
fraction, versus the more optimistic (0.001) reported in Ref.
\citep{Izo07}, due to a very large sample of objects
included in the analysis. For this reason we prefer to average the two main
values as not weighted by their uncertainties, and use the more
conservative error estimate of Ref. \citep{Pei07} at 1-$\sigma$.
In summary, we adopt for the $\He4$ mass fraction \be \yp = 0.250
\pm 0.003 \pp \label{he4est} \ee Note that this range of values
totally contains the one of Ref. \citep{Izo07}, but only partially
covers the determination of Ref. \citep{Pei07} which points out
slightly smaller values. Hence the effects of new $He_{I}$
recombination coefficients \citep{Por05,Por07} on global $\yp$
data analysis is to push slightly upward its estimate, and this is
going to affect at a certain level the determination of
cosmological parameters, as will be discussed in the following
sections.

We would like to briefly point out that other constraints on $Y_p$
can be obtained by indirect methods. For example, in \citep{Sal04}
$Y_p$ was bounded from studies of Galactic Globular Clusters
(GGC). The value they found, $Y_p\lsim
Y_{GGC}=0.250\pm(0.006)_{\rm stat}\pm(0.019)_{\rm sys}$, is
consistent with the above estimate. Finally, CMB anisotropies are
sensitive to the reionization history, and thus to the fraction of
baryons in the form of $^4$He. Present data only allow a marginal
detection of a non-zero $Y_p$, and even with PLANCK the error bars
from CMB will be larger than the present systematic spread of the
astrophysical determinations
\citep{Tro03,Hue04,Ich06,Ich07,Ham08}. On the other hand, imposing
a self-consistent BBN prior on $Y_p$ would improve the diagnostic
power of CMB data on other parameters, thus representing another
nice synergy of CMB and BBN, besides the concordance test provided
by $\eta$.

\subsection{Lithium-7}
\label{sec:lithium7}

Lithium's two stable isotopes, $\lisix$ and $\li7$, continue to
puzzle astrophysicists and cosmologists who try to reconcile their
primordial abundance as inferred from observations with the BBN
predictions. From the astrophysical point of view the questions
mainly concern the observation of lithium in cold interstellar gas
and in all type of stars in which lithium lines are either
detected or potentially detectable \citep{Asp06}.

A chance to link primordial $\li7$  with the BBN abundance was first
proposed by Spite \& Spite (1982) \citep{Spi82}, who showed that the
lithium abundance in the warmest metal-poor dwarfs was independent of
metallicity for [Fe/H]$< -1.5$. The constant lithium abundance defining
what is commonly called {\it the Spite plateau} suggested that this may be
the lithium abundance in pre-Galactic gas provided by the BBN. The very
metal-poor stars in the halo of the Galaxy or in similarly metal-poor GGC
thus represent ideal targets for probing the primordial abundance of
lithium. Even though lithium is easily destroyed in the hot interiors of
stars, theoretical expectations supported by observational data suggest
that although lithium may have been depleted in many stars, the overall
trend is that its galactic abundance has increased with time
\citep{Ste07}.

There is quite a long tale of $\li7$ determinations appearing in
the literature, starting from the Spite \& Spite (1982) value of
$[\li7$/H]$= 2.05 \pm 0.15$ (by definition $[\li7$/H]$\equiv 12 +
\log_{10}(\li7/$H)). For the sake of brevity we will restrict our
analysis roughly to the determinations of the last decade. The
implicit assumption is that, hopefully, the more recent papers
have reached a better understanding of the systematics involved in
the inference of the primordial $\li7$ abundance.

In Ref.s \citep{Rya99,Rya00} a study of a set of very metal-poor
stars showed a very small intrinsic dispersion in the $\li7$
abundance determinations. Moreover, the authors found evidence for
a decreasing trend in the $\li7$ abundance toward lower
metallicity indicating that the primordial abundance of $\li7$ can
be inferred only after allowing for nucleosynthesis processes that
must have been at work in the early stages of the Galaxy. The
primordial $\li7$ abundance reported is $ [\li7/$H]$= 2.09
\pm^{+0.19}_{-0.13}$ ($\li7/$H$= \left(1.23^{+0.68}_{-0.32}\right)
\times 10^{-10}$). Different studies of halo and GGC stars
provided higher lithium plateau abundance $[\li7/$H]$= 2.24 \pm
0.01$ \citep{Bon97a,Bon97b}. A similar analysis is contained in
Ref. \citep{Bon02}, where high resolution, high signal-to-noise
ratio spectra of 12 turn-off stars in the metal-poor globular
cluster NGC 6397 were used. The author conclude that, within the
errors, they all have the same lithium abundance $[\li7/$H]$= 2.34
\pm 0.06$.

In Ref. \citep{Mel04}, a study of $\li7$ abundance in 62 halo dwarfs was
performed by using accurate equivalent widths and a temperature scale from
an improved infrared flux method. For 41 plateau stars (those with
$T_{eff} > 6000\,{\rm K}$) the $\li7$ abundance is found to be independent
of temperature and metallicity, with a star-to-star scatter of only 0.06
dex over a broad range of temperatures ($6000\, {\rm K} < T_{eff} <
6800\,{\rm K}$) and metallicities ($-3.4 <$[Fe/H]$< -1$). Thus they report
a mean $\li7$ plateau abundance of $[\li7/$H]$= 2.37 \pm 0.05$. In Ref.
\citep{Cha05} the authors underwent a very detailed reanalysis of
available observations; by means of a careful treatment of systematic
uncertainties and of the error budget, they find $[\li7$/H]$= 2.21 \pm
0.09$ for their full sample and $[\li7$/H]$= 2.18 \pm 0.07$ for an
analysis restricted to unevolved (dwarf) stars only. They also argued that
no convincing/conclusive evidence for a correlation between $\li7$ and
metallicity can be claimed at present. More recently, the authors of Ref.
\citep{Asp06} have studied a set of 24 very high quality spectra
metal-poor halo dwarfs and subgiants, acquired with ESO's VLT/UVES. The
derived one-dimensional, non-Local Thermodynamical Equilibrium (non-LTE)
$\li7$ abundances from the Li$_I$ 670.8 nm line reveal a pronounced
dependence on metallicity but with negligible scatter around this trend.
The estimated primordial $\li7$ abundance is $\li7$/H $\in (1.1 - 1.5)
\times 10^{-10}$ ($[\li7$/H]$= 2.095 \pm 0.055$). Recently Ref.
\citep{Kor06} has reported the spectroscopic observations of stars in the
metal poor globular cluster NGC6397 that reveal trends of atmospheric
abundance with evolutionary stage for various elements. These
element-specific trends are reproduced by stellar-evolution models with
diffusion and turbulent mixing. They compare their observations of lithium
and iron to models of stellar diffusion, finding evidence that both
lithium and iron have settled out of the atmospheres of these old stars.
Applying their stellar models to the data they infer for the unevolved
abundances, [Fe/H] $= –2.1$ and $[\li7/\textrm{H}] = 2.54 \pm 0.10$.

The list of the last ten years estimates for $\li7$ abundance is then the
following:
\begin{itemize}
\item[i)] $[\li7/{\rm H}] = 2.24 \pm 0.01$ \citep{Bon97a,Bon97b};
\item[ii)] $ [\li7/{\rm H}] = 2.09^{+0.19}_{-0.13}$
\citep{Rya99,Rya00};
\item[iii)] $[\li7/{\rm H}] = 2.34 \pm 0.06$ \citep{Bon02};
\item[iv)] $[\li7/{\rm H}] = 2.37 \pm 0.05$ \citep{Mel04};
\item[v)] $[\li7/{\rm H}] = 2.21 \pm 0.09$ \citep{Cha05};
\item[vi)] $[\li7/{\rm H}] = 2.095 \pm 0.055$ \citep{Asp06};
\item[vii)] $[\li7/{\rm H}] = 2.54 \pm 0.10$ \citep{Kor06}.
\end{itemize}

All but (marginally) the latter value are inconsistent with
standard BBN predictions for the preferred range of $\eta$ singled
out by CMB data, which fits remarkably well deuterium abundance.
It is unclear how to combine the different determinations in a
single estimate, or if the value measured is truly indicative of a
primordial yield. A conservative approach (similar to the one used
for $\He4$) is to quote the simple (un-weighted) average and
half-width of the above distribution of data as best estimate of
the average and ``systematic'' error on $\li7/$H, obtaining
\be
\left[ \frac{\li7}{\rm H}\right] = 2.27 \pm
0.23\Longrightarrow \left( \frac{\li7}{\rm H}\right) =
\left(1.86^{+1.30}_{-1.10} \right) \times 10^{-10}\,.
\label{li7ave2}
\ee
Note that the statistical error is much smaller (of the order of
0.01), but we will not need it since, due to its uncertain status
as tracer of the primordial value, $\li7$ is typically excluded in
``conservative'' BBN statistical analyses (or rather invoked to
support particular non-standard BBN scenarios).

\subsection{Lithium-6 and ``The lithium problems"}
\label{sec:lithium6}
It is clear from the above discussion and from the substantial
disagreement of almost all of the $\li7$ observations with the
standard BBN predicted value (by about 0.4 dex, assuming the
central value of the quoted average) that some piece of
(astro)physics is missing. For a detailed discussion of possible
causes we address the reader to the excellent review given in Ref.
\citep{Asp06} (see also \citep{Lam05a}). Here we want to remark
that: i) a $\sim 1.5\div 2$ lower value of $\eta$ at the BBN time
with respect to the best fit deduced from CMB data is excluded by
the agreement between deuterium observations and CMB value of
$\eta$, but also by the inferred upper limit of the primordial
$\he3$ abundance; ii) underestimated errors in the adopted nuclear
reaction rates are now excluded: the laboratory measurements of
the crucial $\he3(\alpha,\gamma)\bers$ cross-section
\citep{Nar04,Bem06,Con07,Bro07}, its inferred rate from solar
neutrino data \citep{Cyb03a}, and the measurement of the proposed
alternative channel for $\bers$ destruction $\bers(d,p)2\,\alpha$
\citep{Ang05} all point to the conclusion that nuclear
uncertainties cannot explain the discrepancy between observed and
predicted primordial $\li7$ abundances; iii) systematic errors in
the abundance analysis, although in principle still possible, seem
very unlikely. The introduction of 3D model atmospheres, even
accounting for non-LTE, has not resulted in a significant upward
revision of the lithium abundance obtained from more primitive 1D
atmospheres.

The most likely causes of the ``problem" are thus: a) either some
modification to the BBN scenario; b) or, perhaps more likely, that
the lithium abundance of very metal-poor stars is not the one of
the primordial gas. We will illustrate later some scenarios of the
type a). Here, however, we want to point out that explanations of
the type b) are probably not trivial and might involve both
reprocessing {\it in situ} (the observed stars) and earlier
lithium synthesis/depletion in the pre-galactic, young universe
environment. Indeed, even assuming that some diffusion and
turbulent mixing mechanism like the one pointed out in
\citep{Kor06} can explain the $\li7$ problem, still an issue
remains with $\Li6$. The presence of the fragile $\Li6$ isotope,
which is produced during the BBN at the level of $\Li6/{\rm H}
\sim 10^{-15} - 10^{-14}$, has been recently confirmed in a few
metal-poor halo stars, with some hint of a plateau vs.
metallicity\footnote{However, taking into account predictions for
$\Li6$ destruction during the pre-main sequence evolution tilts
the plateau suggesting a $\Li6$ increase with metallicity.
Basically, fairly uncertain stellar pre-main-sequence destruction
of $\Li6$ could be responsible for an apparent plateau
\citep{Ric02,Ric05,Asp06}.} with abundance as high as $\Li6/{\rm
H} \sim 6 \times 10^{-12}$ \citep{Asp06}. These data (at least
partially) confirm the first observations reported in literature
already fifteen years ago \citep{Smi93b,Smi98,Hob97,Cay99,Nis00}.
In particular, in Ref. \citep{Asp06} $\Li6$ is detected in 9 of 24
analyzed stars at the $> 2 \sigma$ significance level. However, it
is worth stressing that these observations have been questioned in
Ref. \citep{Cay07,Cay08}. According to these papers the convective
asymmetry generates an excess absorption in the red wing of the
$\li7$ absorption feature that could mimic the presence of $\Li6$
at a level comparable with published values. This would mean that
only an upper limit on $\Li6/\li7$ can be derived at present.

Both $\li7$, $\Li6$ can be produced by fusion
($\alpha+\alpha\to$Li) and spallation (p+CNO $\to $ LiBeB)
reactions by ordinary cosmic ray primaries impinging on nucleons
and nuclei in the intergalactic medium (see e.g. \citep{Van00}).
Additionally, observations are performed not in ``inert" gas
clouds, but in stellar atmospheres, where thermonuclear burning
depletion (of particularly fragile nuclei) is a crucial effect. It
follows that primordial production mechanisms and later
astrophysical effects might be competing. Viable astrophysical
candidates for the acceleration of cosmic rays, able to account
for the $\Li6$ observed at low metallicity include: the massive
black hole in the Galactic center \citep{Pra05}, radio-loud AGNs
(one of which could have been present in our Galaxy in the past)
\citep{Nat06}, and PopIII stars \citep{Rol05,Rol06} which may also
explain the depletion of $\li7$ \citep{Pia06}, but have been
recently challenged on the light of further constraints adopted in
the model of \citep{Evo08}. The shocks developed during structure
formation \citep{Ino03}, which however conflict with astrophysical
constraints \citep{Pro07}, are not powerful enough, since in the
early times of Galaxy formation, the masses of the assembling dark
haloes were still quite small and the corresponding virial
velocities insufficient \citep{Pra05}.

Perhaps, the most convincing explanation proposed until now is
that $\Li6$ may be produced {\it in situ} from stellar flares
within the first billion years of the star's life \citep{Tat07}.
In particular, the anomalously high $\he3/\He4$ ratio found in
solar flares ($\sim$0.5) and the kinematically much more favorable
process $\He4(\he3, p)\Li6$ compared to $\alpha\,\alpha\to \li7$
fusion reactions provide a physical mechanism for producing large
quantities of $\Li6$ without overproducing $\li7$. The main issue
with this or other scenarios might be the difficulty to test them.

A great help in solving this issue might come from detecting
lithium in a different environment. In Ref. \citep{Pro04}, it was
proposed to independently test the pre-Galactic Li abundance by
looking at high velocity gas clouds falling onto our Galaxy, with
metallicities as low as 10\% of the solar one. If these
low-metallicity clouds have a mostly pre-Galactic composition,
with a small contamination from the Galaxy, they might allow
probing the lithium abundance at least free of the possibility of
thermonuclear depletion {\it in situ}. Another proposal to detect
the cosmological recombination of lithium via its effect on the
microwave background anisotropies \citep{Sta02} has been proved
not to be viable \citep{Swi05}.

We will not use $\Li6$ to derive constraints in this review. However,
using the observed $\Li6$ as an upper limit to its primordial value turns
out to be a powerful constraint in some regions of parameter space for
exotic models. While reporting them in the following, we warn the reader
that their robustness relies on the assumption that no destruction or
major reprocessing of the $\li7$ and $\Li6$ observed in the halo stars has
happened, which might be overly optimistic.

\section{Standard BBN theoretical predictions versus data}
\label{sec:BBNanalysis}

The goal of a theoretical analysis of BBN is to obtain a reliable
estimate of the model parameters, once the experimental data on
primordial abundances are known. In this Section we will consider
only the case of the standard BBN, where the only two free
parameters are the value of the baryon energy density parameter
$\Omega_B h^2$ (or equivalently the baryon to photon number
density, $\eta$) and possibly, a non-standard value for the
relativistic energy content during the BBN. The latter, after
$e^\pm$ annihilation can be parameterized in terms of the
effective number of neutrinos we have recalled in Section
\ref{sec:neutrinodecoupling},
\be \rho_R = \left( 1 + \frac{7}{8} \left( \frac{4}{11}
\right)^{4/3}  \neff \right) \rho_\gamma \pp \label{defneff}
\ee
Similar analyses have been recently presented by various groups,
which might be slightly different depending on the adopted values
of $Y_p$ and/or $^2$H/H experimental determination, see e.g.
\citep{Lis99,Bur99b,Bar03a,Cuo04,Cyb03c,Cyb04,Cyb05,Han02,Man07,Sim08b}.

\begin{figure}[t]
\begin{center}
\epsfig{file=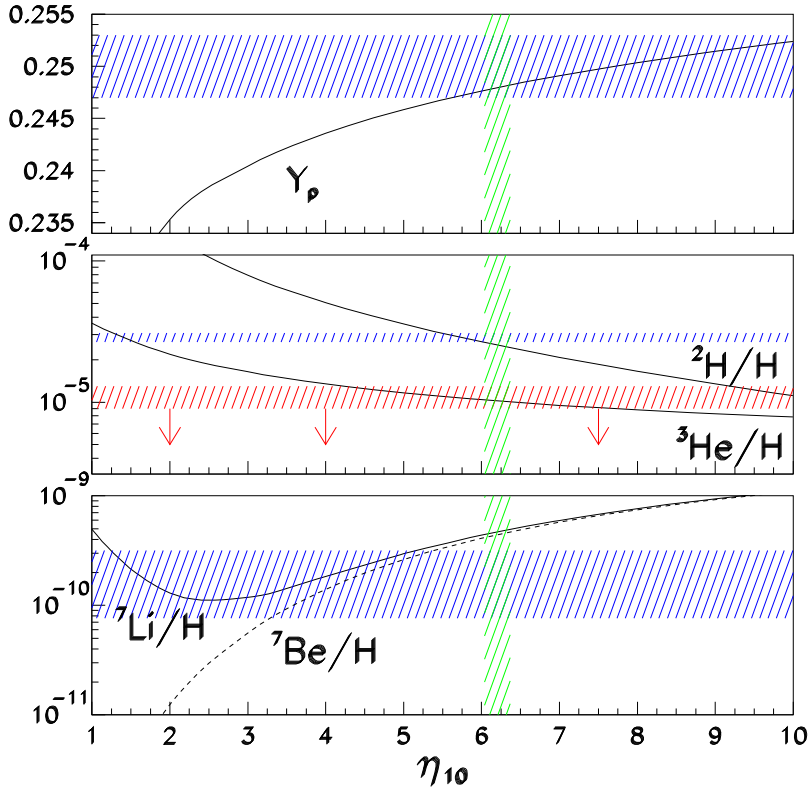,width=0.8\textwidth}
\end{center}
\caption{Values of the primordial abundances as a function of
$\eta_{10}$, calculated for $\Delta \neff = 0$. The hatched blue
bands represent the experimental determination with $1-\sigma$
statistical errors on $Y_p$, $^2H$, and $^7$Li, while the red band
is the upper bound obtained in Ref. \citep{Ban02}. Note that for
high value of $\eta_{10}$  all $^7$Li comes from $^7$Be
radioactive decay via electron capture. The vertical green band
represents the WMAP 5-year result $\Omega_B h^2 = 0.02273 \pm
0.00062$ \citep{Dun08}.} \label{f:abund_parth}
\end{figure}

In the minimal scenario the parameters reduces to the baryon
density only, since $\Delta \neff$ is assumed to vanish. Figure
\ref{f:abund_parth} shows the dependence on $\eta_{10}$ of the
final value of the primordial yields, calculated using
{\texttt{PArthENoPE}, along with the experimental values of the
abundances and their corresponding uncertainties, as discussed in
the previous Section.

To get confidence intervals for $\eta$, one constructs a
likelihood function
\be {\mathcal{L}}(\eta)\propto \exp\left( -\chi^2(\eta)/2\right)
\vv \ee with \be \chi^2(\eta) = \sum_{ij} [ X_i(\eta) - X_i^{obs}
] W_{ij}(\eta) [ X_j(\eta) - X_j^{obs} ] \pp \ee
The proportionality constant can be obtained by requiring
normalization to unity, and $W_{ij}(\eta)$ denotes the inverse
covariance matrix, \be W_{ij}(\eta) = [ \sigma_{ij}^2 +
\sigma_{i,exp}^2 \delta_{ij} + \sigma_{ij,other}^2 ]^{-1} \vv \ee
where $\sigma_{ij}$ and $\sigma_{i,exp}$ represent the nuclear
rate uncertainties and experimental uncertainties of nuclide
abundance $X_i$, respectively (we use the nuclear rate
uncertainties as in Ref. \citep{Ser04b}), while by
$\sigma_{ij,other}^2$ we denote the propagated squared error
matrix due to all other input parameter uncertainties ($\tau_n$,
$G_{\rm N}$, etc.). We use the following values for the
experimental yields of $^2$H and $^4$He (see previous Section):
\be ^2\textrm{H}/\textrm{H}=\left(2.87^{+0.22}_{-0.21}
\right)\times 10^{-5},~~~~ \yp = 0.250 \pm 0.003\pp \ee
\begin{figure}[t]
\begin{center}
\epsfig{file=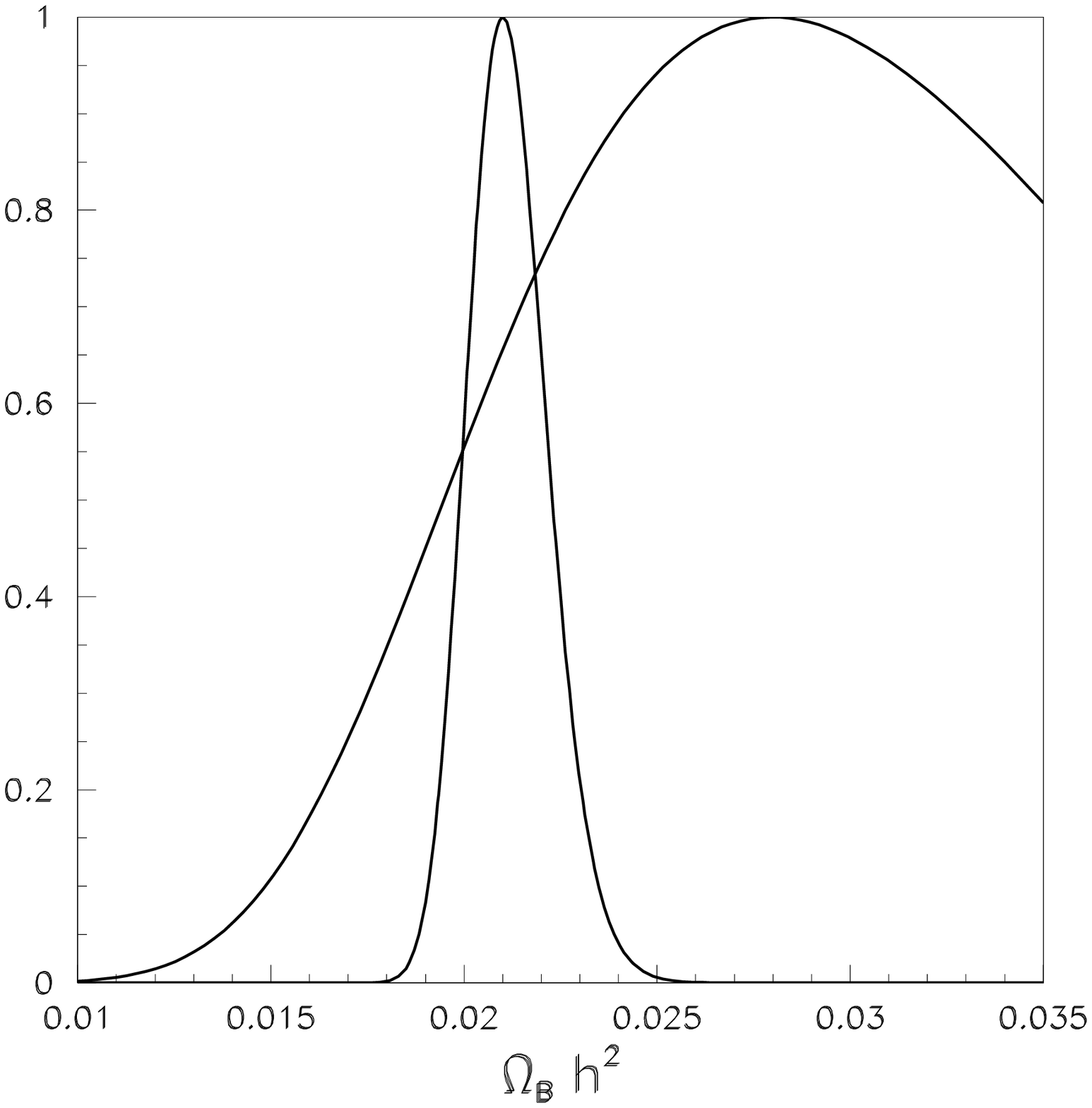,width=0.6\textwidth}
\end{center}
\caption{Likelihood functions for $^2$H/H (narrow) and $Y_p$
(broad).} \label{f:like_uni}
\end{figure}
\begin{table}[b]
\begin{center}
\caption{The theoretical values of the nuclear abundances for some
value of $\Omega_B h^2$.} \label{table:sbbn}
\begin{tabular}{cccccc}
\hline\hline $\Omega_B h^2$        & 0.017  & 0.019  & 0.021  &
0.023 & 0.028 \\ \hline\hline $Y_p$ & 0.245 &
0.246 & 0.247 & 0.248 & 0.250 \\
\hline $^2$H/H  $\times 10^{5}$   & 4.00 & 3.36 & 2.87   & 2.48 &
1.79
\\ \hline $^3$He/H $\times 10^{5}$ & 1.22 & 1.14 & 1.07   & 1.01
& 0.903\\ \hline
$^7$Li/H $\times 10^{10}$ & 2.53 & 3.22 & 3.99   & 4.83 & 7.08  \\
\hline $^7$Be/H
$\times 10^{10}$ & 2.15   & 2.89   & 3.69   & 4.56 & 6.88  \\
\hline $^6$Li/H
$\times 10^{14}$ & 1.72 & 1.45 & 1.25   & 1.08  & 0.791 \\
\hline\hline
\end{tabular}
\end{center}
\end{table}

We first consider $^2$H abundance alone, to illustrate the role of
deuterium as an excellent baryometer. In this case the best fit
values found are $\Omega_B h^2 = 0.021\pm 0.001$ ($\eta_{10} = 5.7
\pm 0.3$) at 68\% C.L. , and $\Omega_B h^2 = 0.021 \pm 0.002$  at
95\% C.L. . A similar analysis can be performed using $^4$He. In
this case we get $\Omega_B h^2 = 0.028^{+0.011}_{-0.007}$
($\eta_{10} = 7.6^{+3.0}_{-1.9}$)\footnote{Note that $\eta_{10}$
reported in this section is conventionally defined, according to
\pth, as $\eta_{10} \equiv 273.49 \, \Omega_B h^2$ which is
slightly different from the definition of Eq. (\ref{eta10def}),
but simply connected to it.} at 68\% C.L. , and $\Omega_B h^2 =
0.028^{+0.024}_{-0.012}$ at 95\% C.L. . Figure \ref{f:like_uni}
shows, as from our discussion in the previous Section, that the
determination of $\Omega_B h^2$ is mainly dominated by deuterium.
In any case, the result is compatible at 2-$\sigma$ with the WMAP
5-year result $\Omega_B h^2 = 0.02273 \pm 0.00062$ \citep{Dun08}.
The slight disagreement might have some impact on the
determination from CMB anisotropies of the primordial scalar
perturbation spectral index $n_s$, as noticed in \citep{Pet08},
where the BBN determination of $\Omega_B h^2$ from deuterium is
used as a prior in the analysis of the five year data of WMAP.

In Table \ref{table:sbbn} we report the values of relevant
abundances for some different baryon densities, evaluated using
\texttt{PArthENoPE} \citep{Par08}. Notice the very low prediction
for $^6$Li (see discussion in Section \ref{sec:obsabund}) and
that, for these values of baryon density, almost all $^7$Li is
produced by $^7$Be via its eventual electron capture process.

\begin{figure}[t]
\begin{center}
\epsfig{file=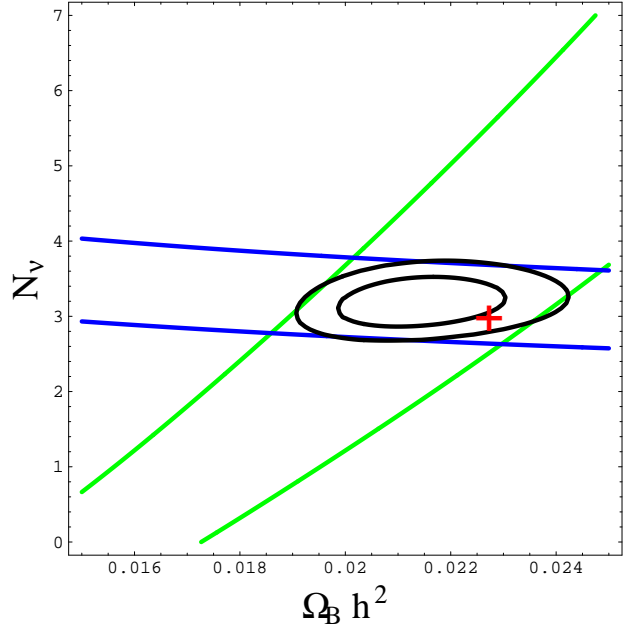,width=0.6\textwidth}
\end{center}
\caption{Contours at 68 and 95 \% C.L. of the total likelihood
function for deuterium and $^4$He in the plane ($\Omega_B
h^2$,$\neff$). The bands show the 95\% C.L. regions from deuterium
(almost vertical) and Helium-4 (horizontal). The red cross
corresponds to the standard $\neff$ and $\Omega_B h^2 = 0.02273$ as indicated
by WMAP 5-year results.}\label{f:omega_nnu_contour}
\end{figure}

If one relaxes the hypothesis of a standard number of relativistic
degrees of freedom, it is possible to obtain bounds on the largest
(or smallest) amount of radiation present at the BBN epoch, in the
form of decoupled relativistic particles, or non-standard features
of active neutrinos (but see our previous discussion in Section
\ref{sec:neutrinodecoupling}). Figure \ref{f:omega_nnu_contour}
displays the contour plots 68\% and 95\% C.L. of the total
likelihood function, in the plane ($\Omega_B h^2$,$\neff$). After
marginalization one gets $\Omega_B h^2 = 0.021\pm 0.001$ and
$\neff = 3.18^{+0.22}_{-0.21}$ at 68\% C.L., and $\Omega_B h^2 =
0.021\pm 0.002$ and $\neff = 3.18^{+0.44}_{-0.41}$ at 95\% C.L. .
Hence the global analysis results to be compatible with $\neff =
3.046$ and $\Omega_B h^2 = 0.02273$ found by WMAP at 1-$\sigma$
level (see Figure \ref{f:omega_nnu_contour}).

\section{BBN and Neutrino physics}\label{sec:BBN_nuphys}

We have already stressed how large is the impact of neutrino
physics and neutrino properties on BBN. In fact, the discovery of
neutrino masses via oscillations, combined with the stringent
bounds on the effective electron neutrino mass via tritium decay
experiments, has had a profound impact on the phenomenology of
active neutrinos in the early universe. At the moment we know
that: (at least two) neutrinos are massive; {\it all} the masses
are small ($m_\nu\lsim$eV, possibly much smaller); individual
lepton numbers are violated (with mixing angles much larger than
in the quark sector), although it is not known if the overall
lepton number, ${\sf L}$, is conserved. We do not know yet if the
neutrino mass term in the Lagrangian is of the Dirac
($\sim[{\mathcal M}\,\bnu_L\nu_R+h.c.]$, flavor indexes omitted)
or Majorana ($\sim[{\mathcal M}\,\bnu_L^c\nu_R+h.c.]$, where
$\nu_L^c$ is the charge conjugate field) type, but in either case
new physics is required. In the Dirac case, one is forced to
introduce the yet undetected right-handed neutrino fields,
$\nu_R$. In the Majorana case, one assumes (differently from the
SMPP) that the lepton number ${\sf L}$ is violated and introduces
a Majorana mass operator, which is allowed for neutrinos, being
the only neutral fermions in the SMPP. The important news for BBN
are that even the incomplete knowledge of the mass matrix
${\mathcal M}$ that we have at present is enough to conclude that
the phenomenology of active neutrinos in the BBN is greatly
simplified. A plethora of cases once popular in the literature are
now excluded. Among the ones which were of remarkable interest
only a decade ago we can mention: (i) a lower-than-three effective
number of neutrinos due to a ``massive $\nut$'' (improper, not
being $\nut$ a mass eigenstate) ; (ii) a decaying ``$\nut$'';
(iii) the thermalization of right-handed neutrinos (for the Dirac
mass case), which is inhibited by the smallness of the neutrino
masses by which they are coupled to the active states. We do not
treat these issues further and address the interested reader to
these historical topics to the review \citep{Dol02b}. In the
following, we focus on the bounds on electromagnetic interactions
of neutrinos in Section \ref{nuem}, while Section \ref{nonemnu}
treats other exotic interactions. The topic of neutrino asymmetry
is briefly reviewed in Section \ref{nuasymm}, while Section
\ref{sterile} treats the impact on BBN of sterile neutrino states.

\subsection{Bounds on electromagnetic interactions of neutrinos}
\label{nuem}
Dropping flavor indexes, the most general structure of effective neutrino
electromagnetic interactions is
\be
\Lag_{\rm int}=-e_\nu\,\bar\nu\,\gamma_\mu\,\nu\,
A^\mu-a_\nu\,\bar\nu\,\gamma_\mu\,\gamma_5\,\nu\,
\partial_\lambda F^{\mu\lambda}-\frac{1}{2}\bar{\nu}\,\sigma_{\alpha\beta}\,(\mu+\epsilon\,
\gamma_5)\,\nu F^{\alpha\beta}\label{MagMom}
\ee
where $F^{\alpha\beta}$ is the electromagnetic field tensor,
$\sigma_{\alpha\beta}=[\gamma_\alpha,\gamma_\beta]$, and the form factors
$\{e_\nu,a_\nu,\mu,\epsilon\}$, which are functions of the transferred
squared momentum $q^2$, in the limit $q^2\to 0$ correspond to the electric
charge, anapole moment, magnetic and electric dipole moment, respectively.

In principle, Dirac neutrinos may have a very small electric
charge $e_\nu$.  BBN bounds may be derived by requiring both that
right-handed partners are not populated and that neutrinos are not
kept in equilibrium too long after the weak freeze-out, which
would alter the photon-neutrino temperature relation. However, for
the range of masses presently allowed, the BBN bounds are never
competitive with other astrophysical or laboratory constraints, as
for instance the red giant bound of $e_\nu\lsim 2\times 10^{-14}$
\citep{Raf99}. Actually, the indirect bound coming from the
neutrality of matter is stronger ($e_\nu\lsim10^{-21}$,~
\citep{Foo90}; see also \citep{Dav00} and reference therein for
details), so we ignore in the following a possible neutrino
charge.

The possibility of a neutrino charge radius\footnote{In complete
analogy, one can define an anapole radius $\langle r_a^2\rangle$
from $a_\nu$. For Majorana neutrinos, symmetries require some of
the e.m. couplings to vanish. Since the astrophysical/cosmological
bounds do not typically distinguish between $\langle r^2\rangle$
and $\langle r_a^2\rangle$, or electric and dipole magnetic
moments, we shall quote bounds on $\langle r^2\rangle$ and $\mu$
in this loose sense.} (which can be negative), \be \langle
r^2\rangle =\frac{6}{e}\left(\frac{\partial e_\nu(q^2)}{\partial
q^2}\right)_{q^2=0}\,, \ee has been considered in the literature.
After a long debate, it has been finally established that $
\langle r^2\rangle$ is a well-defined (gauge-independent) quantity
\citep{Ber00,Ber02,Ber04a}. The Standard Model expectations are in
the range $\langle r^2\rangle\simeq 1\div4\,$nb. Differently from
the case of the magnetic moment, the charge radius does not couple
neutrinos to on-shell photons, so stellar cooling arguments are
not very sensitive to $\langle r^2\rangle$. Yet, for Dirac
neutrinos the channel $e^{-}e^{+}\to \nu_R\bnu_R$ is still
effective, provided that new physics (NP) violates the cancelation
between vector and axial contribution that otherwise applies in
the SMPP. If this cancelation does not take place, the
corresponding bounds from SN 1987 A \citep{Gri89} and
nucleosynthesis \citep{Gri87} read \be |\langle r^2\rangle|^{\rm
NP}\lsim 2\, (7)\,{\rm nb,\: from\:SN1987A\: (BBN)} \pp \ee For
Majorana neutrinos, the previous bounds do not apply. In this
case, however, even in the $\nu_\tau$-sector where BBN may have a
sensitivity comparable to or better than laboratory experiments,
even a change of one order of magnitude above the SMPP level in
the channel $e^{-}e^{+}\to \nu_\tau\bnu_\tau$ due to new physics
would only bring changes at the 0.1\% level in $Y_p$, so that only
laboratory bounds are meaningful. For a further discussion of this
point, see \citep{Hir03}.

Another consequence of the existence of neutrino masses is that a neutrino
magnetic moment is naturally present (and it is usually expressed in terms
of Bohr magnetons, $\mu_B$), although for Majorana particles only
off-diagonal elements in flavor space are non-vanishing. There are two
possible processes of interest for BBN: (i) the thermalization of Dirac
neutrinos via e.g. the process $\nu_L\,e\to \nu_R\,e$; (ii) a radiative
decay of the kind $\nu_L^i\to \nu_L^j\,\gamma$, the latter being possible
also for Majorana neutrinos. In absence of primordial magnetic fields, the
BBN bound on the diagonal elements coming from the thermalization of
right-handed neutrinos is not as restrictive as the one coming from red
giant cooling argument (from plasmon decay $\gamma^*\to \nu\bnu$,
$\mu\lsim 3\times 10^{-12}\,\mu_B$ \citep{Raf99}). The radiative decay
rate for a transition $i\to j$ is
\be
\Gamma^\gamma_{ij}=\frac{|\mu_{ij}|^2+|\epsilon_{ij}|^2}{8\pi}\left(\frac{
m_i^2-m_j^2}{m_i}\right)^3\simeq 5.3\, {\rm s}^{-1}\, \left(
\frac{\mu}{\mu_B} \right)^2\left(\frac{ m_i^2-m_j^2}{m_i\,\times 1\,{\rm
eV}}\right)^3\, ,
\label{kappa}
\ee
and typical bounds lead to a lifetime definitively too long to affect the
BBN cosmology. In any case, the very cosmological bounds on the neutrino
lifetime in \citep{Mir07} exclude any effect of the radiative neutrino
decay at the BBN epoch. However, in presence of very strong primordial
magnetic fields, the BBN bound via spin-precession may be as stringent as
$\mu\lsim 10^{-20}\,\mu_B$ \citep{Enq95}. Although very model-dependent
(see \citep{Enq95}, in particular Eq.s (50,51)), this is to our knowledge
the only bound probing the level of intensity expected for the dipole
moment in the SMPP enlarged with a Dirac neutrino mass term, which is
$\mu=3\,e\,G_F\,m_\nu/(8\sqrt{2}\pi^2)=32\times 10^{-20}\,(m_\nu/{\rm
eV})\mu_B$.

\subsection{Bounds on other exotic interactions of neutrinos}
\label{nonemnu}
Besides anomalous electromagnetic interactions, neutrinos might
undergo non-standard-interactions (NSI) with electrons. Ref.
\citep{Man06} considered low-energy four-fermions interactions of
the kind \be {\mathcal L}^{\alpha\beta}_{\rm NSI}=-2\sqrt{2}G_F \,
[\epsilon^L_{\alpha\beta} \, (\bar{\nu}_L^\alpha \gamma^\mu
\nu_L^\beta)(\bar{e}_L\gamma_\mu e_L)+\epsilon^R_{\alpha\beta}\,
(\bar{\nu}_L^\alpha \gamma^\mu \nu_L^\beta)(\bar{e}_R\gamma_\mu
e_R)]\vv \label{NSI} \ee with the NSI parameters
$\epsilon^L_{\alpha\beta}\, \epsilon^R_{\alpha\beta}$ constrained
by laboratory measurements to be at most of ${\mathcal O}(1)$. It
was found that, for NSI parameters within the ranges allowed by
present laboratory data, non-standard neutrino-electron
interactions do not essentially modify the density of relic
neutrinos nor the bounds on neutrino properties from cosmological
observables. Qualitatively, this depends on the fact that a large
modification of the neutrino spectra would only be achieved if the
decoupling temperature were brought below the electron mass.  The
presence of neutrino-electron NSI within laboratory bounds may
enhance the entropy transfer from electron-positron pairs into
neutrinos, up to a value of $\neff=3.12$ (and $\Delta Y_p\simeq
6\times 10^{-4}$), which are almost three times the corrections
due to non-thermal distortions that appear for standard weak
interactions, but still probably too small to be detectable in the
near future, even for PLANCK.

Another scenario is the one where neutrinos couple to a scalar or
pseudoscalar particle with a Yukawa-type interaction of the kind
\be {\mathcal L}_{\nu\phi}= \frac{1}{2} \partial_\mu \phi
\partial^\mu \phi - \frac{1}{2} m_\phi^2 \phi^2 - \phi
\sum_{\ell\jmath} \overline{\nu}_L^\ell
\lambda_{\ell\jmath}\nu_L^\jmath \vv \label{lagr} \ee where
$\ell,\jmath$ are flavor indexes and, in case of pseudoscalar
coupling, $\lambda_{\ell\jmath}\to\gamma_5\lambda_{\ell\jmath}$.
We shall denote these couplings simply as $\lambda$ if we ignore
flavor effects and only refer to constraints within a factor of
${\mathcal O}(1)$. A famous case of this kind is the Majoron model
\citep{Chi81,Gel81,Geo81} where the Majoron $\phi$ is the
Goldstone boson associated to the breaking of the lepton number
symmetry (and thus $m_\phi\to 0$).  Although laboratory bounds
rule out the original model as an explanation of the small
neutrino masses, still it represents a prototype of  ``secret
neutrino interactions", where neutrinos interact with a sector
precluded to other standard model particles, and may thus have
stronger interactions among themselves at low energies, than
predicted by the SMPP. In the early universe, a large enough
$\lambda$ would allow to populate thermally the species $\phi$.
For $m_\phi\ll 1\,$MeV,  it was found in \citep{Cha94} that
$\lambda\lsim 10^{-5}$, if one considers one additional boson
($\Delta \neff=3/7$) to be incompatible with the observations. If
this is instead considered viable, the same process
$\bar\nu\nu\leftrightarrow\phi\phi$ may be responsible for a
``neutrinoless universe" well after the BBN epoch, provided that
$m_\phi\ll m_\nu$ \citep{Bea04}.

Relatively less attention has been paid to the case where the
associated (pseudo) boson is massive (although a particle of this
kind might have other ``cosmological virtues'', as providing a
warm dark matter candidate, see e.g. \citep{Lat07}). In
\citep{Cuo05} the authors assumed MeV-scale $\phi$ particles
produced at  early epochs via additional couplings with other SMPP
particles and later decaying into neutrinos in out-of-equilibrium
conditions (with a rate $\Gamma(\phi \rightarrow
\overline{\nu}\nu) = 3\lambda^2 m_\phi/(8 \pi)$); when this
happens before the photon last scattering epoch, the produced
neutrino burst directly influences the CMB anisotropy spectrum, as
well as the late LSS formation. They show that current
cosmological observations of light element abundances, Cosmic
Microwave Background (CMB) anisotropies, and Large Scale
Structures (LSS) are compatible with very large deviations from
the standard picture. They also calculate the bounds on
non-thermal distortions which can be expected from future
observations, finding that the present situation is likely to
persist with future CMB and LSS data alone. On the other hand, the
degeneracy affecting CMB and LSS data could be removed by
additional constraints from primordial nucleosynthesis or
independent neutrino mass scale measurements.

In \citep{Han04},  the particle $\phi$ was considered to be the
inflaton, only coupled to neutrinos and with a mass
$m_\phi>>1\,$MeV, to determine via BBN and other cosmological
observation the constraint on the lowest possible reheating
temperature $T_{RH}$. In particular, the author derives
constraints from partial thermalization as well as neutrino
spectral distortions. Barring fine-tuning, the resulting bound is
$T_{RH}\gsim 4\,$MeV. A factor $\sim2$ lower bound was found in
\citep{Ich05} if no $\phi$ boson is included but oscillations are
taken into account. It is worth noticing that in scenarios  with
late-time entropy production a constraint arises anyway due to the
incomplete background neutrino thermalization, even if the
neutrinos do not couple to $\phi$ directly, as already noted
in~\cite{Kawasaki:1999na,Kawasaki:2000en}. In \citep{Ser04a},
light scalar particles annihilating into neutrinos were considered
to analyze the effect on BBN of MeV-scale dark matter particles,
invoked to explain the excess of 511 keV photons from positron
annihilation from the Galactic Center \citep{Boe04}. If such
particles only couple to neutrinos, they need to be heavier than
$\sim 1\,$MeV to be consistent with the $\yp$ constraints.  If
they have an additional coupling to $e^{+}\,e^{-}$ at the level
required to explain the Galactic Center positrons, the bounds may
be more stringent (but then depend on the details of the model).

We have seen that a Dirac mass term could in principle be responsible for
the production of right--handed neutrinos in the primordial plasma. While
this possibility is excluded by the smallness of neutrino masses, it is
still possible to populate $\nu_R$ via direct right-handed currents
mediated by $W_R$ bosons, of the kind $\bnu_R\slW_R\nu_R$ (or analogous
coupling with right--handed charged leptons). These are possible in some
extensions of the standard electro-weak model. If one assumes that the
right-handed interaction has the same form as the left-handed one but with
heavier intermediate bosons, one can obtain from BBN a lower limit on
their mass of the order of $m_{W_R}\gsim 75 m_{W}$ \citep{Ste79, Oli00,
Bar03b}, which depends however on the exact particle spectrum of the
physics beyond the SMPP up to $\sim 75\, m_W$.

\subsection{Neutrino asymmetry}
\label{nuasymm}
The origin of the most fundamental parameter in BBN, the baryon asymmetry
$\eta_B=(n_B-n_{\bar{B}})/n_\gamma$ (or simply $\eta$ at late times), is
not known. While SMPP and SMC contain all the ingredients required to
generate it dynamically from an initially symmetric universe (B, C, and CP
violating interactions, departure from thermal equilibrium) \citep{Sak67},
the amount of CP violation and the strength of the electro-weak phase
transition are insufficient to account for an asymmetry as large as
$\eta\sim 6\times 10^{-10}$. The usual theoretical attitude towards the
cosmic lepton asymmetry $\eta_L$ is that sphaleron effects before/at electroweak
symmetry breaking equilibrate the cosmic lepton and baryon asymmetries to
within a factor of order unity (the relations in the limit of
ultra-relativistic SMPP particles can be found in \citep{Har90}). If this
is the case, for all phenomenological purposes $\eta_L$ is vanishingly
small. Sphaleron effects are a crucial ingredient in most baryogenesis
scenarios \citep{Din04,Rio99}, including leptogenesis \citep{Buc05,Dav08}.
Yet, no experimental evidence for or against these effects exists, and
(even barring the---phenomenologically viable--- alternative that $\eta_B$
and its leptonic counterpart $\eta_L$ are simply ``initial cosmological
conditions'') models have been envisioned where the lepton asymmetry is
large, as for example via Affleck-Dine mechanism or Q-balls
\citep{Cas99,Mar99,McD00a,Kaw02}.  Since charge neutrality implies that
the electron density matches the proton one, we do know that a large
lepton asymmetry could only reside in the neutrino sector. This asymmetry
can be parameterized in terms of the chemical potentials of the different
flavor species, $\mu_{\nu_\ell}$, or better in terms of the degeneracy
parameter $\xi_\ell=\mu_{\nu_\ell}/T_{\nu_\ell}$ which is constant in
absence of entropy releases. For neutrinos distributed as a FD with
temperature $T_{\nu_\ell}$, the asymmetry in each flavor is given by
\be
\eta_{\nu_\ell}=\frac{n_{\nu_\ell}-n_{\bar {\nu_\ell}}}{n_\gamma} =
\frac{1}{12\zeta(3)} \left( \frac{T_{\nu_\ell}}{T_\gamma}\right)^3 \left(
\pi^2 \xi_\ell + \xi_\ell^3 \right)\pp
\ee
Without further input, the quantities $\xi_\ell$ are not determined within
the Standard Model, and should be constrained observationally. Over the
years, BBN with a lepton asymmetry has been studied by many authors and in
different scenarios
\citep{Wag67,Fre83,Kan92,Kohri:1996ke,Whi00,Esp00a,Esp00b,Esp01,Han02,Ori02,Sti02,Bar03a,Cuo04,Kne04,Ser04b,Pop08a,Pop08b,Sim08a}.
There are several effects of $\xi_\ell\neq 0$ on BBN. The most important
one (at least for relatively small $\xi_e$) is a shift of the beta
equilibrium between protons and neutrons, which is however insensitive to
$\xi_\mu$ and $\xi_\tau$.
The leading flavor-blind effect amounts to a mere modification of the
radiation density entering the Hubble expansion rate equation by the amount
\be
\Delta \neff=\sum_\ell \bigg[ \frac{30}{7} \bigg( \frac{\xi_\ell}{\pi}
\bigg)^2 + \frac{15}{7} \bigg( \frac{\xi_\ell}{\pi} \bigg)^4 \bigg]\,.
\label{eq:deltanu}
\ee
Moreover, for sufficiently large $\xi_\ell$ the neutrino decoupling
temperature is higher than in the standard case \citep{Fre83,Kan92}, so
that in principle one could get a non-standard $T_\nu(T)$ evolution.  A
non-zero $\xi_\ell$ slightly modifies the partial neutrino reheating
following the $e^{+}e^{-}$ annihilation, too \citep{Esp00b}. Both effects
are however typically negligible for the ranges of $\xi_\ell$ presently
allowed. The greater sensitivity to $\xi_e$ than $\xi_{\mu,\tau}$ made the
constraints on the latter quantities looser, allowing on the other hand a
richer phenomenology within a quasi-standard scenario.

Again, the new knowledge on neutrino mixing parameters has rescued
the simplicity of the standard cosmological scenario. A few years
ago it was realized that the measured neutrino oscillation
parameters imply that neutrinos reach approximate chemical
equilibrium before the BBN epoch. This is due to the effects of
the background medium on the evolution of the neutrino matrix
density. In presence of neutrino asymmetry the medium term in the
Hamiltonian becomes \be H_1={\rm
diag}(V_e,V_\mu,V_\tau)\pm\sqrt{2}G_F(\varrho-\bar\varrho)\vv \ee
with the $+$ sign for $\nu$, the $-$ sign for $\bar\nu$. In
particular neutrino self-interactions synchronize the neutrino
oscillations and drive all the potentials to the same value
\citep{Dol02a,Won02,Aba02}. Assuming the standard value for
$\neff$, from the $Y_p$ range follows the bound
$\xi_e=-0.008\pm0.013$\footnote{For some regions in parameter
space, the equalization may not be complete,
see~\citep{Pastor:2008ti}. Also, at least one way around the
equalization of the chemical potentials has been proposed in
\citep{Dol04b}: an hypothetical neutrino-Majoron coupling of the
order $g\sim 10^{-6}$ can suppress neutrino flavor oscillations in
the early universe, in contrast to the usual weak interaction
case. }. In Figure \ref{figure1DBBN} we show the predictions for
the primordial light-element abundances as a function of the
neutrino degeneracy parameter $\xi$, taken to be equal for all
flavors \citep{Ser05b}. The gray band is the $1\,\sigma$ predicted
range, including both the uncertainty on $\eta$ of Ref.
\citep{Spe03} and the nuclear reactions and uncertainties adopted
in Ref. \citep{Ser04b}.

\begin{figure}
\begin{center}
\includegraphics[width=0.65\textwidth]{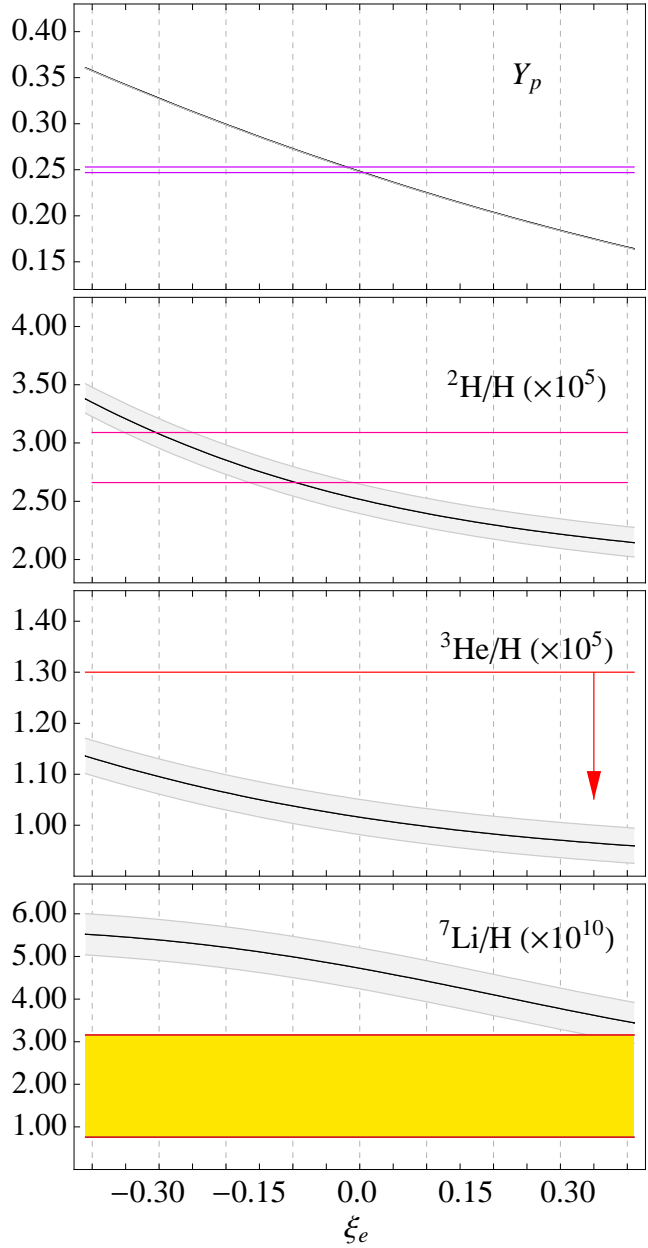}
\caption{Light-element abundances as a function of the neutrino
degeneracy parameter. The top panel shows the primordial $^4{\rm
He}$ mass fraction $Y_{\rm p}$, whereas the other panels show the
$^2{\rm H}$, $^3{\rm He}$, and $^7{\rm Li}$ number fractions
relative to hydrogen. The gray $1\,\sigma$ error bands include the
uncertainty of the WMAP determination of the baryon abundances of
Ref. \citep{Spe03} as well as the uncertainties from the nuclear
cross-sections of Ref.\citep{Ser04b}. Updated, from
\citep{Ser05b}.} \label{figure1DBBN}
\end{center}
\end{figure}
The bound relaxes by a factor of 2$\div$3 (depending on other
priors used) if additional degrees of freedom are present in the
plasma, i.e. $\neff$ is allowed to vary
\citep{Bar03a,Cuo04,Sim08a}. At present, the BBN is by far the
best cosmic ``leptometer'' available, and is virtually the only
one sensitive to the sign of $\xi_e$. Most of its sensitivity
derives however from the sensitivity of $\yp$ to the variation of
the weak $n-p$ rates, so in order to improve these bounds the
systematic error in the determination of primordial helium remains
the major obstacle. Yet, even lacking further progress in this
direction, in the near future the BBN role will be still important
in combination with other cosmological observables to break
degeneracies among different parameters, as in the case of the
PLANCK CMB mission \citep{Ham08,Pop08a,Pop08b}.

\subsection{Sterile Neutrinos and BBN}
\label{sterile}
Sterile neutrinos are, by definition, standard model gauge group singlet
fermions. Their only coupling to SM particles arises via their mass and
mixing parameters with active neutrinos and, provided their mass is not
too high and their mixing sufficiently small, they are long-lived
particles\footnote{For a typical mixing element with the active sector of
the order $\sin\theta_s$ one expects a lifetime for decay into a neutrino
and a photon of the order $\tau_s\simeq 2048\,\pi^4/(9\,\alpha
G_F^2\sin^22\theta_s\,m_4^5)$, $m_4$ being the sterile mass scale
\citep{Mar77}.}. Here, we shall only refer to the case where there is only
one additional neutrino mass eigenstate $\nu_4$, with mass $m_4$. Even in
this case, the resulting $4 \times 4$ neutrino mass matrix $\mathcal{U}$
is described by 4 masses, 6 mixing angles and 3 CP-violating phases (and
possibly, other 3 phases, not entering oscillations, if neutrinos are
Majorana particles). Namely, one mass, three mixing angles and two more
phases with respect to the $3\times 3$ matrix for active Dirac neutrinos.
Since not even the mass pattern of active neutrinos (or their complete
mixing matrix) is known, it comes with no surprise that sterile neutrinos
can manifest quite a rich phenomenology. This is especially true in
cosmology, due to the relevance of medium effects. Basically, sterile
neutrinos with a typical mixing element with the active sector of the
order $\sin\theta_s$ can be populated via incoherent scattering with a
rate which, under ``normal'' conditions, writes
\be
\Gamma_s\simeq \sin^22\theta_s\, \Gamma_a\,,
\ee
$\Gamma_a\sim G_F^2 T^5$ being the active neutrino scattering rate.
However, in the presence of matter with a potential $V$, the effective
mixing angle can be efficiently enhanced whenever the resonance condition
$\delta m_{sa}^2/2\,p\simeq V$ is fulfilled, giving rise to potentially
large effects even when a small vacuum mixing would make the sterile
neutrino undetectable in laboratory experiments. BBN is sensitive to
sterile neutrinos through the following three effects: (i) the partial and
total population of a sterile state induces $\neff> 3$ and thus affects
the Hubble expansion rate; (ii) if $\nu_s$ are produced only after the
decoupling of the active neutrinos from the cosmological plasma, they lead
in general to a depletion of $\nue$ and $\bar\nue$, thus affecting the
weak $n-p$ rates; (iii) the depletion can be $\nu-\bar\nu$ asymmetric,
again affecting in particular, the weak rates.

The interest in the physics of sterile neutrinos in BBN has a long
history (see e.g.
\citep{Kir88,Bar90,Enq90,Kai90,Bar91,Enq92a,Cli92,Shi93,Lis99,Aba03a}).
Due to the otherwise large parameter space, it has roughly
followed the appeal that, from time to time, sterile neutrinos
have had in explaining anomalies in the neutrino phenomenology. An
incomplete account includes $m_4\simeq $17 keV in the beta decay
\citep{Fra95,Wie96}, the KARMEN anomaly ($m_4\simeq $33.9 MeV)
\citep{Arm95}, and in the last few years, the LSND anomaly
($m_4\simeq 1\,$eV) \citep{Agu01}. The recent Miniboone results
\citep{Agu07}, although not ruling out the possibility of more
exotic physics, strongly disfavor or rule out the simplest sterile
neutrino models to explain the LSND signal. At the moment, there
is no clear theoretical or experimental argument suggesting the
existence of (sufficiently light) sterile neutrinos, yet their
rich physics continues to attract a lot of interest (for a review,
see \citep{Cir05}). However, this implies that in absence of
theoretical or experimental prejudice, one has to scan over a
large parameter space.

Most of the old literature referred to mixing between a sterile
and an active state, neglecting mixing among active neutrinos.
This allowed for several simplifications, but the results are
clearly unphysical given the fact that we know that neutrinos do
mix, and the mixing is large. A density matrix formalism is
necessary, which differs from the one we introduced in Section
\ref{sec:neutrinodecoupling} in several points. The vacuum
Hamiltonian $H_0$ includes now four mass eigenstates and a
4$\times$4 mixing matrix. The refractive term $H_1$ writes now in
flavor basis (considering only diagonal elements) \be H_1= {\rm
diag}(V_e,V_\mu,V_\tau,0)~. \ee

The matter potentials $V_\ell$ for each flavor is (with
$\eta_\nu\equiv\eta_{\nue}+\eta_{\num}+\eta_{\nut}$)
\bea
V_e  &=& \pm \sqrt{2} G_F n_\gamma \left[ \eta_e-\frac{\eta_n}{2}
+\eta_\nu+\eta_{\nue} \right]
   - \frac{8 \sqrt{2} G_F\,p}{3}\left(\frac{\rho_{\nue+\bar\nue}}{M_Z^2} +\frac{\rho_{e^-+e^+}}{M_W^2}  \right)\\
V_\mu & =&  \pm \sqrt{2} G_F n_\gamma \left[  -\frac{\eta_n}{2}+  \eta_\nu+\eta_{\num} \right]
   - \frac{8 \sqrt{2} G_F\,p}{3 M_Z^2} \rho_{\num +\bar\num}   \\
V_\tau &=& \pm \sqrt{2} G_F n_\gamma \left[ -\frac{\eta_n}{2}  + \eta_\nu+\eta_{\nut} \right]
   - \frac{8 \sqrt{2} G_F\,p}{3 M_Z^2} \rho_{\nut+\bar\nut}
\eea
where $+$ applies to $\nu$, $-$ to $\bar\nu$.

As initial conditions one assumes usually thermal populations for the
active neutrinos and a vanishing one for the sterile neutrinos (this
assumption is relaxed in some papers, as \citep{Kir04,Kir07,Kir06}, where
a partial filling of the initial sterile state has been considered). In
general, assuming that the active-sterile mixing angles are small (to be
consistent with the phenomenology in the laboratory) is the only
reasonable simplification. Also, as long as $T\ll m_\mu$, in the limit
$\theta_{23}=\pi/4$ and $\theta_{13}\to 0$ there is a $\mu-\tau$ symmetry
which further simplifies the structure of $\mathcal{U}$. A quite thorough
analysis has been performed in \citep{Dol04a}, at least for the range
$\delta m_{4i}^2\lsim 1\,$eV$^2$ which gives rise to the majority of
phenomenologically distinct cases. We summarize here the main features,
while addressing to the original literature for details.
\begin{itemize}
\item If the terms in the square brackets are very small or have
the natural value of the baryon asymmetry, $\eta_i\sim 10^{-9}$,
they are dynamically negligible compared to the  terms of order
$\mathcal{O}(G_F/M_{W,Z}^2)$ at high temperatures and to the
vacuum term $H_0$ at low temperatures. This is also the situation
considered for the standard decoupling in Section
\ref{sec:neutrinodecoupling}. In this case, the matter potential
is always negative, and the existence of resonance conditions
depends only on the mass-square differences $\delta m_{4i}^2$. If
$\delta m_{4i}^2>0,\:\forall\: i$, the sterile-active mixing  is
never resonant, and the analysis simplifies considerably. If
however $\delta m_{4i}^2<0$ for some value of $i$, the system may
undergo one, two or three resonances, and many sub-cases are
possible. \item If a large neutrino asymmetry is present, the
square brackets terms may be large enough to change the impact of
sterile neutrinos on BBN, typically weakening the constraints, as
first noted in \citep{Foo95}. The reason is that the dominance of
the flavor-diagonal medium term compared to the off-diagonal term
due to mixing  suppresses the active-sterile oscillations,
producing sterile neutrinos less efficiently than in a symmetric
background. Apart for larger value of the potential at a given
temperature,  another peculiarity of this case is that its {\it
sign} is opposite for $\nu$ and $\bar\nu$: a pattern of resonances
appears independently of the sign of $\delta m_i^2$, and each one
only in the $\nu$ (or $\bar\nu$) sector. In \citep{Chu06}, some
attention has been paid to strategies to lift the cosmological
bound on sterile neutrinos invoked to explain the LSND anomaly
with a moderate asymmetry (say, $\eta\sim 10^{-4}$). The weakening
of the BBN bounds for a growing asymmetry is represented by the
shift from the solid purple line to the dashed ones in Figure
\ref{fig:lsnd}. Apart for the asymmetry, the distortion of the
neutrino momentum distributions is negligible in the cases studied
in \citep{Chu06}. It is well-known, however, that significant
deviations from a pure FD distribution could occur during the
evolution. Typically they can be relevant either for relatively
small mass splittings, $\delta m_{4i}^2\lsim 10^{-8}\,$eV$^2$
\citep{Kir98} or, for eV scale masses, via matter resonances post
weak decoupling, which might leave both active and sterile
neutrinos with a highly nonthermal spectrum for some choices of
neutrino parameters and energies. This however requires larger
initial asymmetries, $\eta_\nu\gsim 0.01$ \citep{Aba05,Smi06}.
\end{itemize}
\begin{figure}[t]
\begin {center}
\includegraphics[width=0.7\textwidth]{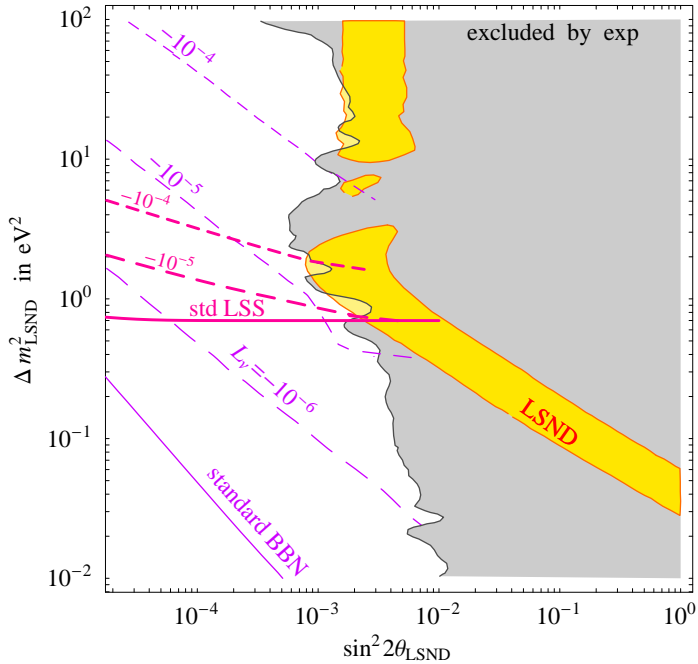}
\end{center}
\caption{The allowed LSND region at 99\% C.L. (yellow/light shaded area)
compared to the cosmological bounds from BBN and LSS in the presence of
primordial asymmetries. The darker shaded area is already excluded at 99\%
C.L. by other experiments. The regions below and to the left of the thin
lines are allowed by BBN because they correspond to $Y_p \le 0.258$. The
regions below and to the left of the thick lines are allowed by LSS
because they correspond to $\Omega_\nu h^2 \le 0.8\ 10^{-2}$. From
\citep{Chu06}.}
\label{fig:lsnd}
\end{figure}

An interesting aspect implicitly omitted above is that, even if
the neutrino asymmetry is vanishingly small before the onset of
the active-sterile oscillations, a large $\nu-\bar\nu$ asymmetry
(up to $\eta\sim 0.1$) might be dynamically generated in the
active neutrino sector (and compensated by an opposite one in the
sterile sector) via a resonant matter effect.  This scenario
requires quite special choices of the parameters: the mixing must
be sufficiently small, $\sin^2 2\theta_s\lsim 10^{-4}$, $|\delta
m_i^2|<0$ for some $i$,  and   $|\delta m_i^2|\,\sin^42\theta_s
\lsim {\rm few}\times 10^{-9}{\rm eV}^2$. For a review we address
to \citep{Dol02b} and reference therein (see also \citep{Dol04a}
for some more details). Perhaps more general is the following
consideration, which is often overlooked in the literature: once a
sterile neutrino is introduced, at least two additional
CP-violating phases are {\it naturally} introduced in the neutrino
mixing matrix. The {\it vacuum} oscillation probability between an
active and the sterile state are naturally CP-violating, even when
$\theta_{13}\to 0$ and no CP violation happens in the
active-active oscillations. One might generate a flavor-dependent
lepton-asymmetry as large as $O(\sin^22\theta_s)$---which in a
large part of the allowed parameter space is much bigger than
$\eta_B$---without resonances.

Finally, some BBN constraint also arises for more massive sterile
neutrino states, in the keV-MeV range, by requiring that the
energy density stored in the sterile states populated via mixing
does not exceed e.g. $\Delta \neff=1$. As reviewed in
\citep{Aba01}, the bounds are not very competitive compared with
others. For significantly heavier states, the most interesting
bounds may come from cascades and dissociation of light elements
from the sterile neutrino decays, which we address in Section
\ref{s:massive-particle}.

\section{Inhomogeneous nucleosynthesis}
\label{sec:IBBN}
In the standard scenario of nucleosynthesis  all constituents are
homogeneously and isotropically distributed, in accordance with
the hypothesis of a FLRW universe. Anisotropic models, studied as
early as in the 60's, or models with adiabatic fluctuations in the
radiation energy density may affect nucleosynthesis essentially
via a variation in the expansion rate. The emergence of a
concordance cosmology model, supported by CMB and LSS data
essentially confirm {\it observationally} that, apart from small
initial adiabatic fluctuations ($\sim 10^{-5}$) as seed of
structure formation, no significant departure from the homogeneity
or isotropy is required. The logical possibility that large
fluctuations existed at the horizon scales at BBN epoch (size of
the order of $10^{-6}$ deg. in the CMB) is not well motivated
either in the favored inflationary scenario to generate the
perturbations. Thus, this line of research has faded away in the
last decade: we address the reader to \citep{Mal93} for a
historical overview. However, it is perfectly consistent with the
present cosmological scenario to speculate on a varying baryon to
photon and neutron to proton ratio on small scales. These
isothermal (or isocurvature) fluctuations may alter BBN in a
non-trivial way, and these scenarios are known as inhomogeneous
BBN (IBBN) models. Till the 90's, the interest in IBBN was due to
two aspects: (i) several theories can lead to inhomogeneous
distributions of neutrons and protons at the time of
nucleosynthesis: a first order quark-hadron phase transition
\citep{Wit84,Kur88a,Sum90,Mal93}, the CP violating interaction of
particles with the bubble enucleated in the electroweak phase
transition \citep{Ful94,Hec95,Kai99,Meg05}, the phase transition
involving inflation-generated isocurvature fluctuations
\citep{Dol93} or kaon condensation phase \citep{Nel90}; (ii) there
was some hope that IBBN scenarios with $\Omega_B \simeq 1$---thus
consistent with a flat, matter dominated universe, or at least an
open universe without non-baryonic dark matter---were
phenomenologically viable (see e.g. \citep{Sal86}). Both
motivations have lost observational or theoretical support in the
last decade. The agreement between the value of $\Omega_B h^2$
from standard BBN deuterium abundance and CMB anisotropies is
indeed quite a compelling result in favor of the simplest
homogeneous scenario. Theoretically, due to the high mass of the
Higgs, the electroweak phase transition is a smooth cross-over for
the SM particle content, and even the QCD phase
transition---although less well established---appears to be a
crossover or weak first order one, probably insufficient to
produce significant departures from the standard BBN scenario
\citep{Sch03}.

In the following, we limit ourselves to introduce and briefly
describe some recent calculations in IBBN, addressing the reader
to \citep{Mal93} for a throughout overview of the topic. Since
IBBN has to reproduce very closely the SBBN yields while requiring
additional parameters, most of its phenomenological interest has
declined, too. Today, perhaps the only way to discriminate IBBN
vs. homogeneous BBN lies in the different predictions for the
intermediate (CNO) or heavy elements.

\subsection{Baryon inhomogeneous models}
\label{sec:BImodels}
An important point to be considered is the sensitivity of the
predictions to the form of the inhomogeneity itself. Many studies
simply have a 2-phase model with a fraction $f$ at high density
and $1-f$ at low, while in the paper \citep{Bar83} a more
realistic distribution (log normal) is considered, which allows a
bound to be placed on the variance.

If there are large fluctuations in the nucleon density, the
differential transport of neutrons and protons can create
neutron-rich regions where heavy elements can be formed. Neutrons
diffuse by scattering on electrons and protons, protons scatter on
neutrons and Coulomb scatter on electrons, but the mean free path
of protons is about $10^6$ times smaller than that of neutrons.
The diffusion of other species is negligible with respect to
neutron scattering due to their larger masses. Neutron diffusion,
however, was not considered in the earliest codes of IBBN
\citep{Zel75,Eps75,Bar83}, where regions of different nucleon
density were treated as separate homogeneous BBN models. The mass
fractions from each model were then averaged, with a weight given
by the corresponding size. A later generation of codes
\citep{App87,Alc87,Kaj90} introduced in the calculation nucleon
diffusion, but only at early times and high temperatures, before
the starting of nucleosynthesis. This led to the neutron
enrichment of the low-density region but, once the original
protons were consumed, neutrons could form $^4$He only when other
protons were produced by neutron decay. The main consequences of
this situation on the light element abundances were: a) since four
neutrons (two of which decaying in two protons) were needed for
each $^4$He nucleus, the final yield of $^4$He was reduced; b)
nucleosynthesis time scale were tuned by neutron decay rate,
extending the process to cooler temperatures and allowing $^2$H to
survive in larger quantities; c) the high neutron density could
help the production of heavier elements through neutron-rich
isotopes. However, once neutron diffusion during nucleosynthesis
is taken into account, all the three previous effects are
weakened, since when neutrons are rapidly consumed in the high
density region, where nucleosynthesis begins first, the excess
neutrons in the low density region diffuse back. Kurki-Suonio {\it
et al.} made the significant step forward of treating nucleon
diffusion both before and during nucleosynthesis, with planar
symmetric baryon inhomogeneities \citep{Kur88b} or cylindrical and
spherical models \citep{Kur89,Kur90a}. In order to decrease the
number of zones needed to obtain a high accuracy, nonuniform grids
were used \citep{Kur90b,Mat90,Mat96,Ori97,Lar05}. The diffusion
equation is sufficient for describing the motion of particles in
an IBBN model if the background fluid is stationary, as in the
case of neutrons, which are much more mobile than the ions  and
electrons they scatter on. The evolution of ions at low
temperatures is more complicated, due to momentum transfers in
collisions with other ion components, which move with comparable
fluid velocities. In this case, the common diffusion approximation
has to be relaxed and one needs to take into account dissipative
processes through hydrodynamic equations
\citep{Alc90,Jed94a,Kei02}.

In an inhomogeneous code with treatment of neutron diffusion, the
region considered is divided into several zones, $s$, where the
time evolution of the number density of the specie $i$, $n_{i,s}$,
obeys the following evolution \citep{Lar05,Kai99,Mat90} \bea
\frac{\partial n_{i,s}}{\partial t} &=& n_{B,s} \sum_{j,k,l}\, N_i
\left( \Gamma_{kl \rt ij}\, \frac{Y_{k,s}^{N_k}\,
Y_{l,s}^{N_l}}{N_k!\, N_l !} \; - \; \Gamma_{ij \rt kl}\,
\frac{Y_{i,s}^{N_i}\, Y_{j,s}^{N_j}}{N_i !\, N_j
!} \right) \nonumber \\
&& -\, 3\, H\, n_{i,s} + \frac{1}{r^p}~ \frac{\partial}{\partial
r} \left( r^p\, D_n\, \frac{\partial \xi}{\partial r}\,
\frac{\partial n_{i,s}}{\partial \xi} \right) \pp
\label{e:ndiff}
\eea
The first three terms are usual, corresponding to reactions which create
or destroy nuclides and to the expansion of the universe, while the last
one is due to diffusion of isotope $i$ between zones. The parameters which
appear in Eq. \eqn{e:ndiff} are the inhomogeneity distance scale, $r$,
which measures the physical distance between inhomogeneity regions at the
starting temperature, $T\sim 10\,$MeV, the stretching function, $\xi (r)$,
which implements the non-uniform grid, marking the zone edges, and the
neutron diffusion coefficient, $D_n$, which is a function of proton
density and temperature \citep{App87}\footnote{Note that, as remarked in
\citep{Kur90b}, in their Eq.~(21) the factor $\pi/16$ is missing from the
numerical value.}. For small distance scales, $r<1$ light-hour, the
inhomogeneities are smeared out by neutron diffusion before
nucleosynthesis starts, and the IBBN results approach standard BBN
results. The constant parameter $p$ changes with geometry (for example,
for the spherical symmetry $p=2$). Other important parameters are the
density contrast, $R$, which is the ratio between the high and low
densities, taken as high as $10^6$, and the volume factor, $f_v$, that is
the fraction of space occupied by the high density region. Note that the
higher $R$, the larger the number of zones needed for a sufficient
accuracy of the calculation.

\begin{figure}[p]
\begin{center}
\begin{tabular}{c}
\includegraphics[angle=90,width=0.7\textwidth]{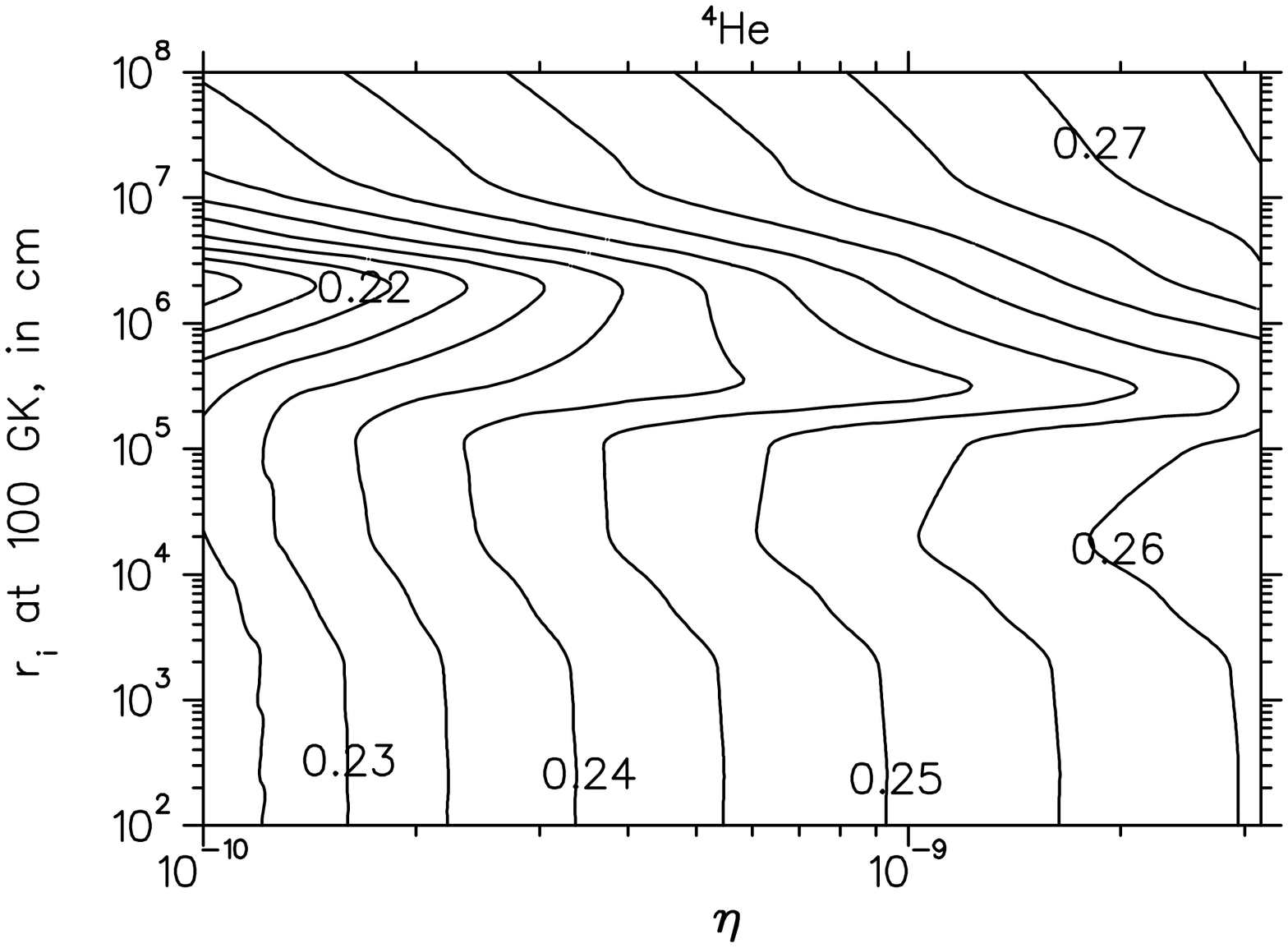} \\
\includegraphics[angle=90,width=0.7\textwidth]{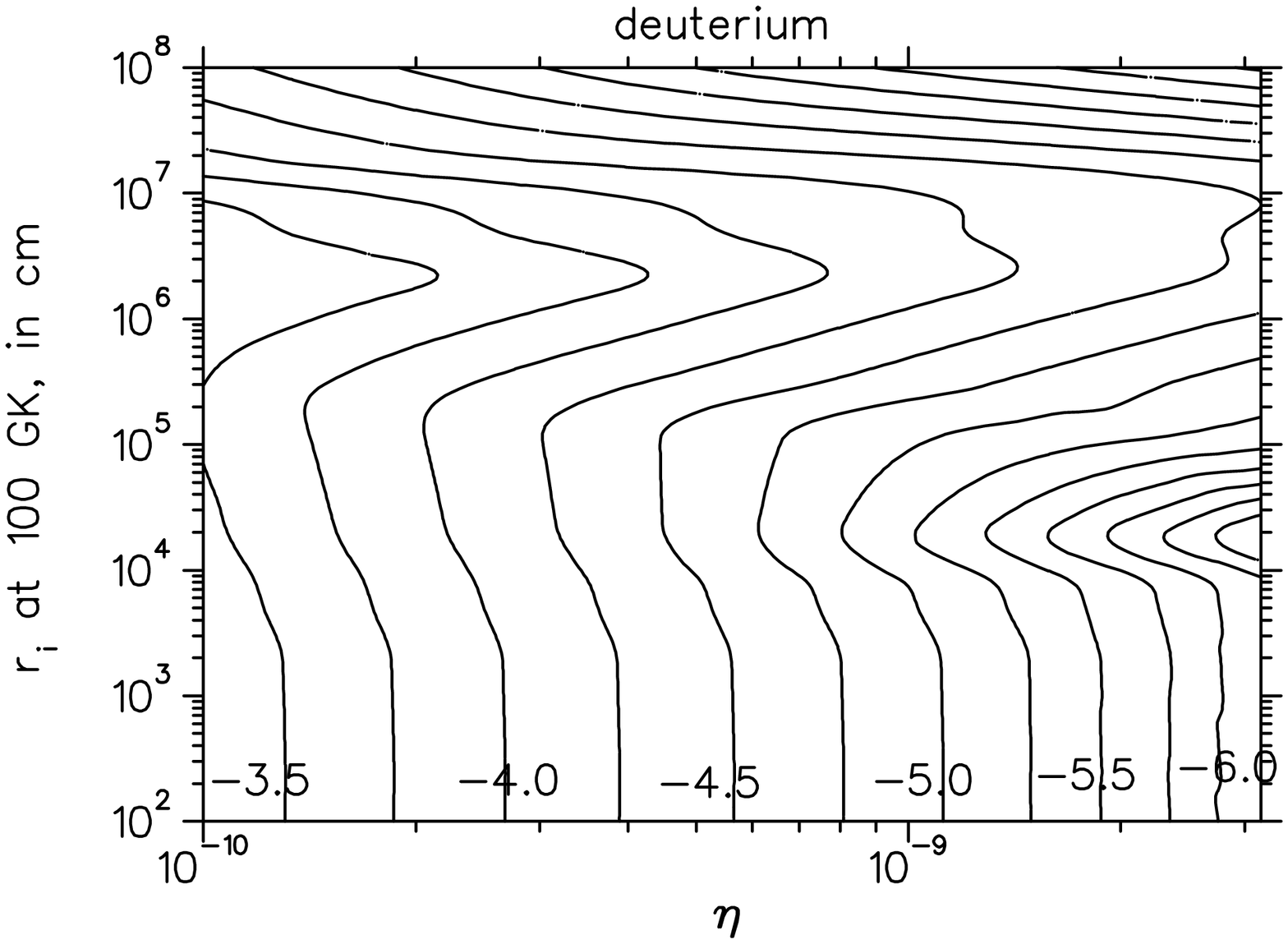} \\
\includegraphics[angle=90,width=0.7\textwidth]{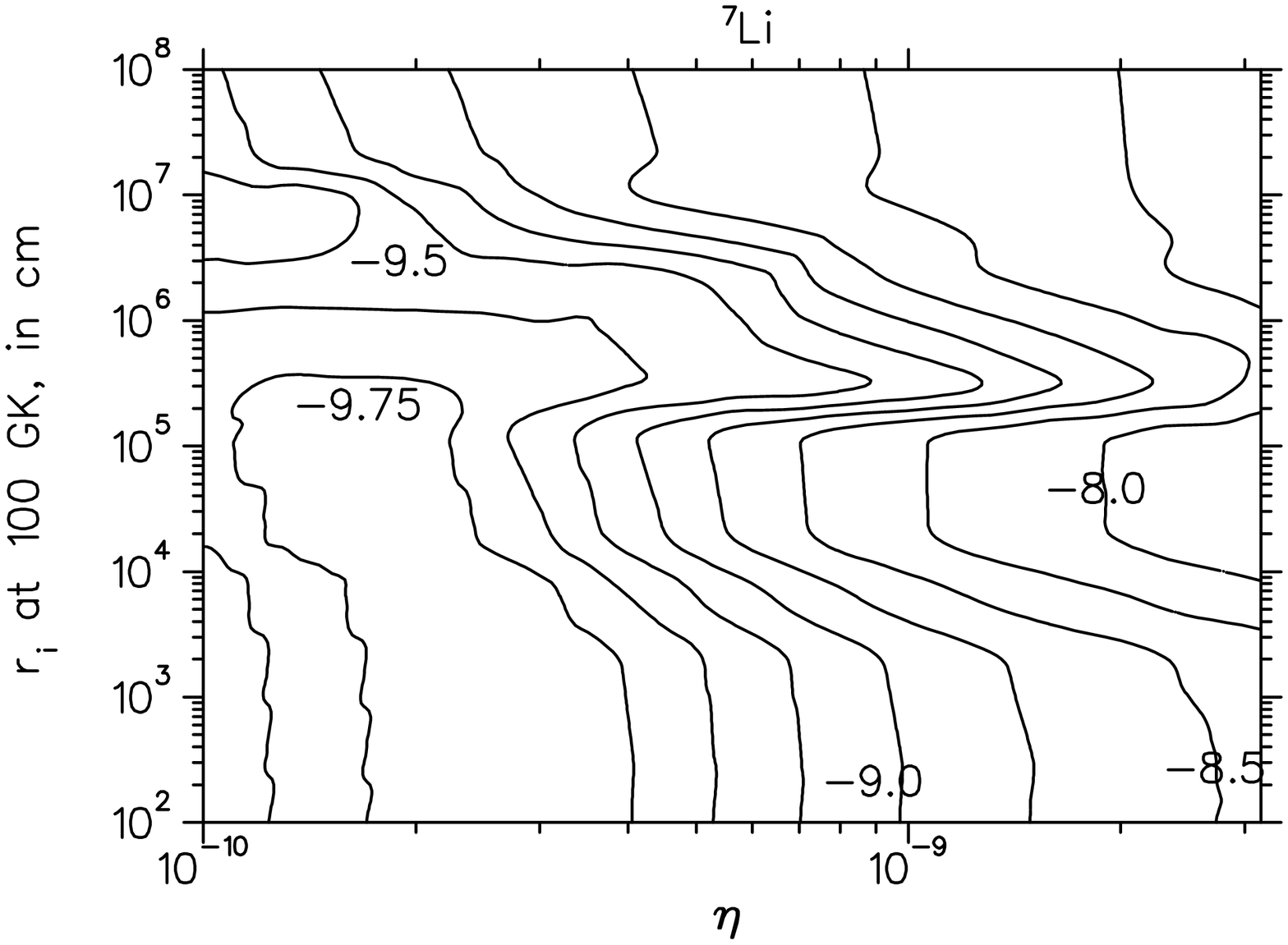}
\end{tabular}
\end{center}
\caption{Mass fraction of $^4$He and $\log$ of the abundances
$Y_i/Y_p$ for $i=^2$H, $^7$Li. From Ref. \citep{Lar05}.}
\label{IBBNabund}
\end{figure}

\begin{figure}[t]
\begin{center}
\begin{tabular}{c}
\includegraphics[angle=90,width=0.7\textwidth]{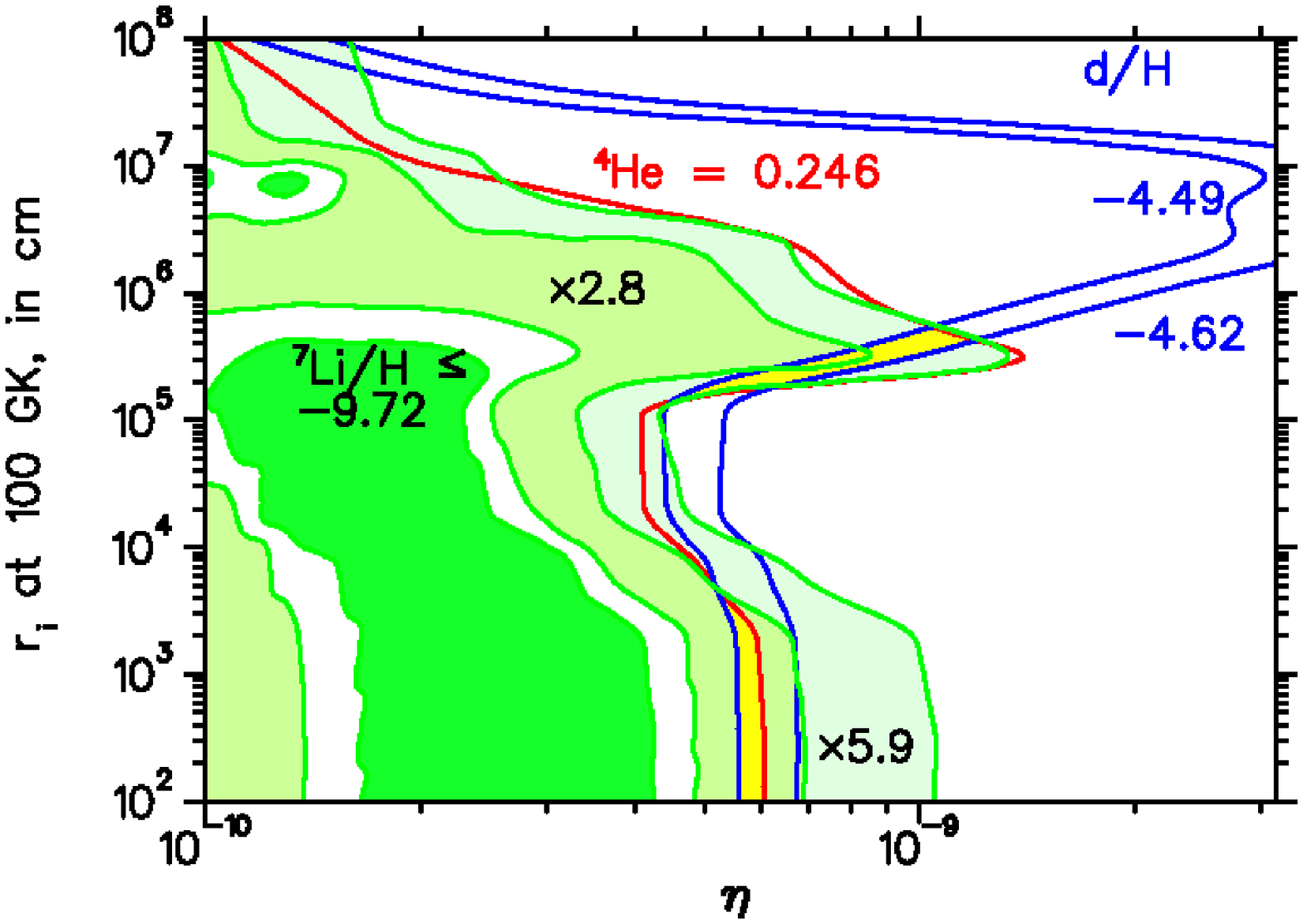} \\
\includegraphics[angle=90,width=0.7\textwidth]{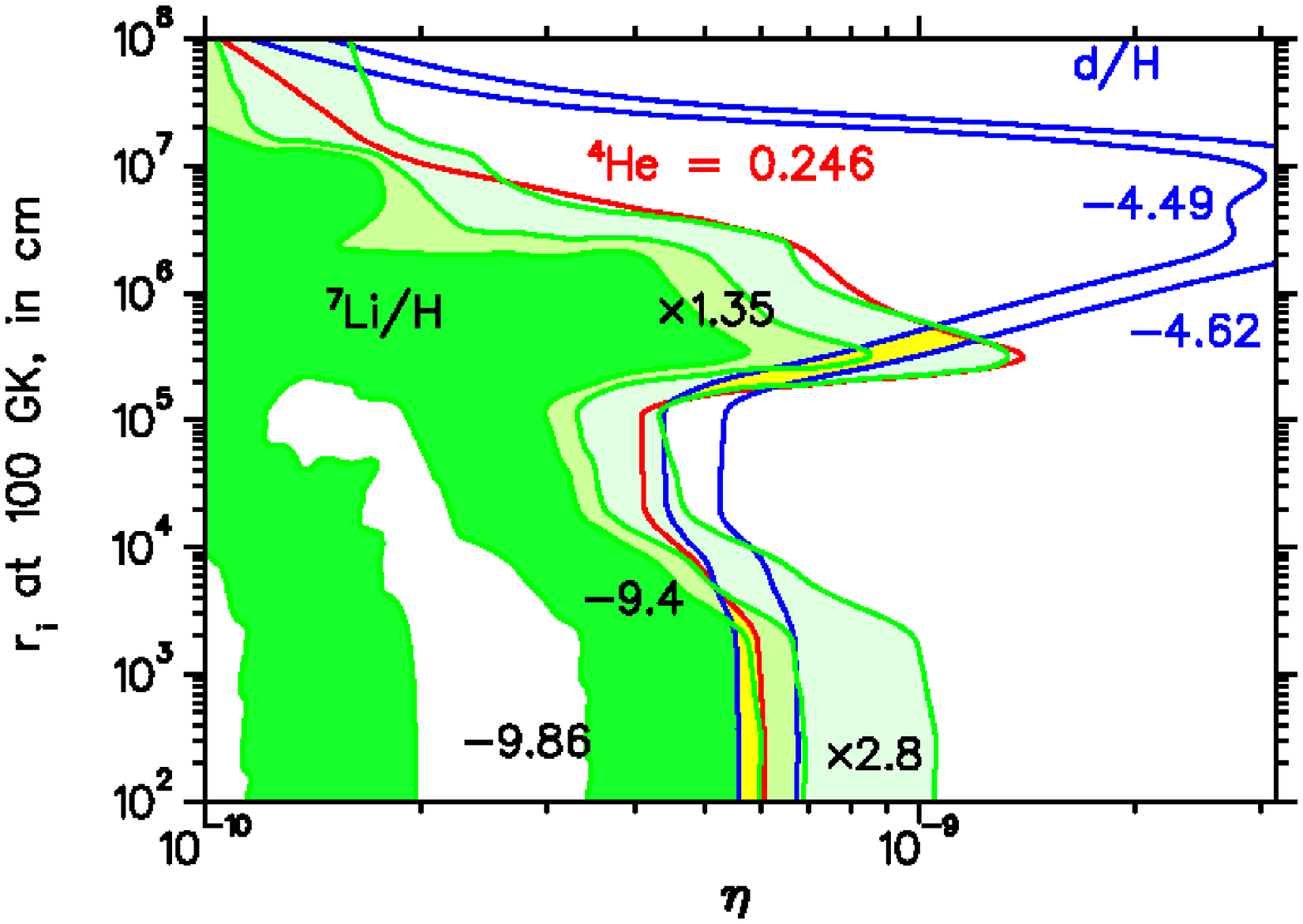}
\end{tabular}{c}
\end{center}
\caption{Concordance between the observational constraints on $^4$He,
$^2$H, and $^7$Li and the model of Ref. \citep{Lar05}. From Ref.
\citep{Lar05}.}
\label{IBBNres}
\end{figure}

The plots in Figure \ref{IBBNabund} are contour maps of the $^4$He
mass fraction, and $^2$H/H and $^7$Li/H, taken from Ref.
\citep{Lar05}, and present the characteristic features of an IBBN
prediction on light element abundances. For small values of $r_i$
neutron diffusion homogenizes neutrons very quickly and protons as
well, since this happens before weak interactions go out of
equilibrium. This means that the final abundances are the same as
a homogeneous model. The shift in the contour lines towards low
values of $\eta$ for distance scales $r_i$ starting from $\sim
2-3\times 10^3$ cm is due to the fact that, for these values of
$r_i$, neutron diffusion and homogenization take a time of the
order of the weak interaction freeze-out. This implies that
protons start to be not efficiently homogenized and
nucleosynthesis occurs before in the high density shells with a
larger proton density and has the characteristics of an earlier
nucleosynthesis (larger $^4$He and $^7$Li and less $^2$H). So, to
recover the same values of the abundances of a model with lower
$r_i$ one needs lower values of $\eta$. Up to $r_i \sim 10^5$ cm
the proton number density is unchanged except for a slight
increase due to neutron decay, and this explains the almost
vertical contour lines in this range. The depletion of neutrons in
the high density shells at temperatures just after nucleon
freeze-out leads to neutron back-diffusion and a more efficient
nucleosynthesis (and a magnification of the production of $^7$Be)
in the high density region. Due to this overproduction of $^7$Be,
which gives $^7$Li after decay, the contour lines in the lowest
plot in Figure \ref{IBBNabund} have a larger shift to lower
$\eta$. When $r_i$ starts to be larger than $10^5$ cm, neutron
back-diffusion does not affect all the shells and nucleosynthesis
is concentrated only in some of them. This leads to a decrease of
the final $^4$He abundance accompanied by an increase in deuterium
production, which corresponds to a shift of the contour lines
towards high $\eta$. Finally, contours turn back to low values of
$\eta$, since for very large $r_i$ diffusion cannot homogenize
neutrons before nucleosynthesis, a large neutron density remains
in the high density region, giving rise again to $^4$He
overproduction and a $^2$H suppressed yield. In this region of
$r_i$, results are equivalent to the average of two separate
homogeneous BBN models, one of high density (with high $^4$He and
$^7$Li and low $^2$H) and one of low density. The basic shapes of
the contour lines of Figure \ref{IBBNabund} are common to all IBBN
models: for different geometries and values of the parameters
there will be regions in $r_i$ where neutron homogenization and
diffusion occur at times between weak freeze-out and
nucleosynthesis or after nucleosynthesis.

Figure \ref{IBBNres} shows the concordance between the
observational constraints on $^4$He, $^2$H, and $^7$Li and the
model of Ref. \citep{Lar05}: upper plot is for the $^7$Li
constraints from Ryan {\it et al.} \citep{Rya00} while lower plot
is for the $^7$Li data of Melendez \& Ramirez \citep{Mel04}. In
both plots the concordance region between $^4$He \citep{Izo04} and
$^2$H data \citep{Kir03} is shown in yellow. While upper plot have
a concordance region for $^7$Li only for a depletion factor
ranging from 2.8 to 5.9, the lower plot does not need any
depletion in $^7$Li if $r_i<5\times 10^3$ (see Ref. \citep{Lar06a}
for a similar analysis using $\tau_n = 878.5\pm 0.7_{\rm stat}\pm
0.3_{\rm syst}$ \citep{Ser05a}).

One interesting consequence of IBBN results is the larger range in
the depletion factor one can obtain for $^7$Li with respect to the
analogous prediction of homogeneous BBN. Figure \ref{IBBNres}
shows, moreover, that IBBN allows for larger values of $\eta$ than
SBBN (up to $\eta\sim 10^{-9}$), requiring at the same time a
depletion factor for $^7$Li to obtain concordance with the
observational limits.

Since lithium is produced quite late in nucleosynthesis, its yield
is particularly sensitive to the late-time transport phenomena
such as hydrodynamic ion diffusion. In this respect, the results
of codes which take into account these phenomena~\citep{Kei02},
which are not considered in~ \citep{Lar05}, might explain the
observed depletion of lithium. This is a consequence of late
separation of elements due to Thomson drag: Thomson scattering of
electrons on background photons makes the diffusion of ions
inefficient which must drag electrons with them to keep charge
neutrality. On the other hand, protons and helium ions, for
instance, are allowed to diffuse in opposite directions. While
protons diffuse out, helium and lithium get concentrated in the
high density regions, leading to enhanced destruction of $^7$Li,
$^2$H, and $^3$He.

Heavy element production in the framework of IBBN was first
investigated in Ref. \citep{Jed94b} and then in
\citep{Mat05b,Mat07}, in the approximation of neglecting baryon
diffusion. The authors claim that there is a parameter region, for
the volume fraction $f$ and density contrast $R$, in which heavy
elements can be produced enough to affect the observation, while
keeping the light element abundances consistent with observation.
The results show that BBN proceeds through both the p-process and
the {\it r}-process, with the transition between the two due to
the Coulomb barriers of proton-rich nuclei and the amounts of
neutrons when heavy elements begin to be synthesized.

\begin{figure}[p]
\begin{center}
\begin{tabular}{c}
\includegraphics[width=0.65\textwidth]{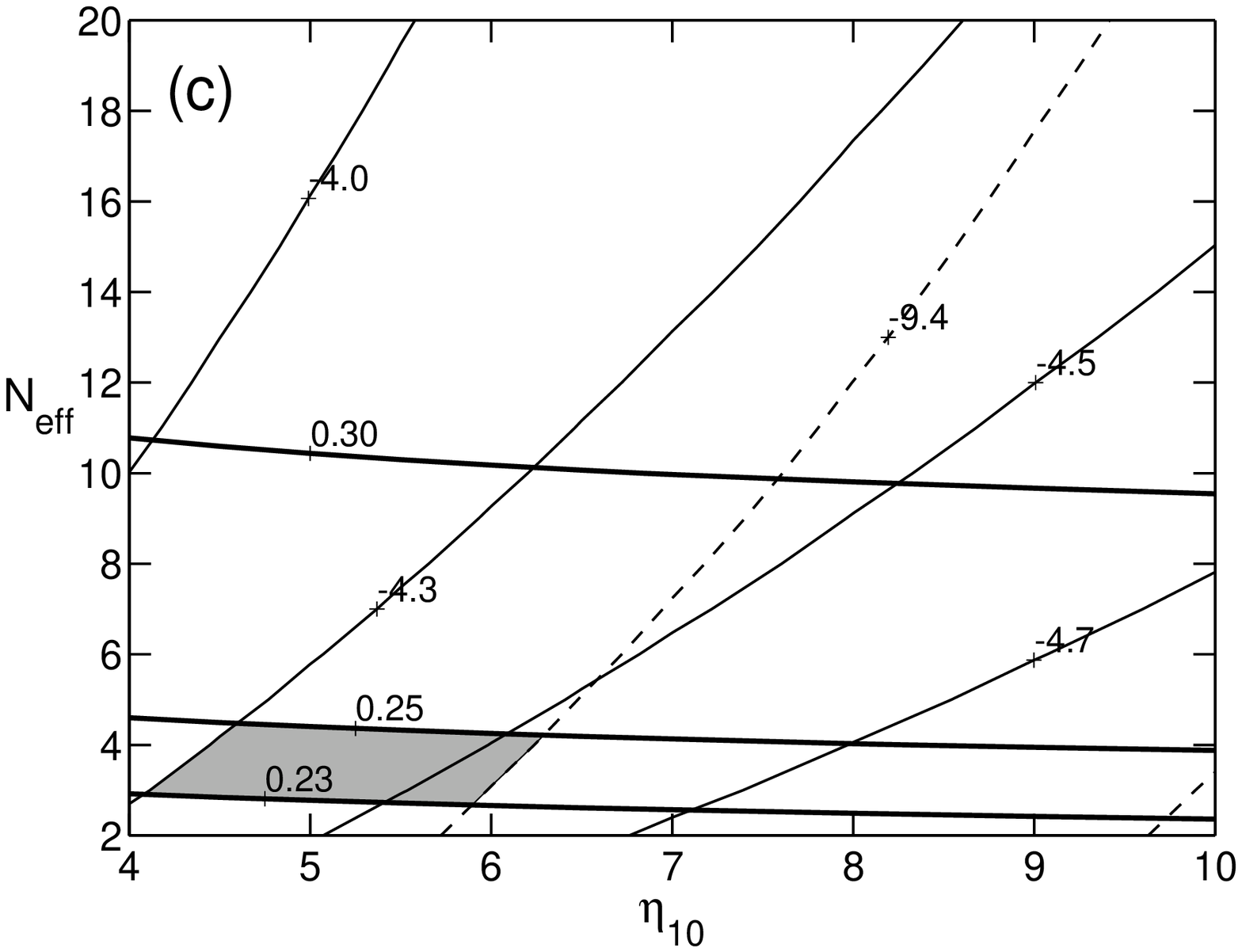} \\
\includegraphics[width=0.65\textwidth]{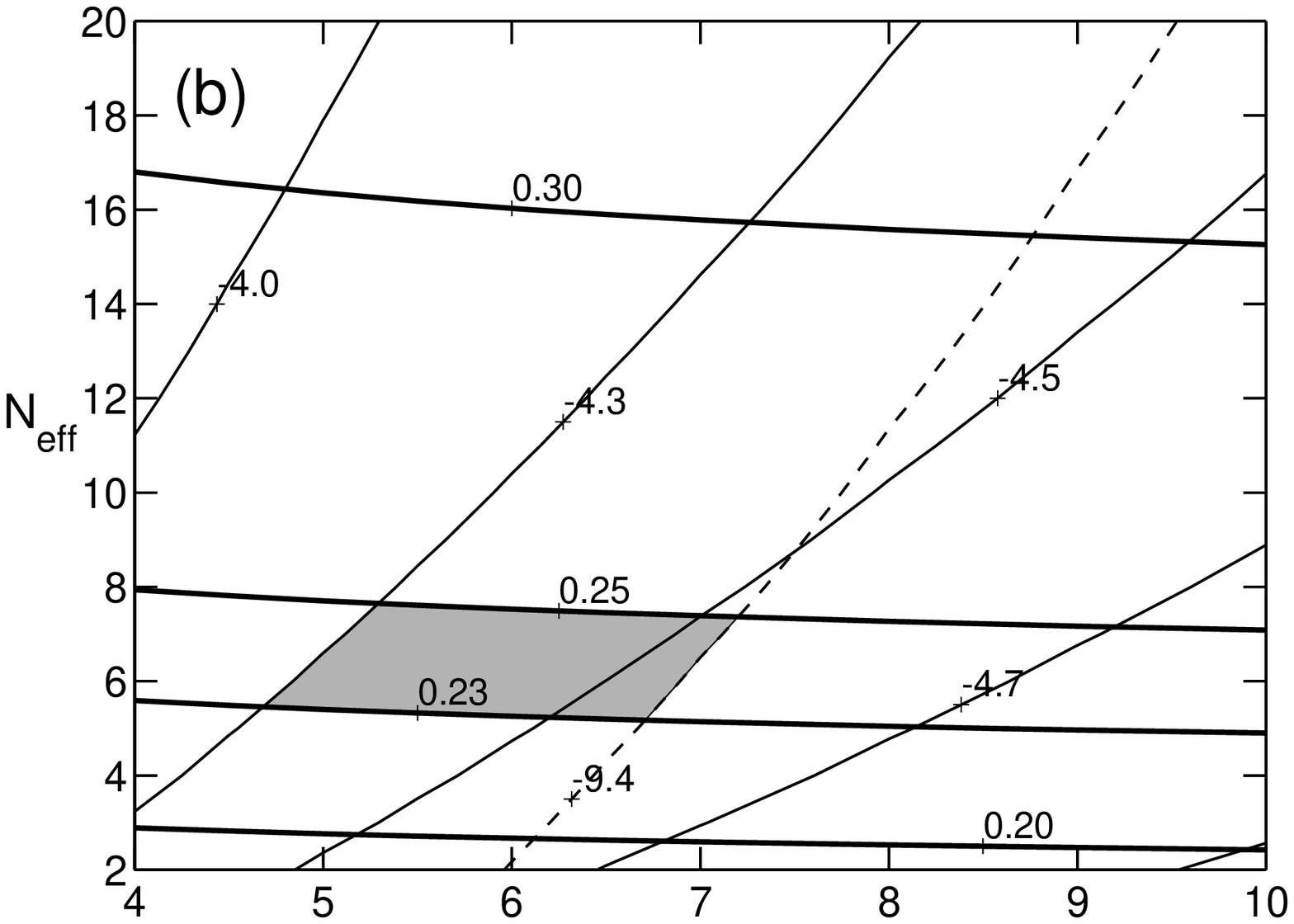} \\
\includegraphics[width=0.65\textwidth]{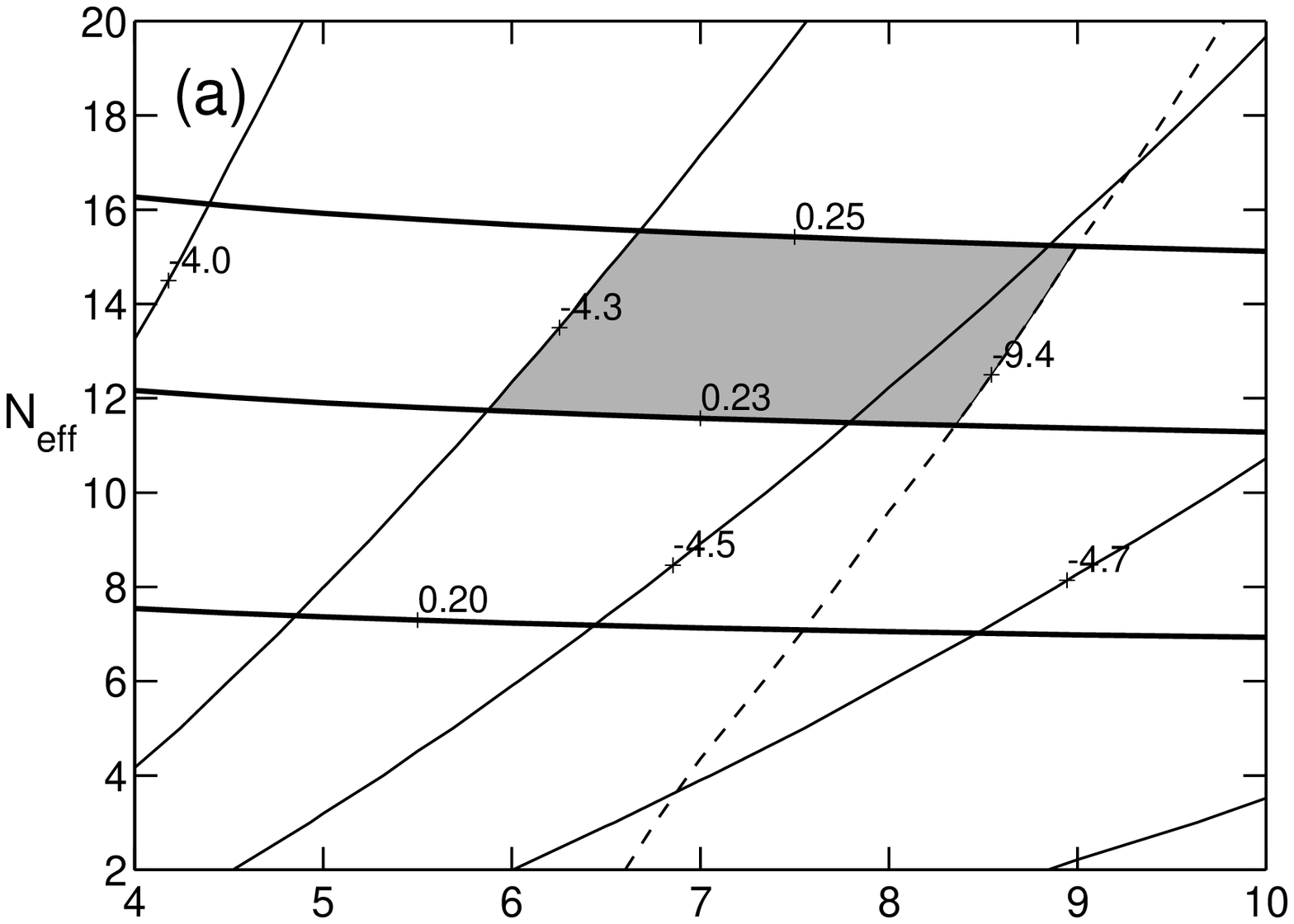}
\end{tabular}
\caption{Light element yields in ABBN as a function of $\eta$ and
$\neff$ for $r_A=10^{6.9}$ m and, from top to bottom, $R=10^{-2},
10^{-1.5}, 10^{-1.2}$. Thick solid lines, thin solid lines, and
dashed lines are for $Y_p$, $log_{10} ^2\textrm{H}/\textrm{H}$,
and $log_{10} ^7$Li/H, respectively. From Ref. \citep{Gio02}.}
\end{center}
\label{ABBNres1}
\end{figure}

\subsection{Matter-antimatter inhomogeneities}
\label{sec:ABBN}

A possible scenario which gives rise to an inhomogeneous
baryon-to-photon ratio is antimatter BBN (ABBN)
\citep{Ste76,Reh98,Reh01,Kur00a,Kur00b,Sih01}. Different
baryogenesis models can give rise to matter-antimatter domains
\citep{Gio98,Dol92,Dol93,Khl00a,Khl00b,Mat04,Dol08}. In the ABBN
scenario, the antimatter regions have radius $r_A$, while $R$ is
the antimatter-matter ratio in the universe. Antimatter regions
should be small enough to be completely annihilated well before
recombination, in order to satisfy CMB constraints. Their size
determines the time when most of the annihilation takes place,
before or after significant amounts of $^4$He are produced by
nucleosynthesis. The first case is realized for typical radii
between $10^5$ and $10^7$ m (comoving distance at T=1 KeV)
\citep{Gio02}\footnote{Smaller antimatter regions would annihilate
before neutrino decoupling without any effect on BBN.}. Thanks to
the different diffusion scale of protons and neutrons, the latter
can more easily move to antimatter regions and annihilate. This
would produce a reduced neutron to proton ratio with respect to
the standard case, which can be compensated by a larger expansion
rate at BBN, provided by more relativistic degrees of freedom,
$\neff>3$, and results in the same $Y_p$. Correspondingly, the
speed-up of expansion shortens the time interval available for
nucleosynthesis, and implies smaller yields for all other light
nuclides, yet this can be compensated by the increase of the
reaction rates due to a higher value of $\eta$. The net result is
thus, a shift of the agreement of theory vs. data towards larger
$\eta$ for large $\Delta \neff$, see Figure \ref{ABBNres1}, but
with the bonus that $\neff$ is no longer constrained, as shown in
Figure \ref{ABBNres2}.

\begin{figure}[t]
\begin{center}
\includegraphics[width=0.6\textwidth]{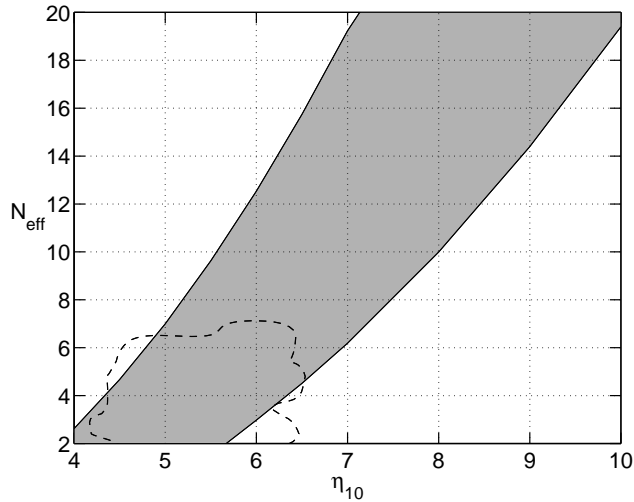}
\end{center}
\caption{Combined allowed region in ($\eta$, $\neff$). The dashed line
is the CMB+SNIa constraint from \citep{Han02}. From Ref. \citep{Gio02}.}
\label{ABBNres2}
\end{figure}

\section{Constraints on fundamental interactions}
\label{sec:constr_fundam}

\subsection{Extra Dimensions and BBN}

\subsubsection{A short journey to extra dimensions}

The idea of introducing extra (spatial) dimensions to generalize
the 4-dimensio--nal theory of fundamental interactions and unify
different forces is quite old. As early as 1919, thus shortly
after the birth of General Relativity, Kaluza considered a
5-dimensional version of Einstein theory which described gravity
and electromagnetism in a unique setting \citep{Kal21}. Shortly
after, in 1926 Oskar Klein stressed the role of having a compact
fifth dimension in order to evade constraints from observations of
large accessible extra dimensions \citep{Kle26}. A nice review on
Kaluza-Klein theories is \citep{Bai87}.

Higher-dimensional theories had perhaps their golden age starting from the
late 70's, after the discovery of the remarkable properties which
superstring and supergravity theories have for particular spacetime
dimensionalities. Quite recently, theories with one or more extra
dimensions with a fundamental scale of TeV$^{-1}$ have been advocated as
possible way to address the long-standing problem of hierarchy between the
electroweak and the much higher Planck scale
\citep{Ark98,Ant98,Ran99a,Ran99b}. In these scenarios, the fundamental
gravity scale is lowered down to the TeV range, and the observed Planck
mass emerges as an effective scale at low energies, smaller than the
Kaluza-Klein (KK) excitation mass scale. This is due to the dilution of
gravitational interactions in the large (millimeter-sized) extra
dimensions (flat scenarios), or the particular configuration of the
gravitational field which provides a static solution to Einstein's
equations (warped extra dimensions). A general feature of these theories
is to assume that ordinary matter is confined to standard 3+1-dimensional
spacetime, a brane embedded in a (4+d)-dimensional manifold, while gravity
can propagate in the whole higher dimensional spacetime.

Interestingly, large (experimentally accessible) extra dimension models
can be tested using collider physics, as for example at LHC, for a review
see e.g. \citep{Giu07,Sun05}. On the other hand, they may have a large
impact on the cosmological evolution of our universe. The issue of
understanding the phenomenological implications of ``brane-cosmology" has
been addressed by several scholars in the last ten years, mainly aimed at
discussing how these scenarios can be constrained by cosmological
observables, CMB and BBN among others. In the following, after a brief
summary of the aspects of the extra dimension models which are relevant
for our discussion, we (mainly) focus on the constraints which can be
obtained by exploiting BBN.

We start by introducing the $D=4+d$ dimensional Einstein and
matter action, which can be written as
\be
S= \int \,d^4 x d^d y \sqrt{-\overline{g}} \, \frac{M_D^{2+d}}{2}
\overline{R}  + \sqrt{-\overline{g}} \, L_m \vv
\ee
where the first term corresponds to the Einstein-Hilbert action,
$\overline{R}$ being the $4+d$-dimensional scalar curvature for the $4+d$
metric $\overline{g}$ and $M_D$ playing the role of the $D$-dimensional
reduced Planck mass, while the second term contains the matter Lagrangian,
with the SMPP fields localized on the 3+1-dimensional brane $y=0$. In the
case of compact extra dimensions (we will consider the specific case of a
$d$-dimensional torus of radius $\delta$) and for a factorized metric,
i.e. if the 4-dimensional part does not depend upon the $d$ extra
coordinates, the action can be reduced to a 4-dimensional action at low
energy, smaller than the inverse compactification radius $\delta$, by
integrating over the $y$ coordinates,
\be
S=  \int \,d^4 x  \sqrt{-g} \,\frac{M_D^{2+d} (2 \pi \delta)^d
}{2} R + \sqrt{-g}\, L_m \vv
\ee
from which we read the expression of the Planck mass $M_P=G_N^{-1/2}= 1.2
\cdot 10^{19}$ GeV in terms of $M_D$ and $\delta$,
\be
(2 \pi\delta)^{-1}= M_D (\sqrt{8 \pi} M_D/M_P)^{2/d} \pp
\label{radius}
\ee
For $M_D \sim$ TeV, the simplest case of one extra factorized
dimension is excluded as it leads to a value of $\delta$ which is
too large, of the order of the scale of the solar system, while
the scenario is viable for $d\geq 2$, as in this case $\delta \leq
1$ mm.

More generally, one can allow for an explicit dependence of the
4-dimensional metric on the extra coordinates. As in
\citep{Bin00a,Chu00}, and usually considered in almost the whole
literature, we consider a (non-factorized) $d=1$ model, where the
extra dimension (the bulk) is compactified on the line segment
$S^1/Z_2$. The metric (preserving 3-dimensional rotation and
translation invariance) can then be written as \be ds^2 =
-n^2(\tau,y) d \tau^2 + a^2(\tau,y) d\vec{x}^2 + b^2(\tau,y) dy^2
\pp \ee The $Z_2$ symmetry identifies the points $y$ and $-y$, so
one can restrict to $0\leq y\leq 1/2$. Two three-branes are placed
at $y=0$ (our ``visible'' brane) and $y=1/2$ (a hidden brane,
which absorbs the gravitational flux lines of the visible brane).
The metric is obtained as usual from Einstein's equations (upper
case latin indexes $A,B=0,1,2,3,5$ run over the 5-dimensional
spacetime), \be \overline{G}_{AB} = M_5^{-3} \overline{T}_{AB} \vv
\label{einstein5} \ee where the stress-energy tensor is the sum of
contributions of ordinary matter on the visible brane, bulk matter
and fields living on the hidden brane, \be \overline{T}^{ AB}=
\frac{T_{vis}^{AB}}{b(\tau,0)} \delta(y) +
\frac{T_{hid}^{AB}}{b(\tau,1/2)} \delta(y-1/2)+ T_{bulk}^{AB} \pp
\label{stress5} \ee Each term corresponds to a perfect fluid,
parameterized as usual in terms of the energy density $\rho$ and
pressure $P$ and specified by the equation of state $P=P(\rho)$,
with furthermore $T_{vis}^{05}=T_{hid}^{05}=0$, so that there is
no flow of matter on the branes along the fifth dimension, and
finally
$T_{bulk}^{AB}=diag(-\rho_{bulk},P_{bulk},P_{bulk},P_{bulk},P_{bulk,5})$.

Before considering the cosmological scenarios corresponding to
this framework, one should look for the static solutions of Eq.
(\ref{einstein5}), the analogous of (empty space) Minkowski
spacetime. Apart for the case $T^{AB}=0$ which leads to a
(factorizable) trivial spacetime, Randall and Sundrum (RS)
\citep{Ran99a,Ran99b} found a new solution by considering a pure
bulk and brane cosmological constant terms. Choosing
$T_{bulk}^{AB}=\Lambda \, diag(-1,1,1,1,1)$, a static solution is
in fact obtained if $\Lambda<0$ and the two brane tensions
$\rho_{vis,hid}=\Lambda_{vis,hid}$ are fine-tuned to the values
\be \Lambda_{vis} = - \Lambda_{hid} = \pm \sqrt{-6 \Lambda M_5^3}
\label{relationrs} \vv \ee so that the effective cosmological
constant in the 3-dimensional space exactly cancels. In this case
one finds \citep{Ran99a} $b(\tau,y)= b_0= const$ and \be a(\tau,y)
= \exp\left(\pm b_0 |y| \sqrt{\frac{-\Lambda}{6 M_5^3}} \right)
\equiv \exp\left(\pm b_0 |y| m \right) \pp \label{warpfactor} \ee
Choosing a negative value for $\Lambda_{vis}$, so that the
solution corresponds to Eq. (\ref{warpfactor}) taken with the
positive sign, leads to a nice solution of the hierarchy problem,
since the fundamental mass scale on the invisible brane $M_5\sim
M_P$ is $red-shifted$ on our visible universe by the conformal
factor $\exp\left(- b_0 m/2 \right)$, which can explain the large
relative ratio of Planck and electroweak scales for a moderate
value of $m b_0 \sim 10^2$. In other words, the non-trivial
dependence of the metric upon $y$ implies that the KK zero-mode of
the graviton wavefunction is peaked around the invisible brane and
has an overlap with the visible brane suppressed by the
exponential "warp" factor $\exp (-m b_0/2)$. Yet the KK tower mass
gap is potentially as low as the TeV scale, and thus these
graviton excitations can lead to testable effects at high energy
colliders as LHC, see e.g. \citep{Giu07} and references therein.

On the other hand RS also observed that choosing our brane with a
positive cosmological constant $\Lambda_{vis}>0$, though does not
solve the hierarchy problem, nevertheless it has the nice
properties of allowing for a non compact fifth-dimension, as one
can take the limit $y\rightarrow \infty$ maintaining consistency
with short-distance force experiments. This scenario is also the
one which is more interesting from the cosmological point of view,
as we will discuss soon.

\subsubsection{Brane cosmology and BBN}

Adding matter on the branes will lead to an evolving universe
analogous to the standard FLRW model. Depending on the choice of
the corresponding reference static solution one starts with, the
prediction for the Friedmann-like equation governing the visible
scale factor, i.e. the value of $a$ in the vicinity of our brane
$y=0$ can be significantly different, leading to testable
predictions for cosmological observables such as BBN, CMB and
structure formation.

The equation governing the evolution of $a(\tau,0)\equiv
a_0(\tau)$ has been worked out in \citep{Bin00a}. If $\rho$ and
$P$ denote energy density and pressure on our visible universe,
thus dropping the index $vis$ in the following, one obtains the
standard conservation equation, \be \dot{\rho} + 3 (\rho+P)
\frac{\dot{a}_0}{a_0} = 0 \vv  \label{cont5} \ee which leads to
the usual power behavior for $\rho \sim a_0^{-3(1+w)}$, as well as
the evolution equation for $a_0$,
\be
\frac{\ddot{a}_0}{a_0} + \left( \frac{\dot{a}_0}{a_0}
\right)^2 = - \frac{1}{36 M_5^6} \rho(\rho+ 3P) - \frac{1}{3 b_0^2
M_5^3} T_{bulk,55} \vv \label{friedmann5} \ee
where the time derivative is with respect to $t$, with $dt =
n(\tau,0) d \tau$ and a flat metric in the ordinary 3-dimensional
space has been assumed for simplicity. This expression shows two
remarkable properties, namely that it is independent of the energy
density and pressure of the second brane, a manifestation of the
local nature of Einstein theory and, furthermore, that the energy
density enters quadratically rather than linearly as in
conventional cosmology.

If the dynamics is dominated by the brane energy density, so that
one can neglect the last term in Eq. (\ref{friedmann5}), using Eq.
(\ref{cont5}) the second order equation (\ref{friedmann5}) can be
put in the form
\be
\frac{d}{dt} (\dot{a}_0^2 a_0^2) = \frac{1}{36 M_5^6}
\frac{d}{dt} ( \rho^2 a_0^4) \vv \ee which gives \be
H^2=\frac{1}{36 M_5^6} \rho^2 + \frac{\mathcal{C}}{a_0^4} \pp
\label{friedmann5bis}
\ee
The second term in the r.h.s of this expression depends upon the
free integration constant $\mathcal{C}$ and behaves as a radiation
term \citep{Bin00b}, thus its popular name of ``dark radiation",
though the sign of $\mathcal{C}$ can be also negative.

This result for the Hubble parameter strongly differs from the
usual Friedmann law, unless one consider the very special case of
a radiation dominated phase driven by a positive $\mathcal{C}$,
while for a negligible value of $\mathcal{C}$ one gets \be a_0(t)
\sim t^{1/(3+3w)} \vv \label{timea} \ee thus a slower expansion
rate compared to the standard result $a_0(t) \sim t^{2/(3+3w)}$.
If we assume that Eq. (\ref{friedmann5bis}) can be applied to the
present universe, writing the energy density as a fraction of the
critical density $\Omega \sim 1$, \be 1 = \Omega^2 \frac{H_0^2
M_P^4}{64 \pi^2 M_5^6} + \frac{\mathcal{C}}{H_0^2} \sim
\frac{\delta^2}{H_0^{-2}} + \frac{\mathcal{C}}{H_0^2} \vv \ee with
the radius of the fifth dimension, $\delta$ (see Eq.
(\ref{radius})), which thus should be of the order of the present
Hubble radius, $H_0^{-1}$ \footnote{Derivation of this result is
presented in a slightly different form in \citep{Bin00a}, where
the second order equation for the scale factor is used.}. This is
ruled out by observations (since we could then observe
five-dimensional gravity directly), as well as the possibility
that dark radiation provides the dominant contribution to the
expansion today.

Strong bounds on this model also come if we assume that it
describes the evolution of the universe at the earliest stage we
can probe in a quantitative manner, namely during BBN. A rough
constraint can be obtained by considering the different behavior
of the Hubble expansion parameter, which changes the neutron to
proton ratio freezing temperature, $T_D$, as well as the
time-temperature relationship in the temperature range from $T_D$
down to the deuterium formation at $T_N \sim 0.1$ MeV
\citep{Bin00a,Chu00}. The value of  n/p at $T_D$ is given by the
standard relation n/p$ = \exp(-\Delta m/T_D)$, with\footnote{We
consider the case $\mathcal{C}=0$. If dark radiation dominates at
BBN, it would be difficult to reconcile later evolution with, say,
CMB and structure formation data.} \be G_F^2 T_D^5 \sim
\frac{\rho^2}{6 M_5^3} \pp \ee On the other hand, the
time-temperature relationship during a radiation dominated epoch
is given by Eq. (\ref{timea}) with $w=1/3$, $t \sim T^{-4}$.
Inserting numerical values and using the expression of the
relativistic degrees of freedom, $g_*$, during BBN one gets \be
T_D \sim 7.5 \left(\frac{\textrm{TeV}}{M_5}\right)^3
\,\textrm{MeV} \vv \ee and \be t \sim 2.8 \cdot 10^{-4}\,{\rm s}
\left(\frac{M_5 }{ \textrm{TeV}}\right)^3
\left(\frac{T}{\textrm{MeV}}\right)^4 \pp \ee Thus, assuming that
all neutrons at $T_N$ are eventually burned into $^4$He nuclei, we
have \be \frac{\rm n}{\rm p} (T_N) \sim \frac{Y_p}{2-Y_p} \sim
\exp\left[-\frac{\Delta m}{7.5
\textrm{MeV}}\left(\frac{M_5}{\textrm{TeV}} \right)^3\right]
\,\e^{-\frac{t(T_N)- t(T_D)}{\tau_n}} \vv \ee which implies that a
correct mass fraction $Y_p \sim 0.25$ requires $M_5 \sim 8$ TeV
and a too large compactification radius.

It should be noted that considering more than one (flat) extra dimension
and an empty bulk could be potentially in better agreement with both BBN
and the requirement of sub-mm extra dimensions, but a careful analysis of
this scenario has not been considered in the literature in details,
evaluating the whole network of light nuclei produced during BBN, in
particular $^2$H and $^7$Li. We also mention that it has been pointed out
that in general, the presence of the KK tower of gravitons may lead to
overclosure of the universe, unless the highest temperature ever achieved
was of the order of MeV, thus with a severe impact on the whole BBN
scenario, as well as on the standard inflationary picture for early
production of perturbations \citep{Han01,Fai01}.

Interestingly, a much more promising brane cosmology can be
obtained by exploiting the RS model. Assuming that the
stress-energy tensor $T_{bulk}^{AB}$ corresponds to a cosmological
constant, $\rho_{bulk}=-P_{bulk}=-P_{bulk,5}$, it has been shown
in \citep{Bin00b} that one can integrate the (0,0) component of
Einstein's equations and obtain the generalized Friedmann equation
in the vicinity of the visible brane, \be
\frac{\dot{a}_0^2}{a_0^2} = \frac{1}{6 M_5^2} \rho_{bulk} +
\frac{1}{36 M_5^6} \rho_{vis}^2 + \frac{\mathcal{C}}{a_0^4} \pp
\label{newfriedmann} \ee This result holds independently of the
metric outside and in particular of the time evolution of the
scale factor $b$. If one assumes that the energy density
$\rho_{vis}$ can be decomposed as the sum of the contribution of
ordinary matter $\rho$ and a (positive) cosmological constant
$\Lambda_{vis}$, and the latter is fine-tuned as in the RS model,
see Eq. (\ref{relationrs}), one recovers a standard cosmology
\citep{Cli99,Csa99,Chu00,Bin00b}, \be \frac{\dot{a}_0^2}{a_0^2} =
\frac{\Lambda_{vis}}{18 M_5^6} \rho + \frac{1}{36 M_5^6} \rho^2 +
\frac{\mathcal{C}}{a_0^4} \vv \label{newfriedmann2} \ee if one
identifies \be 8 \pi G_N =\frac{\Lambda_{vis}}{6 M_5^6} \vv
\label{lambdavisgn} \ee and the limit $\Lambda_{vis} >> \rho$ is
assumed\footnote{The standard behavior is indeed recovered only in
the stronger limit $\rho << \Lambda_{vis}^2/M_P^4$ if the bulk
spacetime is not exactly anti-de Sitter \citep{Shi00}.}. Of
course, choosing the original RS proposal, which provides a
beautiful solution to the hierarchy problem but requires a
negative tension on the visible brane, one would obtain a negative
sign relative to the conventional Friedmann equation, so that the
visible brane behaves as an anti-gravity world \citep{Shi00}. In
particular, this implies that, as soon as the universe becomes
matter dominated, it would collapse on a scale of the order of the
matter-radiation equality time \citep{Csa99}.

From Eq. (\ref{newfriedmann2}) we see that adjusting the value of
$\Lambda_{vis}$ as in Eq. (\ref{lambdavisgn}) for each given
$M_5$, the Hubble rate depends on two new parameters, the
fundamental scale $M_5$ which controls the quadratic term in
$\rho$ and the dark radiation constant $\mathcal{C}$. Bounds on
both these parameters can be obtained from BBN, and have been
discussed by several authors using the simple argument on $^4$He
mass fraction described above \citep{Csa99,Cli99,Bin00b, Maa00,
Lan01, Miz01,Bar02a,Fla06}, namely requiring that both terms be
sufficiently small that an acceptable value for $Y_p$ be produced.
A more careful analysis based on a full numerical integration of
the BBN dynamics and considering the predictions for $^2$H and
$^7$Li as well, has been instead performed in
\citep{Ich02a,Bra02}. In particular \citep{Ich02a} only consider
the dark radiation term and its effect on both BBN and CMB,
neglecting the effect of linear perturbations of dark radiation
during photon decoupling and recombination. The result of
\citep{Ich02a} (and of \citep{Bra02} in the limit of large $M_5$)
can be easily translated in terms of the well-known bound on the
effective number of neutrino species (see Section
\ref{sec:bbn_overview}),
\be \mathcal{C}= \frac{8 \pi}{3} G_N \Delta \neff \rho_{\nu,0} a^4
\pp \ee
The allowed range for $\mathcal{C}$ can be easily obtained from
the result on $\neff$ of Section \ref{sec:bbn_overview}. More
interestingly, if the value of $M_5$ is sufficiently low, the
effect of the $\rho^2$ term can be non-negligible, but it can be
compensated by a large and negative value of $\mathcal{C}$. A
degeneracy is thus expected in the $M_5-\mathcal{C}$ plane which
qualitatively is of the form $a/M_5^6 + \Delta \neff =const$, see
Figure \ref{fig1_Bra02}.
\begin{figure}
\begin{center}
\includegraphics[width=0.7\textwidth]{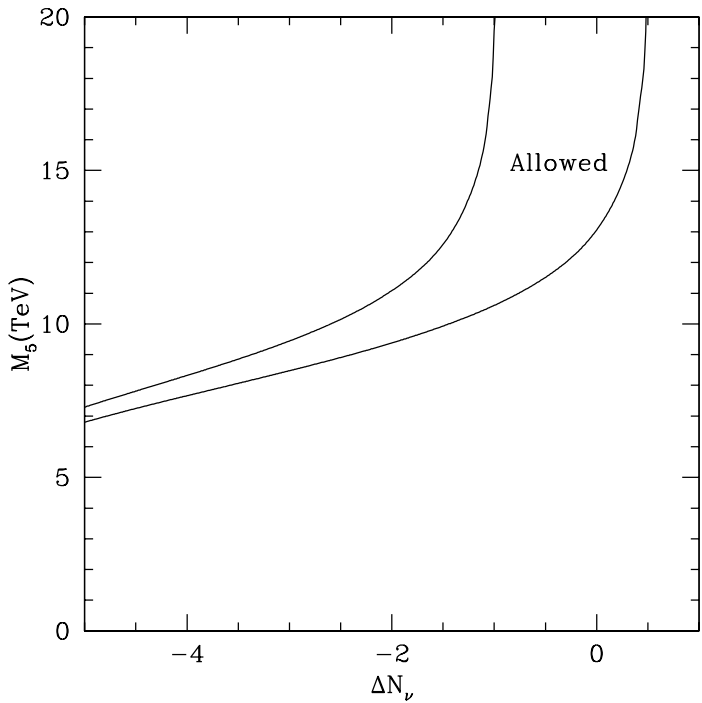}
\end{center}
\caption{The BBN-allowed range in the $M_5-\mathcal{C}$ plane.
From \citep{Bra02}.}
\label{fig1_Bra02}
\end{figure}

We have assumed in the previous discussion that the size of the
extra dimensional space is stabilized by some dynamical mechanism,
as for example discussed in \citep{Gol99}, but indeed it might
have some dynamics during cosmological epochs. In general, for a
homogeneous and isotropic model, the evolution of the extra
dimensions is controlled by a single scalar field, the radion,
with a canonical kinetic term which interacts with ordinary
matter, see e.g. \citep{Li06}. These interaction terms lead to a
time dependent behavior of the Higgs vacuum expectation value,
$v$, which in turn affects fermion masses. The BBN is extremely
sensitive to such variations, since changing $v$ produces a
different Fermi coupling constant, $G_F$, and neutron proton mass
difference, both entering the determination of the freezing
temperature, $T_D$, as well as shifting the pion mass which
influences the nucleon potential and the deuterium binding energy.
We will discuss this issue in details in the next Section.

Finally, BBN can be also used to put constraints on theories which
consider the possibility of bulk neutrinos. As we have mentioned,
in the extra dimension scenarios the SMPP particles are assumed to
be localized on the visible brane, while gravitational
interactions can propagate in the bulk. If the fundamental scale
is of the order of TeV, this leads to a serious problem in
understanding the smallness of neutrino masses, which is usually
thought to be produced via a see-saw mechanism. For sub-eV
neutrino masses, this scheme requires the existence of some new
mass scale of order $10^{11}-10^{12}$ GeV, much larger than the
extra dimension scale. Furthermore, operators as $LHLH/M_D$ ($H$
and $L$ are the Higgs and left-handed fermion doublets,
respectively) could be induced in the low energy Lagrangian which
lead to an unacceptable neutrino mass. This problem was realized
quite early on and several solutions have been proposed
\citep{Die99,Moh99,Moh00,Ark01,Cal01a,Cal01b}, which postulate the
existence of one or more gauge singlet neutrinos in the bulk which
couple to the lepton doublet on the brane. Their corresponding KK
modes give rise to an infinite tower of sterile neutrinos labeled
by an integer $n$, which mix with active neutrinos and have masses
typically of the order of $m_n = n \, 10^{-3}$ eV for an extra
dimension size of mm. This mixing has a relevant effect on both
solar and atmospheric neutrino phenomenology since for each mode
with mass larger than the three active neutrino with Dirac mass
$\mu_i$ ($i=1,2,3$) there will be a corresponding vacuum mixing
angle $\theta \sim \mu_i/m_n$ (see e.g. \citep{Lam01} for a review
and references therein).

Sizeable effects are also expected in cosmology
\citep{Bar00,Aba03b,Goh02}. In particular, KK modes of bulk
neutrinos can be produced in the early universe before BBN by
incoherent scatterings or coherent oscillations, thus contributing
to the total radiation content parameterized by $\neff$. Bounds on
this parameter from BBN amounts to require that the whole tower of
modes should be equivalent to no more than approximately one
active neutrino species. Furthermore, the photoproduction of $^2$H
and $^6$Li by decays of modes after BBN may potentially spoil the
whole standard nuclei production scenario. These effects limit the
possible values for the extra dimension length scale $\delta$ as
function of the Dirac active neutrino mass, see e.g. Figure
\ref{fig1_Aba03} (from \citep{Aba03b}). For illustration, for a
neutrino mass of order $0.1$ eV \citep{Aba03b} find from BBN $0.01
\textrm{MeV} \leq \delta^{-1} \leq 10^3 \textrm{MeV}$.
\begin{figure}
\begin{center}
\includegraphics[width=0.8\textwidth]{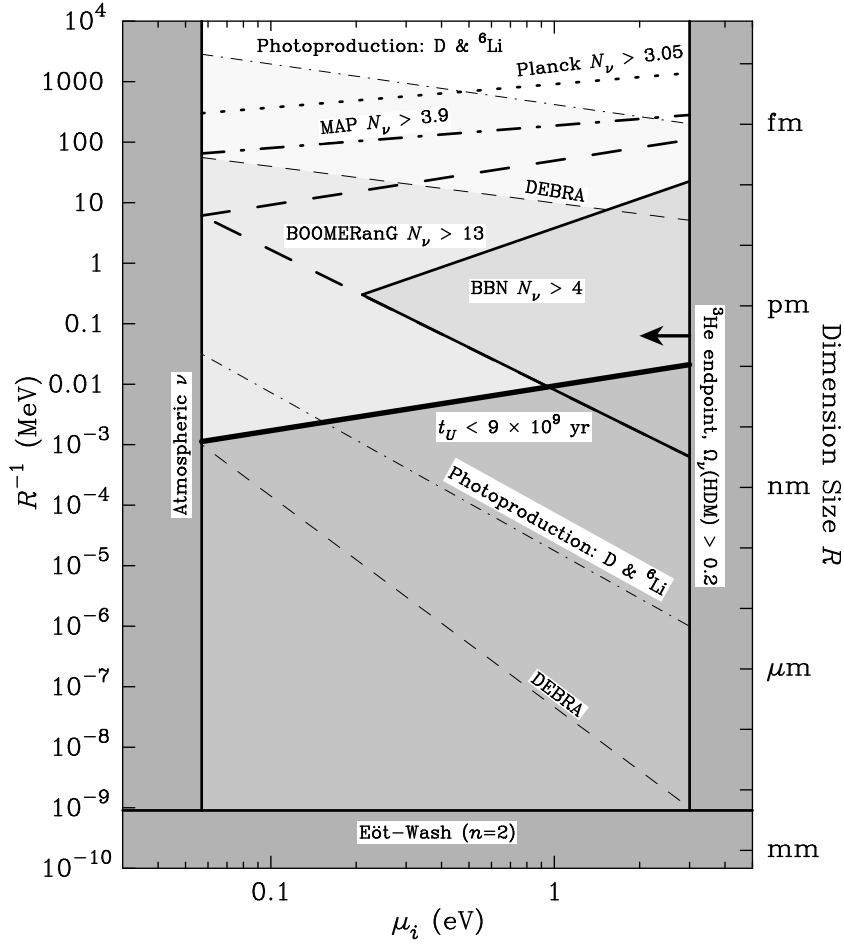}
\end{center}
\caption{Constraints on the extra dimension length scale R versus
active neutrino masses $\mu_i$ for bulk neutrino scenarios. From
\citep{Aba03b}.}
\label{fig1_Aba03}
\end{figure}
%
\subsection{Variation of fundamental constants}

\subsubsection{Introductory remarks}
\label{s:genremarks}
Already in 1937 P.A.M. Dirac first introduced the idea that the
fundamental constants of physics may be indeed variable parameters
characterizing the particular state of the universe \citep{Dir37}.
Physicists have long scrutinized this possibility. On one hand, a
strong effort has been devoted to embed this paradigm into a
definite theoretical framework. For example, theories with extra
dimensions such as Kaluza-Klein or string theories, naturally
predict that 4-dimensional constants may vary (in time and space),
since they represent effective values in the low energy limit, and
are sensitive to the size and structures of extra dimensions. Any
variation of these invisible dimensions, for example over
cosmological times, would lead to varying 4-dimensional constants.
On the other hand, measuring variations of fundamental constants
has been pursued at the experimental level by several groups and
techniques, ranging from short time laboratory based measurements,
to astronomical or geological scale studies and, finally, to
cosmological time variation searches. All these investigations
provided quite strong constraints on the possible time evolution
of e.g. the fine structure coupling, $\alpha$, or the Newton
gravitational constant, $G_N$. Both theoretical and experimental
aspects of this intriguing research issue are beautifully covered
in the review \citep{Uza03}.

In the following we will summarize the impact that fundamental
constant time variations have on BBN, by describing the main
effects on the light nuclei abundances and the bounds which
therefore is possible to obtain using BBN as a probe.
Unless otherwise stated, we will limit to constrain the
departure of these parameters at the BBN epoch compared to
present values, independent of a specific theoretical model for
their evolution (this is instead the approach considered e.g. in
\citep{Bar02b,San02,Sha05}.)
As a general
remark, we would like to stress that all results depend quite
strongly upon the general assumption which are made to perform the
analysis, the {\it priors} in presently fashionable  Bayesian
language. In fact,

1) there are too many constants which enter the BBN physics (the fine
structure coupling, the Newton constant, the strong interaction scale,
$\Lambda_{QCD}$, the Yukawa couplings, the Higgs vacuum expectation
value), so that if they are all considered as free independent parameters
to be fixed by data, one lacks predictive power;

2) there is no unique theoretical framework which allows for an
unambiguous determination of their relative evolutionary history.

For these reasons, there are two typical strategies which have
been exploited. One may assume that only a single fundamental
constant (or a subset of them) is promoted to the role of a free
parameter to be constrained, keeping all the others as fixed.
Several analyses of variation of the fine structure coupling,
which we describe in the next Section are based on this
assumption. On the other hand one can consider a specific
theoretical model, which reduces the number of independent
parameter one starts with, and this typically allows for tighter
constraints. Examples of this approach has been considered in the
last decade in details, based on dilaton inspired theories, or
rather on unified gauge theories, where all gauge coupling of the
SMPP get unified at some high mass scale $M_{GUT}$.

To conclude this short introduction, let us stress that checking
for a time (and space) variation of fundamental constants is only
meaningful for {\it adimensional} quantities, such as $\alpha$.
This is due to the fact that measurement of {\it dimensional}
parameters is strongly intertwined with both the system of units
and the particular measurement technique which is employed, so
that an absolute determination of, say, the time evolution of the
speed of light, $c$, is meaningless.

This can be illustrated with a simple example. Suppose we want to
measure the value of $c$ by using a {\it light clock device}, a
source $S$ and a mirror placed at a distance $D$ away from $S$,
which reflects the light ray back to $S$. The value of $c$ is then
computed as the ratio of the distance $2D$ over the total time
elapsed from emission to light collection, which is expressed in
terms of an adimensional number, {\it once} units of length L and
time T are specified. For example, we can choose an atomic clock
to specify T, using the hyperfine splitting transition rate of
caesium-133 atoms. In this case the particular combination of
fundamental parameters  $m_e^2 c^2 \alpha^4/m_p \hbar$ (the
typical hyperfine frequency) is kept fixed by definition, $m_p$
being the proton mass. If we choose the length unit L as the
distance traveled by light in $k$ units of time T, it is rather
trivial that there is no possibility to detect the time variation
of the speed of light, as also $c$ in this case is kept fixed by
definition. On the other hand we may use the standard prototype
platinum-iridium bar as the value of L, which depends on the
interatomic spacing of the material, and thus ultimately on the
value of the Bohr radius $a_B= \hbar/m_e c \alpha$. In this case
if we find that at different epochs the value of $c$ has changed
in these units, this amounts to say that the ratio \be
\frac{c}{L/T} \propto \frac{m_p}{m_e} \frac{1}{\alpha} \ee is time
dependent. Therefore, either $\alpha$ or the adimensional ratio
$m_p/m_e$ or both, change with time.

As this simple example shows, evidences for time varying fundamental
constants are in all cases evidences for particular combination of
adimensional quantities, which depends upon the particular choice of units
which is adopted. In the following, when time changes of dimensional
parameters are considered, a ratio of two independent and homogenous
constants will be always implicitly understood, as for example the ratio
$\Lambda_{QCD}/m_q$, with $m_q$ some quark mass, or $G_N M_{GUT}^2$.

\subsubsection{Varying the fine structure constant}
\label{s:alpha}
There are several direct measurements in the laboratory on the variation
of $\alpha$ over relatively short time period, using different techniques,
see e.g. \citep{Uza03}. The geological limit from the Oklo natural reactor
is about $|\delta \alpha/\alpha| \leq 10^{-8}$ over a period of few
billion years \citep{Dam96,Fuj00,Oli02}. Astrophysical observations of
high red-shift quasar absorption lines provide the only evidence for a
possible time variation of $\alpha$ \citep{Web99,Mur03,Mur04},
\be
\delta \alpha/\alpha = (-0.57 \pm 0.11) \cdot 10^{-5} \pp
\ee
A result compatible with zero is instead reported by \citep{Sri04,Cha04},
which however has been criticized in \citep{Mur07}.

Over longer time scales, the value of $\alpha$ can be constrained
by cosmological observables, such as CMB and BBN. Indeed, CMB
anisotropies are a very good probe of $\alpha$ since its value is
imprinted in the ionisation history of the universe. The main
effects are a change in the redshift of recombination due to a
change of hydrogen energy levels, the modification of the Thomson
scattering cross-section which is proportional to $\alpha^2$, and
at subleading level a change of $^4$He abundance
\citep{Ave00,Ave01,Mar02}.  If the value of $\alpha$ is increased,
the last scattering surface moves towards larger redshifts, which
correspond to a shift of first Doppler peak towards larger $l$'s.
Moreover, this shift produces a larger Integrated Sachs-Wolfe
effect (i.e. more power around the first peak), while the high
multipole diffusion damping is decreased by a larger $\alpha$,
thus increasing the power on very small scales. With present CMB
data, including those from the first year release from WMAP
Collaboration, one finds the bound $-0.06 \leq \delta
\alpha/\alpha \leq 0.02$ at 95$\%$ C.L., while a Fisher matrix
analysis shows that future experiments such as PLANCK should be
able to constrain variations of $\alpha$ during CMB formation with
an accuracy of $0.3 \%$ \citep{Roc04}.

The value of $\alpha$ enters the physics of primordial nucleosynthesis in
several ways and, remarkably, BBN represents the earliest reliable probe
of possible variation of the fine structure coupling over cosmological
times, though it suffers from being model dependent with respect to CMB
analysis. This issue was first studied in \citep{Kol86,Bar87,Cam95}, where
the focus was on the abundance of $^4$He, while a detailed analysis has
been presented mainly in \citep{Ber99} and \citep{Nol02}, which we follow
for our discussion.

During the early stages of BBN, at the n/p ratio freeze out
temperature, $T_D$, the fine structure coupling affects the weak
n-p rates in two ways. First of all, it changes the neutron-proton
mass difference $\Delta m$, which can be only phenomenologically
parameterized as in \citep{Gas82}, where the authors find \be
\Delta m (\textrm{MeV}) \sim 2.05 - 0.76 \left(1+\frac{\delta
\alpha}{\alpha} \right) \pp \label{massdiffnp} \ee This result is
obtained by studying the the behavior of the nucleon masses versus
$\alpha$, which determines the electromagnetic quark masses and
binding energy. The fact that this parametrization, though quite
reasonable, is not deduced from first principles in a QCD-based
calculation, is the main source of possible uncertainties and
renders all predictions model--dependent. Furthermore, weak rates
also depend upon $\alpha$ when QED radiative and thermal
corrections are included. However, these corrections are at the
level of few percent, see e.g. \citep{Esp99}, so the effect is
very small for moderate variation of $\alpha$, i.e. $\delta
\alpha/\alpha << 1$. Since the $^4$He mass fraction is mainly
sensitive to the n/p ratio at freeze out, while it depends more
weakly on the whole set of nuclear reaction rates, this implies
that at first approximation the whole dependence of $Y_p$ upon
$\alpha$ is through $\Delta m$, which also fixes the decoupling
temperature\footnote{Notice that the Fermi coupling constant $G_F$
does not depend upon the values of gauge couplings in the SMPP,
but on the Higgs vacuum expectation value only.}, since \be Y_p
\sim \frac{2}{1+ \exp(\Delta m / T_D(\Delta m))} \pp \ee

The dependence of other nuclei on $\alpha$ is more involved, and
one should scrutinize the way the fine structure constant appears
in the set of nuclear reaction rates. The leading effect is due to
the change of Coulomb barrier. Charged particle reactions at low
energies take place via tunneling through a repulsive potential,
and this produces most of the energy dependence in the
cross-section,
\be \sigma(E) = \frac{S(E)}{E} e^{-2 \pi \eta}\label{Sfacparam}
\vv
\ee
where $\eta = \alpha Z_1 Z_2 \sqrt{\mu/2E}$, with $Z_1$ and $Z_2$
the charges of the incoming nuclei and $\mu$ the reduced mass. If
one neglects the dependence on $\alpha$ of the astrophysical
$S$-factor and of reduced mass, as in \citep{Ber99}, then only the
Sommerfeld parameter $\eta$ would change linearly with $\alpha$,
and one can obtain the thermally averaged rates versus $\delta
\alpha/\alpha$, see Tables 1 and 2 of \citep{Ber99}, as well as
the fractional variation of light nuclei abundances. Not
surprisingly, the most dramatic changes are those of $^7$Li, due
to the strong Coulomb barrier in its production.
\begin{figure}
\begin{center}
\includegraphics[width=0.8\textwidth]{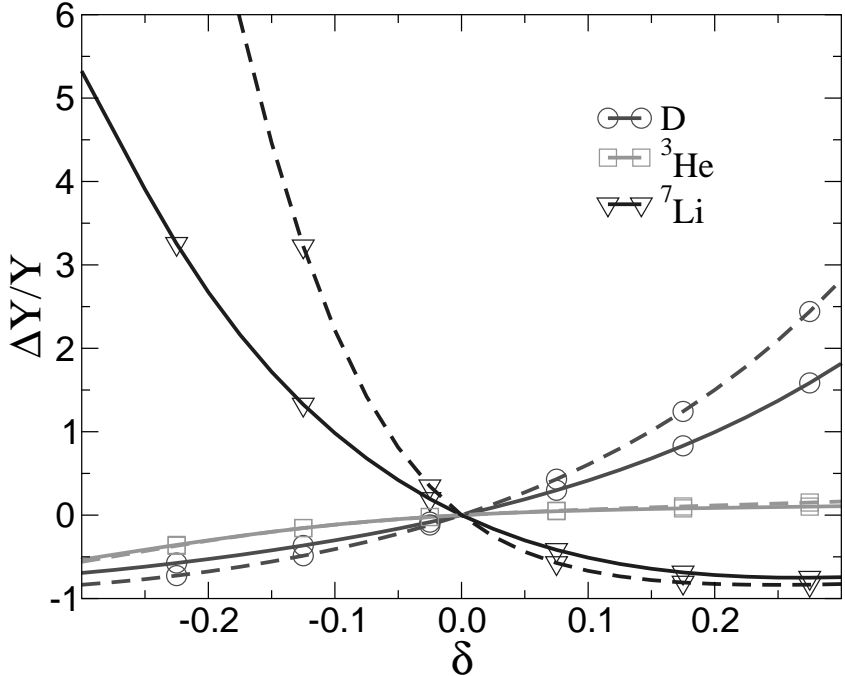}
\end{center}
\caption{The relative variation of $^2$H, $^3$He and $^7$Li produced during
BBN as a function of the variation of the fine structure coupling,
$\delta=\delta \alpha/\alpha$. From \citep{Nol02}.}
\label{fig2_Nol02}
\end{figure}

The analysis of \citep{Ber99} has been improved in \citep{Nol02},
see Figure \ref{fig2_Nol02}, by including several corrections
previously neglected, which accounts for the $\alpha$ dependence
of the $S$-factor, namely the normalization of initial-state
Coulomb penetrabilities, final state charged particle Coulomb
interactions, linear dependence on $\alpha$ of photon coupling to
nuclear currents in radiative processes, photon energy and barrier
penetration in external direct captures and, finally, the
electromagnetic contribution to the nuclear masses. All these
corrections can amount to a third of the total dependence of
$^7$Li on $\delta \alpha/\alpha$, while they are smaller for
$^3$He and $^2$H.

These results can be used to get bounds on $\delta \alpha/\alpha$
by comparing data with the results of a fully numerical
computation of light nuclei versus $\alpha$, using a modified
version of public BBN codes. For example \citep{Ave01} reports at
$95 \%$ C.L. $ \delta \alpha/\alpha = -0.007 \pm 0.009$, using the
experimental values of $^4$He, $Y_p=0.244 \pm 0.002$, and of
deuterium, $^2$H/H=$(3.0 \pm 0.4 )\cdot 10^{-5}$, while
\citep{Nol02} for the same experimental values obtain $\delta
\alpha/\alpha= -0.007^{+0.010}_{-0.017}$. If we use the $^2$H/H
and $Y_p$ of Section \ref{sec:obsabund} and the response matrix
formalism of \citep{Den07} we find the updated bound $-0.007 \leq
\delta \alpha/\alpha \leq 0.017$ at 95\% C.L.. It should be
noticed that a varying value of $\alpha$ does not significantly
improve the global fit of data when $^7$Li is also included in the
analysis. If one uses $\log_{10} (^7\textrm{Li}/\textrm{H}) =
-9.91^{+0.19}_{-0.13}$ \citep{Rya00}, combined with the deuterium
experimental result as above, one would find a reasonable
agreement for $\delta \alpha /\alpha \sim 0.23$ (the standard
value  being excluded at more than 3-$\sigma$), but including
$^4$He gives a minimum $\chi^2$ very far from this value but with
a very low likelihood \citep{Nol02}. Stated differently, it seems
difficult that the problem of the lower observed $^7$Li with
respect to theoretical expectation could be alleviated by allowing
a different value of $\alpha$ at the BBN epoch, unless one also
includes $\neff$ as a free parameter in the analysis. In this
case, a larger value of $\alpha$, which suppresses the theoretical
value of $^7$Li, and a slower expansion rate at BBN, $\neff<3$
(which is degenerate with $\alpha$ and can compensate for the
otherwise too large amount of $^4$He for positive $\delta
\alpha/\alpha$) can reasonably fit all nuclear abundances
\citep{Ich04}.

\subsubsection{The role of Higgs vacuum expectation value, fermion masses and $\Lambda_{QCD}$}
\label{s:v}
In the SMPP the vacuum expectation value, $v$, of the Higgs field provides
the mass term to vector bosons $W^\pm$ and $Z$ via the Higgs-Englert-Brout
mechanism, as well as to all fermions via Yukawa couplings. Any variation
of the weak scale, $v$, or better to say of the ratio of the weak to
strong scales, $v/\Lambda_{QCD}$, and weak to gravitational scale over
cosmological times can therefore, have a dramatic impact since it induces
a time dependence of all massive particles after spontaneous symmetry
breaking of the electroweak symmetry.

The main effects on BBN are the change of the Fermi coupling
constant, the neutron-proton mass difference, the electron mass
and finally, the binding energy of deuterium
\citep{Dix88,Sch93,Ich02b,Yoo03}. The change of $G_F=1/\sqrt{2}
v^2$ re-scales all weak reactions which keep neutron and proton in
equilibrium, and thus shifts the decoupling temperature, $T_D$.
Similarly, these rates are also affected by the variation of the
neutron-proton mass difference, due to the $u$ and $d$ quark mass
contribution to $\Delta m$. This contribution is proportional to
the difference between the two corresponding Yukawa couplings and
can be estimated once the electromagnetic contribution to $\Delta
m$ is singled out. Since the latter can be calculated with
relatively little uncertainty, being the Born term of the
Cottingham formula \citep{Gas82}, one obtains, see Eq.
(\ref{massdiffnp}), \be \Delta m (\textrm{MeV})= 1.29 + 2.053
\frac{\delta v}{v} \vv \ee with the assumption that in this case
only $v$ is considered as a free parameter, while both $\alpha$
and the Yukawa couplings are fixed to their standard values. This
estimate is consistent with recent lattice QCD calculation with
dynamical quarks \citep{Bea07}. Finally, the electron mass is
shifted linearly with $v$, $\delta m_e/m_e = \delta v/v$. The
changes in $G_F$, $\Delta m$ and $m_e$ fixes the new value of
$\Delta m /T_D$, which is a key parameter for $Y_p$, as we
stressed already several times\footnote{The change in electron
mass also produces a different value for the $e^+-e^-$ energy
density, and thus of the Hubble parameter, but the effect is
sub-leading \citep{Yoo03}.}. For example, if $v$ is increased, the
lower value of $G_F$ and the larger $m_e$ (which reduces the
available phase space) implies a less efficient n-p rates, so that
their freeze out occurs earlier (more $^4$He), but the larger
effect is due to the increased $\Delta m$, which reduces the n/p
ratio, and therefore the final value of $Y_p$.

Finally, changing $v$ alters the deuteron binding energy
$B_D$\footnote{Regardless of its expression in terms of
fundamental parameters, the effect of changing $B_D$ has been
studied for example in \citep{Dmi04}. Combining a full BBN data
analysis, which also includes $^7$Li abundance with the WMAP prior
on $\eta$, they find $\delta B_D/B_D =-0.019\pm 0.005$, and point
out that this 4-$\sigma$ shift of the D binding energy might
reconcile the whole BBN picture with data, in particular the low
value of observed $^7$Li.} through the change in the pion mass
$m_\pi$. Since the pion is a Goldstone boson, its mass scales as a
geometric mean between weak and strong scales as found by
Gell-Mann, Oakes and Renner \citep{Gel68}, $m_\pi^2 \sim (m_u+m_d)
\Lambda_{QCD} \propto v$. It is worth noticing that a change of
$B_D$ can be induced by a variation of $v$, as we are presently
discussing, or a time evolving values of Yukawa couplings, $y_i$,
with $i=u,d$, or finally by a change of the strong interaction
scale, $\Lambda_{QCD}$, which we consider later on, \be
\frac{\delta m_\pi}{m_\pi}= \frac{1}{2}\left( \frac{\delta
\Lambda_{QCD}}{\Lambda_{QCD}} + \frac{\delta v}{v} +\frac{\delta
y_u+ \delta y_d}{y_u+y_d}\right) \pp \ee

Therefore, the effects which we now describe of a modified value
of $B_D$ will be relevant when any of these parameters is assumed
to be time-dependant.

The value of $B_D$ fixes both the initial conditions for BBN,
given by the Nuclear Statistical Equilibrium, as well as the
cross-sections that burn deuterium into heavier nuclei. As $B_D$
increases, deuterium becomes more stable, and BBN would start at a
higher temperature epoch $T_N$. Interaction rates of the BBN
network occur more rapidly at higher temperature (though the
cross-sections have changed) leading to a more efficient $^2$H
processing. One thus expect in this case a decreased $^2$H/H ratio
and an increased $Y_p$ \citep{Kne03b,Dmi04}, see also Table I in
\citep{Den07}.

Static properties of nuclei and, in particular, binding energies
depend strongly on $m_\pi$, which sets the length scale of
attractive nuclear forces, and gives the dominant contribution to
two- and three-body potential via the one and two pion--exchange
terms. Though accurate calculations have been performed which
reproduce several experimental properties
\citep{Pud97,Pie01,Fla07}, a numerical determination of the
functional dependence of binding energies upon $m_\pi$ is still
lacking. On the other hand, this dependence has been studied
extensively in the framework of low energy effective theory that
respect the approximate $SU(2)_L \otimes SU(2)_R$ of $QCD$
\citep{Bul97,Bea02,Bea03,Epe03}. Depending on the values of the
coefficients (fixed by data) which weight the s-wave four-nucleon
operator expansion of the lagrangian density, the result seems to
suggest that deuterium remains loosely bound for a wide range of
$m_\pi$, or rather that the value of $B_D$ strongly changes with
$m_\pi$ \citep{Bea03}. Despite the large uncertainties, the
results are compatible with a linear dependence for small
variations around the current value, \be B_D (\textrm{MeV}) = 2.22
\left( 1+r  \frac{\delta m_\pi}{m_\pi} \right) \vv \ee with $-10
\leq r \leq -6$ \citep{Bea03,Epe03}.

One further caveat is represented by the fact that the role of
strange quark mass in nuclear quantities such as binding energy is
not fully understood, being so close to the non--perturbative
scale, $\Lambda_{QCD}$ \citep{Fla02,Den07}. The sensitivity of
$B_D$ to $m_s$ has been estimated in \citep{Fla03}, by inspecting
the role of $\sigma$ meson mass to nuclear potential. They find
$\delta B_D/B_D \sim -17 \delta m_s/m_s$. Some authors have also
pointed out that large pion mass variation could also render
deuterium an unbound system, and discussed the values of $m_\pi$
which would lead to stable di-neutron (or di-proton) systems
\citep{Fla02,Den03,Kne04}. This would have dramatic consequences
on the standard BBN picture, leading to a stable or long--lived
$^8$Be, and by-passing the $A=5$ bottleneck through formation of
$^5$He. Both effects would produce a large enhancement of lithium
or metallicity production in the early universe.

Implementing all changes described so far in a BBN numerical code,
and assuming $v$ as a free parameter, one can typically constrain
$\delta v/v$ at the level of percent. In particular, one can
estimate $\delta Y_p \sim \delta v/v (\%)$. Using both $^2$H/H
with an error of 30$\%$ and a rather low value for $Y_p=0.238 \pm
0.005$, which seems presently disfavored, Yoo and Sherrer find
$-0.7 \% \leq \delta v/v \leq 2.0 \%$ \citep{Yoo03}.
Interestingly, a looser bound (10$\%$) is obtained from (pre-WMAP)
data on CMB, by considering the effect of a varying electron mass
in both the Thomson scattering cross-section and hydrogen binding
energy. We have updated the bound from BBN of \citep{Yoo03}, using
the data on $^2$H/H and $Y_p$ discussed in Section
\ref{sec:obsabund} and find $-0.005\leq \delta v /v  \leq 0.012$
at 95 $\%$ C.L.

The scenario with a varying strong interaction scale, $\Lambda_{QCD}$, can
be described in quite a complete analogy. In this case, assuming that the
weak scale is kept fixed, while $v/\Lambda_{QCD}$ is time dependent, the
effects are due to the change of neutron-proton mass difference, since the
electromagnetic contribution is weighted by the strong scale,
\be
\Delta m (\textrm{MeV})= 1.29 - 0.76 \frac{\delta
\Lambda_{QCD}}{\Lambda_{QCD}} \vv
\ee
and the shift in deuterium and heavier nuclei binding energy
\citep{Fla02,Kne03b}. In particular, in \citep{Kne03b} results are
obtained by exploiting a simplified BBN network up to nuclei with A=3.
This introduces some systematics in the result, but it has the benefit of
reducing the dependence on $\Lambda_{QCD}$ only through the two parameters
$\Delta m$ and $B_D$. The result is a bound on $\Lambda_{QCD}$ at the
level of 10$\%$, if one corrects for such a systematic effect.

The effect of $^2$H binding energy on BBN is also considered in
\citep{Fla02}, where the authors quote $|\delta
\Lambda_{QCD}/\Lambda_{QCD}| \leq 0.06 $ as a conservative bound.
Interestingly, they also point out that even when both the strong scale
and quark masses are modified by the same amount, so that all binding
energies remain unchanged (and so does the reference temperature, $T_N$),
nevertheless the freezing temperature is changed, since it also involves
the Planck mass scale which is assumed to be fixed. In this case one
obtains a bound of the order of $(\delta \Lambda_{QCD}/ M_P ) /
(\Lambda_{QCD}/ M_P) \leq 0.1$ \citep{Fla02}.

\subsubsection{Correlated variation of fundamental constants in unified scenarios}

A summary of bounds on fundamental parameters considered in the
previous Sections is presented in Table
\ref{table:constrfundconst}. As we mentioned already, they refer
to a single parameter analysis, i.e. assuming that all but a
single fundamental constant are held fixed.
\begin{table}[t]
\caption{A summary of BBN constraints on fundamental parameters
discussed in Sections \ref{s:alpha} and \ref{s:v}.}
\label{table:constrfundconst}
\begin{tabular}{lccc}
\hline & \textrm{Data} & \textrm{Range} & \textrm{Ref.} \\
\hline $\delta \alpha/\alpha$ & $Y_p = 0.244 \pm 0.002$ & &\\ &
$^2$H/H $=(3.0 \pm 0.4)\times 10^{-5}$ & -0.016 $\div$ 0.002 (95 \% C.L.)& \citep{Ave01} \\
\hline $\delta \alpha/\alpha$ & $Y_p = 0.244 \pm 0.002$ & &\\ &
$^2$H/H
$=(3.0 \pm 0.4)\times 10^{-5}$ & -0.024 $\div$ 0.003 (95 \% C.L.)& \citep{Nol02} \\
\hline $\delta \alpha/\alpha$ & $Y_p = 0.250 \pm 0.003$ & & this paper\\
& $^2$H/H $= \left(2.87^{+0.022}_{-0.021}\right) \times 10^{-5}$ &
-0.015 $\div$
0.014 (95 \% C.L.) & (using \citep{Den07}) \\ \hline $\delta v/v$ & $Y_p = 0.238 \pm 0.005$ & & \\
& $^2$H/H $=\left(3.0^{+1.0}_{-0.5}\right) \times 10^{-5} $ &
-0.007 $\div$ 0.02 & \citep{Yoo03} \\ \hline  $\delta v/v$ & $Y_p
=
0.250 \pm 0.003$ & & \\
& $^2$H/H $= \left(2.87^{+0.022}_{-0.021}\right) \times 10^{-5}$ &
-0.005 $\div$ 0.012 (95 \% C.L.) & this paper \\ \hline $\delta
\Lambda_{QCD}/\Lambda_{QCD} $ & $Y_p = 0.238 \pm 0.005$
& & \\
& $^2$H/H $=(2.6 \pm 0.4)\times 10^{-5}$ & $\sim$ -0.1 $\div$
$\sim$ 0.1 & \citep{Kne03b}
\\ \hline
$\delta \Lambda_{QCD}/\Lambda_{QCD} $ &  $^2$H/H $= (1 \div
10) \times 10^{-5}$ & $\sim$ -0.06 $\div$ $\sim$ 0.06 & \citep{Fla02} \\ \hline \\
\end{tabular}
\end{table}
On general grounds, one could expect that this assumption is
rather ad hoc. For example, in models with extra dimensions, or
based on embedding of the SMPP into a larger gauge symmetry group,
it is quite natural that more than a single coupling or
fundamental scale, if not all, would be time dependent during the
(homogeneous) expanding history of the universe. When using BBN to
constrain such a variation (and CMB anisotropies to a lesser
extent as they are not sensitive to strong interactions), one
immediately faces the problem of several degeneracies among these
parameters. This results in a much less predictive power, unless a
specific theoretical framework is assumed in the analysis, which
reduces the number of free parameters of the theory, and relates
possible variations of different fundamental constants.

The purpose of this Section is to describe the impact on the BBN
of simultaneous variation of all involved fundamental parameters
and discuss the constraints on these variations which can be
obtained when some particular model is assumed (GUT theories,
string-theory inspired dilaton scenario).

The first steps in this programme are basically the following: to
identify those combination of fundamental parameters which
influence the BBN dynamics and quantify how light nuclei
abundances change when this parameters are left as a free input.
In general, this task is very involved, as large variation of,
say, the strong scale parameter, $\Lambda_{QCD}$, or quark masses
can change the standard picture of BBN in quite a dramatic way
(e.g. by rendering deuteron an unbound state or predicting bound
(pp) or (nn) states which may add new paths to nucleosynthesis).
On the other hand, it is easier to study the problem
perturbatively, in the neighborhood of the standard values that
all these couplings, mass scales, etc., take today as measured in
laboratory. A comprehensive analysis of this sort was lacking
until quite recently, and has been mainly pursued in a very
detailed study of Dent, Stern and Wetterich \citep{Den07} (see
also a semi-analytical {\it tour de force} of \citep{Lan06} in the
chiral limit of $QCD$). We will follow their approach in the
following, and choose to work in the units for which the strong
interaction scale is held fixed (see our discussion in Section
\ref{s:genremarks}). One can then identify the following set of
fundamental constants playing a role at BBN epoch, which we
collectively denote by $\varphi_k$, $k=1,...7$: the Newton
constant, $G_N$\footnote{The results of this Section have some
overlap with the analysis presented later of scenarios when one
varies the gravitational action, including a time-evolving Newton
constant.}, the fine structure coupling, $\alpha$, the Higgs
vacuum expectation value, $v$, the electron mass,
$m_e$\footnote{Differently than in our previous discussion, $m_e$
is assumed an independent parameter with respect to $v$. This is
equivalent to consider the lepton Yukawa couplings as varying
parameters.}, the light quark mass difference, $\Delta_q=m_d-m_u$,
the averaged light quark mass, $M_q=(m_d+m_u)/2 \propto m_\pi^2$
\citep{Den07}. Finally, the baryon density parameter is also
included. Notice that in this analysis variation of the strange
quark mass, $m_s/\Lambda_{QCD}$, is not accounted for.

The leading linear dependence of nuclear abundances X$_i$
($i=^2$H, $^3$H, $^4$He, $^6$Li, $^7$Li) upon small changes of
these parameters is then encoded in the response matrix $R$
defined as
\be
R_{ik}=\frac{\varphi_k}{\textrm{X}_i } \frac{\partial
\textrm{X}_i}{\partial \varphi_k} \pp
\ee
This matrix can be evaluated by numerically integrating the BBN equations.
In particular, in \citep{Den07} this has been performed in two steps, by
first varying a set of nuclear physics parameter $r_j$ which includes
binding energies of light nuclei up to $^7$Be, nucleon mass,
neutron--proton mass difference, neutron lifetime and a subset of the
$\varphi_k$, ($\alpha$, $m_e$, $G_N$ and $\eta$) and computing the matrix
\be
C_{ij}=\frac{r_j}{\textrm{X}_i } \frac{\partial \textrm{X}_i}{\partial
r_j} \vv
\ee
and then relating the variation of the $r_j$ to small changes of the
$\varphi_k$,
\be
F_{jk}=\frac{\varphi_k}{r_j } \frac{\partial r_j}{\partial \varphi_k} \pp
\ee
The response matrix $R$ is therefore given by the matrix product $R=CF$.
This decomposition turns out to be useful as the determination of $C$ is
rather robust, while computing $F$ requires some theoretical assumption,
as for example the dependence of binding energies on the pion mass.
\begin{table}[t]
\begin{center}
\caption{The response matrix $R$ for $^2$H, $^3$He, $^4$He, $^6$Li
and $^7$Li (from \citep{Den07}). The reference value for the
baryon fraction is $\Omega_B h^2=0.022$. \label{table:dent2007}}
\begin{tabular}{lccccc}
\hline & $^2$H & $^3$He & $^4$He & $^6$Li & $^7$Li \\ \hline \\
$G_N$ & 0.94 & 0.33 & 0.36 & 1.4 & -0.72 \\
$\alpha$ & 3.6 & 0.95 & 1.9 & 6.6 & -11 \\
$v$ & 1.6 & 0.60 & 2.9 & 5.5 & 1.7 \\
$m_e$ & 0.46 & 0.21 & 0.40 & 0.97 & -0.17  \\
$\Delta_q$ & -2.9 & -1.1 & -5.1 & -9.7 & -2.9 \\
$M_q$ & 17 & 5.0 & -2.7 & -6 & -61 \\
$\Omega_B h^2$ & -1.6 & -0.57 & 0.04 & -1.5 & 2.1\\ \hline \\
\end{tabular}
\end{center}
\end{table}
In Table \ref{table:dent2007} the entries for the $R$ matrix are
reported \citep{Den07}. The last row is the logarithmic variation
for small changes of the baryon fraction around the reference
value $\Omega_B h^2=0.022$. We have checked that the result is
very weakly dependent on the reference point in the range $0.015
\div 0.025$, the variation being of the order of $5\%$ for $^2$H
and even smaller for $^4$He.

Using these results one can construct the $\chi^2$ function (we do not use
$^7$Li in this analysis)
\be
\chi^2 = \frac{\left( ^2\textrm{H}^{(th)} (\varphi)-^2 \textrm{H}^{(exp)}
\right)^2}{\sigma^2_{^2 H}} + \frac{\left( Y_p^{(th)}(\varphi) -
Y_p^{(exp)} \right)^2}{\sigma^2_{Y_p}} \vv
\label{linearchi2}
\ee
where the theoretical abundances are computed using the linear expansion
in terms of $R$ (but we have retained an exact dependence on the baryon
fraction). In this way, one gets an estimate of the kind of constraints
which can be obtained on all $\varphi_j$, and more importantly, it serves
to illustrate the degeneracies which appear between the several pairs of
fundamental constants, at least as long as we consider small variations
with respect to the standard results, so that the linear expansion used in
Eq. (\ref{linearchi2}) is legitimate.

\begin{table*}[t]
\caption{The bidimensional 68 and 95 \% C.L. contours illustrating
the correlation of fundamental parameters in a BBN analysis
($^2$H/H and $Y_p$ as in Section \ref{sec:obsabund}).}
\label{degeneracies}
\begin{center}
\begin{tabular}{ccc}
\includegraphics[width=0.3\textwidth]{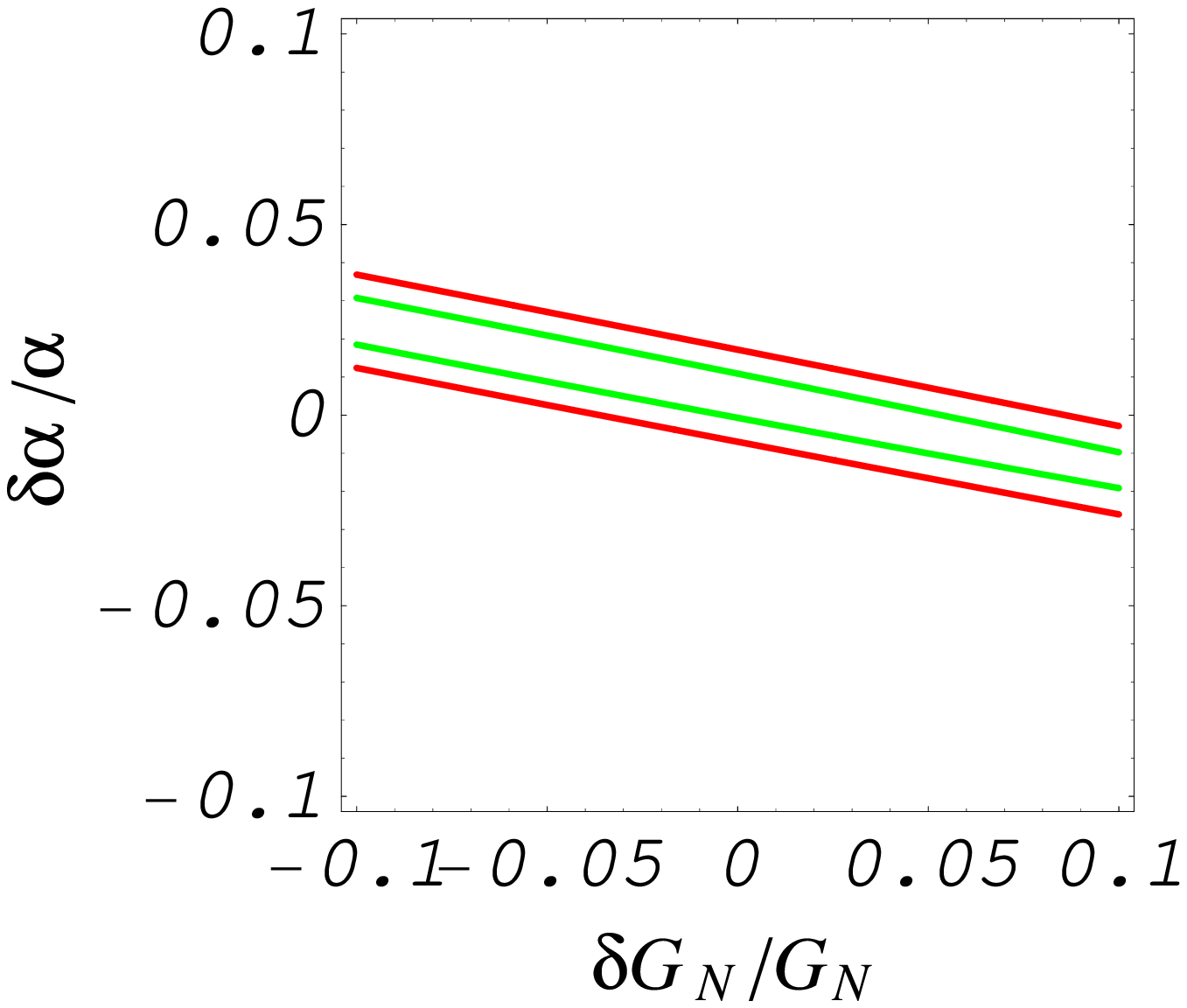}
&
\includegraphics[width=0.3\textwidth]{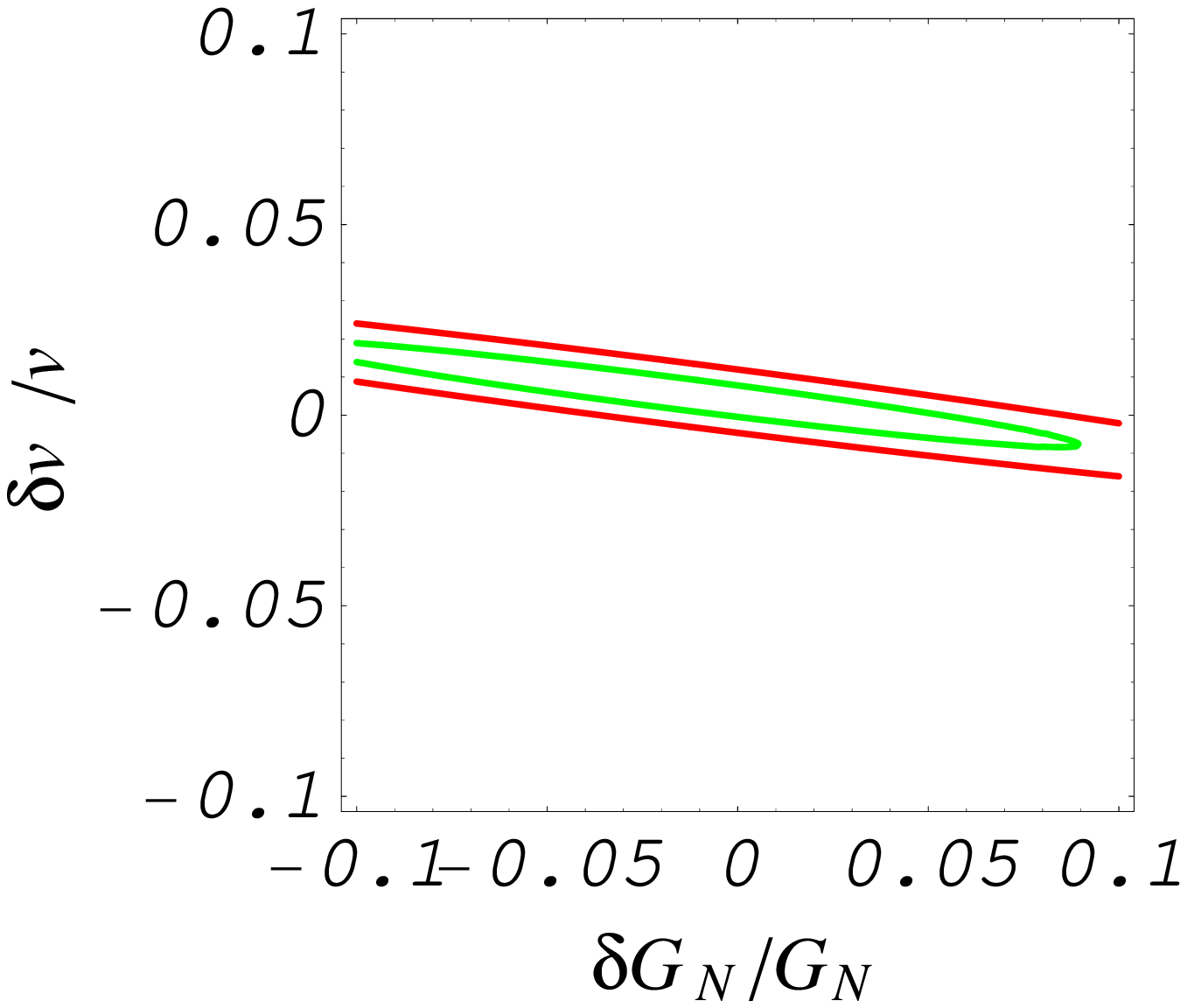}
&
\includegraphics[width=0.3\textwidth]{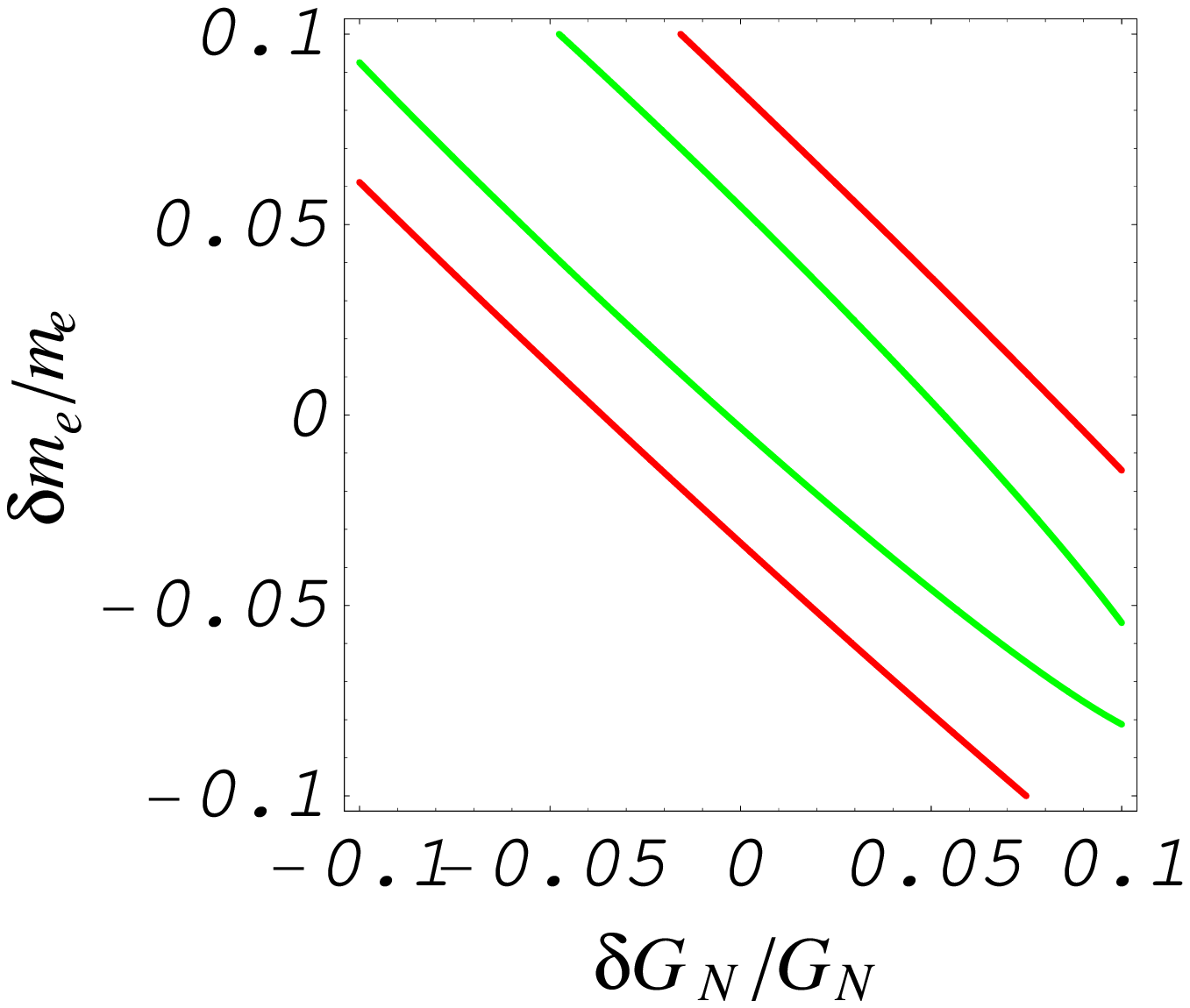}
\\
\includegraphics[width=0.3\textwidth]{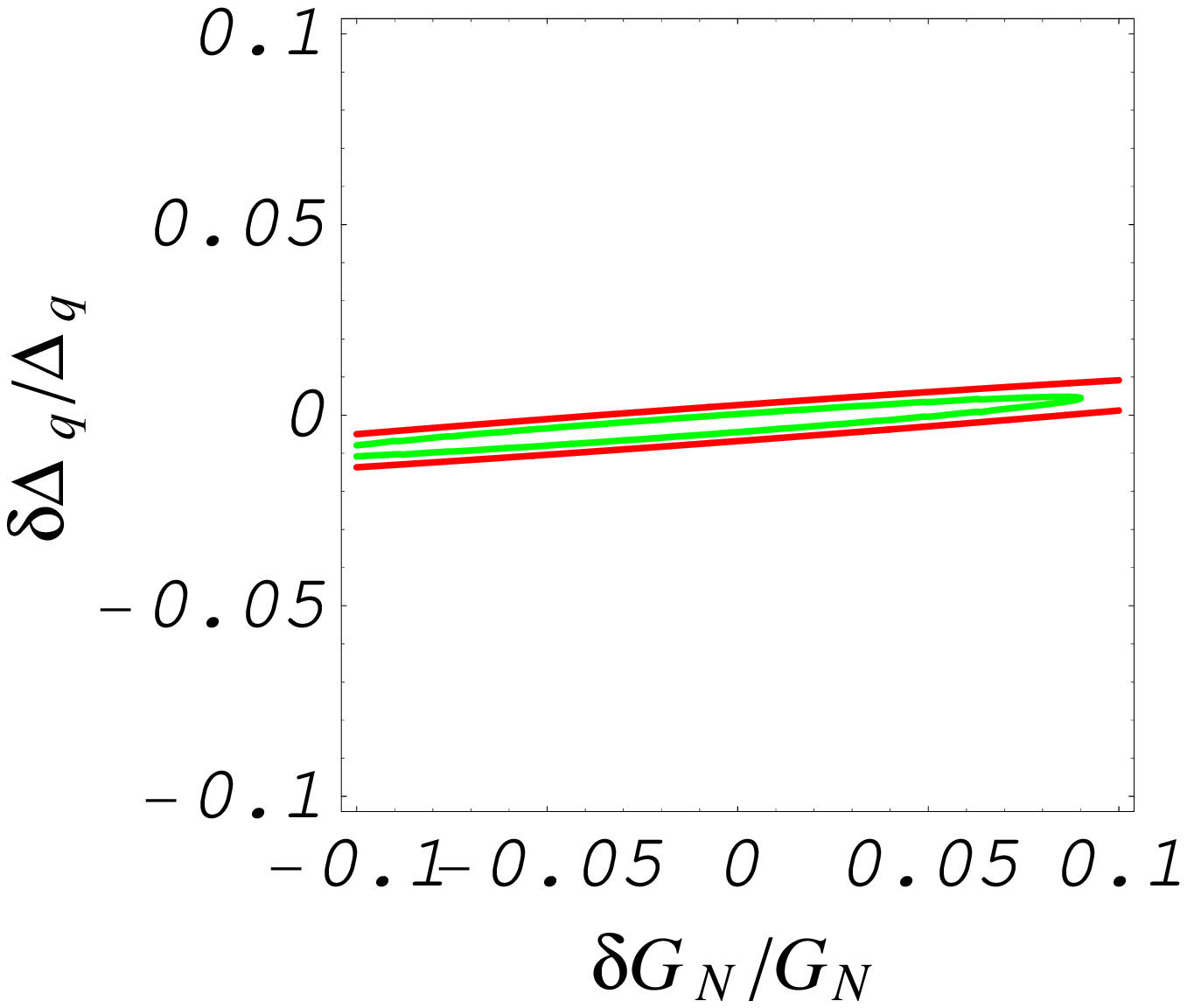}
&
\includegraphics[width=0.3\textwidth]{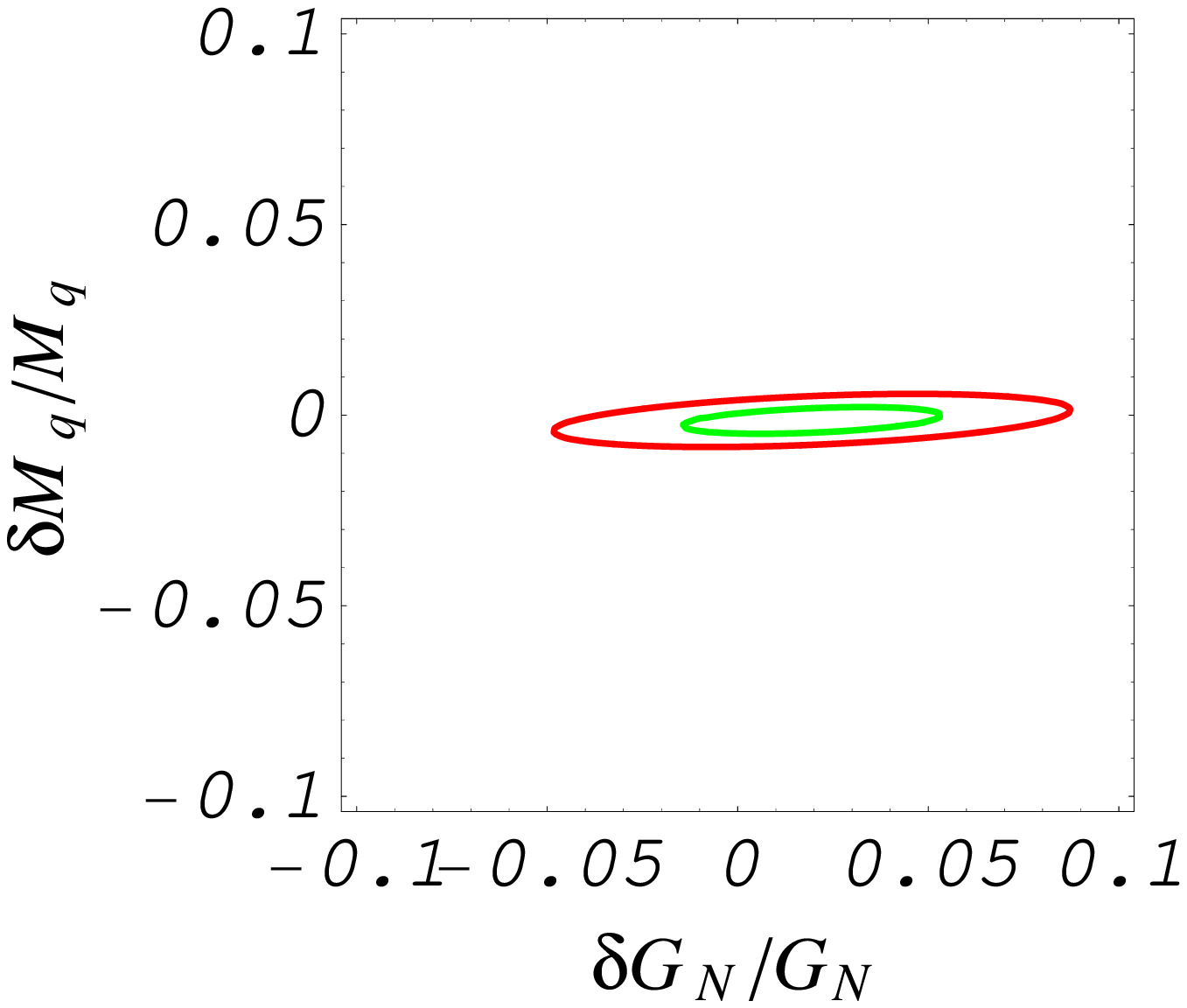}
&
\includegraphics[width=0.3\textwidth]{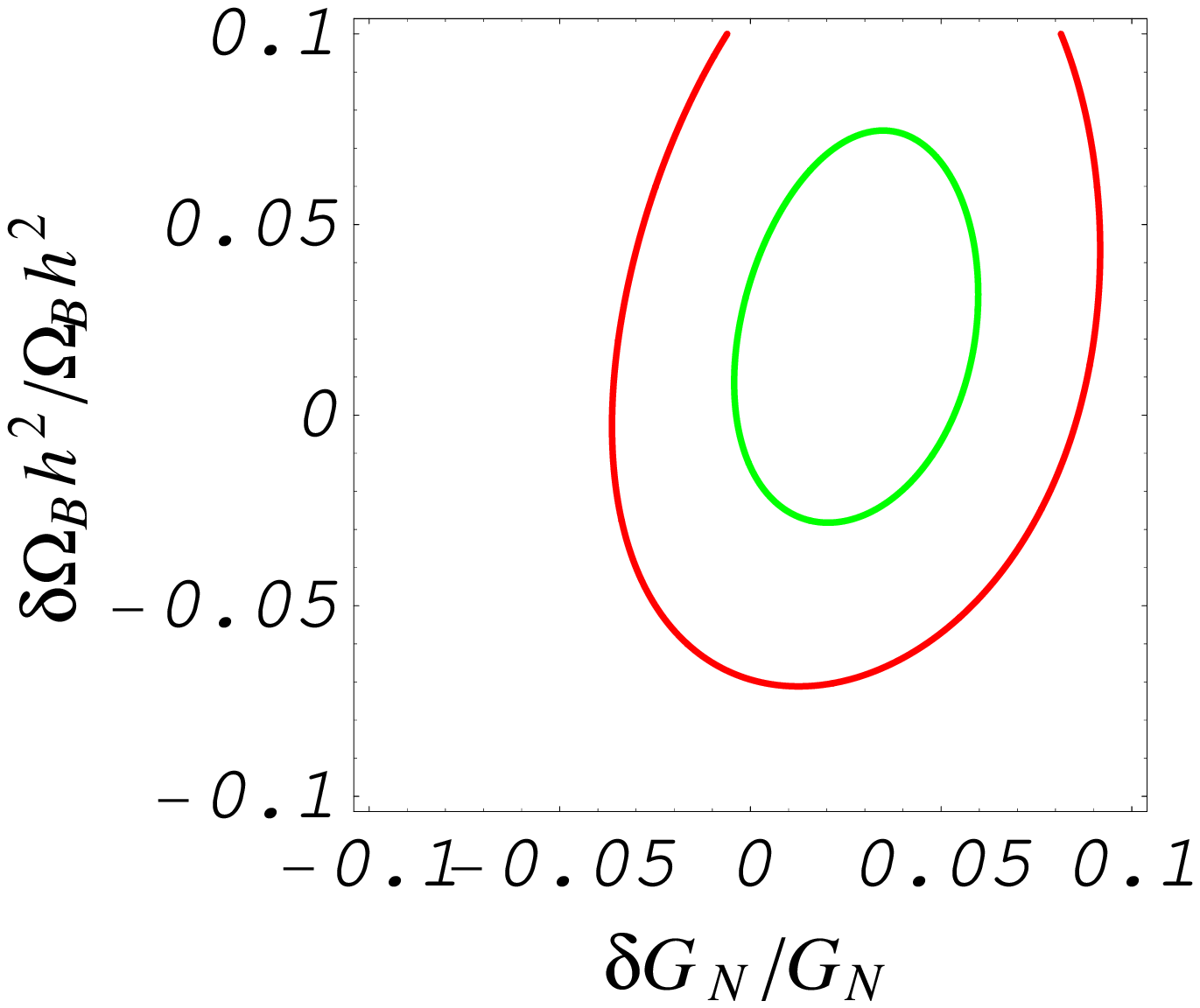}
\\
\includegraphics[width=0.3\textwidth]{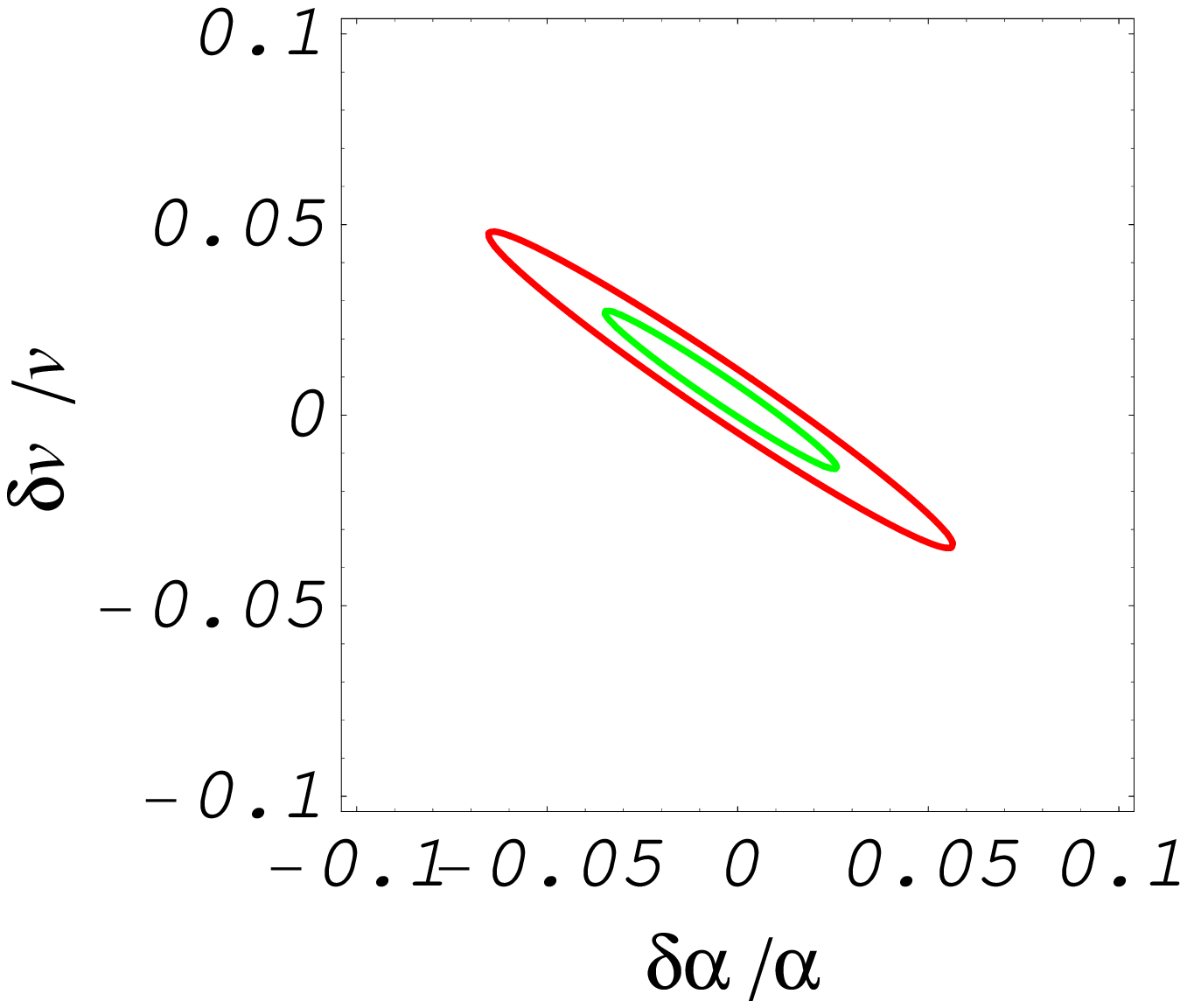}
&
\includegraphics[width=0.3\textwidth]{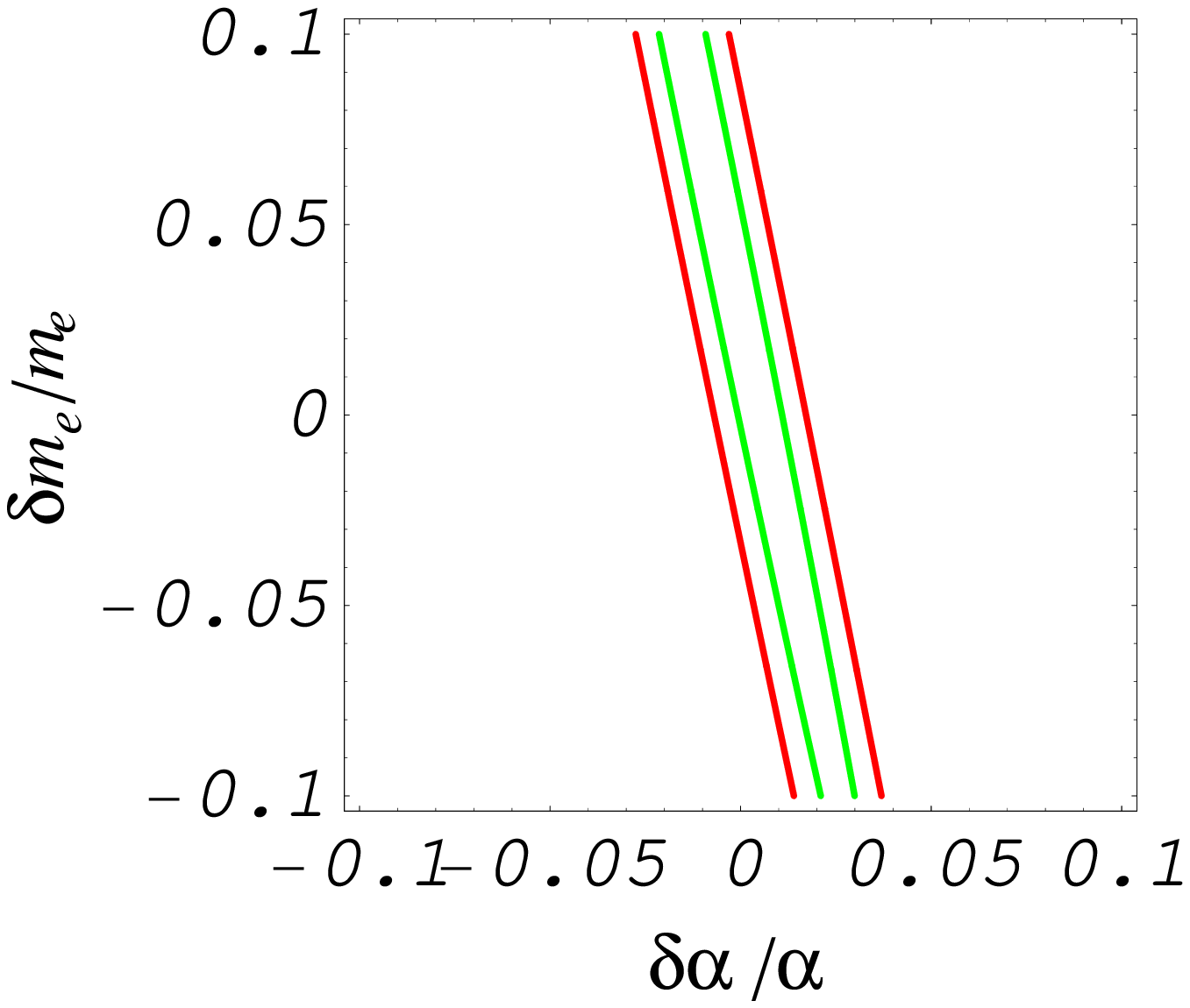}
&
\includegraphics[width=0.3\textwidth]{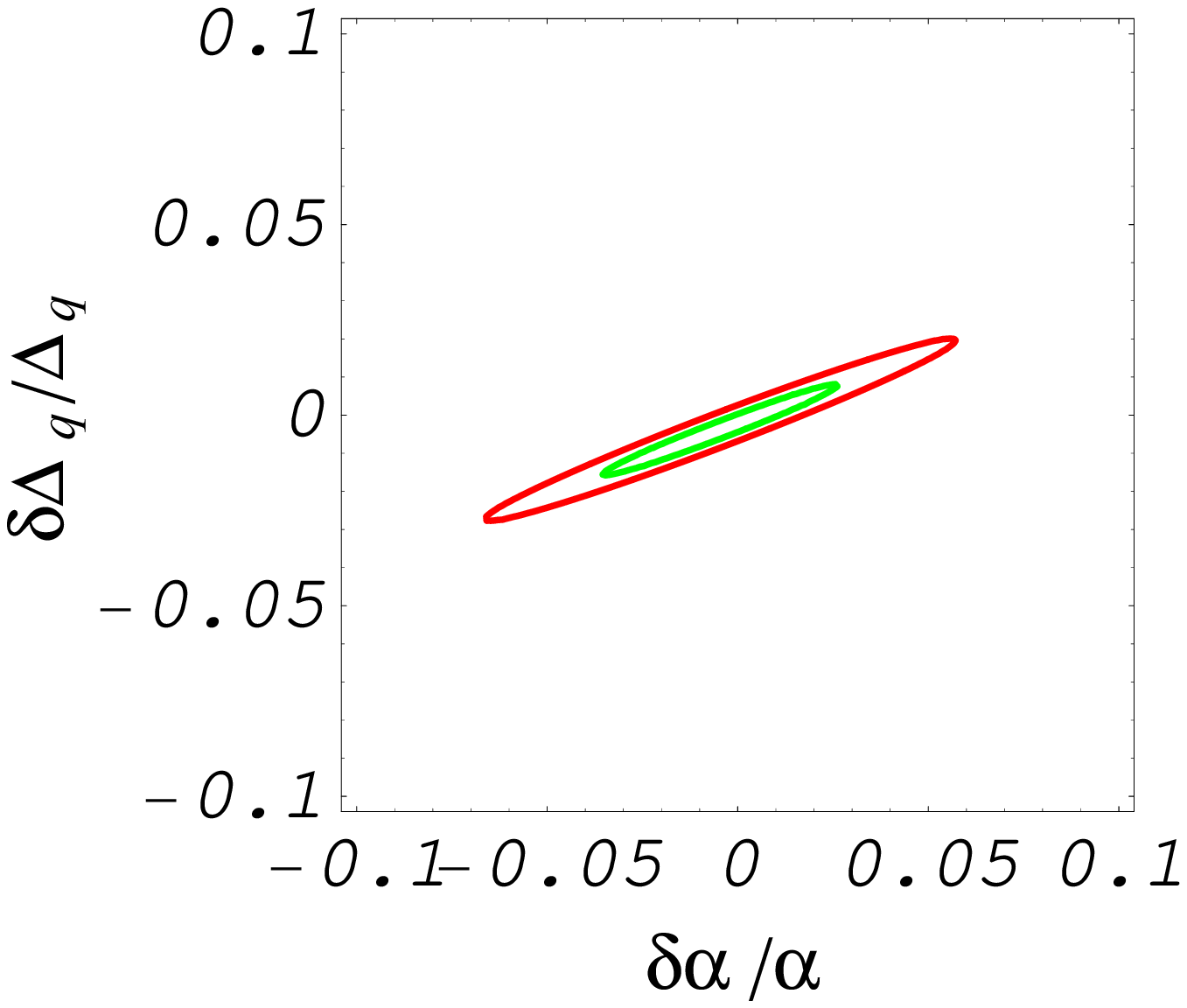}
\\
\includegraphics[width=0.3\textwidth]{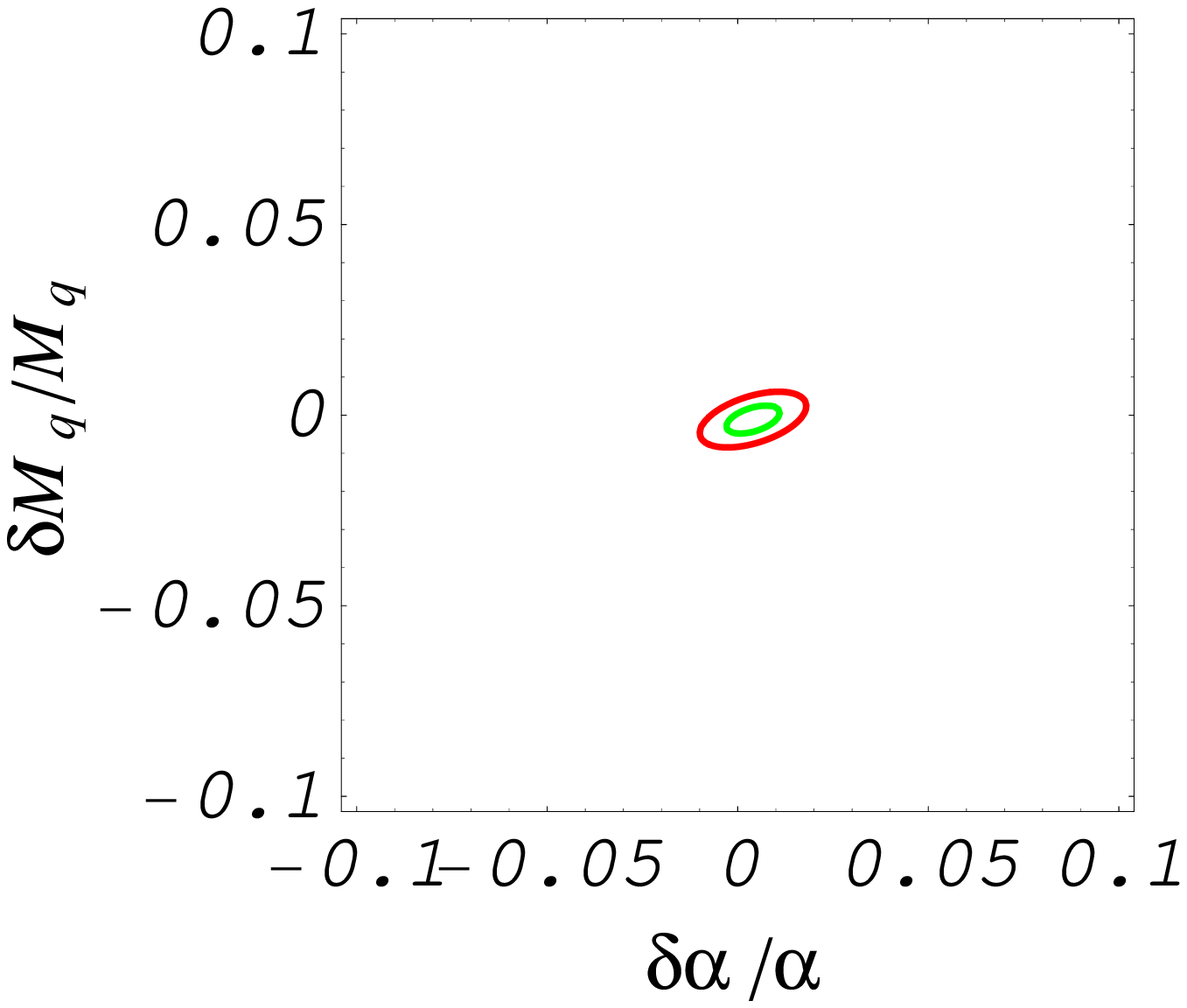}
&
\includegraphics[width=0.3\textwidth]{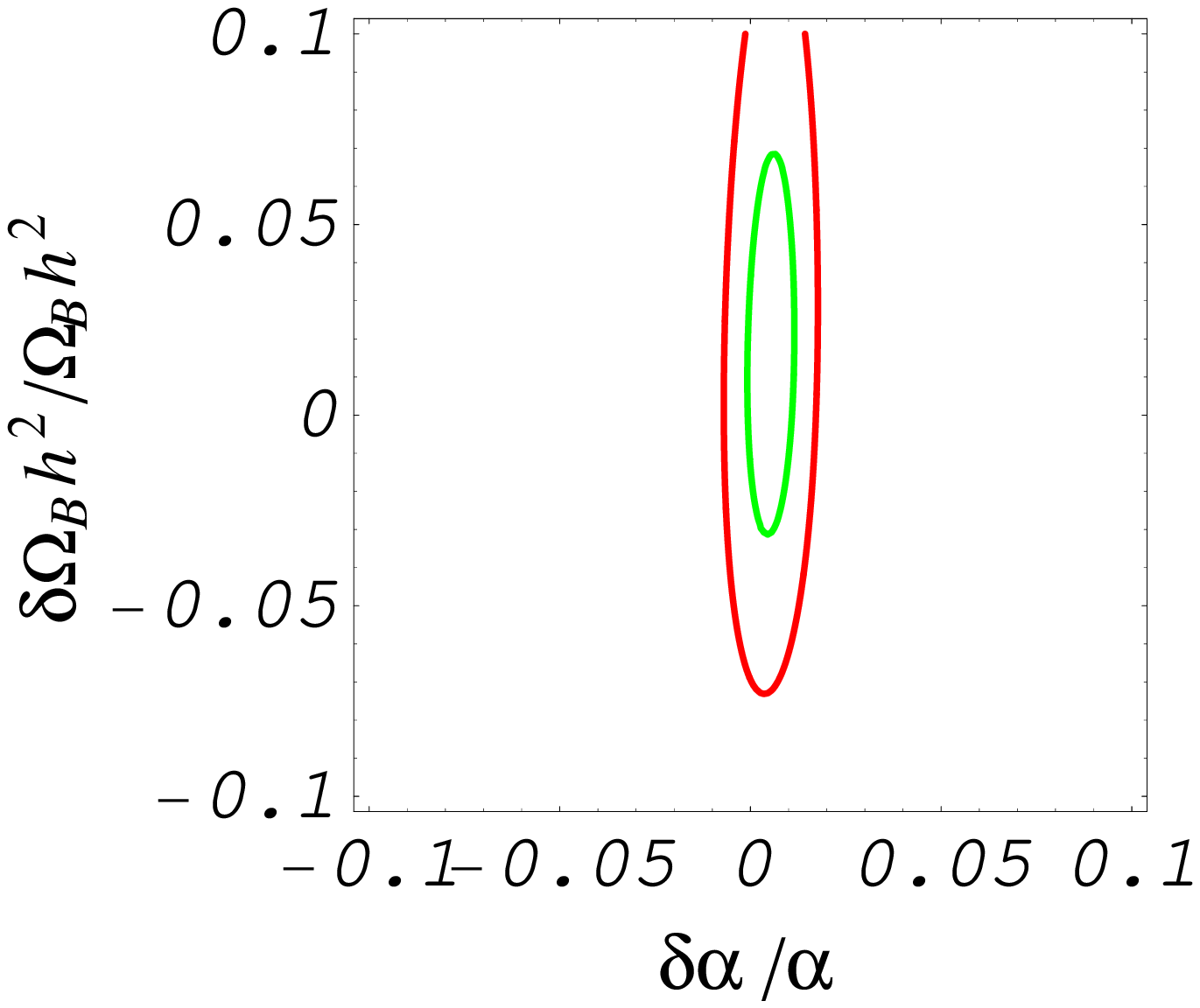}
&
\includegraphics[width=0.3\textwidth]{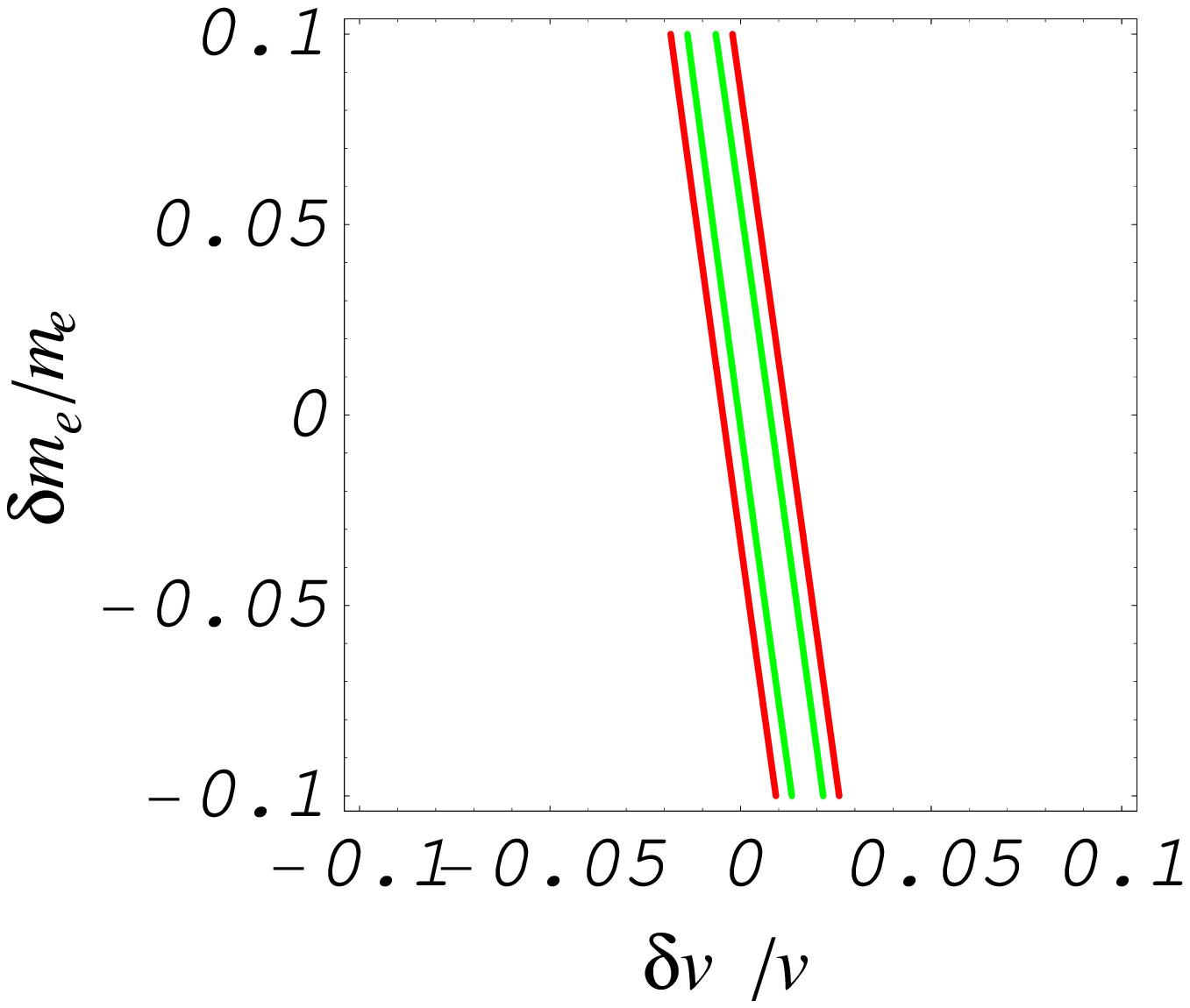}
\\
\end{tabular}
\end{center}
\end{table*}

As an example, we have considered the experimental values and
errors of Section \ref{sec:obsabund}. The results are shown in the
figures of Tables \ref{degeneracies} - \ref{degeneracies2}, where
the various curves correspond to a Fisher matrix analysis around
the reference model with $\Omega_B h^2 \sim 0.0209$ and all other
parameters at their standard values (this is indeed, locally, the
minimum of the total $\chi^2$). When one of the two selected
parameters is the baryon density, the plots correspond to the case
where only one fundamental constant is varied while all others are
fixed to the standard value. In this case, one can read from the
contours the typical order of magnitude of the constraint which
can be obtained in this single-parameter analysis, which is at the
few percent level for $\alpha$, $v$, $\Delta_q$ and $M_q$, while
it is almost one order of magnitude larger for $G_N$ and $m_e$.
More interestingly, the other bidimensional contours show the
degeneracies among pairs of parameters. In particular, notice the
strong correlations of $G_N$ with $\alpha$ and $m_e$ and between
the pairs $v-\alpha$, $m_e-\alpha$, $\Delta_q-\alpha$, $v-m_e$,
and $v-\Delta_q$.

\begin{table*}[t]
\caption{Figures of Table \ref{degeneracies} continued.}
\label{degeneracies2}
\begin{center}
\begin{tabular}{ccc}
\includegraphics[width=0.3\textwidth]{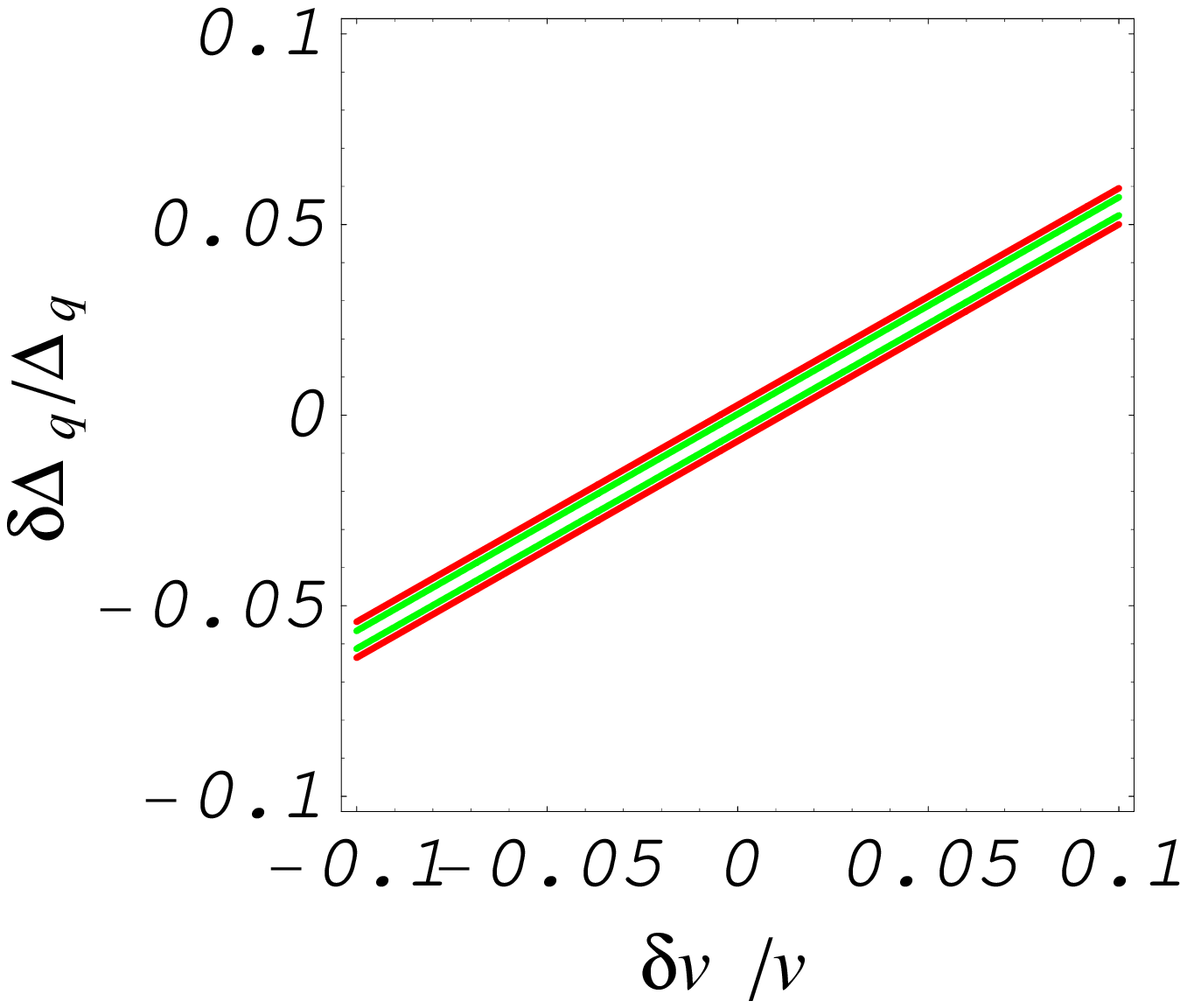}
&
\includegraphics[width=0.3\textwidth]{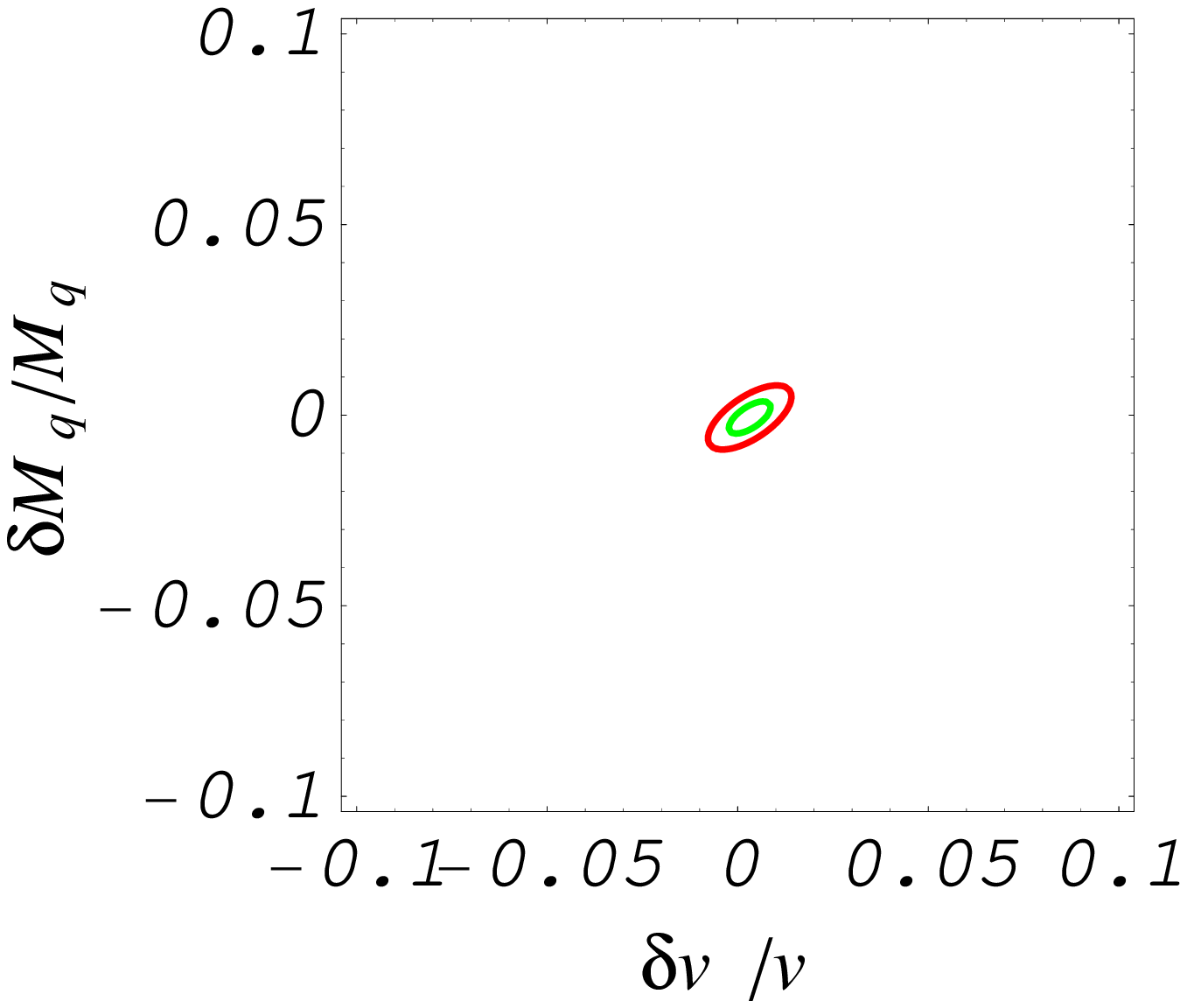}
&
\includegraphics[width=0.3\textwidth]{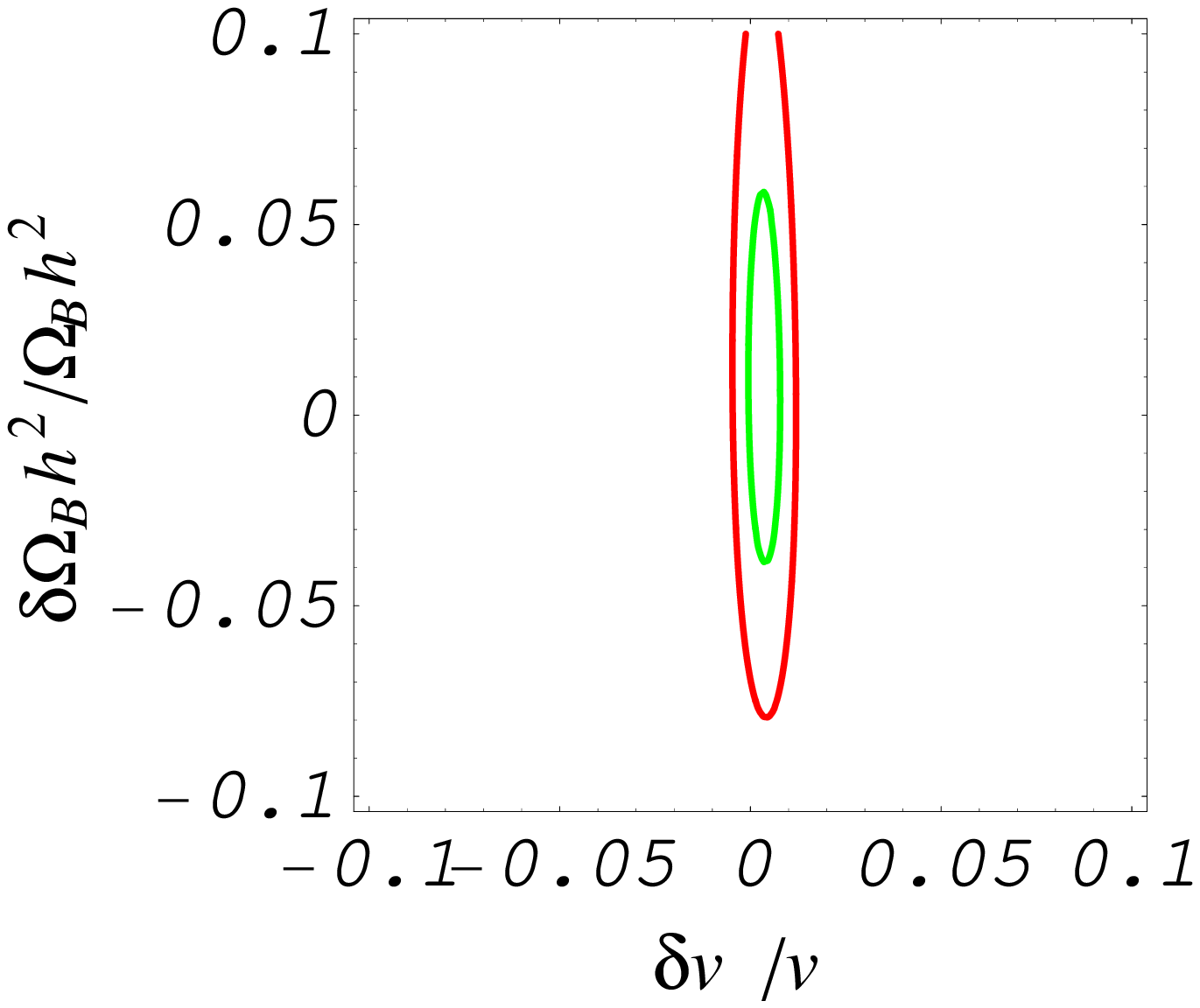}
\\
\includegraphics[width=0.3\textwidth]{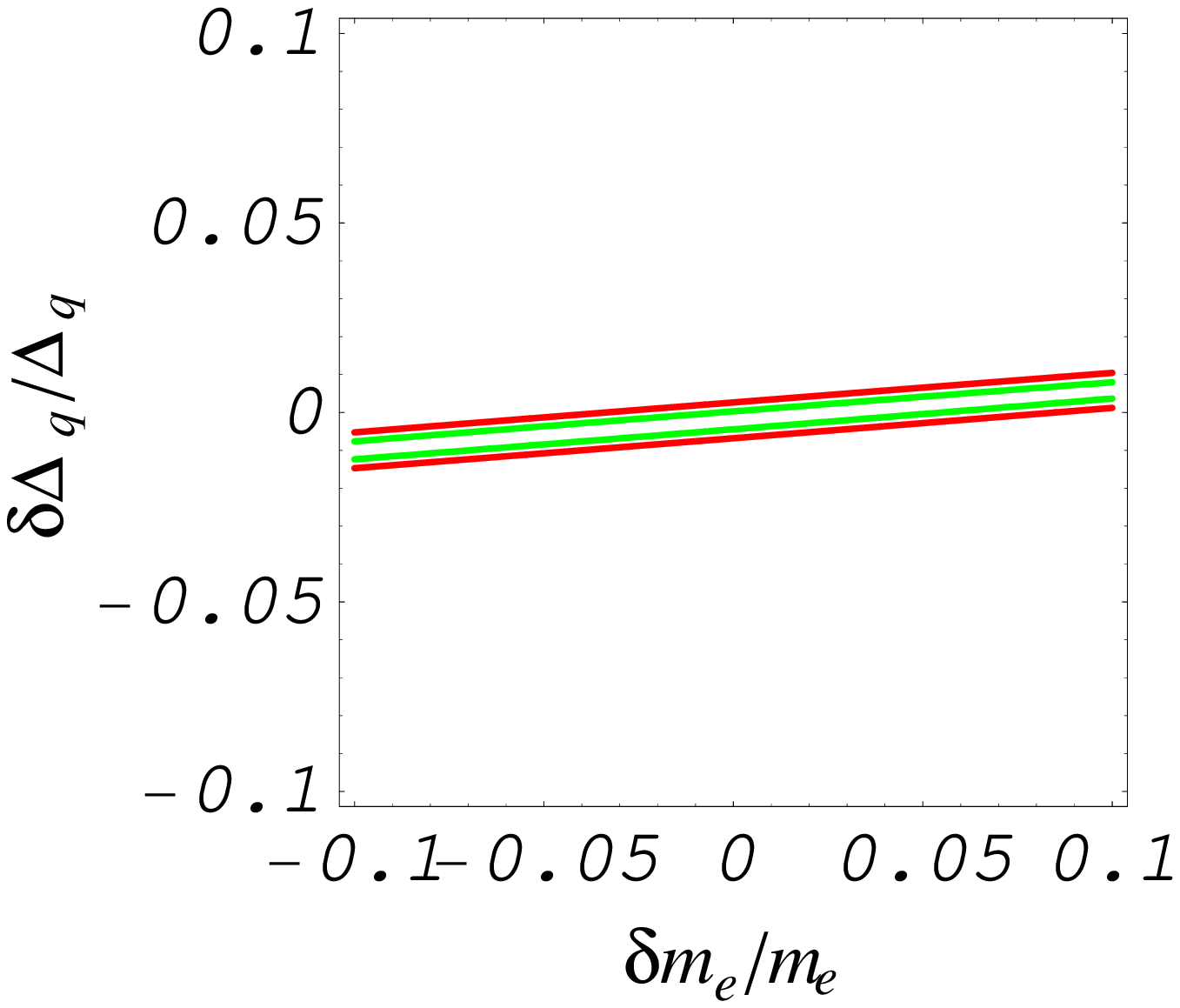}
&
\includegraphics[width=0.3\textwidth]{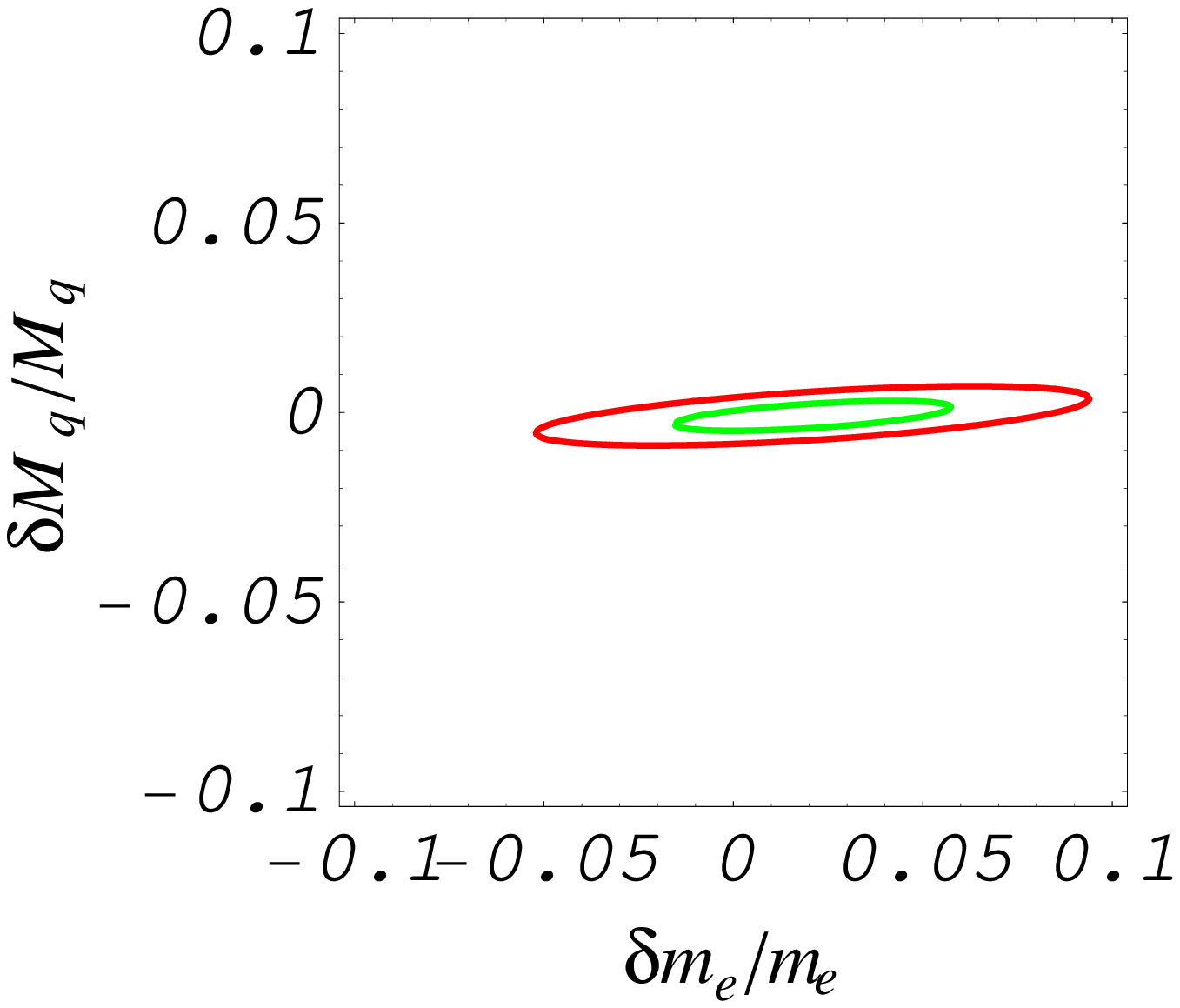}
&
\includegraphics[width=0.3\textwidth]{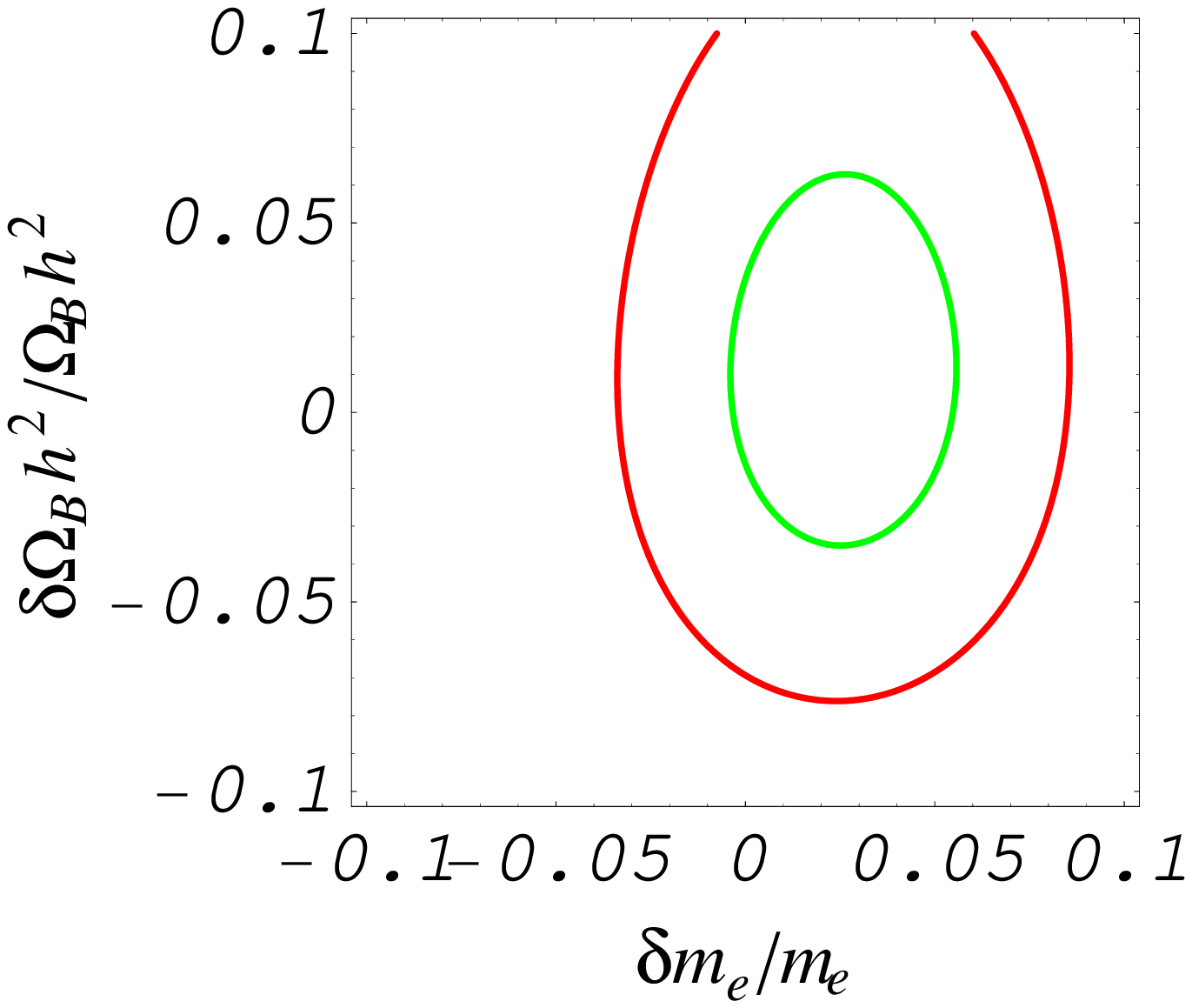}
\\
\includegraphics[width=0.3\textwidth]{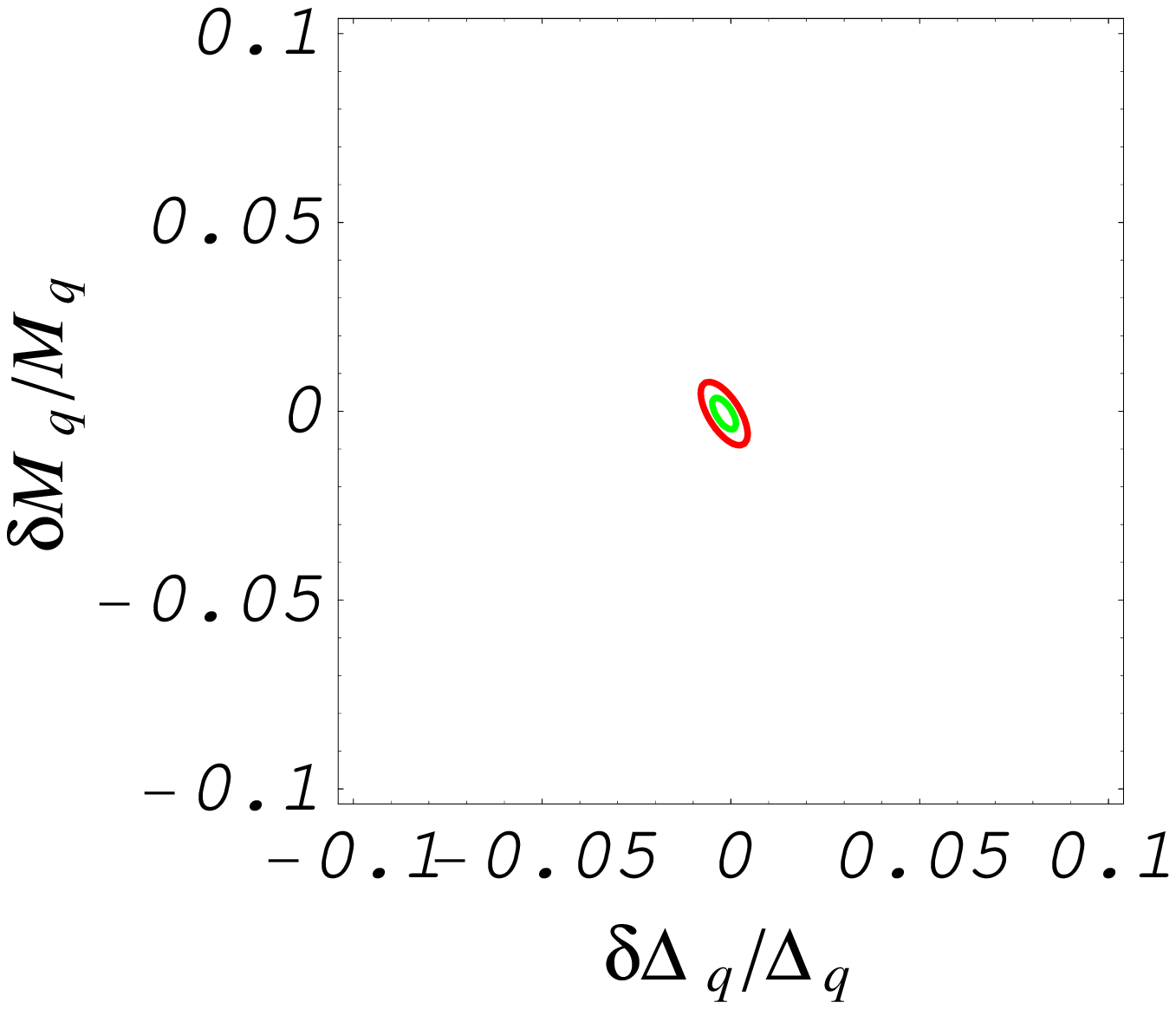}
&
\includegraphics[width=0.3\textwidth]{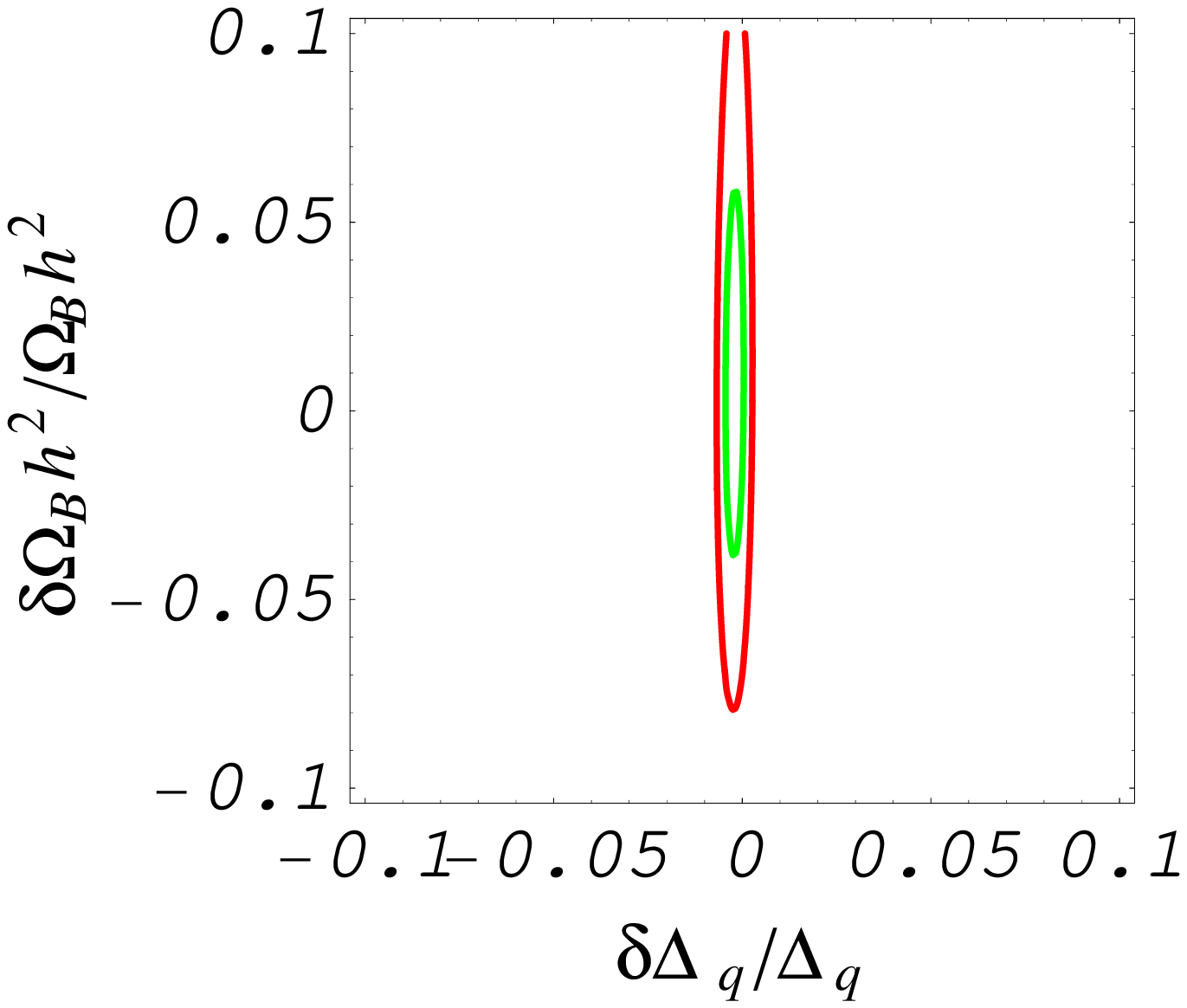}
&
\includegraphics[width=0.3\textwidth]{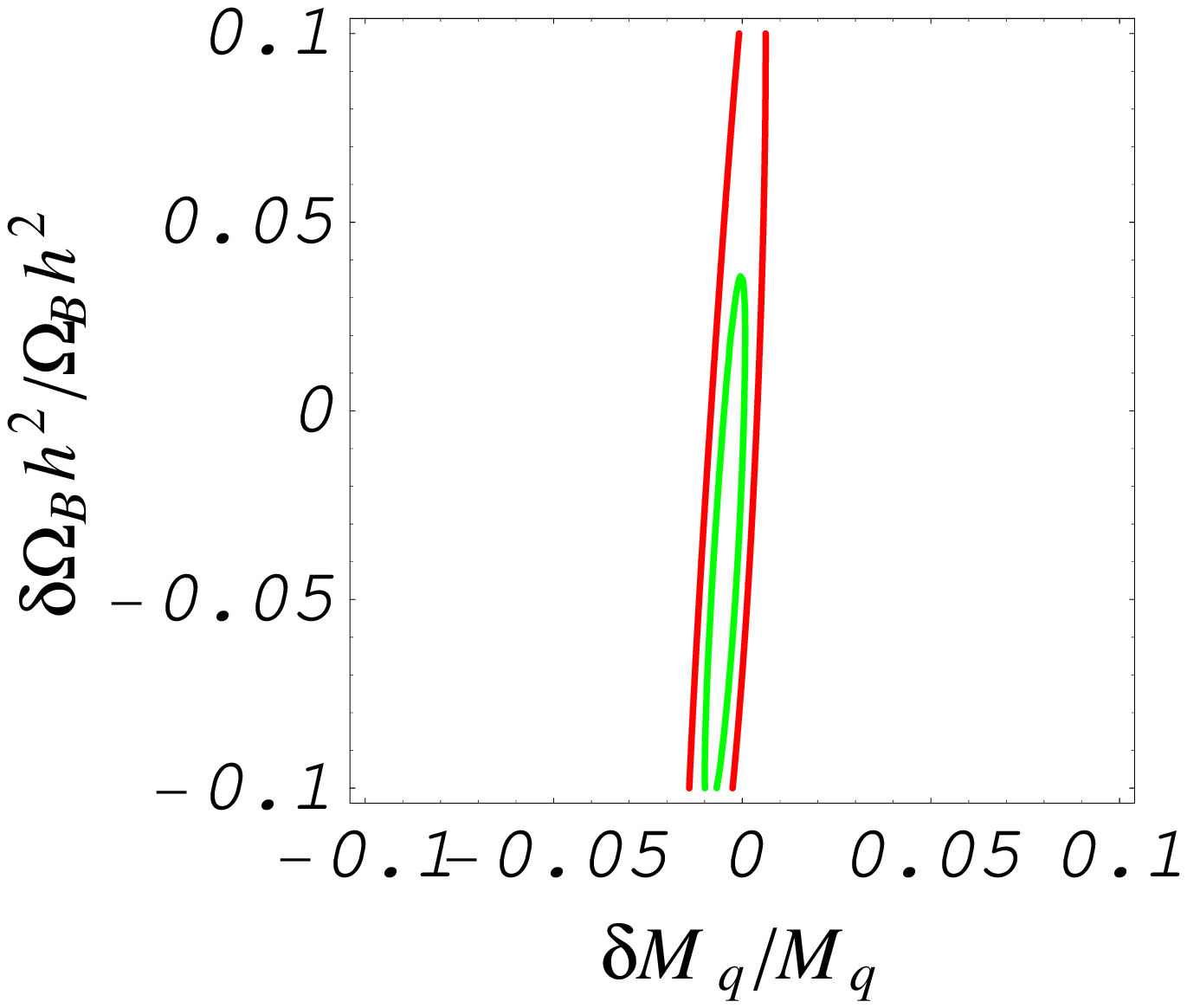} \\
\end{tabular}
\end{center}
\end{table*}

The difficulty of treating simultaneous variations of more than a single
fundamental parameter is somehow alleviated if one works in the framework
of a definite theoretical model beyond the SMPP, which imposes further
relationships among (some) of them. A typical example is given by Grand
Unified Theories (GUT) which assume that the three SMPP gauge couplings
get unified at some high mass scale, $M_G$, where the dynamics is dictated
by a larger (simple) symmetry group. This scenario and its implications
for BBN has been worked out in details in
\citep{Cal02a,Lan02,Cal02b,Den03,Den07,Coc07}. The general argument is
that in GUT theories unification of couplings implies that the various
fundamental constants are likely to vary simultaneously and, for example,
a change of the fine structure coupling implies a much larger variation of
the strong interaction scale, $\Lambda_{QCD}$, at low energies.

If we denote by $\alpha_G$ the (common) value of the three
$SU(3)_c\times SU(2)_L\times U(1)_Y$ couplings, $\alpha_i$, at
$M_G$ (typically $M_G\sim 10^{16}$ GeV), then their running is
given by Renormalization Group equations, \be \alpha_i^{-1}(M_Z) =
\alpha_G^{-1} + \frac{b_i}{2 \pi} \log \left( \frac{M_G}{M_Z}
\right) \vv \ee where the coefficients $b_i$ depend on the
particle multiplets. If $\alpha_G$ undergoes time variation at a
cosmic time above the unification scale, then one can traces the
correlated variation of the SMPP couplings at low energies.
Neglecting the threshold effect one obtains \citep{Cal02a,Lan02}
\be \frac{\delta \Lambda_{QCD}}{\Lambda_{QCD}} \sim (34 \div 40)
\frac{\delta \alpha}{\alpha} \pp \ee Also Yukawa couplings and the
Higgs vacuum expectation value $v$ (which is for example, tied to
the supersymmetric breaking scale in most supersymmetric models)
are expected to vary and to be related to time evolution of
$\alpha$. However, the size of the effect is model dependent, or
simply unknown, so it is usually parameterized in terms of an
unknown constant, $\delta y_i/y_i = c \delta \alpha/\alpha$, and
similarly for $v$. In general, in addition to the time evolution
of $\alpha_G$, one could also consider other possibilities, where
the GUT mass scale $M_G$ also varies \citep{Cal02b}.

The BBN constraints in these scenarios on the variation of $\alpha$ (and
all other related parameters) have been first analyzed in a qualitative
manner in \citep{Lan02,Den03} and then worked out using a full BBN
numerical study in \citep{Coc07,Den07}, with two main results. On one
hand, the constraints become typically tighter, $|\delta \alpha/\alpha|
\leq 10^{-5} - 10^{-4}$. Moreover, in some cases the theoretical value of
$^7$Li is lowered and can be rendered compatible with the experimental
result. This is the case, for example, when the weak scale, $v$, is
determined by dimensional transmutation, so that variation of the largest
top quark Yukawa coupling induces changes of $v$ \citep{Cam95,Coc07}, or
assuming that the variation of fundamental constants are triggered by an
evolving dilaton scalar field \citep{Ich02b,Coc07}.

\subsubsection{Varying the Newton constant and scalar-tensor theories of gravity}
The effect of a varying gravitational constant, $G_N$, on the BBN
is through the expansion law, $H \propto \sqrt{G_N}$, which
determines both the neutron-proton density ratio at freeze-out
($Y_p$), and the efficiency of $^2$H burning abundance when
nucleosynthesis starts. We have already seen in the previous
Section the typical constraints which can be put on $\delta
G_N/G_N$, of the order of few percent. In particular, using the
values of $Y_p$ and $^2$H/H already discussed, we find $-0.036
\leq \delta G_N/G_N \leq 0.086$, at 95 \% C.L., which is dominated
by the effect on $^4$He. In this case, the limit can be read off
directly from the bound on $\neff$, since a change of $G_N$ is
equivalent to the change \be \frac{\delta G_N}{G_N} = \frac{7}{43}
\Delta \neff \,.\pp \ee

Recently, the effect of a varying $G_N$ has been considered by
several authors, with similar results, depending on the
experimental values for light nuclei adopted in their analysis
(for earlier studies see e.g. \citep{Yan79,Acc90}). In
\citep{Cop04} it is stressed the sensitivity of $^2$H to the value
of the Newton constant. In the pessimistic case of a large
systematic error on $Y_p$, deuterium can provide a 20 \% bound. On
the other hand, the results of \citep{Cyb05}, which adopt $Y_p =
0.249 \pm 0.009$, suggests that $Y_p$ still gives the strongest
possible constraint. If one makes the assumption of a monotonic
behavior for the time dependence of the gravitational constant,
$G_N \sim t^{-x}$, the bound on $\delta G_N$ can also be cast in a
constraint on the exponent $x$ and thus on the present value of
$\dot{G}_N/G_N$. In \citep{Cyb05} it is found $-0.0029 < x <
0.0032$, and $-2.4 \times 10^{-13} \textrm{yr}^{-1} <
\dot{G}_N/G_N < 2.1 \times 10^{-13} \textrm{yr}^{-1}$. Our
estimate on $\delta G_N/G_N$ translates into $-0.6 \times 10^{-13}
\textrm{yr}^{-1} < \dot{G}_N/G_N < 1.6 \times 10^{-13}
\textrm{yr}^{-1}$ using $t_0=13.7\,$Gyr for the lifetime of the
Universe.

An interesting class of models which allows for a time dependent
gravitational coupling is represented by scalar-tensor theories of gravity
\citep{Jor49,Bra61}, where the latter is mediated, in addition to the
usual spin-2 gravity field, by a spin-0 scalar $\varphi$ which couples
universally to matter fields. Its dynamics define the evolution of an
effective Newton constant for ordinary matter, to which $\varphi$ is
``universally" coupled, as one can read by the action of the model in the
``Einstein" frame,
\be
S = \int \frac{d^4 x}{16 \pi G_*} \, \sqrt{-g} \left[ R - 2 g_{\mu \nu}
\partial^\mu \partial^\nu \varphi -V(\varphi) \right] + S_m\left[
F^{-2}(\varphi) g_{\mu \nu }; \Psi \right] \vv
\label{bdj}
\ee
where $G_*$ is the bare gravitational coupling, $V(\varphi)$ the scalar
field potential, and $F(\varphi)$ a positive function, which enters the
coupling of ordinary matter fields $\Psi$ to the metric. Strong
constraints on the present (denoted by the index 0) values of the
(generally) $\varphi$ post-Newtonian parameters \citep{Wil01},
\be
\gamma-1 = -2\frac{\alpha_0^2}{1+\alpha_0^2} \vv \, \,\,\,\, \beta-1 =
\frac{1}{2} \frac{\beta_0 \alpha_0^2}{(1+\alpha_0^2)^2} \vv
\ee
where
\be
\alpha(\varphi) = \frac{d \log F^{-1/2}}{d \varphi} \vv \, \,\,\,\,
\beta(\varphi) = \frac{d \alpha}{d \varphi} \vv
\ee
are set by solar system experiments, as the shift of the Mercury
perihelion or the Shapiro delay of radio signals from the Cassini
spacecraft as it passes behind the Sun, $\gamma_0 -1=(2.1 \pm 2.3 ) \times
10^{-5}$ \citep{Ber03a}\footnote{This bound is also usually presented in
terms of a lower limit on the Brans-Dicke parameter $\omega_0 \geq 40000$,
the constant which weights the scalar field kinetic term in the Jordan
frame and in the minimal model with $V(\varphi)=0$,
$F(\varphi)=\varphi$.}. The value of $\alpha_0$ should be thus, very
small, while $\beta_0$ can still be large.

\begin{figure}
\begin{center}
\includegraphics[width=0.7\textwidth]{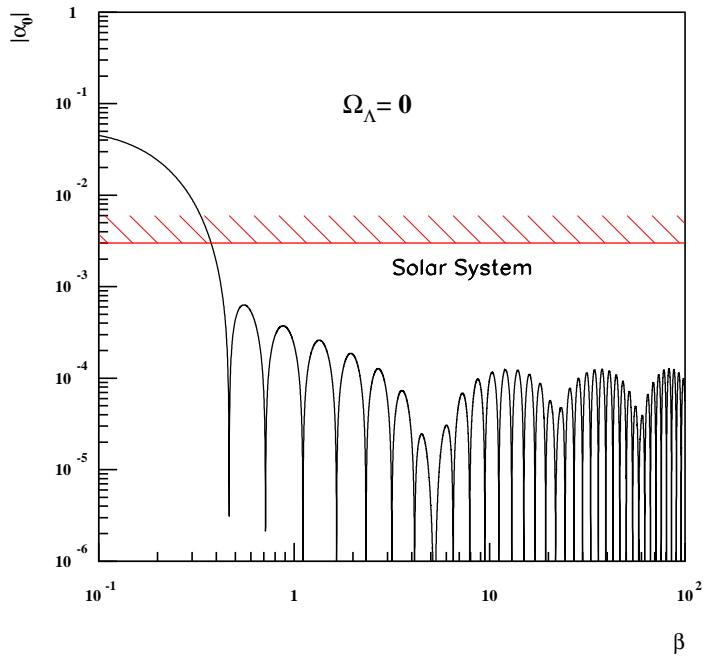}
\end{center}
\caption{Bounds on the post-newtonian parameters $\alpha_0$ and
$\beta$ from $^4$He in scalar-tensor theories in the case of a massless
dilaton with quadratic coupling, $F(\varphi)= \exp(-\beta \varphi^2)$. The
value of baryon density is $\Omega_B h^2= 0.0224$. From \citep{Coc06}.}
\label{fig_Coc06}
\end{figure}

The implications of models described by the action in Eq.
(\ref{bdj}) have been discussed by several authors with different
choices of the arbitrary functions $F(\varphi)$ and $V(\varphi)$,
see e.g.
\citep{Cas92a,Cas92b,Ser92,Ser96,Eto97,San97,Dam99,Alv01,Che01,Ser02,Kne03a,Nag04,Cli05,Pet05,Coc06,DeF06,Lar06b,Nak06}.
In the case of a standard Brans-Dicke theory the value of
primordial $^4$He produced during BBN can be evaluated
semi--analytically as a function of the Brans-Dicke parameter
$\omega$ (which weights the scalar field kinetic term in the
Jordan frame) \citep{Alv01}, in case one consider the particular
solution $\varphi=const$ during the radiation dominated epoch,
giving the bound $\omega_0 \geq 100$. The fact that the BBN
constraints on the model depend strongly on the particular
scalar-tensor theory considered in the analysis, thus providing
quite different bounds on $\omega_0$, have been stressed in
\citep{Ser02}, while a quite general numerical code for BBN in
scalar-tensor models is described in \citep{Cli05,Coc06}. In
particular, in \citep{Coc06} it is studied in details the case of
a massless dilaton with a quadratic coupling, $V(\varphi)=0$,
$F(\varphi)= \exp(-\beta \varphi^2)$, including the mass threshold
effects when the universe cools down. The latter is due to the
variation of the trace of the energy momentum tensor of ordinary
matter-radiation, $\rho-3 P$, whenever a single specie becomes
non--relativistic, which changes the source term in the Klein
Gordon equation for $\varphi$. The results of the study of
\citep{Coc06} show that the strongest bound comes from $^4$He and
indeed it is stronger than from solar system experiments,
$\alpha_0\leq 10^{-3}$ for $\beta \geq 0.3$, see Figure
\ref{fig_Coc06}. Finally, we mention that particular choices of
the scalar field potential $V(\varphi)$ (and of the initial
conditions for $\varphi$ before the BBN) can solve the lithium
problem, yet leading to values of $^2$H/H and $Y_p$ compatible
with the experimental values. One example is illustrated in Figure
\ref{fig_Lar06}, from \citep{Lar06b}, where the potential is
chosen to be $V(\varphi)= \Lambda^2 \varphi^4$, and
$F(\varphi)=\exp(-\beta \varphi^2)$. We see that there is a region
in the parameter space which corresponds to primordial abundances
of all light nuclei in agreement with experimental data.
\begin{figure}
\begin{center}
\includegraphics[angle=90,width=0.8\textwidth]{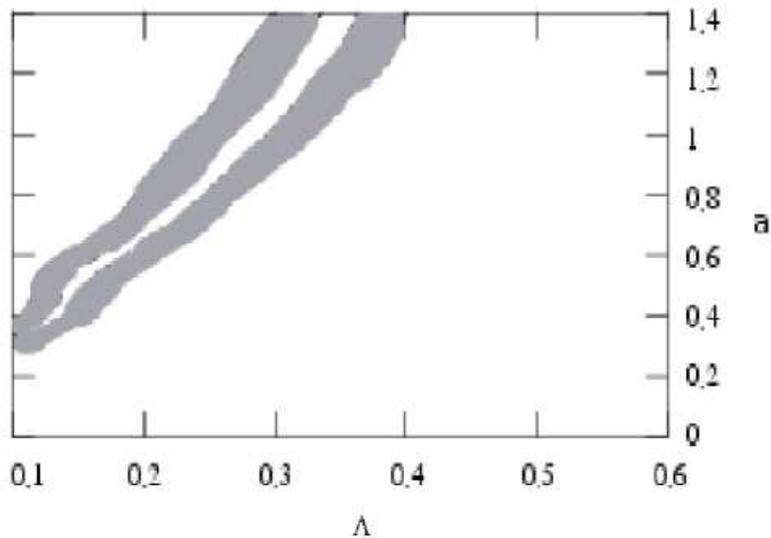}
\end{center}
\caption{Bounds on the parameters $a$ and $\Lambda$ which give
acceptable results for $^2$H/H, $Y_p$ and $^7$Li/H for a
scalar-tensor theory defined by $\alpha=a \varphi^2$ and
$V(\varphi)= \Lambda^2 \varphi^4$. The initial conditions
(pre-BBN) for $\varphi$ are $\varphi_{in}= -1.3$,
$\dot{\varphi}_{in}=0$. From \citep{Lar06b}.}
\label{fig_Lar06}
\end{figure}

\subsubsection{A varying Cosmological Constant: Quintessence models}
Observations of type Ia supernovae \citep{Ast05,Rie04}, structure
formation \citep{Col05,Teg06} and CMB \citep{Dun08} provide quite a strong
evidence for an accelerated expansion of the Universe at recent times. The
simplest explanation of this result is to invoke a new component of the
total energy--momentum tensor in the form of a cosmological constant
$\Lambda$ ($\Lambda$CDM model), which is a nice fit of data, yet it gives
rise to two related fine tuning problems. On one hand, the value of
$\Lambda$ happens to be extremely small with respect to the typical energy
scale of any fundamental physics model, as the Planck mass (122 orders of
magnitude larger), or supersymmetry breaking scale, or finally, the
electroweak mass scale (54 orders of magnitude off). Moreover, it seems
really a coincidence that the energy density stored in the form of
$\Lambda$ and matter appears to be of the same order of magnitude {\it
just} today, $\Omega_\Lambda \sim 0.75$ and $\Omega_{B+DM}\sim 0.25$. If
the cosmological constant had been larger it would have changed the whole
structure formation history, while a smaller value would have been
irrelevant for observations.

To solve (or to alleviate) these two related problems, many
attempts have been made in the last two decades and based on the
idea that the ``dark energy'' whose present value is given by
$\Omega_\Lambda$, is dynamically changing and is due to the
evolution of a scalar field $Q$ (the quintessence, k-essence, etc.
field) see e.g.
\citep{Wet88,Rat88,Cal98,Cop98,Fer98,Lid99,Ste99,Zla99,Bra00,Chi00,Arm00,Arm01}.
In particular, depending on the choice of the field potential,
typically modeled as an exponential, $V(Q) = V_0 \exp(-\lambda
Q/M_P)$ or an inverse power, $V(Q) = \lambda \Lambda^{4+a}/Q^a$,
the field $Q$ evolves according to an attractor-like solution of
the equation of motions. This means that for a wide range of
initial conditions at early times, the evolution of $Q$ rapidly
converges towards these solutions (tracker solutions), which lead
naturally to a crossover from the radiation dominated epoch to a
dark energy dominated era at late times.

If the $Q$-field provides a non-negligible contribution to the total
energy density during radiation dominated regime, it follows that its
early dynamics can be constrained by nucleosynthesis, and later on by CMB
power spectrum. A first rough estimate can be obtained by requiring that
at the neutron-proton decoupling temperature $T_D\sim$ MeV, the scalar
field energy density $\Omega_Q$ should be small enough not to disturb the
eventual amount of frozen neutrons, i.e. of the final $^4$He mass
fraction, see e.g. \citep{Fer98}. This bound can be cast in terms of the
largest acceptable value for deviations of the effective number of
neutrino from its standard value {\it at that particular temperature}, as
in general the $Q$ energy density would not scale simply as radiation
\be
\Omega_Q(T_D) \leq \frac{7 \Delta \neff /4}{10.75 + 7 \Delta \neff /4}
\leq 0.09 \vv
\ee
where we have used our bound at 95 \% C.L. on $\neff$ of Section
\ref{sec:BBNanalysis}.

More detailed analysis have been presented in
\citep{Yah02,Kne03a}. In \citep{Kne03a}, the authors consider as a
model the particular potential of \citep{Alb00}, \be V(Q) =
\left[(Q-Q_0)^2+A \right] \exp(-\lambda Q) \vv \ee and obtain the
bound on the quintessence contribution to the total energy density
during BBN $\Omega_Q \leq 0.12$, $\lambda \geq 5.7$ at 99 \% C.L..
In \citep{Yah02} the analysis is performed for both the originally
proposed inverse power potential of Ratra and Peebles
\citep{Rat88}, \be V(Q)= M^{4+\alpha} Q^{- \alpha} \vv \ee and a
modified version of it based on the hypothesis that the
quintessence field be a part of supergravity models. In this case,
for a flat K$\ddot{\textrm{a}}$hler potential, the potential
receives an extra factor $\exp(3 Q^2/2 M_P^2)$. Their results are
shown in Figure \ref{fig_Yah02}, for the following choices of
$Y_p$ and $^2$H/H \bea 0.226 \leq & Y_p & \leq 0.247 \vv \nonumber \\
2.9 \times 10^{-5} \leq & ^2\textrm{H}/\textrm{H} & \leq 4.0
\times 10^{-5} \pp \eea The plots show the bound on the initial
ratio of the $Q$ field to background (B) (ordinary radiation and
matter) energy density at initial redshift $z=10^{12}$, assuming
equipartition of $Q$ energy at that epoch, $\dot{Q}^2/2 = V(Q)$,
versus the potential parameter $\alpha$. The regions denoted by
"Tracker Solution at BBN" denotes models in which the evolution of
$Q$ has already reached the tracker solution. The main effect of
BBN is to exclude a large family of possible kinetic-dominated
solutions with a $Q$-energy density exceeding that of relativistic
species prior or during nucleosynthesis.
\begin{figure}
\begin{center}
\includegraphics[angle=270,width=0.9\textwidth]{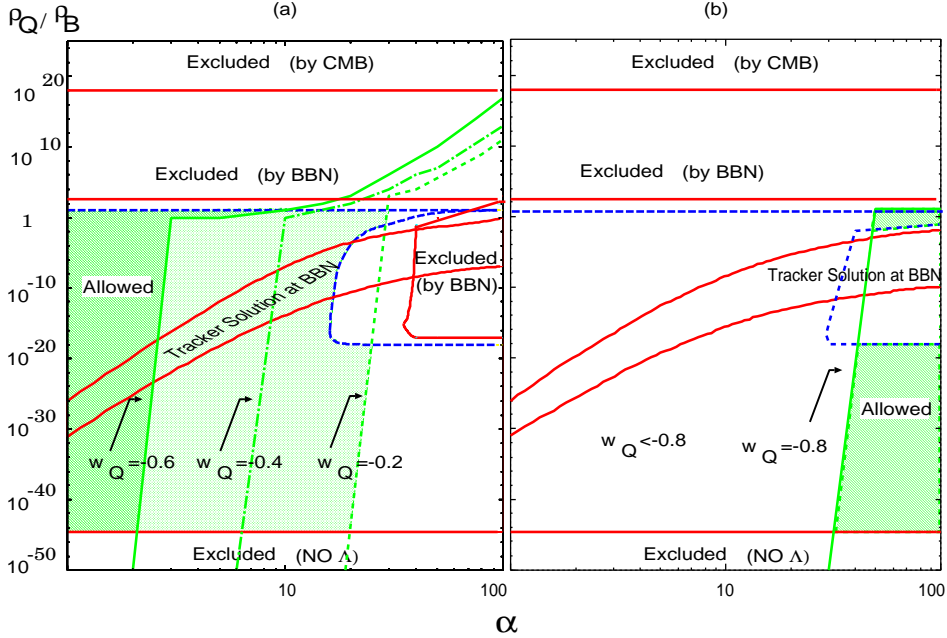}
\caption{Allowed values of the $Q$ field potential parameter
$\alpha$ and initial $\rho_Q/\rho_B$ at $z=10^{12}$ (B stands for
"background", i.e. ordinary radiation and matter). The plots refer
to power law (a) and SUGRA corrected potentials (b). Models in
which the tracker solution is obtained by the BBN epoch are those
in the band denoted by "Tracker Solution at BBN". Values of
$\alpha$ to the right of the lines labeled $w_Q=-0.6,-0.4,-0.2$
are excluded by requiring that the present equation of state be
sufficiently negative. The BBN constraint for a maximum energy
density of the quintessence field of 0.1 \% and 5.6 \% are shown
as dotted and solid lines, respectively. From \citep{Yah02}.}
\label{fig_Yah02}
\end{center}
\end{figure}

\subsection{Miscellanea}
\label{s:miscellanea}

\subsubsection{Testing Friedmann equation}

We have already stressed several times that changing the expansion
history in the early universe has a big impact on the BBN
predictions for light nuclei abundance, in particular $^4$He.
Several modifications to the standard Friedmann equation have been
already discussed, arising in a variety of contexts such as
non-standard theory of gravity, time evolving Newton constant, or
brane-world models. One may also try to perform somehow a $blind$
test of the validity of Friedmann expansion law without a
particular theoretical framework behind, and using a suitable
parametrization of possible deviations from the standard behavior.
This has been considered for example in \citep{Car02}, where the
Hubble factor is expressed in terms of two parameters, \be H(T) =
H_1 \left( \frac{T}{1 \textrm{MeV}}\right)^\alpha \vv
\label{exotich} \ee where $H_1$ and the exponent $\alpha$ should
be constrained by data, once we fix the baryon density parameter,
$\eta$. A similar analysis has been performed in \citep{Mas03},
with a slight different definition of $H(T)$. Using Eq.
(\ref{exotich}) one can compute the values of the three leading
quantities which enter BBN dynamics: the freeze-out temperature,
$T_D$, the time elapsed from $T_D$ and the onset of BBN at $T_N$,
and the value of the expansion rate at $T_N$. Interestingly, $H_1$
and $\alpha$ show a large degeneracy. The same Helium mass
fraction and $^2$H abundance can be obtained increasing at the
same time the Hubble rate at every temperature (i.e. $H_1$), which
raises the freezing temperature and thus leads to a larger initial
neutron to proton density ratio, and the value of $\alpha$ (for a
definite $H_1$ a larger $\alpha$ means a relatively lower
expansion rate at $T_N$). Furthermore, for each nuclide there are
two branches in the $H_1$-$\alpha$ plane for which $Y_p$ or
$^2$H/H are the same, see Figure \ref{figs_Car02}: a lower branch,
where also the standard result, $\alpha=2$, $H_1 \sim$ MeV, is
lying, and an upper branch, for higher values of $H_1$, which
however cannot simultaneously fit both $^4$He mass fraction and
deuterium. The whole compatibility region is thus characterized by
a single (almost) linear behavior, $\log(H_1) \propto \alpha$,
with $1.5 \leq \alpha \leq 3$. This bound can also be used to
constrain possible non-universal coupling of gravity to the three
neutrino generations as considered in \citep{Mas03}, where the
study is also extended to the case of a degenerate BBN, or the
possibility that matter and antimatter may have different
couplings to gravity \citep{Mas04}. A different gravitational
coupling to bosons and fermions, $G_{N,B}$ and $G_{N,F}$
respectively, has been instead considered in \citep{Bar04b}, with
the result $0.33 \leq G_{N,B}/G_{N,F} \leq 1.10$ at $2-\sigma$.

\begin{figure}
\begin{center}
\begin{tabular}{c}
\includegraphics[width=0.7\textwidth]{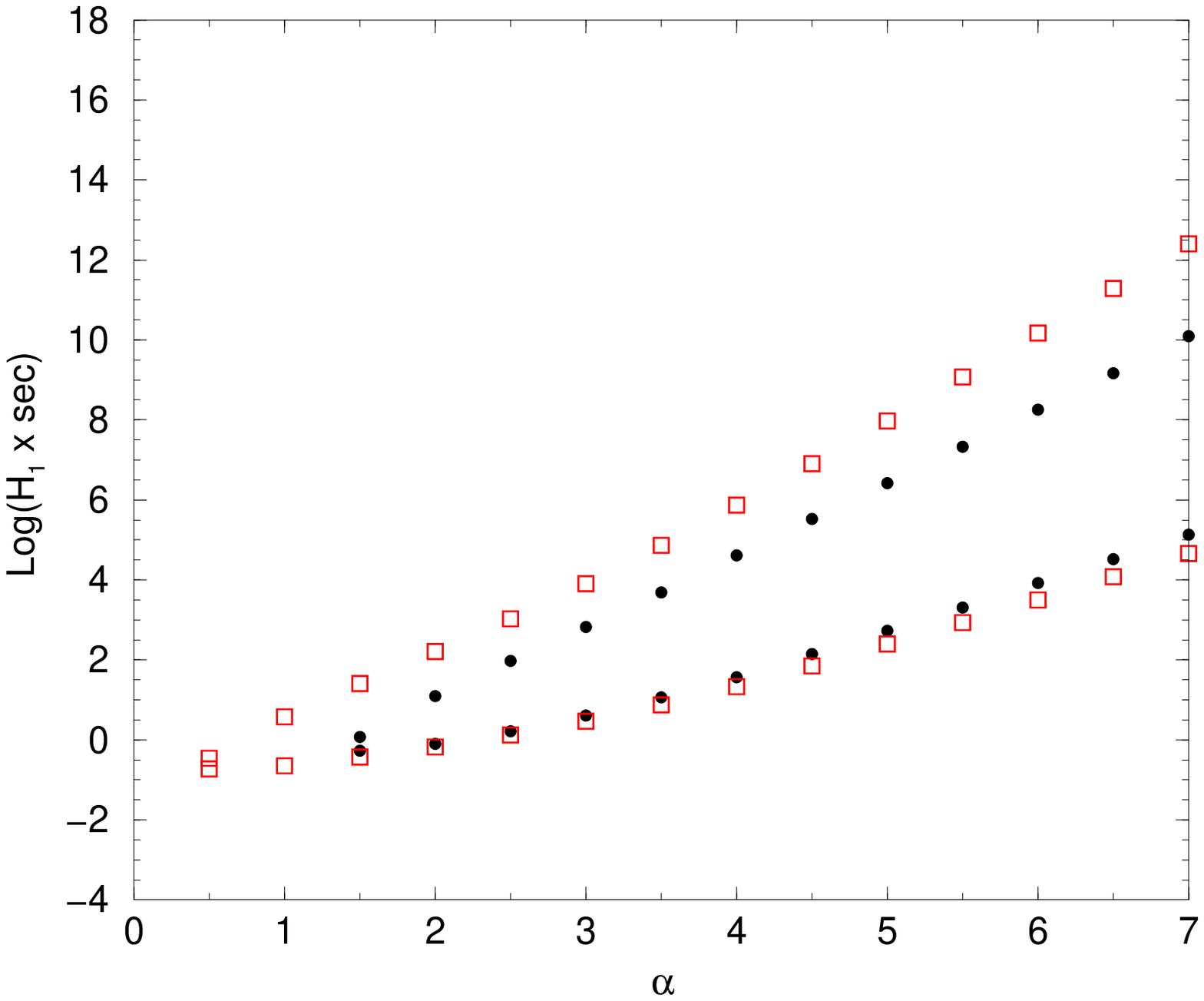} \\
\includegraphics[width=0.7\textwidth]{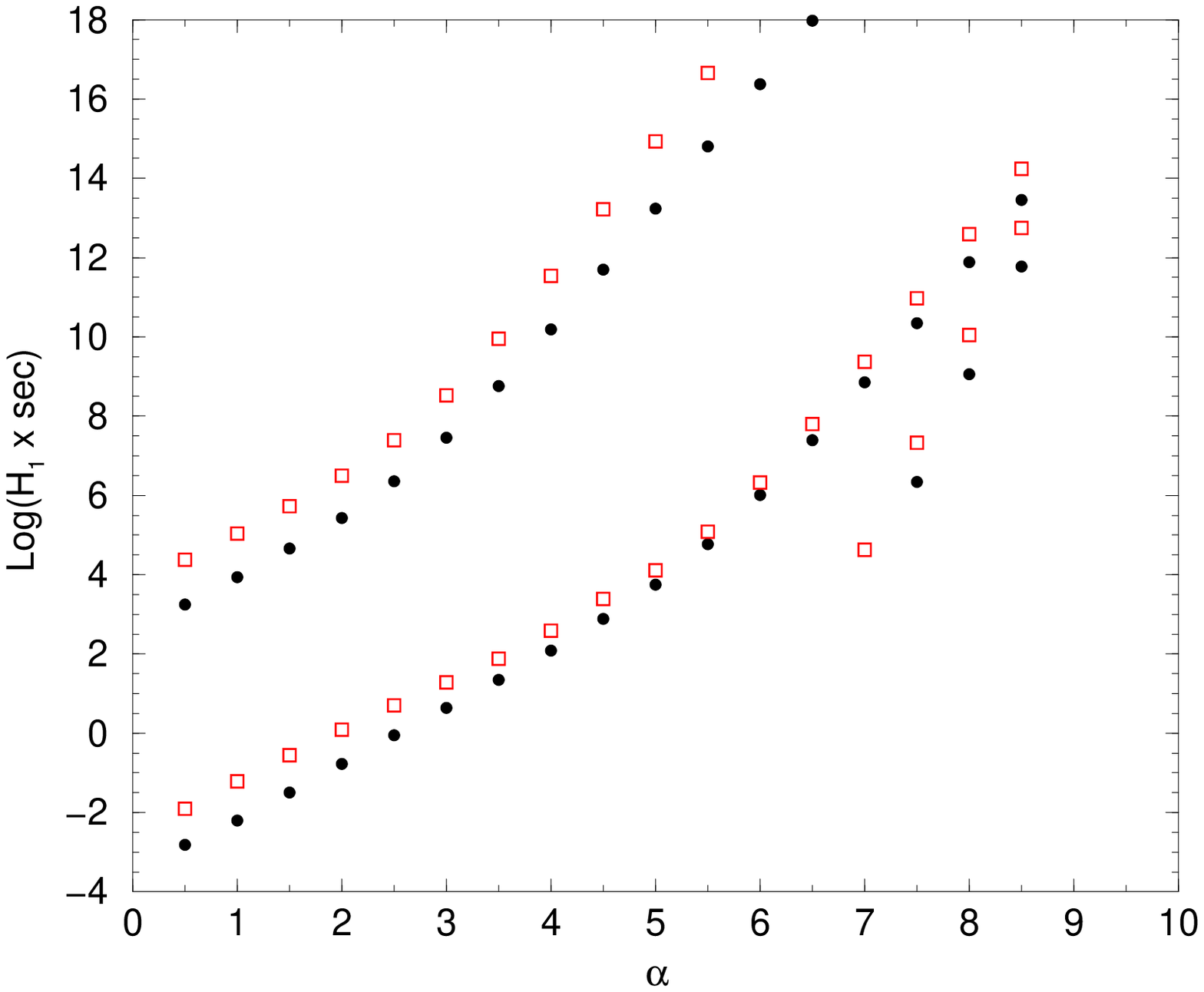}
\end{tabular}
\end{center}
\caption{Contours of constant helium (top) and deuterium (bottom)
in the $\alpha-H_1$ plane. Chosen values are $Y_p=0.24$ and $^2$H/H=$3
\cdot 10^{-5}$. Squares and filled circles correspond to $\eta= 10^{-9}$
and $\eta=10^{-10}$, respectively. From \citep{Car02}.}
\label{figs_Car02}
\end{figure}

Similar analyses have been also applied in the framework of
metric-affine gravity models, where the non-Riemaniann effects are
encoded in a fictitious fluid with equation of state $P=\rho$
\citep{Kra05}, or which account for the quantum gravity
corrections for matter fields computed in a loop quantum gravity
approach, with a non-canonical equation of state \citep{Boj08}.
Finally, one can bound the values of parameters which measure
departure from Lorentz invariance \citep{Lam05b}, see e.g.
\citep{Kos04} for a review of the Lorentz- and CPT-violating
extension of the SMPP.

The sensitivity of BBN to any modification of the Friedmann
equation can be also exploited to test a specific prediction of
general relativity, which has no analog in Newton theory, namely
the fact that pressure contributes to the acceleration of the
scale factor,
\be
\frac{\ddot{a}}{a} = - \frac{4 \pi G_N}{3} \left( \rho + 3P
\right) \pp \label{selfgravityp}
\ee
Indeed the Friedmann equation $knows$ about the pressure
contribution in Eq. (\ref{selfgravityp}) via the Bianchi identity.
If one allows for deviation from this distinctive feature of
Einstein theory and introduces, as in \citep{Rap08}, a free
constant parameter, $\chi$, which modifies the spatial component
of Einstein's equations, \be \frac{\ddot{a}}{a} = - \frac{4 \pi
G_N}{3} \left( \rho + 3 \chi P \right) \vv \label{selfgravityp2}
\ee one obtains during a radiation dominated expansion (neglecting
the spatial curvature term), \be H^2 = \frac{1+ \chi}{2} \frac{8
\pi G_N}{3} \rho \pp \ee Bounds on $\chi$ can then be obtained
from the BBN. Not surprisingly, the results of \citep{Rap08} show
an excellent agreement with the standard expectation, $\chi=1$.
Notice that variation of this parameter is completely degenerate
with a change of the Newton constant during the BBN, or a
non-standard effective number of neutrinos $\neff$.

\subsubsection{Primordial Black Holes and BBN}

Primordial black holes may form in the early universe in presence
of sub-horizon density perturbations of order unity
\citep{Zel67,Haw71,Car75}. Assessing cosmological constraints upon
their density over some mass ranges can provide important
information on the primordial density fluctuations. For a recent
review see e.g. \citep{Khl08b}. These constraints have been
discussed by many authors in the 1970s \citep{Haw74,Car76,Zel76,
Zel77,Vai78a,Miy78,Vai78b,Lin80}, while updated analysis using BBN
\citep{Koh99} and CMB data \citep{Ric07,Tas08} have been presented
quite recently. Primordial black holes evaporating by the BBN
epoch ($t\leq 10^3$ s) have a mass lying in the range $M\leq
10^{10}$ g, since there is a simple relation between mass and
lifetime, $\tau_{bh}$, see e.g. \citep{Koh99}, \be M \sim 10^9
\left( \frac{\tau_{bh}}{\textrm{sec}} \right)^{1/3} \textrm{g} \pp
\ee They emit various particles such as neutrinos/antineutrinos,
photons, quark-gluon jets which produce hadrons through
fragmentation processes. All these (high energy) particles
interact with species already present in the thermal bath, and
induce several effects which can change the standard picture of
light nuclei production. For example, production of high energy
neutrinos and antineutrinos change the weak interaction freeze-out
temperature, and thus the neutron to proton ratio. Similarly,
large hadron injection after the weak process decoupling
temperature $T_D$ may revival chemical equilibrium between
nucleons, thus leading to quite different amounts of $^4$He and
deuterium with respect to the standard case.

\begin{figure}
\begin{center}
\includegraphics[width=0.9\textwidth]{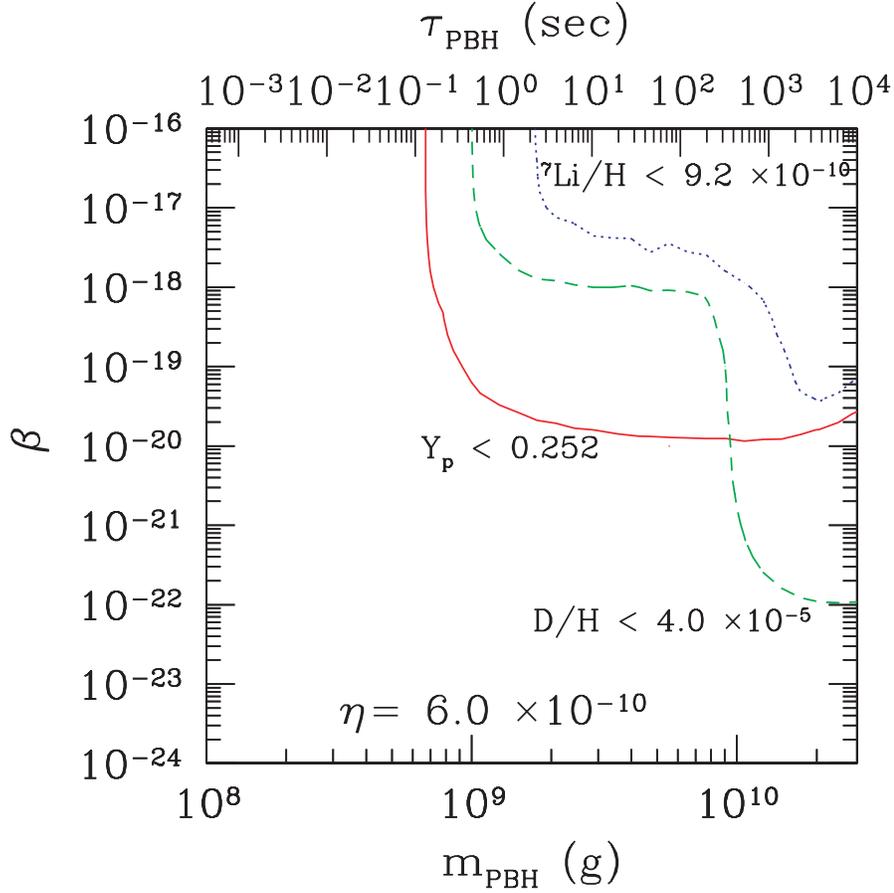}
\end{center}
\caption{The BBN bounds on the primordial black hole fraction
versus their mass. Regions to the right and above the curves are excluded
by primordial nucleosynthesis. From \citep{Koh99}.}
\label{fig2_Koh99}
\end{figure}

In Figure \ref{fig2_Koh99} we report the results of \citep{Koh99}
for $\beta(M)$, the primordial black hole initial fraction to the
total energy density as function of their mass. The bound for
$10^{9}$ g $\leq M \leq 10^{10}$ g, $\beta(M) \leq 10^{-20}$ is
strongly constrained by the upper bound on $Y_p$ (they use $Y_p
\leq 0.252$ in their analysis), since in this mass range the
effect of black hole evaporation is to delay the freeze out of n-p
chemical equilibrium. On the other hand, for larger masses and
lifetimes, the extra produced neutrons cannot be burn into $^4$He
and thus contribute to deuterium abundance, which provides a
stronger constraint in this region, $\beta(M) \leq 10^{-22}$.

\subsubsection{Mirror world}

The theory of a hidden mirror world, an exact duplicate of our visible
world, is based on the product of two identical gauge symmetry group $G
\times G'$, with $G$ the SMPP group, $SU(3)_c \times SU(2)_L \times
U(1)_Y$, in the minimal case. The two sectors communicate through gravity
and, possibly, by other interactions (kinetic mixing of photons and mirror
photons\footnote{BBN and CMB bounds on the mixing of photon with a hidden
light abelian gauge boson have been recently considered in
\citep{Jae08}.}, neutrino mixing, common gauge symmetry of flavor) and are
exchanged by the action of a discrete symmetry, the mirror parity, which
implies that both particle sectors are described by the same action, and
are characterized by the same particle content. For a review and a
comprehensive list of references we address the reader to \citep{Ber03b}.
The mirror world would influence the whole evolution history of the
universe through its contribution to the expansion rate, and can represent
a natural candidate for dark matter, as well as providing a mechanism for
baryogenesis. Of course, if mirror particles populate the primordial
plasma with the same densities of ordinary particles, they would
contribute to the effective neutrino number for a too large value $\neff
\sim 6.14$ at the onset of BBN, which is excluded by light nuclei
abundances. Thus, the mirror particle density should be reduced, and
characterized by a plasma temperature, $T'$, lower than the ordinary
photon temperature, in order to be compatible with the bound on $\Delta
\neff$. This can achieved if the inflationary reheating temperature is
different in the two sectors, and all possible interactions between the
two worlds are weak enough to ensure that they do not come into mutual
thermal equilibrium.

The ratio $T'/T$ can be computed by using entropy conservation.
Use of the definition of Section \ref{sec:cosm_overview} leads to
\be \frac{T'(t)}{T(t)} =  \left( \frac{s'}{s} \right)^{1/3} \left(
\frac{g_{*s}(T)}{g_{*s}(T')} \right)^{1/3} \equiv x \left(
\frac{g_{*s}(T)}{g_{*s}(T')} \right)^{1/3} \vv
\label{mirrortratio}
\ee
 with $s$ ($s'$) the entropy density of
ordinary (mirror) species. During the radiation dominated epoch we
have \bea H(T) &=& \left[\frac{8 \pi G_N}{3} \frac{\pi^2}{30}
\left( g_*(T)T^4 + g_*(T')T'^4 \right) \right]^{1/2}
\nonumber \\
&\sim& \left[\frac{8 \pi G_N}{3} \frac{\pi^2}{30} g_*(T)(1+ x^4)
\right]^{1/2} T^2 \vv \eea where the last approximate equality
holds as long as the value of $x$ is not too small. On the other
hand the parameter $x$ can be re-expressed in terms of $\Delta
\neff$. Since photons, electron/positron pairs and active
neutrinos correspond to $g_* = 10.75$ we have \be \Delta \neff =
6.14 x^4 \vv \ee and thus the BBN bound $\Delta \neff \leq$ 0.4
gives $x\leq$ 0.51.
\begin{figure}
\begin{center}
\includegraphics[width=0.7\textwidth]{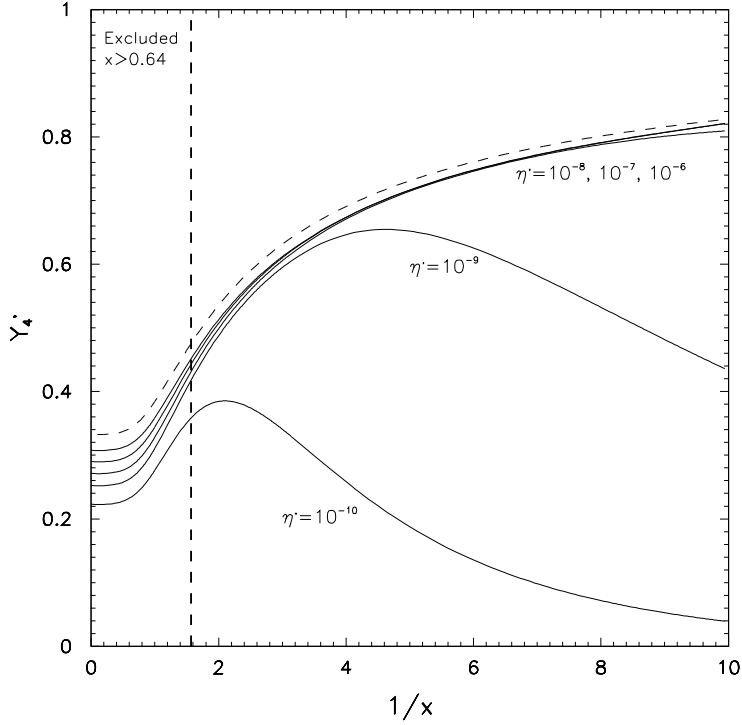}
\end{center}
\caption{The primordial mirror $^4$He mass fraction as a function of
$x$ (see Eq. \ref{mirrortratio}). The dashed curve is the approximate
result of Eq. (\ref{mirrorhe}). The solid curves are obtained via exact
numerical calculation. From \citep{Ber01}.}
\label{fig3_Ber01}
\end{figure}

It is interesting to sketch how nucleosynthesis proceeds in the
mirror world, in particular to compute the final yield of mirror
$^4$He.  As usual one can use the limit that all (mirror) neutrons
get bound in $^4$He and compute the decoupling and deuterium
formation time. In this case we have \citep{Ber01} \be Y'_p \sim
\frac{2 \exp\left[-t(T_N)/\tau_n (1+x^{-4})^{1/2}\right]}{1+ \exp
\left[ \Delta m /T_D(1+x^{-4})^{1/6} \right]} \vv \label{mirrorhe}
\ee where it has been used the fact that the weak interaction
decoupling temperature in the mirror world is larger than $T_D$ by
a factor $(1+x^{-4})^{1/6}$ and that, since $T'_N \sim T_N$ unless
the mirror baryon density parameter $\eta'$ is very different than
$\eta$, the time of nucleosynthesis is $t(T'= T_N)
=t(T_N)/(1+x^{-4})^{1/2}$. As noticed in \citep{Ber01}, the
estimate (\ref{mirrorhe}) is not valid for small $x$ and $\eta' =
10^{-10} - 10^{-9} \sim \eta$, since in this case deuterium
production can become ineffective, thus inhibiting the whole BBN
reaction network. The result of $Y'_p$ versus $x$ as numerically
computed in \citep{Ber01} is shown in Figure \ref{fig3_Ber01}.
Notice that for large $\eta' \geq 10^{-8}$ the value of $Y'_p$ is
in good agreement with the estimate provided by Eq.
(\ref{mirrorhe}). Indeed, this large $\eta'/\eta$ regime is
particularly interesting, since it corresponds to a sizeable
contribution of mirror baryons to the present energy density of
the universe, \be \frac{\Omega'_B}{\Omega_B} = x^3
\frac{\eta'}{\eta} \pp \ee In this case mirror baryons might
constitute the dark matter or one of its components\footnote{The
growth of perturbations and the power spectrum in presence of a
sizeable mirror baryon density has been studied in
\citep{Ber01,Ign03} and in more detail
in~\cite{Ciarcelluti:2003wm} and reference therein.}. In view of
the bound on $x$, this requires a large value for the ratio
$\eta'/\eta \geq 10 $, so that the mirror world would contain a
considerably bigger fraction of $^4$He than the visible world.
Further details on the thermodynamics of the early Universe with
mirror dark matter are available in~\cite{Ciarcelluti:2008vs}.

\section{Massive Particles \& BBN}
\label{s:massive-particle}
The existence in the primordial plasma of one or more species of
massive particles  ($m_X\gg m_e$)  besides baryons is at very
least a likely possibility, given the numerous pieces of evidence
in favor of the existence of dark matter (DM). A wide variety of
observations suggests that most of the matter in the universe is
not in a visible form (with $\Omega_{DM}\simeq 5\Omega_B$), and
several of them also imply that $DM$ is non-baryonic and cold,
i.e. made by particles with a non-relativistic momentum
distribution. Direct and indirect evidence, starting more than
seventy years ago with the orbital velocities of galaxies within
clusters \citep{Zwi33}, now includes also the rotational speeds of
galaxies \citep{Bor00}, gravitational lensing \citep{Dah07,Clo06},
the cosmic microwave background \citep{Dun08}, the large scale
structure \citep{Teg06}, and the light element abundances
themselves \citep{Oli00}.

At first sight, it might appear that $DM$ has no impact on BBN
since the expansion of the universe at the BBN epoch is dominated
by radiation, and the presence of additional massive particles at
the BBN epoch is thus dynamically irrelevant.  We recall that, for
example, the analysis of \citep{Car02,Mas03} without assuming
strong priors on $\eta$ show that the behavior of the Hubble
parameter should be very close to the radiation dominated regime
during the BBN, see Section \ref{s:miscellanea}. Once the WMAP
prior on $\eta$ is used, this result would narrow even further.
Thus, even from an observational point of view, any effect due to
massive particles at the BBN must be of non-gravitational nature.

Additional couplings for the $DM$ particles are far from being a remote
possibility: there are instead strong motivations which point to a
connection between dark matter and the electroweak scale. A stable
particle with an electroweak scale mass and couplings would naturally be
produced in the thermal bath of the early universe in an amount similar to
the observed matter abundance \citep{Lee77}. From a particle physics
perspective, the hierarchy problem appears to require new physics at or
above the electroweak scale. Furthermore, stringent constraints from
electroweak precision measurements (and from the stability of the proton)
indicate that these new particles respect symmetries which limit their
interactions. Such symmetries can also lead to the stability of one or
more of the new particles, such as the lightest exotic particle in the
spectrum of R-parity conserving supersymmetry, K-parity  in universal
extra dimensional models, or T-parity in little Higgs models. For a review
of $DM$ models, see e.g. \citep{Ber04b}.

Dark matter cannot bring electric charge by definition (otherwise
it would be visible) and---although the reason is less
obvious---it appears that it cannot be strongly interacting (see
\citep{Mac07} and reference therein for a recent overview of the
stringent bounds on such scenarios). However, $DM$ might be
produced as a decay product of charged (or strongly interacting)
progenitors, and there are also viable particle physics scenarios
where this can take place, the most popular of which being the
so-called super-WIMP scenario \citep{Fen03}. Charged or strongly
interacting metastable particles at the BBN epoch may form bound
systems with baryons, altering the nuclear reactions pattern and
thus the yields of light elements. The analysis of this scenario,
which goes under the name of ``catalyzed nucleosynthesis'', is
addressed in Section \ref{CataBBN}. On the other hand, particles
which decay (or annihilate) into Standard Model products, even if
only weakly interacting, may affect BBN via the cascades they
induced in the plasma: the injection of secondaries triggers
directly or indirectly  non-thermal reactions, altering again
standard BBN predictions. Of course, if the lifetime satisfies
$\tau_X\ll 1\,$s, the particles have decayed well before the BBN
and no meaningful bound can be derived.

In general, annihilations have a similar effect as decays, the
main difference being the time (or redshift) distribution of the
injection: for decays the rate per unit volume is $n_X/\tau_X$,
while for annihilations is given by $n_Xn_{\bar X}\langle \sigma
v\rangle$. If we parameterize $\langle \sigma v\rangle\propto
\sigma_n(T/m_X)^n$ ($n=0,1$ correspond then to s-wave and p-wave
annihilation respectively), in the early universe when homogeneity
is a good approximation, the former gives an injection rate
scaling as $(1+z)^3$, or $(1+z)^{6+n}$ for annihilations. In the
following, unless otherwise stated, we will generically refer to
``decays'' to indicate any mode of injection of energetic
particles in the plasma, thus implicitly including the
annihilating case. Some specific features of the annihilation
injection mode will be discussed at the end of
Section~\ref{hadcasc}.

\subsection{Cascade Nucleosynthesis}\label{DMdecays}
The phenomenology of non-thermal BBN bounds strongly depends on
the branching ratios of the secondaries the $X$ particle decays
into: the hadronic, electromagnetic (photons, e$^\pm$), neutrino
or inert (exotic invisible) ones. The last case requires two or
more (meta)stable particles, which does not happen very naturally
in most realistic models of physics beyond the SMPP. Perhaps, one
exception is provided by the case of the axino decaying into a
gravitino and an axion, see e.g. \citep{Chu93}. In general, the
only constraint in this case comes from the requirement that the
universe at $T\sim 0.1\,$MeV is radiation-dominated, which implies
\be \rho_X=m_X\,n_X\lsim \rho_\gamma \Leftrightarrow
\left(\frac{m_X}{0.1\,{\rm
GeV}}\right)\left(\frac{n_X}{n_\gamma}\right) \lsim 10^{-4}\,. \ee
We do not treat this scenario and its constraints further. For
more details, see for example \citep{Sch88}.

It is worth noting that even a decay into neutrinos $X\to Y+\nu$
has effects on the BBN. The effects are two-fold: i) energetic
neutrinos can create charged leptons by annihilation onto the
thermal neutrino background\footnote{In principle, they can also
upscatter $e^{\pm}$ in the plasma, but decays are only relevant
when $T\ll 1\,$MeV, when almost no charged particle populate the
plasma.}; although with a suppressed efficiency, sufficiently high
energy neutrinos can also produce pions via $\nu+\bar\nu_{th}\to
\pi^{+}+\pi^{-}$ which affect the $p-n$ equilibrium.
ii)electromagnetic (and possibly hadronic) showers are induced by
3 or 4 body decay channels via a virtual or real weak boson
propagator, like  $X\to Y+\nu+e^{+}+e^{-}$; the previous final
state is always considered as kinematically allowed, when a tree
level $X\to Y+\nu$ is included in the analysis, since---given the
binding energies of nuclei---to induce any change to BBN the phase
space available must be $\gg1\,$MeV anyway. The importance of
these sub-leading channels for phenomenological constraints from
astrophysical arguments has also been emphasized in other contexts
\citep{Kac07,Mac08}. We address the reader to \citep{Kan07b} for a
more through analysis. In the remaining of this Section we
describe electromagnetic cascades (Section \ref{emcasc}) and
hadronic ones (Section \ref{hadcasc}). It is worth noting that,
for a very broad range of lifetimes involved ($0.1\,{\rm s}\lsim
\tau_X\lsim 10^{12}\,{\rm s}$), the thermalization of the
secondaries takes place in a time which is negligible with respect
to the Hubble time, and therefore redshifting of particles can be
safely neglected.

\subsubsection{Development of the electromagnetic cascade}
\label{emcasc}
There are some features of the  electromagnetic cascades in the
primordial plasma which significantly simplify the treatment with
respect to hadronic ones. When the injected $e^{+},e^{-},\gamma$
are energetic enough, the cascade develops very rapidly by a
combination of two processes: the pair-production on the CMB
thermal distribution ($\gamma + \gamma_{\rm CMB}\to e^+ + e^-$)
and the inverse Compton scattering of the non-thermal electrons
and positrons ($e^{\pm}+\gamma_{\rm CMB}\to e^{\pm} + \gamma$) off
the CMB photons \citep{Aha85,Zdz89,Sve90,Pro95,Kaw95}. There is a
critical energy, $E_C(T)$, above which the non-thermal spectrum is
quickly cutoff by pair-production, leaving virtually no photon
available to photo-dissociate the few available nuclei\footnote{No
$e^{+},e^{-}$ remain instead available for
electro-disintegrations, since the cross-section with the abundant
CMB photons is effective to quickly cool them down to thermal
energies.}. Although the typical energy for pair-producing on the
bulk of the CMB distribution is  $m_e^2/\langle E_{\rm
CMB}\rangle\simeq m_e^2/2.7\,T$, since reactions on the energetic
tail of the distribution are also important, $E_C$ turns out to be
smaller: Numerical calculations estimate $E_C\simeq m_e^2/22\,T$
\citep{Kaw95} or  $E_C\simeq m_e^2/23.6\,T$ \citep{Pro95}. By
equating $E_C$ to the $\hi2$ and $^4$He binding energies of 2.2
MeV and 19.8 MeV respectively, one infers the keV-scale
characteristic temperatures below which a small but non-negligible
fraction of $\gamma$'s  (at the percent level) is available to
photodisintegrate the light nuclei. This is why meta-stable
particles with a too short lifetime cannot significantly affect
thermal BBN yields. Meaningful bounds on electromagnetic cascades
are only achieved if $\tau_X\gsim 10^5\,$s for deuterium
dissociation and $\tau_X\gsim 10^7\,$s for the more tightly bound
$^4$He (in a radiation-dominated universe $t\propto T^{-2}\propto
E_C^2$).

\begin{figure}[t]
\begin{center}
\includegraphics[width=0.8\textwidth]{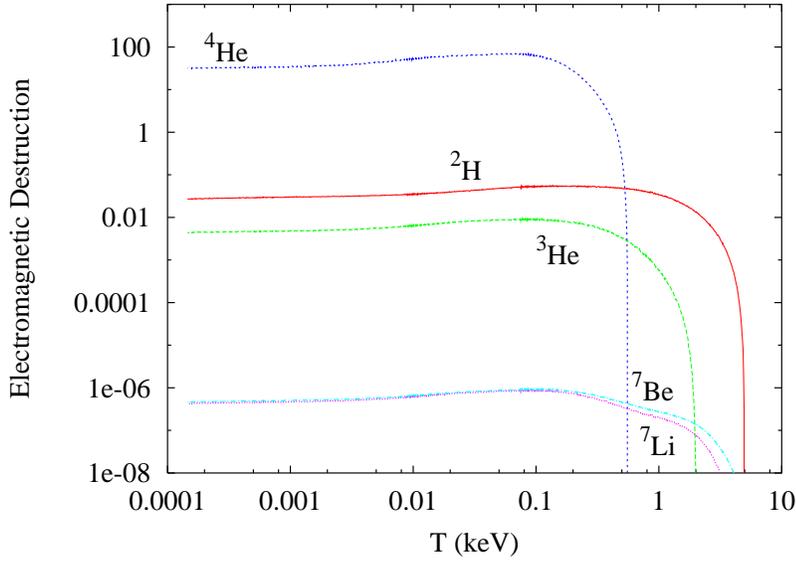}
\end{center}
\caption{Number of destroyed nuclei per TeV of electromagnetically
interacting energy injected in the plasma at temperature $T$. From
\citep{Jed06b}.}
\label{DestrEM}
\end{figure}

Another interesting feature is that the e.m. cascade develops so rapidly
that most of the effect  depends only on the total amount of injected
energy $E_0$ and the time of injection, rather than the nature and energy
of the primary.  It is customary to define a parameter representing the
energy stored in meta-stable particles before the cascade begins (hence
the zero subscripts), for example $\zeta_X\equiv m_X\,
n_{X,0}/n_{\gamma,0}$, which together with the decay time $\tau_X$
characterizes almost completely the process. A quasi-universal shape of
non-thermal photons is reached very quickly. Numerical simulations have
found a spectrum below $E_C$ well approximated by \citep{Pro95,Kaw95}
\be
\frac{{\rm d}N_{\gamma}}{{\rm d}E_{\gamma}} = \left\{
  \ba{lcl}
\displaystyle
K_0\biggl(\frac{E_{\gamma}}{E_X}\biggr)^{-3/2} &{\rm for}  & E_{\gamma}< E_X
\\*[4mm]
\displaystyle
K_0\biggl(\frac{E_{\gamma}}{E_X}\biggr)^{-2}&{\rm for}  &  E_X \leq E_{\gamma}\leq E_C
  \ea \right.
\quad ,
\label{Eq:spectrum}
\ee
where $K_0 = E_0/(E_X^2[2+{\rm ln}(E_C/E_X)])$ is a normalization constant
such that the total energy in $\gamma$-rays below $E_C$ equals the total
energy $E_0$ injected. One has $E_X\approx 0.0264E_C$ according to Ref.
\citep{Pro95}, or $E_X\approx 0.03E_C$ according to \citep{Zdz89}. A more
accurate calculation of the evolution of this spectrum should include
interactions of these ``break-out'' photons via photon-photon scattering
$\gamma +\gamma_{\rm CMB}\to \gamma + \gamma$, (mainly redistributing the
energy of energetic $\gamma$-rays right below energy $E_C$), Bethe-Heitler
pair production $\gamma + p({}^4{\rm He})\to p({}^4{\rm He}) + e^- + e^+$,
Compton scattering off thermal electrons $\gamma + e^-\to \gamma + e^-$
(with the produced energetic $e^-$ in turn inducing inverse Compton
scattering and thus further low-energy $\gamma$'s) and, of course, nuclear
photodisintegration, which directly affects BBN.

\begin{figure}[t]
\begin{center}
\includegraphics[width=0.8\textwidth]{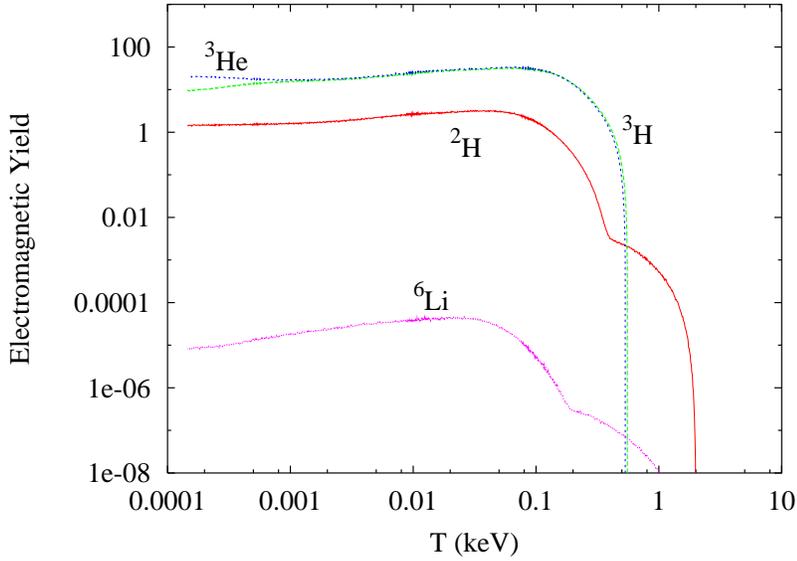}
\end{center}
\caption{Number of produced nuclei per TeV of electromagnetically
interacting energy injected in the plasma at temperature $T$, due to
photodisintegration and fusion reactions. From \citep{Jed06b}.}
\label{YieldsEM}
\end{figure}

In Figure \ref{DestrEM} we show the amounts of destroyed nuclei
per TeV of injected electromagnetic energy, $E_0$, at a
temperature $T$. The sharp rises at characteristic temperatures
are due to the pair production threshold effect for energetic
$\gamma$, which depend on the binding energy. In Figure
\ref{YieldsEM} we show a similar plot for the nuclei {\it
produced} as a result of the dissociations. It is evident that,
although a minor amount of $\hi2$ starts to be produced from
spallation of $\he3$ already at $T\simeq 1\,$keV, a major
secondary production of $\hi2,\,\h3$ and $\he3$ happens at $T\lsim
0.5\,$keV, when $^4$He photodisintegration becomes relevant. In
this range, not only does $\hi2,\,\h3$ and $\he3$ production
over-compensate their photodisintegration (which takes place at a
slower rate) but a significant synthesis of $\lisix$ is induced by
the non-thermally produced $\h3$ and $\he3$, with a  subleading
contribution from direct photodisintegration of $\li7$ and
$\bers$. For a compilation of the reactions involved in the
cascade nucleosynthesis calculations, we address the reader to the
excellent review \citep{Jed06b}. Also, some relevant new nuclear
astrophysics data and their implications for BBN have been
presented in~\cite{Kusakabe:2008kf}.

\begin{figure}[t]
\psfig{file=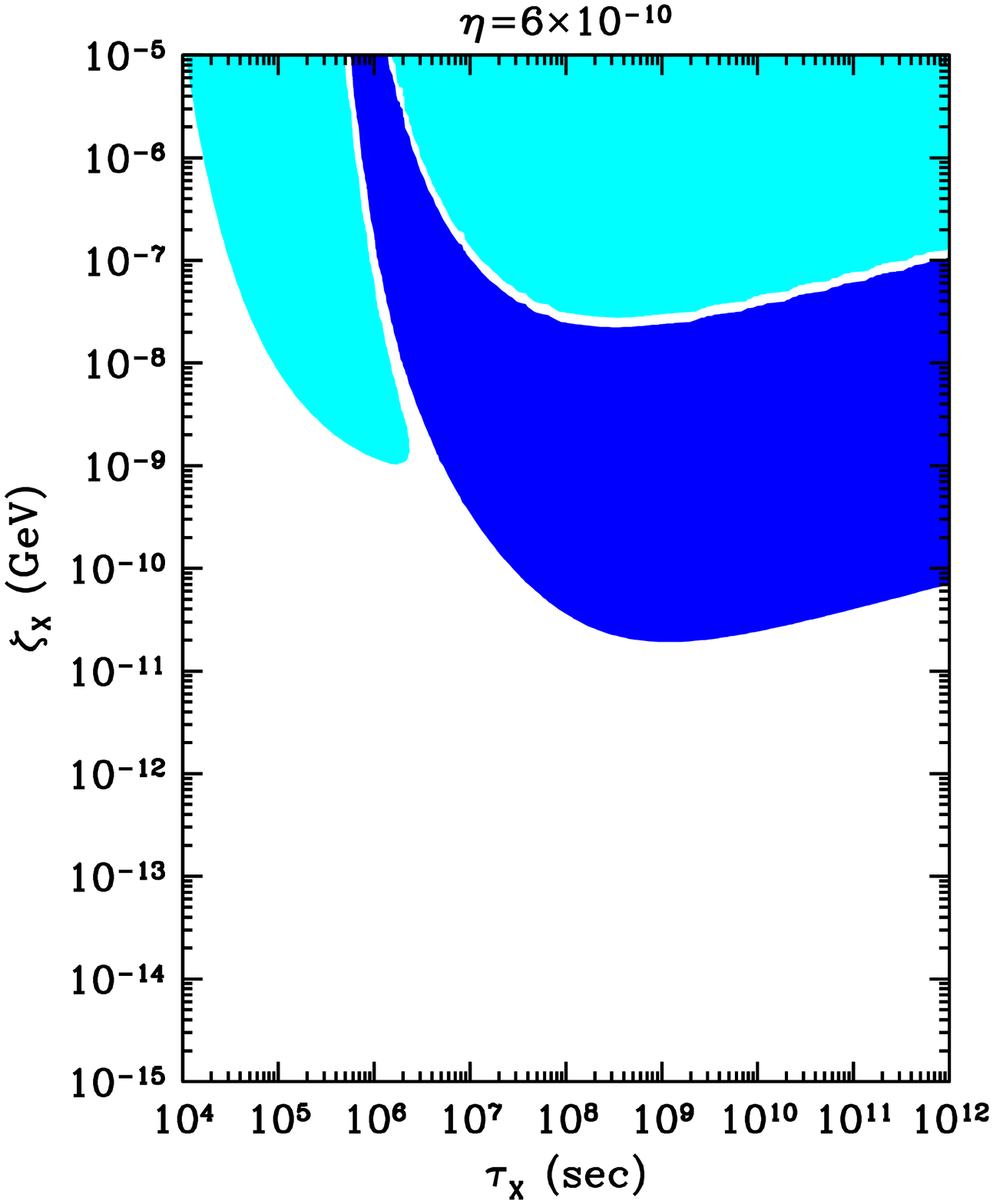, width=0.50\textwidth} \hfill
\psfig{file=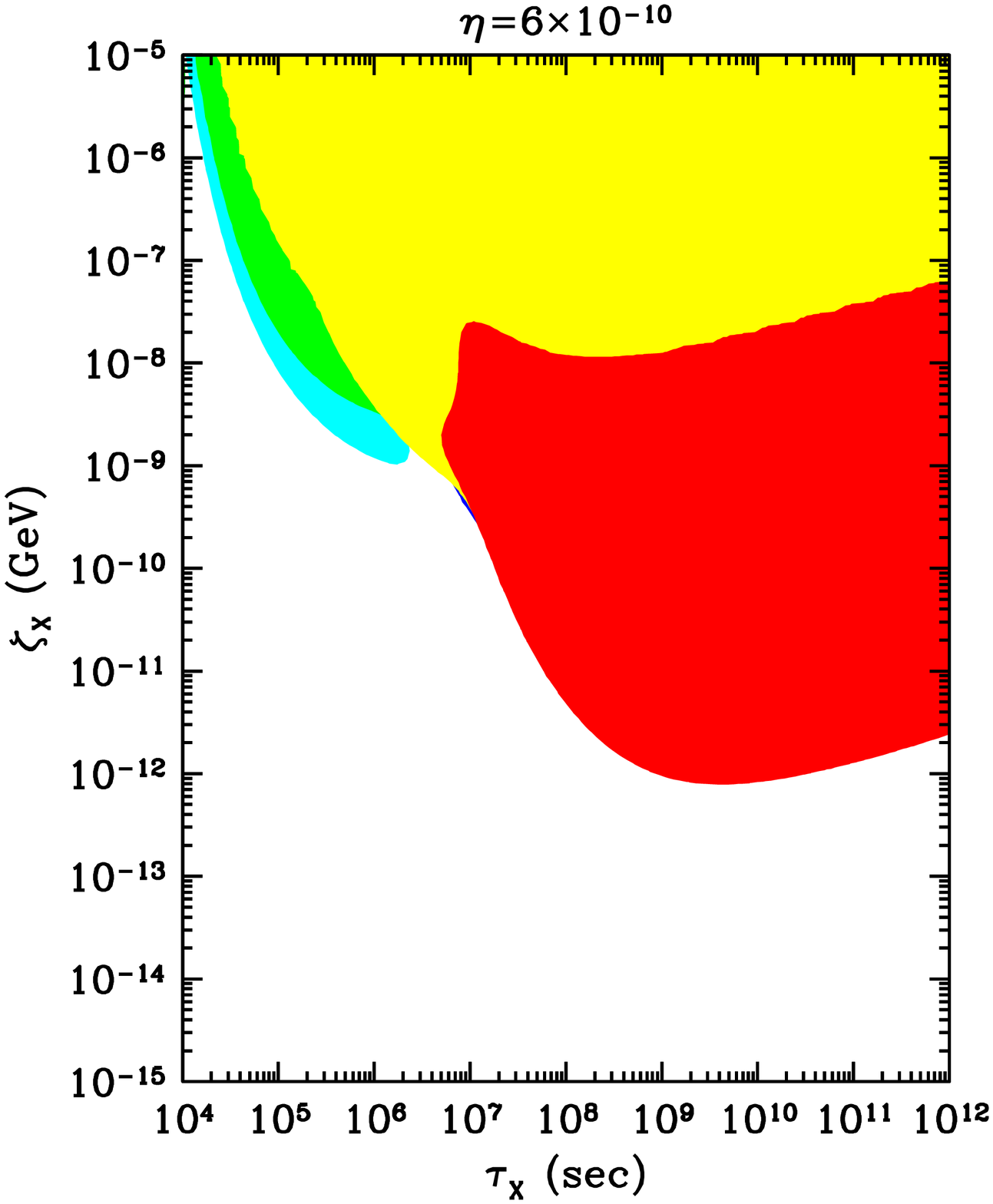, width=0.50\textwidth} \caption{Regions of
the parameter space   $\zeta_X$-$\tau_X$ excluded by: left panel:
$\hi2$ overproduction (dark blue) or $\hi2$ under-production
(light blue); right panel: including also the constraints from
overproduction of $\lisix$ (dark red), shown on the top of the
$\li7$ (yellow) and $^4$He (green) constraints. From
\citep{Cyb03a}.} \label{emConstraints}
\end{figure}

In Figure \ref{emConstraints} we show a typical result for the
excluded regions in the plane $\zeta_X$-$\tau_X$ (where $0$ refers
to a time $t\ll \tau_X$), in this case taken from \citep{Cyb03a}.
In the left panel only deuterium constraints are considered, in
the right panel constraints from all elements are included. The
light blue regions are excluded due to underproduction of
deuterium: in the left shoulder at relatively small $\tau_X$ this
is due to the fact that direct photodestruction of $\hi2$
dominates. For $\tau\gsim 10^7\,$s $^4$He destruction is
important, and leads typically to over-production of $\hi2$,
unless the injected energy is so high that in turn even this
secondary deuterium is destroyed (upper-right corner of the plot).
In this parameter range, typically even $^4$He and $\li7$ are
destroyed to a level inconsistent with observations, but these
additional constraints basically overlap with the deuterium ones.
On the other hand, if one considers $\lisix$ production from
non-thermally produced $A=3$ nuclei, one can also disfavor the red
region shown in the right panel due to over-production of
$\lisix$, which is however less robust due to the likely
reprocessing of this fragile isotope in the observed stellar
systems. Note that if the metastable particle is the progenitor of
the $DM$ candidate, it must fulfill the condition $m_X\,n_X\gsim
4\,m_p\, n_B$, i.e. a typical value for $\zeta_X$ fulfills
$\zeta_X\gsim 2\times 10^{-9}\,$GeV. Larger values imply a very
large mass difference between the $X$ particle and the $DM$
candidate, smaller values refer to metastable particles which
account at most for a sub-leading fraction of the $DM$ today. So,
a way to summarize previous constraints is to say that BBN
excludes electromagnetically decaying particles as progenitors of
$DM$ when their lifetime is longer than $\tau_X\gsim {\rm
few}\times 10^{5}\,$s.

\subsubsection{Including hadronic channels}
\label{hadcasc}
\begin{figure}[t]
\begin{center}
\includegraphics[width=0.7\textwidth]{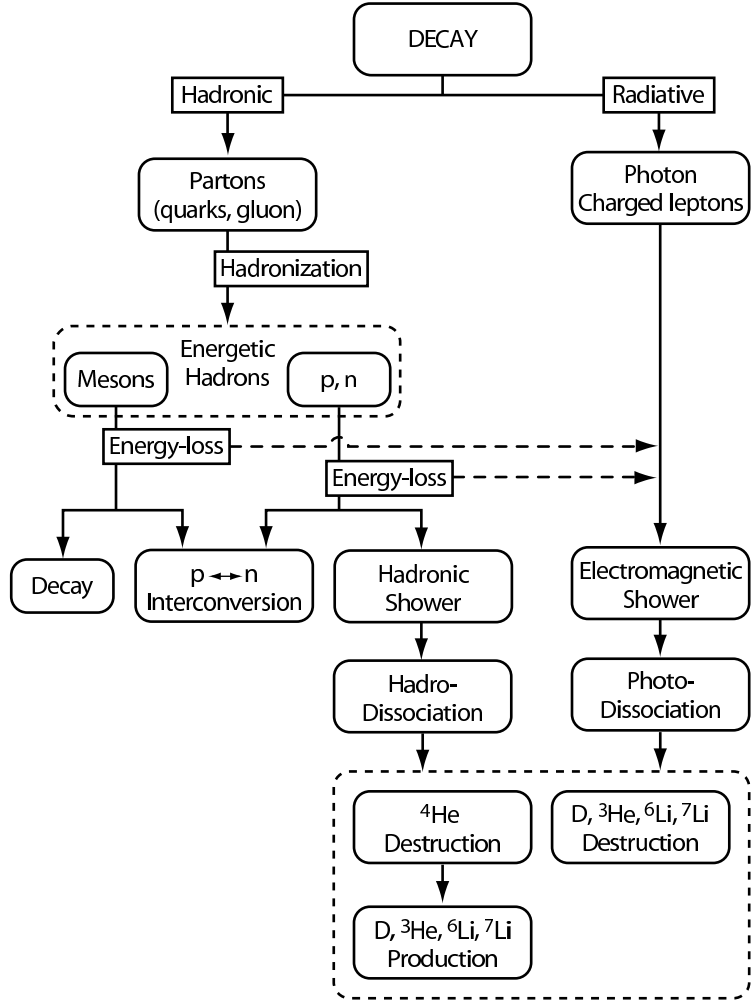}
\end{center}
\caption{Flow--chart of the decay effects on BBN. From
\citep{Kaw05a}.}
\label{scheme}
\end{figure}

Generally, in the decay of the $X$ particles one expects both
hadronic and electromagnetic channels. The treatment of cascades
in presence of hadrons is significantly more involved, since it
introduces more particles, more timescales and more processes to
take into account. The flow-chart in Figure \ref{scheme}
summarizes the decay scheme and physical processes involved. Note
that a hadronic branching ratio unavoidably leads to secondary
electromagnetic showers, which affect BBN along the lines
described in Section \ref{emcasc}. To have a first qualitative
understanding of the effects of hadronic particles, it is worth
recalling that different hadronic species interact in the plasma
via a few reactions, which assume different relevance at different
times:
\begin{itemize}
\item
\emph{mesons}, mostly $\pi^{\pm}$ and kaons (with rest frame lifetime in
the $10^{-8}\,$s range), have only an effect at times $t \approx 1-10\,$s,
when ordinary weak interactions are not efficient anymore, but they still
have time to interact before decaying. They mostly act by enhancing the
n/p ratio and thus the final value of $\yp$.
\item
\emph{antinucleons}, by the preferred tendency to annihilate onto protons,
have a similar final effect of increasing n/p. Compared with mesons, they
also have an additional peculiar effect at later times
($t\simeq$10$^2\,$s) by annihilating onto $\yp$ and leaving $\hi2$, $\h3$,
and $\he3$ among secondaries. \item \emph{nucleons}: at early times,
nucleons thermalize via electromagnetic processes: magnetic moment
scattering off  $e^\pm$ for neutrons, Coulomb stopping  off $e^\pm$ and
Thomson scattering off thermal photons for protons. However, at late times
other energy loss mechanisms start to dominate for high energy nucleons,
namely nucleon--nucleon collision and nuclear spallation reactions. Due to
the different electric charge, these nuclear processes are already
dominant for neutrons at $t\gsim $200 s, while for protons only  at
$t\gsim$10$^4$ s. When they are effective, a cascade nucleosynthesis can
take place: each nucleon--nucleon scattering will produce another
energetic nucleon (a single 100 GeV nucleon can produce several tens of 10
MeV nucleons) and their effect of spallation over $^4$He will produce many
$\hi2$, $\h3$, and $\he3$ nuclei. The total effect will be more efficient
than for antinucleons: a single 100 GeV antimatter particle has a much shorter
mean free path before annihilating with a $^4$He nucleus and produce the
mentioned secondaries, and nucleon induced production is therefore much
more efficient. Non-thermal nucleon injection lead to an increased $\yp$
abundance at $t\leq$200 s, increased $\hi2$ abundance at 200 s$\leq
t\leq$10$^4$ s, or decreased $\li7$ abundance at $t\approx$10$^3$s.
Spallation of $^4$He to produce $\hi2$, $\h3$, and $\he3$ may have as a
secondary effect the synthesis of $\lisix$.
\end{itemize}

To illustrate this point, in Figure \ref{YieldsHadr}, we report
the light element yields produced by the decay $X\rightarrow q\bar
q$ of a single metastable particle with mass $m_X$=1 TeV, as a
function of the photon temperature at which the injection takes
place. It is interesting that initially, after hadronization of
the quark--antiquark state, on average only 1.56 neutrons result;
all others are created at $T\leq$90 keV by the thermalization of
injected neutrons and protons due to inelastic nucleon--nucleon
scattering and $^4$He spallation. Similarly, all the $\hi2$,
$\h3$, and $\he3$ nuclei are due to $^4$He spallation processes
and n--p nonthermal fusion reactions (for $\hi2$) induced by the
thermalization of the injected energetic nucleons.

\begin{figure}[t]
\begin{center}
\includegraphics[width=0.8\textwidth]{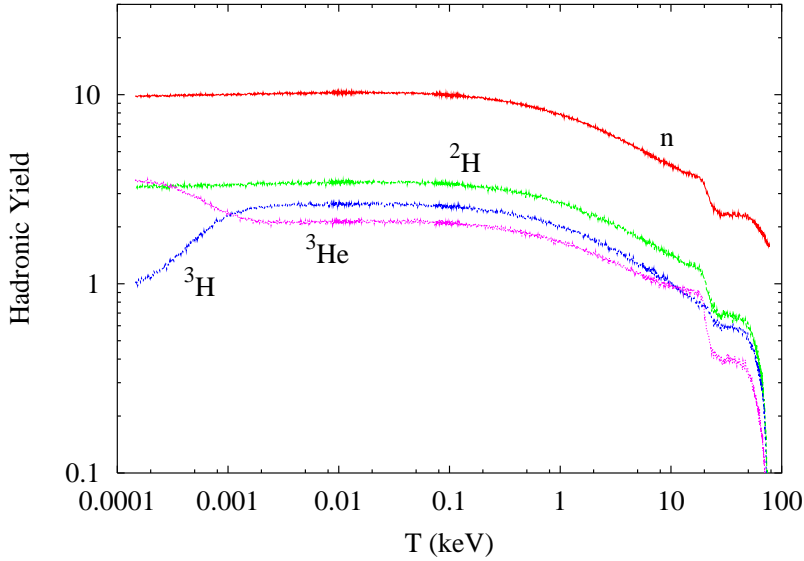}
\end{center}
\caption{Light element nuclei yields for a particle of mass $m_X$=1 TeV
decaying as $X\rightarrow q\bar q$ as a function of the temperature $T$ of
the cascade injection in the plasma.  From \citep{Jed06b}.}
\label{YieldsHadr}
\end{figure}

For a more quantitative analysis, a numerical treatment is
required. Technically, the problem of solving cascade
nucleosynthesis is highly non-trivial, especially in presence of
significant hadronic branching ratios. One major difficulty is
that one cannot study the non-thermal effects independently of the
earlier standard BBN stage, as it is possible (at least as first
approximation) in the case of electromagnetic cascades. Although
already in the 80's several authors have estimated the effects of
injecting a non-thermal population of particles in the plasma at
the BBN epoch, only recently several authors  have followed
\emph{self-consistently} the reactions taking place during their
thermalization and coupled them to the simultaneously ongoing
thermal nuclear network \citep{Kaw05b,Jed04b,Cyb06,Jed06b}. The
latest treatments typically adopt a Monte Carlo technique to
calculate the interaction probability of the particle shower, and
codes such as PYTHIA are employed to determine the energy
distribution of the shower particles given the initial branching
ratios in the decay. Nucleon energy losses must be taken into
account as well. For a more detailed overview of the scheme and
techniques of the numerical treatment, see e.g.
\citep{Kaw05a,Jed06b}.

\begin{figure}[p]
\begin{center}
\includegraphics[width=0.8\textwidth]{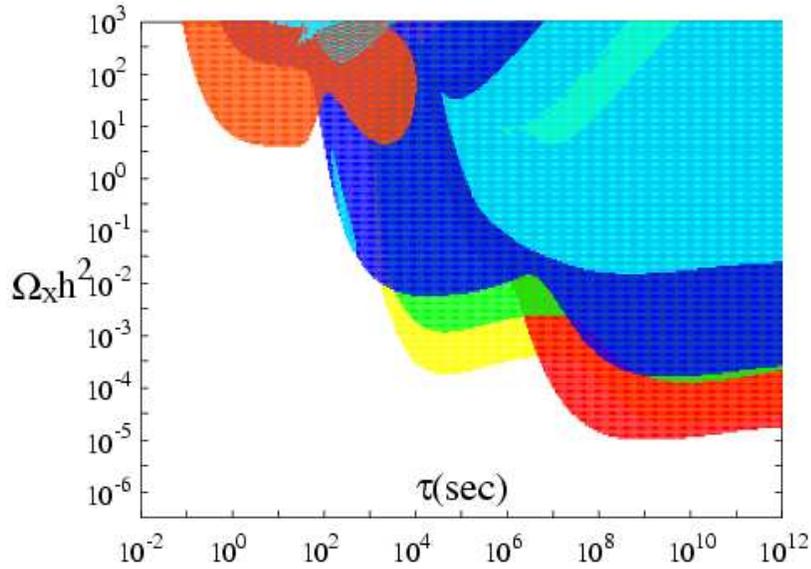}
\end{center}
\caption{Conservative BBN constraints on the abundance of relic
decaying neutral particles as a function of their lifetime; particle mass
is $m_\chi$=1 TeV and the hadronic branching ratio
$B_h$=3.33$\times$10$^{-2}$. $\Omega_\chi h^2$ is the contribution neutral
particles would have given to the total relic density today, would they
have not decayed. Colored regions are excluded and correspond to
constraints imposed by observations (see text). From \citep{Jed06b}.}
\label{HadrBrat-elem}
\end{figure}
\begin{figure}[p]
\begin{center}
\includegraphics[width=0.8\textwidth]{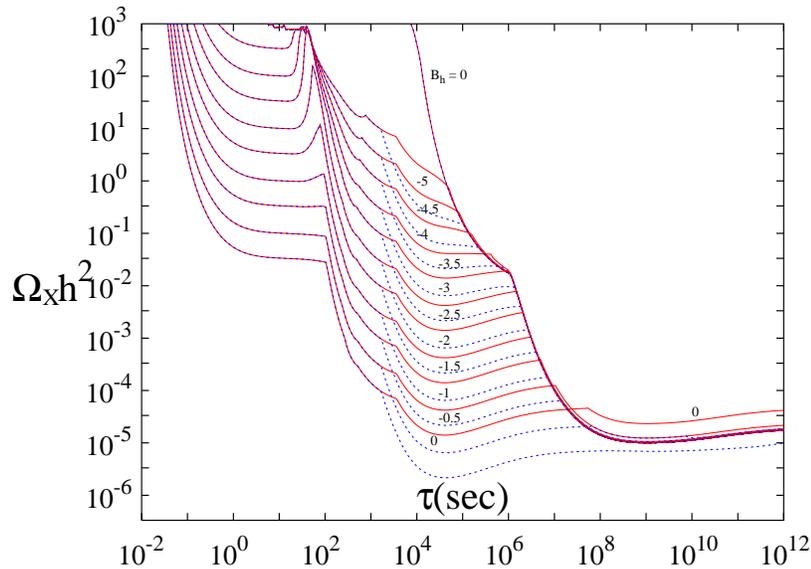}
\end{center}
\caption{Constraints on the relic abundance of neutral decaying particles
as a function of their decay time, for a particle mass $m_\chi$=100GeV and
varying branching ratio, $B_h$. Numbers in the picture refer to the solid
line above and stand for the corresponding $\log_{10}B_h$. Dotted lines
are the actual constraints if one considers the less conservative
$\lisix$/$\li7$. From \citep{Jed06b}.}
\label{BHconstraints}
\end{figure}
The resulting constraints can be presented similarly to the ones
for e.m. decays, still it is crucial to specify not only the
injected energy, but also the hadronic branching ratio, $B_h$. For
example, in Figure \ref{HadrBrat-elem} we report the BBN
constraints derived in \citep{Jed06b} for the decay of particle of
mass $m_X$=1 TeV and hadronic branching ratio,
$B_h$=3.33$\times$10$^{-2}$. The quantity $\Omega_X h^2$ is the
contribution that neutral particles would have given to the total
energy density today, would they have not decayed, and is
proportional to the parameter $\zeta_X$ introduced in the previous
Section ($\Omega_X\,h^2=(M_X\,n_{X}^{0} h^2)/\rho_{cr}=
(n_{\gamma}^{0}h^2/\rho_{cr}) \zeta_X\simeq 3.9\times
10^7\zeta_X/$GeV). Colored regions are excluded and correspond to
constraints imposed by upper limit for $\yp$ (orange area), upper
limit on $\hi2$ (blue), upper limit on $\he3$/$\hi2$(red), and
lower limit on $\li7$ (light blue). The yellow region violates the
less conservative bound from $\lisix$/$\li7$. References for
observational abundances used to get these constraints, which
partly differ from the ones we have compiled in Section
\ref{sec:obsabund}, can be found in the original paper
\citep{Jed06b}.

As a consequence of the different mechanisms dominating at
different times, the most stringent constraints are given by: the
overproduction of $^4$He at early times
($\tau_\chi\leq$10$^2\,$s), overproduction of $\hi2$ for
10$^2\,$s$\leq \tau_\chi\leq$ 10$^3\,$s, overproduction of
$\lisix$ for 10$^3\,$s$\leq \tau_\chi\leq$ 10$^7\,$s, and an
overproduction of the $\he3$ /$\hi2$ ratio for
$\tau_\chi\geq$10$^7\,$s. Figure \ref{BHconstraints} illustrates
how the constraints depend on the branching ratio, $B_h$, this
time for a particle mass, $m_\chi$=100 GeV, with the allowed
region below the lines. In the limit $B_h\to 0$, one recovers
bounds due to the e.m. channels, while even for $B_h\sim 1\%$
metastable particles with a relic abundance comparable to today's
$DM$ one are excluded if $\tau_X\gsim 10^{2}\,$s. Of course, one
can turn the argument around and explore the possibility that
decays may {\it explain} some discrepancies existing between the
SBBN model and observation. In particular, both ``lithium
problems'' mentioned in Section~\ref{sec:lithium6} may be solved
by hadronically decaying particles with lifetime $\sim 2000\,$s
(see e.g. Figure 2 in~\citep{Jed06a,Jed08a}). It is worth
mentioning that in some of these scenarios ``low-energy''
observables (like the properties of gravitino dark matter) would
be linked to very high energy physical scales, as the reheating
temperature and the leptogenesis scale, see for
example~\cite{{Cer06},Kan07a,Steffen:2008bt}.

In case of an annihilation mode, the major peculiarity is that
in order to have the same injection rate of a decay
at an epoch $z_{\rm inj}$, a very large annihilation rate at epochs
$z>z_{\rm inj}$ is required.
Typically, in order to produce significant modifications of the light element yields,
often an unphysically large $\langle \sigma v\rangle$ is required.
Stronger bounds follow then from non-BBN considerations, unless a relatively
light particle ($m_X\lsim $few GeV) is considered (see e.g. \citep{McD00b}).
A possible exception is provided by the
$\lisix$ nuclide. It has been shown that, for typical electroweak-scale WIMP masses,
the primordial $\lisix$ yield from DM annihilation exceeds the SBBN abundance
for any  $\langle \sigma v\rangle > 10^{-27}\,$cm$^3$/s~\citep{Jed04a,Jed04b}.

This is illustrated in Figure~\ref{li6annihl}: interestingly for
the annihilation rate required to produce the correct WIMP DM
thermal relic abundance, an $\mathcal{O}$(100) GeV DM particle
produces an amount of $\lisix$ in the same range required to
explain observational claims of $\lisix$.

\begin{figure}[t]
\begin{center}
\includegraphics[width=0.8\textwidth]{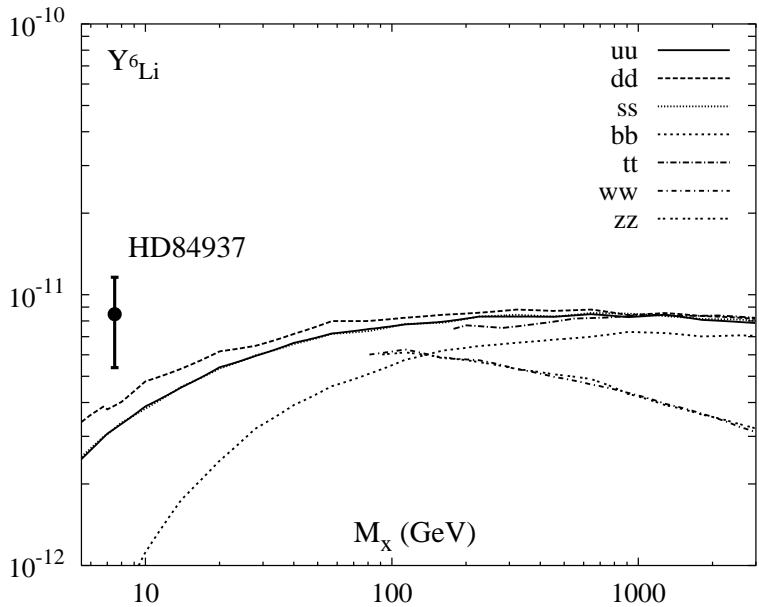}
\end{center}
\caption{The abundances of $\lisix$ as produced in scenarios
with hadronic cascades injected by neutralino annihilation, for different
channels, as labeled.
The annihilation rate is taken to be
 $\langle \sigma v\rangle$= 10$^{-25}$cm$^3$/s.
From \citep{Jed04b}.}
\label{li6annihl}
\end{figure}

\subsection{Catalyzed BBN}
\label{CataBBN}
The constraints derived in the previous Section assume that,
before decaying, the meta-stable particle $X$ is ``inert''. While
$DM$ can not be charged or strongly interacting, there is no a
priori reason why its parent particle should bring no electric or
strong charge. Whenever this happens, the possibility arises that
$X$ particles may form bound systems with baryons, altering the
nuclear reaction pattern and thus the yield of light elements.
This scenario is currently known as catalyzed BBN (CBBN). The
cosmological role of charged massive particles (CHAMPs) was
already considered in the late 80's \citep{DeR89,Dim89,Raf89}, but
the influence of bound states in BBN has only been fully
appreciated recently.  In 2006, within a few weeks three papers
appeared pointing out the importance of CBBN
\citep{Pos06,Koh07,Kap06} and identifying the main physical
mechanisms responsible for the alteration in the nuclear network.
In the last two years, several articles have followed, clarifying
the physical ingredients regulating this complex scenario,
including  refinement in the calculation of catalyzed reactions,
late-time nucleosynthesis, etc.
\citep{Cyb06,Ham07,Bir07,Kaw07,Jit07,Jed08a,Cum07,Jed08b,Kam08,
Kus07a,Kus07b,Jit08}. A quite complete review of the physics can
be found in \citep{Jed08a}, which we shall mainly follow for the
present summary. We limit ourselves to consider singly charged
CHAMPs, although it has been argued that negative doubly charged
and {\it stable} CHAMPs bound to $^4$He$^{++}$ may be viable as
$DM$ candidates in walking technicolor theories \citep{Khl08a}. It
is worth noting that, at least qualitatively, one expects similar
catalytic mechanisms if the  particle $X$ is strongly interacting,
rather than being electrically charged. One physically motivated
scenario of this kind is the long-lived gluino in split-SUSY
\citep{Ark05a,Giu04,Ark05b}. Some calculations of the primordial
nucleosynthesis in presence of massive, strongly interacting
particles have been performed in the past (see
\citep{Dic80,Pla95}), where bound states with ``ordinary'' nuclei
were considered.  The main difficulty with this scenario is that
the nuclear physics of such bound states is very hard to treat
reliably. For example, analogies with toy-models used to describe
hypernuclei are employed \citep{Moh98} and the description of the
bound systems is at best parametric.  An accurate treatment is
made highly non-trivial by the effects of non-perturbative
physics, and we shall not consider them further. We want to
remark, however, that the BBN cascade bounds that are sometimes
considered in the literature for such particles (see
\citep{Arv05}) may be altered, since catalytic effects are
completely neglected.

\subsubsection{Early time CBBN: formation of bound states and catalysis}\label{CataPhys}
\begin{table}[b]
\caption{Binding energies $E_b$ in the Bohr approximation
$|E_b|\simeq  Z^2\, \alpha\, m_A/2$ and photo-dissociation
decoupling temperatures $ T_{ph}$ in keV (as calculated in
\citep{Pos06}) for exotic bound states $AX$. }
\label{PospeBoundstates}
\begin{center}
\begin{tabular}{|c|c|c|c|c|c|c|c|c|}
\hline
bound state & $p\X$ & $\hi2\X$& $\h3\X$& $\he3\X$&     $^4$He$\X$&  $\li7\X$&
$\bers\X$&  $\bero\X$\\ \hline\hline
 $|E_b|$ & 25 & 50 & 75 & 299 & 397 & 1566 & 2787 & 3178 \\ \hline
$T_{ph}$ & 0.6 & 1.2 & 1.8 & 6.3 & 8.2 & 21 & 32 & 34 \\ \hline
\hline
\end{tabular}
\end{center}
\end{table}

A negatively charged, long lived particle $X$ (or CHAMP) with mass
$M_X \gg m_p$  would form bound--states with nuclei and alter the
network of reactions leading to the synthesis of light elements.
Compared with a generic nucleus $A$ of charge $Z$ and mass $m_A$,
its corresponding bound state with a CHAMP, $AX$, has a  mass
higher by $\sim m_X$, a charge lower by one unit ($Z-1$), and it
is characterized by an $AX$ binding energy given by $E_b\simeq
Z^2\,\alpha\, m_A/2$ in the limit $m_A\ll m_X$. Another
interesting quantity is the photo-dissociation temperature
$T_{ph}$ of the $AX$ system, defined as the temperature at which
the photodissociation rate, $\Gamma_{ph}(T)$, for the bound
nucleus becomes smaller than the Hubble rate, $H(T)$.  Roughly
speaking, below $T_{ph}$ the $AX$ system is stable against
photodestruction. In Table \ref{PospeBoundstates} we report
binding energies in the Bohr-like atom approximation (corrections
are of order $\sim 10\%$ for $\He4 X$, up to 50\% for $^7$Be$X$)
and the $T_{ph}$, as originally calculated in \citep{Pos06}. A
precise calculation of the fractional abundance of bound nuclei
requires solving the Boltzmann equation, including all relevant
reactions for the $AX$ state. The inadequacy of simple approaches
based on the Saha equation to determine the evolution of the $AX$
abundance was realized immediately in \citep{Pos06} and further
analyzed e.g. in \citep{Koh07,Cyb06}. However, even the latter
analysis had to rely on strong approximations on the reaction
rates for bound nuclei (that the bound state nucleus would be
destroyed in the interaction, and that standard BBN is over at the
time CBBN takes place, etc.). Figure \ref{JedaBoundState}, which
we take from the more recent and accurate analysis in
\citep{Jed08a}, shows the evolution of bound state fractions
$f_A^b\equiv n_{A X}/n_{A}^{tot}$ of nuclei $A$ bound to $X$ as a
function of temperature $T$, for a model with $m_X = 100\,$GeV and
$\Omega_X h^2 = 0.1$. It confirms that the behavior of $f_A^b$ is
significantly different than that expected from simple estimates
by the Saha equation due to nuclear destruction of bound states
and slow recombination rates  (with respect to the Hubble time).
This is particularly relevant in $f_{\rm ^7Li}^b$  due to the
$\li7X(p,X)\,2\,^4$He reaction.

\begin{figure}[t]
\begin{center}
\includegraphics[width=0.8\textwidth]{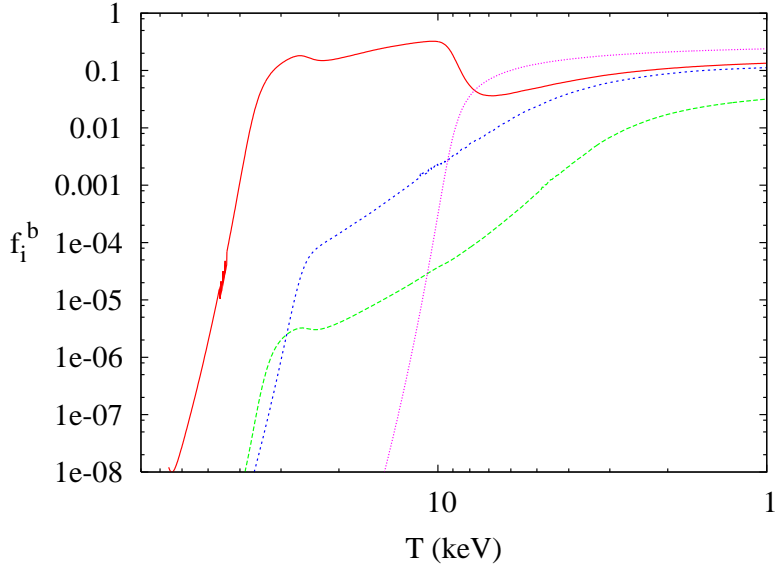}
\end{center}
\caption{Bound state fractions $f_A^b\equiv n_{AX}/n_{A}^{tot}$ of
nuclei $A$ bound to CHAMP $X^-$ as a function of temperature $T$,
for a model with $M_X = 100\,$GeV and $\Omega_X h^2 = 0.1$. Shown
are $f_A^b$ for $\bers$ solid (red), $\li7$ long-dashed (green),
$\lisix$ short -dashed (blue), and $^4$He dotted (purple),
respectively. From \citep{Jed08a}. } \label{JedaBoundState}
\end{figure}

Once CHAMP--nuclei states  $XA$ are formed in the plasma, each nuclear
reaction will have its CHAMP homologue:
\bea
\textrm{standard} \,\,BBN: && A + A_1 \rightarrow A_2 + A_3,
\nonumber
\\ CBBN: &&  AX+ A_1\rightarrow A_2 + A_3 +X\,.
\eea
At first sight, the main advantage of CBBN reactions is the smaller
Coulomb barrier. However, there is a more subtle effect---whose importance was
already recognized in \citep{Pos06}---which acts on homologues of
radiative captures , i.e. of reactions of the kind $A(A_1,\gamma)A_2$. If
we denote by $\lambda_\gamma$ the wavelength of the emitted photon, in
general of the order of 100 fm,  electric dipole (E1) reaction rates scale
as $\lambda_\gamma^{-3}$, whereas electric quadrupole (E2) ones scale as
$\lambda_\gamma^{-5}$. The introduction of the photonless state in the
CHAMP-mediated reaction replaces $\lambda_\gamma$ with the Bohr radius of
the bound system, approximately 5 fm for $^4$He. In terms of the usual parameterization
of low-energy nuclear reaction (see Eq.~(\ref{Sfacparam})), the former Coulomb effect enters
the exponential barrier-factor, the latter---when present---influences the $S$-factor, which can be indeed enhanced by orders of magnitude, as reported in Table~\ref{CyburtModiCHAMP}.
Another, usually sub-leading effect in the direction of decreasing $S$ is a decrease
of phase-space due to the reduction of the reaction Q-value.

\begin{table}[t]
\caption{Dipole amplitude, Q-values, and catalyzed S-factor
enhancement in the cross-section for relevant ($\alpha,\gamma$)
reactions.  Modified from \citep{Cyb06}, taking into account the
values recently reported  by \citep{Kam08}.}
\label{CyburtModiCHAMP}
\begin{center}
\begin{tabular}{ccccc}
\hline\hline
 & EM & $N(\alpha,\gamma)C$ & ${\rm N}\X(B,C)X^-$ & Enhancement \\
Reaction  &  Transition  & $Q_{\rm SPN}$ (MeV) &  $Q_{\rm CBBN}$ (MeV) & $S_{\rm CBBN}/S_{\rm SBBN}$  \\ \hline\hline
$\hi2(\alpha,\gamma)\lisix$     &  E2 & 1.474   & 1.124  & $\sim10^7$ \\
$\h3(\alpha,\gamma)\li7$   &  E1 & 2.467   & 2.117  & $\sim$30\\
$\he3(\alpha,\gamma)\bers$  &  E1 & 1.587   & 1.237  & $\sim$30\\
\hline
\end{tabular}
\end{center}
\end{table}
The most important alteration to the standard scenario in
catalyzed BBN is mainly due to the  enhancement of the single
$\lisix$ producing process, \be \textrm{standard} \,\,BBN: \yp +
\hi2 \rightarrow \lisix + \gamma;\,\,Q=1.47\,{\rm MeV}\vv
\label{Sdgammali6} \ee replaced by the process \be CBBN:  \He4 X
+\hi2\rightarrow\lisix+X;\,\, Q\simeq 1.13\,{\rm MeV}\vv
\label{Cdgammali6} \ee which in \citep{Pos06} was identified to be
very effective even for a small fraction of $X$ particles bound
with nuclei. The usual BBN process of Eq.~(\ref{Sdgammali6}) is
indeed only allowed at the quadrupole level (due to the almost
identical mass to charge ratio of $\hi2$ and $\yp$), which is the
reason for the very small value of $\lisix/\li7$, as already
discussed in Section \ref{sec:obsabund}. Another possible path to
enhance the $\lisix$ yield in CBBN was proposed in \citep{Kap06},
who noticed that the decay of $X$ when still in a bound state with
$\yp$ could result in a break--up of the $\yp$ nucleus, producing
$\he3$ and $\h3$ that would eventually fuse into $\lisix$ when
reacting with $\yp$. However, the possibility that this would
happen is estimated to be very low and the $\hi2(\He4
X,\,\lisix)X$ appears to be in all cases still dominating the
production of $\lisix$ \citep{Kap06,Cyb06}. The enhancement of the
$\lisix$ yield in CBBN due to the process in
Eq.~(\ref{Cdgammali6}) has been confirmed by all the published
analysis as the most remarkable effect of CBBN and in particular
by \citep{Kam08}, who performed an accurate quantum three--body
calculation of the cross-section of the reactions involved in
CBBN. The observational hints of a plateau in $\lisix$ at a value
well above standard BBN predictions, as well as the persisting
discrepancy between $\li7$ observations in the Spite Plateau and
the (apparently overproduced) $\li7$ yield has motivated several
authors to explore the CBBN scenario further, trying to also
explain the $\li7$ ``problem''. A mechanism to address this issue
was pointed out in \citep{Koh07}. Since significant fractions of
$\li7$ and (mostly) $\bers$ are in bound states with CHAMPs,
$\bers$ can be depleted by the enhancement of the CBBN analogues
of $\li7$(p,$\alpha$)$^4$He, $\bers$(n,p)$\li7$, and
$\bers$(n,$\alpha$)$^4$He. The authors of Ref. \citep{Cum07}
performed a CBBN analysis also adding the effect of $X$ decay
cascades. They concluded that in presence of strong showers from
decaying relic particles, bound-state effects on nucleosynthesis
are negligible, and both Li problems are solved (if at all) in a
way very similar to the cascade BBN case in absence of catalysis.
\citep{Bir07} proposed a more elaborate solution of the $\li7$
problem: the $\bers X(p,\,\gamma) ^8{\rm B} X$, and the subsequent
beta--decay of $^8{\rm B}\rightarrow \bero$+$e^+$+$\nu_e$ would
deplete the final $\li7$ abundance, with little consequence on
$\yp$. The $\bers X(p,\,\gamma)$ reaction would happen through a
shifting of the resonance as an effect of the $X$ presence, which
would lead to a huge rate enhancement.

To a large extent, the reason why settling these issues is far
from trivial is that significant uncertainties remain in the
estimates of binding energies and CBBN reaction rates, due to the
use of very simplified nuclear models and of the Born
approximation. An account of the situation has been given in
Section III of \citep{Jed08a}, which we address the reader to for
further technical details.  \citep{Jed08a} also contains the most
systematic and up-to-date analysis of CBBN, including calculations
of rates of $A X$ recombination--photodisintegration and CBBN
analogues of BBN nuclear reaction rates. As long as the range
$10\,{\rm keV}>T>0.8\,{\rm keV}$ is concerned, the author confirms
that the most relevant role is played by
reaction~(\ref{Cdgammali6}), even when using the detailed
calculation for its rate obtained in \citep{Ham07}. Moreover, in
this regime only nine reactions (reported in Table II of
\citep{Jed08a}) are sufficient to describe the physics of CBBN.
The evolution of light nuclide abundances in presence of
bound--state reactions is shown in Figure \ref{JedaParaCPNconst},
where the enhancement of $\lisix$ abundance is clearly visible
(note however that the role of $X$ decays has been ``switched
off'').

\begin{figure}[!thb]
\begin{center}
\includegraphics[width=0.8\textwidth]{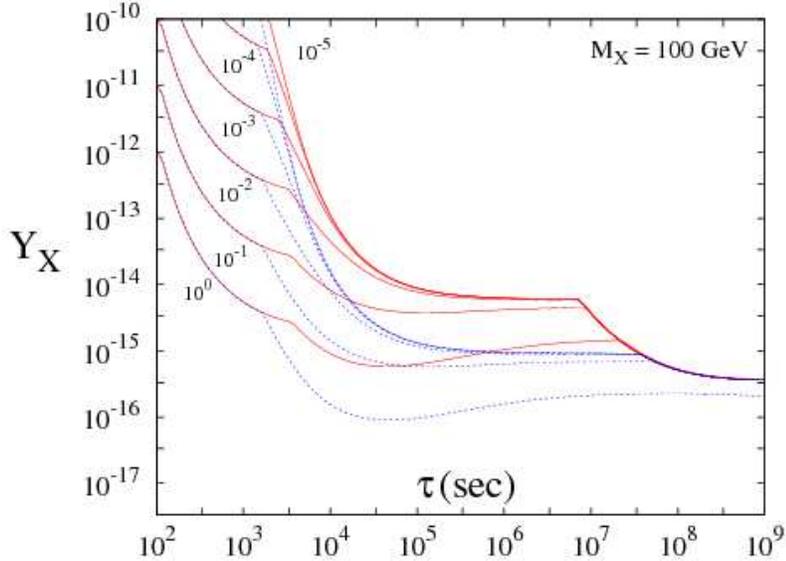}
\end{center}
\caption{Limits on the primordial CHAMP to entropy ratio $Y_x = n_X/s$
for CHAMPs with intermediate lifetimes. Shown are constraint lines for
CHAMPs of mass $m_X $=100 GeV and a variety of hadronic branching ratios,
$B_h = 10^{-5}-1$, as labelled in the Figure. Solid (red) lines correspond
to the conservative limit $\lisix/\li7$ $< $0.66, whereas dashed (blue)
lines correspond to $\lisix/\li7$ $< $0.1. From \citep{Jed06b}, updated (courtesy of K. Jedamzik).}
\label{JedaParaCPNconst}
\end{figure}

\subsubsection{Late time CBBN: H$X$ bound states, CHAMP-exchange,  and decays}

The authors of \citep{Koh07} suggested the possibility that the
formation of bound state nuclei of CHAMPs with $\h3$, $\hi2$ and
$p$ at very late times might induce the suppression of synthesized
$\lisix$.  Bound states of $Z$=1 nuclei with $X$ form at 1--2 keV
temperatures,  see Table \ref{PospeBoundstates}. These bound
states behave essentially as ``long-lived'' neutrons, which can
dissociate Li and $\bers$ Coulomb-unsuppressed. \citep{Jed08a}
carried on an extensive analysis of these effects, making use of
reaction rates derived in the Born approximation, identifying
nineteen reactions as relevant when late time CBBN effects are
taken into account. These early results indicated that bound
states of CHAMPs with $Z=1$ WOULD induce at late times the
destruction of most of the synthesized $\lisix$ and some $\li7$.
Initially, it appeared that the stringent constraints initially
put on the abundance of CHAMPs, e.g. in \citep{Pos06} and
\citep{Kaw07} loosen significantly. To add another layer of
complication, an additional late time effect was pointed out in
\citep{Jed08a}, which somewhat compensates the previous one. It is
due to exothermal transfer of CHAMPs from $Z=1$ nuclei  into
heavier nuclei (CHAMP-exchange reactions), i.e. ${\rm H} X(\He4,
{\rm H})\He4 X$. By lowering the abundance of neutral states ${\rm
H}X$, more of the $\lisix$ and $\bers$ produced can survive. The
detailed study of bound state reaction rates performed by
\citep{Kam08} has finally shown that CHAMP-exchange reactions
dominate the late time CBBN, thus ``protecting'' the abundances of
Li and $\bers$ produced at earlier times, as it can be seen in
Figure \ref{JedaCPNevol1}. The late-time drop in $\bers$ is due to
electronic capture decays.

\begin{figure}[t]
\begin{center}
\includegraphics[width=0.8\textwidth]{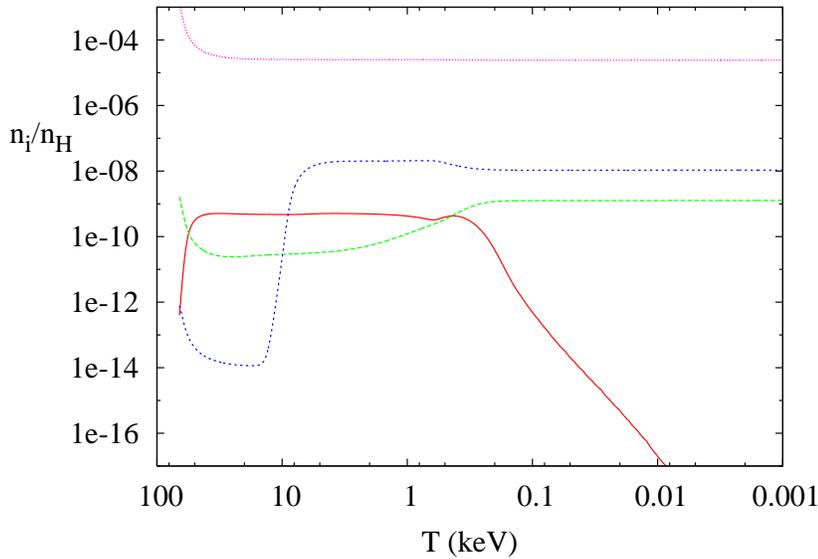}
\end{center}
\caption{Evolution of light-element number ratios $\bers$/H (solid
- red), $\li7$/H (long-dashed - green), $\lisix$/H (short-dashed -
blue), and $\hi2$/H (dotted - purple), for a CHAMP model with $M_X
= 100\,$GeV, $\Omega_Xh2 = 0.01$, and $\tau_X = 10^{10}$s.
Neither effects due to electromagnetic and hadronic energy release
during CHAMP decay nor charge exchange effects have been taken
into account. From \citep{Jed08a},  updated (courtesy of K.
Jedamzik).} \label{JedaCPNevol1}
\end{figure}

\subsubsection{$^9$Be from CBBN}
\label{Be9CBBN}
Finally, we want to point out here that modifications to the
yields of heavier than $\li7$ elements can take place in CBBN. As
we have summarized in Section \ref{NucNet}, one robust prediction
of the standard BBN is the absence of sizeable yields of ${\rm
A}>$7 elements. Roughly speaking, this is due to the lack of
stable ${\rm A}$=8 elements, and to the inefficiency of the
$3\,\alpha\rightarrow ^{12}C$, which would respectively allow a
slow light element chain to produce heavier elements or ``bridge''
it, as it happens in stars. The very short lifetime of $\bero$ is
a problem that can be overcome in CBBN, where sizeable amounts of
meta--stable bound states $\bero X$ can be created through the
mechanism $\He4 X$($\alpha$,$\gamma$)$\bero X$ at
T$\lesssim$30~keV, as showed by \citep{Pos07}. Once this bound
state has been produced, the neutron capture reaction $\bero
X$(n,$\bern$)$X$ can take place and $\bern$ be efficiently
produced in CBBN, whereas the analogue SBBN mechanism is
suppressed as a consequence of the lifetime of $\bero$, shorter
than a femtosecond. Although the absolute abundance of $\bern$
produced by this mechanism is sensitive to the CHAMP abundance in
the plasma, $Y_X$, \citep{Pos07} argued that an enhanced
production of $\bern$ and $\lisix$ are peculiar signatures of
CBBN, and found that their ratio is independent of $Y_X$. The
author derives the conclusion that the primordial ratio
$\bern/\lisix\sim$10$^{-3}$ should therefore be a ``signature'' of
CBBN, as it dramatically differs from $\bern/\lisix\sim$10$^{-5}$
($\bern/$H$\sim$10$^{-19}$, \citep{Ioc07}) which one obtains in
SBBN. In their detailed study, \citep{Kam08} have challenged this
result by arguing that the rate estimate of $\bero
X$(n,$\bern$)$X$ by \citep{Pos07} is too high; although strong
conclusions would require a careful examination of all the nuclear
effects involved, they hint toward a much lower efficiency of the
process proposed by \citep{Pos07}. While this issue is yet to be
clarified, the kind of constraints to supersymmetric scenarios
that might follow from a peculiar $\bern$ production in CBBN has
preliminarily been studied in \citep{Pra08,Bai08}.

\subsubsection{Constraints in CHAMP parameter space}
If the lifetime of the CHAMP is $\tau_X\leq$3$\times$10$^2\,$s,
i.e. before that bound states can form, clearly its only effect on
BBN is equivalent to injecting electromagnetically and
hadronically interacting particles into the plasma. For
intermediate lifetimes, 3$\times$10$^2\,$s$\leq \tau_X
\leq$5$\times$10$^5\,$s, the main novel constraint is due to the
possible overproduction of $\lisix$ via $\He4 X(\hi2,\,X)\lisix$.
Fortunately, as this reaction is now known within a factor three
in light of the dedicated calculations of \citep{Ham07}, this
range of lifetimes is relatively well constrained. However the
decay of a CHAMP at this time induces again a cascade
nucleosynthesis; in hadronic cascades from a neutral particle
decaying  at the same time, it is well known that $\lisix$
overproduction is the main effect. Thus, unexpectedly, for
hadronic branching ratios $B_h\geq$0.01 (and e.g. $M_X$=1 TeV) the
effects of a charged particle decay do not differ much from those
of a neutral one \citep{Jed08b}. The results of the previous
Section apply since the $\lisix$ produced by the hadronic shower
induced reactions dominates over the effect of catalyzed $\lisix$
production. Only for $B_h\lsim 10^{-2}$  and for sufficiently
large decay times $\tau_X$ the CBBN mechanism provides new bounds.

Finally, for longer lifetimes, $\tau_X\geq$5$\times$10$^5\,$s, the
conservative  limits on charged decaying particles initially
derived in \citep{Jed08b} appeared to be no stronger than those on
neutral particles, due to the large uncertainties in
nucleosynthesis at T$\leq$3 keV. However, the nuclear rate
calculations for CBBN presented in \citep{Kam08} strongly reduce
those uncertainties. A recent application of these updated
calculations has been performed in \citep{Bai08} with an analysis
restricted to a minimal SUGRA model with heavy gravitino, which we
address for further details.

\section{Conclusions}\label{sec:conclusions}
In this review we have reported the current status of BBN,
focusing in the first part on precision calculations possible in
the standard scenario, which provide a tool for current
cosmological framework, in the second part on the constraints to
new physics, which become particularly important in the
forthcoming LHC era. The ``classical parameter" constrained by BBN
is the baryon to photon ratio, $\eta$, or equivalently the baryon
abundance, $\Omega_B h^2$. At present, the constraint is dominated
by the deuterium determination, and we find $\Omega_B h^2=0.021\pm
0.001$(1 $\sigma$). This determination is consistent with the
upper limit on primordial $^3$He/H (which provides a lower limit
to $\eta$), as well as with the range selected by $^4$He
determinations, which however provides a constraint almost one
order of magnitude weaker. The agreement within 2 $\sigma$ with
the WMAP determination, $\Omega_B h^2=0.02273\pm 0.00062$,
represents a remarkable success of the Standard Cosmological
Model. On the other hand, using this value as an input, a factor
$\gsim 3$ discrepancy remains with $^7$Li determinations, which
can hardly be reconciled even accounting for a conservative error
budget in both observations and nuclear inputs. Even more puzzling
are some detections of traces of $^6$Li at a level far above the
one expected from the Standard BBN. If the observational
determinations are solid, both nuclides indicate that either their
present observations do not reflect their primordial values, and
should thus be discarded for cosmological purposes, or that the
early cosmology is more complicated and exciting than the Standard
BBN lore. Neither a non-standard number of massless degrees of
freedom in the plasma (parameterized via $\neff$) or a lepton
asymmetry $\xi_e$ (all asymmetries assumed equal) can reconcile
the discrepancy. Current bounds on both quantities come basically
from the $^4$He measurement, $\neff=3.2\pm 0.4\,(1\,\sigma)$ and
$\xi_e=-0.008\pm 0.013\,(1\,\sigma)$.

On the other hand, other exotic proposals have been invoked to
reconcile this discrepancy. Typically they involve massive
meta-stable particles with weak scale interactions, which should
be soon produced at the LHC. In Supersymmetric scenarios,
long-lived particles are possible whenever the Next to Lightest
Supersymmetric Particle (NLSP) decays into the Lightest
Supersymmetric Particle (LSP) are gravity-mediated, or
``disfavored'' by phase space arguments, with a modest mass
splitting between NLSP and LSP. Cases frequently considered in the
recent literature are neutralino $\to$ gravitino decays, for
example, or stau $\to$ gravitino. The phenomenology associated
with the catalysis of reactions due to bound states of charged
particles (as the stau) with ordinary nuclei is a particularly new
topic in recent investigations. Also, the importance of a possible
primordial origin of the $\lisix$ measured in a few systems of the
$\li7$ plateau has been recognized: first,  the bounds in
parameter space tighten significantly if lithium constraints are
used, especially $\lisix$ \citep{Hol96,Jed00,Hol99,Jed04a};
second, because these exotic BBN scenarios may accommodate for a
cosmological origin for $\lisix$ while solving the $\li7$ excess
problem as well, their phenomenology is very appealing. Although
these links among primordial nucleosynthesis, dark matter, and
perhaps SUSY phenomenology are quite fascinating, it is worth
stressing that BBN bounds on cascade decays or annihilations of
massive particles apply well beyond the restricted  class of
SUSY-inspired models. For example, the electromagnetic cascades
following heavy sterile neutrino decays are constrained by these
kinds of  arguments, as well as decays of massive pseudo
Nambu-Goldstone bosons, as considered in \citep{Mas97,Mas04}.

There are two directions along which we can expect the BBN field to
develop in the future. On one hand, BBN is an important tool for precision
cosmology, especially if its priors are used in combination with other
cosmological observables. Already BBN provides the best bounds on
parameters as $\neff$ and $\xi_e$ (and bounds on $\eta$ comparable to the
CMB); yet, since theoretical uncertainties are at the moment well below
observational ones, there is surely room to refine its power, provided
that significantly greater efforts are devoted to determine light element
abundances, and in particular $Y_p$. It is instructive in this sense to
look back to what S. Sarkar wrote in his review \citep{Sar96} thirteen years
ago:

{\it Thousands of person years of effort have been invested in obtaining
the precise parameters of the $Z^0$ resonance in $e^{+}-e^{-}$ collisions,
which measures the number of light neutrino species (and other particles)
which couple to the $Z^0$. In comparison, a modest amount of work has been
done, by a few small teams, on measuring the primordial light element
abundances, which provide a complementary check of this number as well as
a probe of new superweakly interacting particles which do not couple to
the $Z^0$.}

Despite the improvements reported in this article, we feel that
unfortunately insufficient attention has been devoted to this
problem, if compared to other areas of observational cosmology. In
particular, the $\He4$ determination is still plagued by
systematic uncertainties. Although their importance has been
recently recognized and assessed more carefully, the fact that
this reanalysis was triggered after the independent determination
of $\eta$ from CMB (and its agreement with the ``low deuterium
determinations" in QSO spectra) shows that there is still a long
way to go towards a precision era for primordial elements. On the
other hand, a significant improvement has taken place in assessing
and reducing theoretical uncertainties, mostly related to nuclear
reaction data. BBN has benefit from a wealth of new nuclear
astrophysics measurements at low energies and covering large
dynamical ranges. Given the much larger observational
uncertainties, in this sector an effort in reassessing the
systematic errors in older datasets might be more useful in
reducing remaining discrepancies in the nuclear rates error
budget. This is in particular the case for reactions involving
$^7$Be.

The other direction of development follows from the interplay with
Lab experiments. Neutrinos have reserved many surprises, and it is
not excluded that exotic properties may show up in future
experiments with important implications for BBN, as we illustrated
in Section \ref{sec:BBN_nuphys}. However, it is in particular from
LHC that one expects a better understanding of high energy scales,
and thus of the cosmology at earlier times and higher
temperatures. Most theories that go beyond the Standard Model of
Particle Physics require new states to appear at or above the
electroweak scale and, as already reported, they might have
implications for the phenomenology at the BBN epoch. If the LHC
should provide indication for the existence of the SMPP Higgs and
nothing else, there will be no natural scale to explore. In this
case, albeit sad, BBN and other cosmological tools might be the
only practical means to explore very high energy phenomena leaving
their imprint on the cosmos. One example treated here is the
effect of variations of fundamental ``constants"  on cosmological
time-scales that emerge in extra dimensional scenarios, possibly
embedded in grand unified theories or string theories. If, as
hopefully more likely, the LHC will reveal  new dynamics above the
electroweak scale, we might be able to infer from the empirical
evidence the presence of cosmological effects before the BBN
epoch. A new Standard Cosmological Model would emerge as well,
perhaps making the BBN one more step in the ladder back to the Big
Bang, rather than the first one.

\ack We would like to thank C. Abia, A.D. Dolgov, J. Lesgourgues,
S. Pastor and G.G. Raffelt for valuable comments and suggestions,
and G.L. Fogli for having particularly encouraged this work. We
also thank  M. Kamimura, and especially K. Jedamzik, for
suggestions and clarifying remarks which much improved the
manuscript, and K. Jedamzik for providing also the updated version
of some Figures. F. Iocco is supported by MIUR through grant
PRIN-2006, and acknowledges hospitality at Fermilab during some
stage of this work. G. Miele acknowledges supports by Generalitat
Valenciana (Grant No.\ AINV/2007/080) and by the Spanish MICINN
(grants SAB2006-0171 and FPA2005-01269). G. Mangano, G. Miele, and
O. Pisanti acknowledge supports by INFN - I.S. FA51 and by PRIN
2006 ``Fisica Astroparticellare: Neutrini ed Universo Primordiale"
of Italian MIUR. P.D. Serpico is supported by the US Department of
Energy and by NASA grant NAG5-10842. Fermilab is operated by Fermi
Research Alliance, LLC under Contract No.~DE-AC02-07CH11359 with
the United States Department of Energy.

\end{document}